\documentstyle[12pt,epsf]{report}

\parskip=\medskipamount                 
\thicklines                         
\def\cc{{\cal C}}
\def\cd{{\cal D}}

\def\al{\alpha}
\def\be{\beta}
\def\ga{\gamma}
\def\de{\delta}
\def\lm{\lambda}
\def\om{\omega}

\def\veps{\varepsilon}
\def\eps{\epsilon}

\def\Ga{\Gamma}

\def\d#1/d#2{ {\partial #1\over\partial #2} }
\def\pdr{\partial}

\def\half{{1\over 2}}

\def\non{\nonumber}
\def\beq{\begin{equation}}
\def\eeq{\end{equation}}
\def\beqa{\begin{eqnarray*}}
\def\eeqa{\end{eqnarray*}}
\def\beqs{\begin{eqnarray}}
\def\eeqs{\end{eqnarray}}
\def\n{\global\advance \eqnumber by 1\eqno(\the\eqnumber)}
\def\puteqno{\global\advance \eqnumber by 1 (\the\eqnumber)}

\newcommand{\opr}[3]{~{Phys. Rev.}~{#1}~{(19#2)}~{#3.}}
\newcommand{\npb}[3]{~{Nucl. Phys.}~{B#1}~{(19#2)}~{#3.}} 
\newcommand{\prc}[3]{~{Phys. Rev.}~{C#1}~{(19#2)}~{#3.}}
\newcommand{\prd}[3]{~{Phys. Rev.}~{D#1}~{(19#2)}~{#3.}}
\newcommand{\prl}[3]{~{Phys. Rev. Lett.}~{#1}~{(19#2)}~{#3.}}
\newcommand{\phlt}[3]{~{Phys. Lett.}~{B#1}~{(19#2)}~{#3.}}

\newcommand{\etc}{{\it et al }}
\newcommand{\bib}{\bibitem}
\newcommand{\beqar}[1]{\begin{eqnarray} \label{#1}}
\newcommand{\eeqar}{\end{eqnarray}}
\newcommand{\bra}[1]{\langle {#1}|}
\newcommand{\ket}[1]{|{#1}\rangle}
\newcommand{\perb}[1]{\underline{#1}}
\newcommand{\nid}{\noindent}
\newcommand{\ovl}[1]{\overline{#1}}
\newcommand{\T}{{\rm T}}
\newcommand{\W}{{\rm W}}
\newcommand{\mn}{{\mu \nu}}
\newcommand{\dt}{\! \cdot \!}
\def\Lm{\Lambda}

\begin{document}

\pagestyle{empty}
\begin{centering}
\huge \bf Virtual Compton Scattering \\
at High Energy \\
\vspace{1.0in}
\Large Zhang Chen \\
\vspace{1.0in}
{\em Advisor:} Professor Alfred H.~Mueller\\
\vspace{1.0in}
\large
Submitted in partial fulfillment of the \\
requirements for the degree \\
of Doctor of Philosophy \\
in the Graduate School of Arts and Sciences \\
\vspace{0.5in}
COLUMBIA UNIVERSITY \\
1999 \\
\end{centering}
\clearpage

\clearpage
\vspace*{-2cm}
\begin{centering}
{\large
ABSTRACT \\
}
\vspace*{2ex}
{\Large
Virtual Compton Scattering at High Energy\\
}
\vspace*{2ex}
{\large
ZHANG CHEN\\
}
\vspace*{6ex}
\end{centering}

In this dissertation we develop a theoretical framework in the context
of perturbative QuantumChromoDynamics (pQCD) for studying non-forward 
scattering processes. In particular, we investigate a non-forward
unequal mass virtual Compton scattering amplitude by performing the 
general operator product expansion (OPE) and the formal
renormalization group (RG) analysis. 

We discuss the general tensorial decomposition of the amplitude to
obtain the invariant amplitudes in the non-forward kinematic region.
We study the OPE to identify the relevant operators and their reduced
matrix elements, as well as the corresponding Wilson coefficients. We 
find that the OPE now should be done in double moments with new moment 
variables. There are in the expansion new sets of leading twist 
operators which have overall derivatives. They mix under 
renormalization in a well-defined way. We compute the evolution 
kernels from which the anomalous dimensions for these operators can 
be extracted. We also obtain explicitly the lowest order Wilson coefficients. 

In the high energy limit we find the explicit form of the dominantly 
contributing anomalous dimensions. We are then able to solve the resulting 
renormalization group equations (RGE) and give a prediction of the
high energy behavior of the invariant amplitudes. We find that it is the 
same as is indicated by the conventional double leading 
logarithmic analysis. 

\clearpage

\pagestyle{plain}
\pagenumbering{roman}
\setcounter{page}{1}
\tableofcontents
\clearpage
\addcontentsline{toc}{chapter}{List of Figures}
\listoffigures
\clearpage
\chapter*{Acknowledgments}
\addcontentsline{toc}{chapter}{Acknowledgments}

I am most thankful to Professor Alfred H.~Mueller, my advisor,
for his guidance throughout my studies at Columbia as a
graduate student. 

I wish to thank Professor Allan S. Blaer, for his advice and help in
the development of my career, not only as a researcher, but also as an
effective teacher.

I would like to thank Professor Norman Christ, Professor Frank Sciulli
and Professor Michael Tuts for their support and help. Many thanks to
Professor Xiangdong Ji at University of Maryland for stimulating physics
discussions specially on Virtual Compton Scattering and to Anne Billups,
Lalla Grimes and Roy Jerome for their administrative work.

I also wish to thank Professor T.~D.~Lee and all other faculty
members of the Physics Department of Columbia University.  Thanks 
to all members of the theory group during my study for the fun and 
stimulating atmosphere in which I am able to pursue my research.  
And special thanks all my friends for the happiness they bring to me.

Finally, with all my heart, I wish to thank my parents for their 
guidance, support and encouragement throughout my life.

\clearpage

\pagestyle{myheadings}
\markright{}
\pagenumbering{arabic}
\setcounter{page}{1}

\chapter*{Introduction}
\thispagestyle{myheadings}
\markright{}

\addcontentsline{toc}{chapter}{Introduction}

It is generally believed that {\it Quantum Chromodynamics} (QCD) is the 
fundamental theory for the strong interaction and it describes the
properties of nuclear and subnuclear, or hadronic, matter from first
principles. QCD is a quantum field theory that is invariant under a
local non-Abelian gauge symmetry--the SU(3) color symmetry. The basic 
matter particles of QCD, the {\it quarks} (and antiquarks), interact 
with eight massless {\it gluons}, non-Abelian gauge bosons that also 
couple to themselves, to form confined bound states of hadrons.   
The SU(3) gauge symmetry uniquely determines the forms of interactions
between quarks and gluons, and dictates that the theory is both color
confined and asymptotically free. This results in the bifurcation of
QCD into {\it Lattice} QCD at low energy and {\it Perturbative} QCD
(pQCD) at high energy. For more detailed discussion on various 
aspects of Quantum Field Theory and QCD, see, for example,
References \cite{I&Z, Peskin, Sterman, Cheng&Li, 
Collinsbook, Group, QCDbook1, pQCD, AplpQCD}. 

At low energies, because the coupling of QCD is strong, the usual
analytic method of perturbative expansion in quantum field theory
breaks down. As a result many fundamental properties of QCD at low 
energies, such as quark confinement, the dynamical chiral symmetry 
breaking and ultimately the low energy hadron mass spectrum, remain 
unresolved, at least analytically. The lattice formulation of 
QCD \cite{Wilson74} provides us with a new way of studying the theory 
non-perturbatively. By defining the quark and the 
gluon fields on a $4$ dimensional Euclidean space-time lattice, the
theory is regulated by an ultra-violet cut-off, the inverse lattice
spacing.  Physical quantities of interest defined through the path
integral of the quantum fields can now be studied numerically, using
Monte Carlo methods. 

At high energies, on the other hand, 
perturbation theory is valid due to the smallness of the QCD coupling. 
Perturbative QCD has had great success in explaining and 
predicting results of high energy experiments, for example, from the 
ratio of cross sections of $e^+ e^- \to \mu^+ \mu^-$ 
to $e^+ e^- \to hadrons$, to structure functions of the nucleon in 
deeply inelastic scattering (DIS) (e.g., see \cite{DESYreview} 
and the references therein). In particular, pQCD has been successfully
applied to the study of the structure of the nucleon, one
of the most important frontiers in strong interaction physics.
Here by structure we mean not only the traditional nuclear structure
like spin structure and form factors but also, more relevant to the
work in this thesis, the distribution functions of quarks and gluons,
which we collectively call {\it partons}, inside the nucleon.

Physics processes usually involve both hard (high energy) and soft
(low energy) physics. The key to a pQCD analysis is the proper
separation (``factorization'') between the hard part, where analytic 
calculation using perturbation theory is possible, and the soft part, 
where we have to rely on non-perturbative methods like numerical study and
phenomenological modeling. By application of factorization theorems and 
resummation methods, and/or more formal methods like the operator product 
expansion (OPE) and renormalization group equation 
(RGE) \cite{Cheng&Li, Collinsbook, Christ&Mueller72, A&P77, Mueller78},
pQCD has been very successful in extracting the parton distribution functions 
of the nucleon and predicting their behaviors in a wide range of 
kinematic regions of high energy particle-particle scattering. 
However, there is still much work to be done, especially in extreme 
kinematic regions like very small and/or very large Bjorken-$x$ 
(see equation \ref{eq:x-bj}), where we have to not only look for new 
physics but also take greater care in separating the hard scattering 
process where pQCD is applicable and the soft physics parts where 
perturbation theory breaks down.  

Recently there is much interest in the analysis of non-forward scattering 
processes, for example, deeply virtual Compton scattering 
(DVCS) \cite{Ji97, Radyphlt961} and hard diffractive electroproduction
of vector 
mesons \cite{BrodskyVM, Radyphlt962, Collins&FS97, Frankf&Strikman98} 
in DIS. This creates a new territory to explore the quark and gluon
structure of the nucleon besides the traditional inclusive (forward 
parton distributions) and exclusive (form factors) processes. 
A new collection of parton distribution functions called the {\it
skewed} parton distributions (SPDs) \footnote{SPD is the unified
terminology used to replace the many terms like non-diagonal,
non-forward, off-diagonal and off-forward from different
parameterizations.} appear in these exclusive, hard diffractive
processes which contain new information on long distance physics 
due to the non-forwardness (see, for example, 
Ref. \cite{Jiprl97, Ji&Song97, Radyprd97, RadySPD}). At the same time, 
diffractive vector meson production also provides a new way of directly 
measuring the (forward) gluon density in the proton \cite{BrodskyVM}.
The work of this thesis is aimed at developing a 
theoretical frame work to properly study non-forward processes in the
context of operator production expansion and renormalization group
analysis \cite{nfvcs}. The thrust of our research is obtaining the 
high energy behavior of these non-forward processes in general.

\subsubsection{DVCS and Skewed Parton Distributions}

The Compton process, referring to the elastic scattering of a photon 
off a charged object, historically provided one of the early evidences 
of the quantization and particle nature of the electromagnetic
wave \cite{Compton}. Its role in studying the structure of hadrons 
has been explored since the 50s with the derivation of Low's
low-energy theorems \cite{LowE} which assert, for instance, that 
at sufficiently low energy the spin-dependent part of the Compton 
amplitude is determined by the anomalous magnetic moment of a 
composite system. Going to higher-order terms in the
low-energy expansion, one finds the electric and magnetic
polarizabilities. In recent years experimental and theoretical works in
measuring and understanding the polarizabilities of the nucleon and
pion have flourished \cite{polarizability}.

The virtual Compton amplitude for scattering a virtual photon
off a hadron has become one of the basic tools in QCD to understand 
the short-distance behavior of the theory. The process of deeply
virtual Compton scattering (DVCS) is, assuming the virtual photon is 
generated by inelastic lepton scattering, the Bjorken limit, i.e.,
the energy and momentum of the virtual photon going to infinity at 
the same rate, of virtual Compton scattering. The basic mechanism 
for DVCS is a quark absorbing the virtual photon, immediately
radiating a real photon and falling back to the nucleon ground state.
Ji \cite{Ji97,Ji&Song97} was the first to introduce and study DVCS. 
He proposed that we can obtain from DVCS information on the so-called 
off-forward parton distributions (OFPD) which in this case contain 
new information on long distance physics.

The OFPDs have come up in different theoretical studies prior to Ji's 
work. In the late 1980s, Geyer and 
collaborators \cite{interpolating, UMCompton1} studied the relation 
between the Altarelli-Parisi evolution for parton
distributions \cite{A&P77} and the Brodsky-Lepage 
evolution \cite{Brodsky&Lepage} for leading-twist meson wave
functions. The ``interpolating functions'' introduced in 
Ref. \cite{interpolating} are essentially Ji's OFPDs.
In the early 1990s, Jain and Ralston \cite{OFAMP} studied hard 
processes involving hadron helicity flip in terms of an 
``off-diagonal transition amplitude'' that involves off-forward 
matrix elements of bi-quark fields in the nucleon. 
It was shown that the integral of this amplitude over
the quark four-momentum yielded elastic form factors.
 
Ji \cite{Ji97, Jiprl97} introduced OFPDs in the study of the spin 
structure of the nucleon. His main observations were that the
fractions of the spin carried by quarks and gluons can be determined
from form factors of the QCD energy-momentum tensor, and that the
latter can be extracted from the OFPDs. Furthermore, the DVCS process
was proposed \cite{Ji97} as a practical way to measure the new
distributions. Because the spin-$2$ twist-two quark and gluon
operators are part of the QCD energy-momentum tensor and because 
the form factors of the energy-momentum tensor contain information 
about the quark and gluon contributions to the nucleon spin, DVCS 
provides a novel way to measure the fraction of the nucleon spin 
carried by the quark orbital angular momentum, a subject of great 
current interest \cite{qspin}. The more relevant part of his work to
ours is his study on the evolution and sum rules (in terms of form 
factors) of OFPDs. He also obtained certain estimates at low energy 
of the OFPDs and form factors \cite{Ji&Song97} using the MIT bag 
model \cite{bag}. 

Radyushkin also studied the scaling limit of DVCS \cite{Radyphlt961}, and 
generalized the discussion to hard exclusive electroproduction 
processes \cite{Radyphlt962,Radyprd97}. The non-perturbative
information is incorporated in his double distributions $F(x,y;t)$ 
and non-forward distribution functions $F_\zeta(X;t)$. He discussed 
their spectral properties, the evolution equations they satisfy, 
their basic uses and some general aspects of factorization for hard 
exclusive processes. 

A third parametrization for the non-perturbative information was 
proposed by Collins {\it et al} \cite{Collins&FS97,Frankf&Strikman98}. 
In addition to the discussion of factorization theorem and evolution,
they performed a numerical study of their non-diagonal 
parton distributions in leading logarithmic approximation. They found 
that the non-diagonal gluon distribution $x_2G(x_1,x_2,Q^2)$ can be well 
approximated at small $x$ by the conventional (forward) gluon density 
$xG(x,Q^2)$.

It is clearly both interesting and important for us to study in depth
these SPDs, because they will give us information on novel long
distance physics due to the non-forwardness.

\subsubsection{Diffractive Vector Meson Production}

Conventionally, for example, in DIS, the gluon density $xG(x,Q^2)$ is
measured indirectly. The structure function $F_2$ (the same as $\nu
\W_2$ of, eg, equation \ref{eq:fw-operesult}) that comes out
directly from the experimental data is (essentially) the sum of quark 
(and antiquark) distribution functions while the gluonic content can only 
be obtained from a DGLAP type evolution analysis (see Chap. 1 and 
Ref. \cite{DESYreview}). This is because gluons only have color charge
and all probes in nature are color neutral due to color confinement. 
As a result any interaction with the gluonic content inside the nucleon 
has to have at least a quark loop involved and thus has to be of 
higher order. 

On the other hand, at very small Bjorken-$x$, the gluonic sector
dominates the high energy behavior of scattering cross sections (see
Chap.1 for explanation). It is well-known that there is a sharp
rise in the parton densities inside a nucleon at small $x$, which is
driven by the gluons. While this might be an indication of the 
emergence of the so-called BFKL pomeron \cite{BFKL}, 
higher order DGLAP evolution is able to account for most of the 
increase (see, for example, Ref. \cite{DESYreview}). It is obvious
that new ways of measuring the gluonic content in the nucleon,
especially direct measurements, will be extremely useful.

It has been shown that the cross section for diffractive
electroproduction of vector mesons can be predicted in 
pQCD \cite{BrodskyVM, Ryskin, Frankf&Strikman96} 
(see Figure \ref{fig:vm}). This process provides a novel probe 
of the dynamics of diffractive scattering in QCD. The prediction 
that the cross section is proportional 
to the square of the gluon density $xG(x,Q^2)$ in the 
hadron \cite{BrodskyVM} means that we do have a process through 
which direct measurement of the gluon density is possible. 
Experimental data \cite{VMexp} appears to be in accord with the 
predictions, including an enhancement due to the rapid rise of 
the gluon density at small $x$. 

The pQCD calculation leading to the above prediction in
Ref. \cite{BrodskyVM} was done under a double leading logarithmic 
approximation (DLLA)--resumming leading logarithms in both the 
longitudinal momentum (leading $\ln 1/x$) and the transverse momentum 
(leading $\ln Q^2/\Lm^2$, where $\Lm$ is the QCD scale). For further
studies we need to relax (at least one of) these two approximations
and also incorporate consistently the BFKL contribution. One such 
possibility can be found in applying the high energy factorization 
developed by Catani \etc \cite{Catani}, relaxing the leading
logarithmic approximation (LLA) on the transverse momentum
$\perb{k}$ and discussing the cross section in terms of the so-called
un-integrated gluon distribution (see, for example, \cite{Ryskin2}).
However, up to the time of the work in this thesis \cite{nfvcs}, there 
had not been a completely consistent treatment.

One of the basic problems lies in the fact that conventional forward
factorization has been used on a non-forward process--because of the
time-like four-momentum of the final state vector meson, diffractive
vector meson production is necessarily a non-forward scattering 
process. It is therefore not obvious without theoretical investigation 
of non-forward factorization and/or OPE that it should be the forward 
gluon density that comes into the expression of the cross section.

The research of this thesis was indeed originally motivated by the
diffractive vector meson production. The ultimate goal is a better
understanding of the gluon distribution inside the nucleon, especially
at small $x$. However, we realised that some formal theoretical
development on OPE and renormalization group analysis in the 
non-forward cases has to be made before we can consistently attempt
to relax the LLAs. In other words, we have to justify (or prove wrong)
the use of forward factorization under DLLA before we can go further.

\subsubsection{Unequal Mass Virtual Compton Scattering}

\begin{figure}[ptb]
\begin{center}
\epsfxsize=3in
\epsfysize=4in
\leavevmode
\hbox{ \epsffile{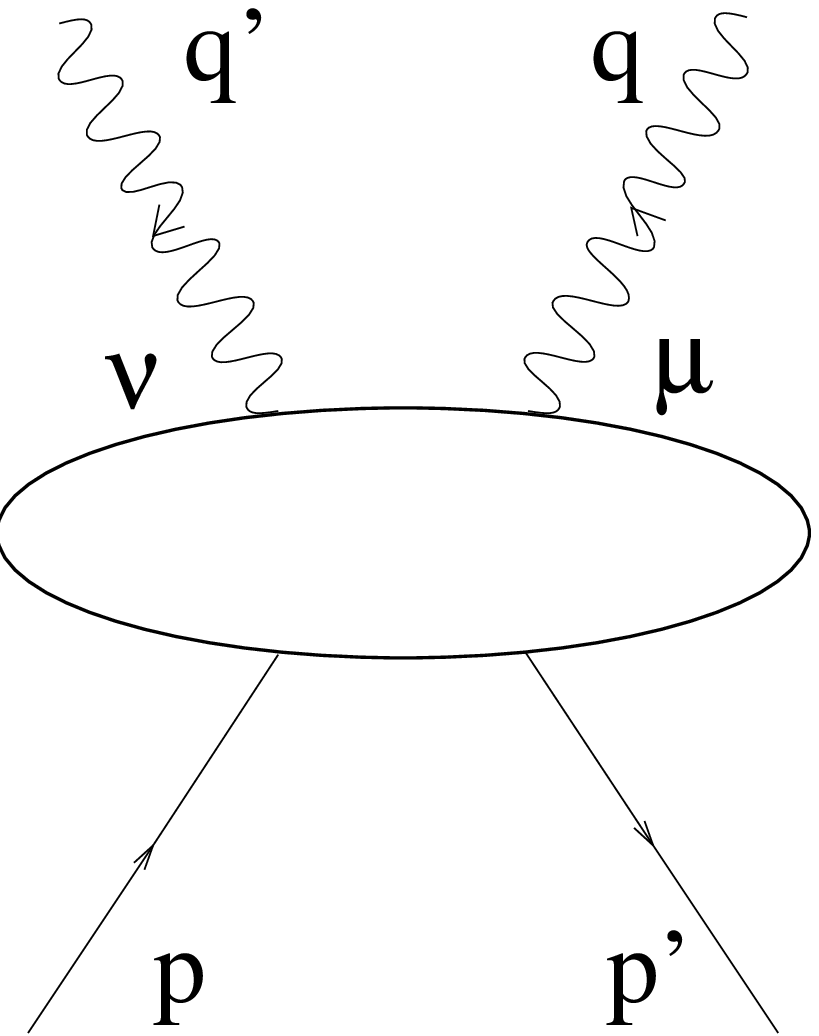}}
\end{center}
\caption{Non-forward Scattering Amplitude}
{This is the unequal mass double virtual Compton scattering diagram. 
Both $q$ and $q'$ are virtual photons and they are of different 
invariant masses. There is a non-zero
four-momentum transfer $r$ from the incoming virtual photon to the 
target proton.}
\label{fig:ampdef}
\end{figure}

This dissertation aims exactly at tackling the above problem.
In this research, we formulate a general operator product expansion 
(OPE) description for a generic non-forward unequal mass virtual 
Compton scattering amplitude(see Figure \ref{fig:ampdef}). In order to
avoid complications from dealing with meson light-cone wave
functions (these have been shown by Collins \etc \cite{Collins&FS97} to
be factorizable), we have replaced the vector meson vertex with
another virtual photon coupling. At this point, the main goal is to
study the influence of the non-forwardness on OPE, rather than the 
phenomenology of the vector meson production itself. DVCS and DIS are 
extreme kinematic cases of this generic unequal mass double virtual 
Compton scattering (see section \ref{sec:kinematics}).

General proofs of the DVCS factorization have been given by
Radyushkin in his approach based on 
$\al$-representation \cite{Radyphlt961, Radyphlt962, Radyprd97} 
and by Ji in an alternative method \cite{Ji&Osb98}. It has been
established that factorization for DVCS is on the same footing as 
other well-known examples like DIS. Collins \etc \cite{Collins&FS97}
have also proved the factorization theorem for general diffractive
vector meson production.

Some early studies of unequal mass Compton processes can be found in
Refs. \cite{UMCompton1, UMCompton2}. The factorization property of 
the general two virtual photon process can be summarized beautifully 
in terms of Wilson's OPE. This expansion now requires operators with 
total derivatives \cite{nfvcs, UMCompton1, UMCompton2} to describe 
the non-forward nature of the process. While it is well-known that 
these derivative operators contribute to the wave functions of 
mesons \cite{Brodsky&Lepage, mesonwf}, and that the scale evolution 
of total derivative operators can best be studied using 
conformally-symmetric operators \cite{conformal},
our work will concentrate on the renormalization group evolutions 
of these operators (and their corresponding Wilson coefficients) 
in the high energy (and small $x$) limit (see later chapters of this 
thesis). 

Because of the non zero momentum transfer $r=p-p'$ from the 
initial state proton to the final state proton
(Fig.~\ref{fig:ampdef}), the scaling behavior must be different from
that in a forward case because of the new kinematic degree of freedom 
(the non-forwardness). Indeed we will show that compared with forward 
scattering one has two scaling variables, or equivalently, two moment 
variables $\omega$ and $\nu$ (see equation \ref{eq:momvardef}). 
The amplitude should now be expanded in terms of double moments with 
respect to these two moment variables. There are associated with these 
double moments new sets of operators that have overall derivatives in 
front (see equation \ref{eq:operators}), which mix among 
themselves under renormalization via an anomalous dimension matrix. 
The reduced non-forward matrix elements of these operators are thus 
double distribution functions. However, they do not seem to have a simple 
probability interpretation. While formally we can go from DVCS to DIS
using the same formalism and in principle we can proceed beyond DLL to
next to leading order, we will focus our attention on the high 
energy behavior of these distribution functions. We solve their 
renormalization group equations at high energy, and show that in the 
high energy limit the gluonic double distribution function actually 
reduces to the conventional (forward) gluon density and the high 
energy behavior is the same as obtained by a conventional double 
leading logarithmic (DLL) analysis. This also means that we have 
justified the DLLA calculations \cite{BrodskyVM}. 

The outline of the thesis is as follows. 
In Chapter 1 we review the standard operator product expansion and
renormalization group equation analysis of forward scattering
amplitudes, using DIS as an example. In Chapter 2 we define the 
process and the kinematics of the non-forward scattering amplitude 
${\rm T}_{\mu \nu}$. A general tensorial decomposition of 
${\rm T}_{\mu \nu}$ into invariant amplitudes is given. In Chapter 3 
we perform a general operator product expansion of ${\rm T}_{\mu \nu}$,
define new moment variables and give their relationship to the more 
conventional Bjorken type scaling variables. In Chapter 4 we write down 
the renormalization group equations, present the equivalence of 
evolution kernels for the double moments, and calculate explicitly the 
lowest order Wilson Coefficients. In Chapter 5 we go to the high energy 
limit to solve the renormalization group equations and make connection 
with the double leading logarithmic analysis. Chapter 6 gives 
conclusions and outlook for future work. 

\chapter{Forward Scattering}
\thispagestyle{myheadings}
\markright{}

In this chapter, we give a review of the traditional operator
product expansion (OPE) and renormalization group (RG) analysis for 
a deeply inelastic scattering (DIS) of lepton and proton (e-p), which 
serves to lay out the framework and conventions that will be used in 
later discussions of non-forward scatterings. This will be the forward 
case that we refer and compare to in those chapters.

We start with the parton model description of DIS to present
the problem and establish the kinematics. We proceed to decompose 
the amplitude into invariant amplitudes and define the 
structure functions. We then perform an OPE of the invariant
amplitude and discuss the RG equation and its solution of the operators 
and Wilson coefficients involved. To set the conventions, we
also include a review of the Light Cone(LC) gauge convention and
explicitly calculate the quark-quark splitting function in the LC gauge.

\nid Most results in this chapter concerning quantum field theory are
more or less standard material and can be found in, for example, 
\cite{Peskin}, \cite{Cheng&Li} or \cite{Collinsbook}, while more
specific details of DIS process can be found in for example,
\cite{DESYreview}. However, we will be presenting them in a different
way \cite{G8070}. We will generally not make explicit
references to the standard literature on well-known results.

\section{Amplitude and Kinematics \label{sec:fw-kinematics}}

\begin{figure}[ptb]
\begin{center}
\epsfxsize=3in
\epsfysize=2.5in
\leavevmode
\hbox{ \epsffile{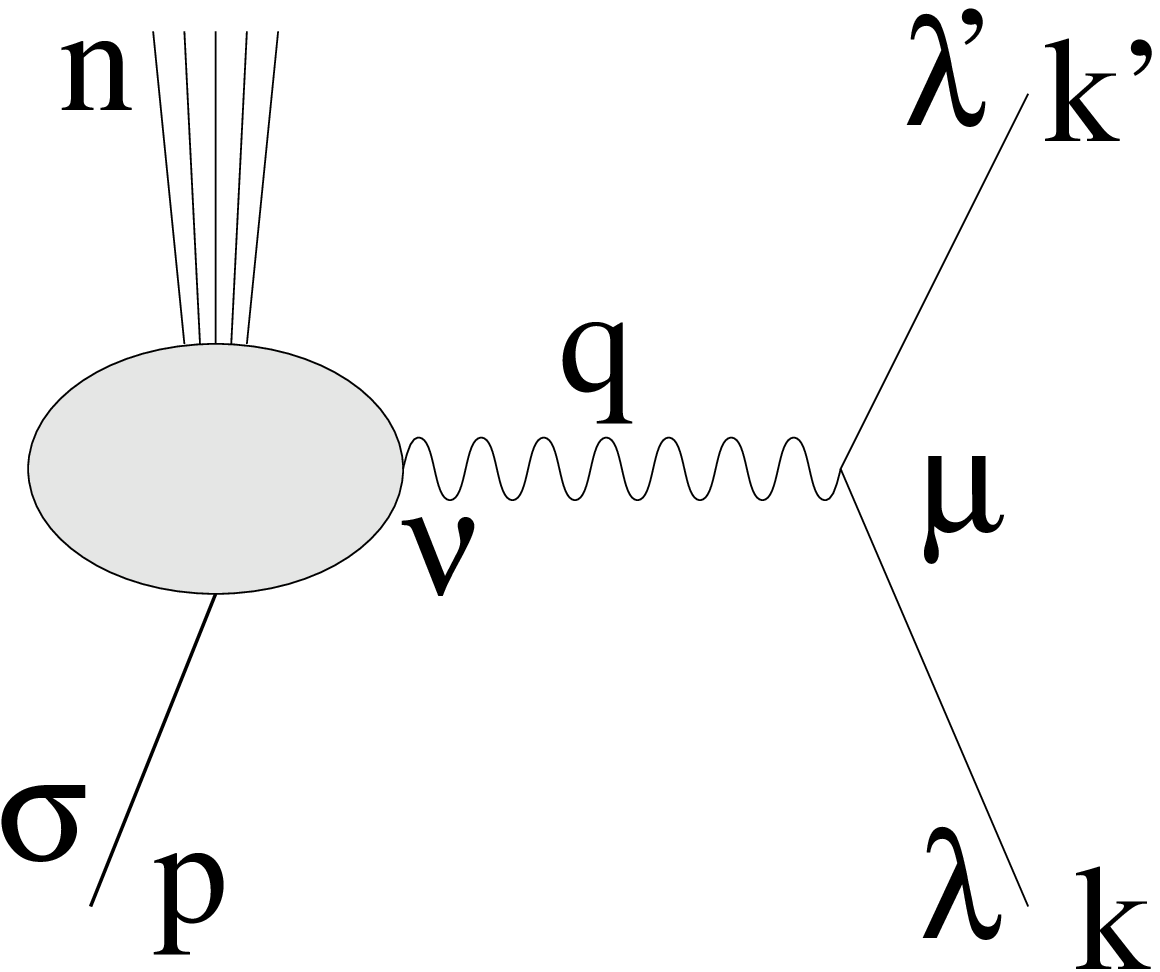} }
\end{center}
\caption{Deeply Inelastic Scattering}
{This is the  diagrammatic representation of the Deeply Inelastic Scattering
process. This is a neutral current process via one photon exchange.
The initial proton is $\ket{p, \sigma}$, the initial and final
electron are $k, \lm$ and $k',\lm'$, respectively, and $n$ labels
collectively the different final states of the proton.}
\label{fig:DIS}
\end{figure}

The process we consider here is an electron-proton scattering at very
high energy, such as those that occur at DESY. Figure \ref{fig:DIS} depicts
one such process, where the initial proton has four momentum $p$ and
spin $\sigma$ while that of the initial electron is $k$ and $\lm$,
respectively. After the scattering, the electron has four momentum
$k'$ and spin $\lm'$ while the final state of the proton is
collectively labeled $n$.

\nid The exchanged virtual particle, with momentum $q = k-k'$, can 
be either a photon, a $Z$ or a $W$ boson (in figure \ref{fig:DIS} it
is a photon). For the purpose of our 
discussion, we will consider only one photon exchange, which is a 
good approximation at the leading level. In the kinematic region we 
will be working, that is, with $Q^2 = -q^2 \ge 1 \; GeV^2$ and energy 
of the electron at tens of $GeV$ (for example, DESY has a electron 
beam energy of about $30 \; GeV$), we can neglect the mass of the electron 
in the following discussion, which means $k^2 = k'^2 = 0$.

\subsection{Invariants of DIS \label{sec:fw-invariants}}

We first work in the rest frame of the proton to define the
kinematic variables used to describe such a process. We have
\beq
\label{eq:dis-ini}
\left\{
\begin{array}{ll}
p = (M,\!\!&\!\! 0, 0, 0) \\
k = (\,\eps, \!\!&\!\! 0,0, \eps)
\end{array} \right. \,.
\eeq
 
\nid If the scattering angle of the electron is $\theta$, then
\beq
k = (\,\eps', \eps'\sin\theta, 0, \eps'\cos\theta) \,,
\eeq

\nid thus the exchanged four momentum square is
\beq
Q^2 = -q^2 = - (k-k')^2 = 2 k \cdot k' - k^2 - k'^2
= 2 \eps \eps' (1 - \cos\theta) \,.
\eeq

\nid The first invariant of a DIS process, $Q^2$, is then given by
\beq
\label{eq:fw-Qsqyare}
Q^2 = 4 \eps \eps' \sin^2 {\theta \over 2} \,,
\eeq

\nid which indicates the hardness of the scattering. We can see that
large $Q^2$ needs large $\eps$, $\eps'$, and $\theta$ not
approaching zero.

A second invariant $\nu$ is defined by
\beq
M\nu \equiv p \cdot q = p \cdot (k-k') = M (\eps - \eps') \,,
\eeq

\nid which gives
\beq
\label{eq:fw-nu}
\nu = \eps - \eps' \,.
\eeq

\nid We can see that $\nu$ is the energy loss of the electron in the
proton rest frame.

A third invariant, the Bjorken-$x$ variable, is
more commonly used instead of $\nu$ and is defined as
\beq
\label{eq:x-bj}
x_{Bj} \equiv {1 \over \om} \equiv { Q^2 \over {2 M \nu}} 
\equiv {Q^2 \over {2 p \cdot q} }\,.
\eeq

\nid We note $0 \leq x_{Bj} \leq 1$.

$Q^2$ and $x_{Bj}$ are the usual variables used to describe
DIS, and sometimes we use equivalently $Q^2$ and $\nu$.  Although 
they are defined above in the photon rest frame, they are by
definition invariant when we go to any other frame.

\subsection{Scattering Amplitude and Differential Cross-section}

The scattering amplitude of the DIS process in figure \ref{fig:DIS}
can be written as
\beq
\label{eq:fw-t}
T^{(n)}_{\sigma \lm'\lm} = -ie \tilde{U}(k') \ga_\mu U(k)
{ {-i g_\mn} \over {q^2} } \bra{n^{(-)}} j_\nu (0) \ket{p\sigma}
i e \,,
\eeq

\nid with the minus sign on the $n$ state meaning outgoing states and
$j_\nu$ the electromagnetic current given by
\beq
\label{eq:emcurrent}
j_\nu = \sum_f e_f \tilde{q}_f \ga_\nu q_f \,.
\eeq

\nid The differential cross-section is then given by
\beq
\label{eq:fw-dsigmadef}
d \sigma = {1 \over 4} \sum_{\lm'\lm} \sum_{\sigma n}
|T^{(n)}_{\sigma \lm'\lm}|^2 (2\pi)^4 \de^4(p+q-p_n) 
{ {d^3 k'} \over {2 \eps 2 \eps'} } \,,
\eeq

\nid where the factor ${1 \over 4}$ comes from averaging over initial
proton and electron spin and the phase space and kinematic factor are
obtained from a plane wave normalization.

To simplify the expression of $d\sigma$ we define the electron factor
$l_\mn$ and the proton factor $L_\mn$ such that
\beq
\label{eq:fw-dsigma}
{ {d^3 \sigma} \over {d k'^3} } = {1 \over {2 \eps 2 \eps'}}
{ e^4 \over q^4} L^\mn l_\mn \,.
\eeq

\nid We have, for the electron part,
\beqs
\label{eq:efactor}
l_\mn &\equiv& \half \sum_{\lm'\lm} 
[\tilde{U}_{\lm'}(k') \ga_\mu U_\lm(k)]^*
[\tilde{U}_{\lm'}(k') \ga_\nu U_\lm(k)] \non \\
&=&  \half \sum_{\lm'\lm} \tilde{U}_\lm (k) \ga_\mu U_{\lm'}(k') 
\tilde{U}_{\lm'}(k') \ga_\nu U_\lm(k) \,.
\eeqs

\nid Because we know that
\beq
\sum_\lm U_{\lm,\al}(k) \tilde{U}_{\lm,\be}(k) 
= (\not{\!k} + m_e)_{\al\be} \,,
\eeq

\nid we arrive at
\beqs
l_\mn &=& \half Tr[\ga_\mu (\not{\!k}' + m_e) \ga_\nu (\not{\!k} + m_e)]
\non \\
&=& \half Tr[\ga_\mu \not{\!k}' \ga_\nu \not{\!k}] \,,
\eeqs

\nid where we have taken the limit $m_e = 0$. Thus we obtain
\beq
\label{eq:efactorfinal}
l_\mn = 2 (k'_\mu k_\nu + k_\mu k'_\nu - g_\mn k \cdot k') 
+ O(m_e^2) \,,
\eeq

\nid which is determined completely from the kinematics of the
process.

On the other hand, for the proton part, we have
\beqs
\label{eq:pfactor}
L^\mn &=& \half \sum_\sigma \sum_n \{ 
\bra{n^{(-)}} j^\mu (0) \ket{p\sigma}^* 
\bra{n^{(-)}} j^\nu (0) \ket{p\sigma} (2 \pi)^4 \de^4(p+q-p_n) \} \non \\
&=& \half \sum_{\sigma,n} \bra{p\sigma} j^\mu (0) \ket{n^{(-)}}
\bra{n^{(-)}} j^\nu (0) \ket{p\sigma} (2 \pi)^4 \de^4(p+q-p_n) \,,
\eeqs

\nid where we have used the fact that $j_\mu$ is Hermitian.

\nid Because the presence of the $\de$-function, the sum of $n$ state
is not a complete set summation, but rather an on-shell summation. We
rewrite the $\de$-function and obtain
\beqs
L^\mn &=& \half \sum_{\sigma,n} \int d^4x e^{i (p+q-p_n) \cdot x}
\bra{p\sigma} j^\mu (0) \ket{n^{(-)}}
\bra{n^{(-)}} j^\nu (0) \ket{p\sigma} \non \\
&=& \half \sum_{\sigma,n} \int d^4x e^{i q \cdot x}
\bra{p\sigma} j^\mu (x) \ket{n^{(-)}}
\bra{n^{(-)}} j^\nu (0) \ket{p\sigma} \,,
\eeqs

\nid where we have used the translation relationship
\beq
\label{eq:translation}
j_\mu(x) = e^{i \hat{p} \cdot x} j_\mu(0) e^{-i \hat{p} \cdot x} \,.
\eeq

\nid This translation operation is exactly analogous to the spacial
translation by the three momentum operator $\hat{\vec{p}}$ or the time
translation by the Hamiltonian $\hat{H}$ in quantum mechanics (see,
for example, \cite{SakuraiQM}). In the above, $\hat{p}$ is the 
$4$-momentum operator where 
\beqs
\label{eq:p-hat}
\left\{
\begin{array}{l}
\hat{p} \cdot x = \hat{H} t - \hat{\vec{p}} \cdot \vec{x} \\
\hat{p} \ket{\, Vacuum \,} = 0 \\
\hat{p}_\mu \ket{n} = p_{n \mu} \ket{n} \,.
\end{array}
\right.
\eeqs

\nid The $n$ state summation is now of a complete set, thus by the
completeness relationship we arrive at the expression of the proton 
factor as
\beq
\label{eq:pfactordef}
L^\mn = \half \sum_\sigma \int d^4x e^{i q \cdot x}
\bra{p\sigma} j^\mu (x) j^\nu (0) \ket{p\sigma} \,.
\eeq

\section{Structure Function}

The structure functions of the proton are defined from the invariant
amplitudes of the proton factor, which in turn are obtained by
a general tensorial decomposition.

\subsection{Tensorial Decomposition \label{sec:fw-decomp}}

\nid $L_\mn$ as a rank-$2$ tensor depends on the kinematic variables
$p$ and $q$ only. Thus the general tensorial decomposition of $L_\mn$
can be written as
\beq
\label{eq:fw-decomp}
L_\mn(p,q) = A g_\mn + B p_\mu p_\nu + C (p_\mu q_\nu + p_\nu q_\mu)
+ D q_\mu q_\nu + E (p_\mu q_\nu - p_\nu q_\mu)
+ F \veps_{\mu\nu\rho\sigma} q^\rho p^\sigma \,, 
\eeq

\nid where $A, B, ...$ are invariant amplitudes depending only on the
invariants of the process, that is, $A = A (p \cdot q, q^2)$ and so
on. We have deliberately grouped the terms into symmetric and
anti-symmetric parts to make the symmetry properties of $L\mn$ more
manifest. The last term does not occur for an electron scattering,
however, we do have to consider it if it is a neutrino scattering.

The conservation of the electromagnetic current gives 
(cf equation \ref{eq:currentconsv})
\beq
\label{eq:fw-crtcsv}
q^\mu L_\mn = q^\nu L_\mn = 0 \,.
\eeq

\nid Thus we have
\beq
q^\mu L_\mn = A q_\nu + B p \cdot q p_\nu 
+ C (p \cdot q q_\nu + q^2 p_\nu) + D q^2 q_\nu 
+ E (p \cdot q q_\nu - q^2 p_\nu) = 0 \,,
\eeq

\nid being true for any $p$ and $q$. We can vary $p$ and $q$ while
keeping $p \cdot q$ and $q^2$ constant and this means that the
coefficients of the individual $p_\nu$ and $q_\nu$ have to vanish. We
have
\beq
\label{eq:qmuresult}
\left\{
\begin{array}{l}
A + C p \cdot q + E p \cdot q + D q^2 = 0 \\
B p \cdot q + C q^2 - E q^2 = 0
\end{array}
\right. \,.
\eeq

\nid Similarly, from
\beq
q^\nu L_\mn =  A q_\mu + B p \cdot q p_\mu 
+ C (p \cdot q q_\mu + q^2 p_\mu) + D q^2 q_\mu 
+ E (q^2 p_\mu - p \cdot q q_\mu) = 0 
\eeq

\nid we have
\beq
\label{eq:qnuresult}
\left\{
\begin{array}{l}
A + C p \cdot q - E p \cdot q + D q^2 = 0 \\
B p \cdot q + C q^2 + E q^2 = 0
\end{array}
\right. \,.
\eeq

\nid Combine equations \ref{eq:qmuresult} and \ref{eq:qnuresult}
we arrive at
\beq
\left\{
\begin{array}{l}
E = 0 \\
A + C p \cdot q + D q^2 = 0 \\
B p \cdot q + C q^2 = 0
\end{array}
\right. \,,
\eeq

\nid which eventually leads to
\beq
\left\{
\begin{array}{l}
C = - { {p \cdot q} \over q^2 } B \\
D = - { 1 \over q^2} ( A - { {(p \cdot q)^2} \over q^2 } B) \\
E = 0
\end{array}
\right. \,.
\eeq

\nid Therefore, only two of the invariant amplitudes survive after
considering current conservation and we have
\beq
L_\mn = A g_\mn + B p_\mu p_\nu 
- B { {p \cdot q} \over q^2 } (p_\mu q_\nu + p_\nu q_\mu)
- { 1 \over q^2} ( A - { {(p \cdot q)^2} \over q^2 } B) q_\mu q_\nu 
\,,
\eeq

\nid or eventually, after collecting terms,
\beq
\label{eq:fw-decompfinal}
L_\mn = A (g_\mn - { {q_\mu q_\nu} \over q^2 })
+ B \left( p_\mu p_\nu 
         - { {p \cdot q} \over q^2 } (p_\mu q_\nu + p_\nu q_\mu)
         + { {(p \cdot q)^2} \over {(q^2)^2} } q_\mu q_\nu
    \right) \,.
\eeq

\subsection{The Structure Functions \label{sec:fw-strfcn}}

The traditional structure structure function $\W_\mn$ of the proton is indeed
just $L_\mn$ with kinematic factors. Explicitly (see \ref{eq:pfactordef}),
\beq
\label{eq:fw-w}
{1 \over {4 \pi^2}} {M \over E_p} \W_\mn (p,q)  \equiv L_\mn 
= \half \sum_\sigma \int d^4x e^{i q \cdot x}
\bra{p\sigma} j^\mu (x) j^\nu (0) \ket{p\sigma} \,.
\eeq

\nid And after the tensorial decomposition, the two proton structure
functions $\W_1$ and $\W_2$, which are only functions of the kinematic
invariants of the problem,  are defined by
\beqs
\label{eq:fw-w12def}
\W_\mn(p,q) &=&  \W_1(Q^2, x) \left( -g_\mn + {{q_\mu q_\nu} \over q^2}
\right)   \non \\
&+& \, { {\W_2(Q^2, x)} \over M^2 }
\left( p_\mu p_\nu 
         - { {p \cdot q} \over q^2 } (p_\mu q_\nu + p_\nu q_\mu)
         + { {(p \cdot q)^2} \over {(q^2)^2} } q_\mu q_\nu
    \right) \,.
\eeqs

\nid Now in terms of the structure functions, the differential
cross-section \ref{eq:fw-dsigma} becomes
\beq
\label{eq:fw-dsigmaW}
{ {d^3 \sigma} \over {d k'^3} } = {1 \over {4 \pi^2}} {M \over E_p}
 {1 \over {2 \eps 2 \eps'}} { e^4 \over Q^4} \W^\mn l_\mn \,.
\eeq

In calculating $\W^\mn l_\mn$ we note that 
$q_\mu l^\mn = 0 + O(m_e^2)$, so we can drop all terms having explicit
factor of $q_\mu$ or $q_\nu$ in $\W_\mn$. We get
\beqs
\label{eq:wl}
l_\mn \W^\mn &=& 2 (k'_\mu k_\nu + k_\mu k'_\nu - g_\mn k \cdot k')
\; (- \W_1 g^\mn + { \W_2 \over M^2 } p^\mu p^\nu) \non \\
&=& 4 \W_1 k \cdot k' + 2 { \W_2 \over M^2 } 
    (2 p \cdot k p \cdot k' - M^2 k \cdot k') \non \\
&=& 4 \eps \eps' (2 \W_1 \sin^2 {\theta \over 2}
+ \W_2 \cos^2 {\theta \over 2} ) \,.
\eeqs
 
\nid Thus the differential cross-section is given by
\beq
{ {d^3 \sigma} \over {d k'^3} } = 4 { {\al_{em}^2} \over Q^4}
{M \over E_p}
(2 \W_1 \sin^2 {\theta \over 2} + \W_2 \cos^2 {\theta \over 2} ) \,,
\eeq

\nid where $\al_{em} \equiv {e^2 \over {4 \pi}}$ is the fine structure
constant.

It is conventional to give $d \sigma /d Q^2 d \nu$ rather than
$d \sigma / d^3k'$, so we need the Jacobian of the variable
transformation. We have
\beqs
\label{eq:Jacobi}
d^3 k' &=& 2 \pi d \cos \theta (\eps')^2 d \eps' \non \\
&=& 2 \pi (\eps')^2 \left|
\begin{array}{cc}
\d{\cos \theta}/d {Q^2} & \d{\cos \theta}/d{\nu} \\
\d{\eps'}/d{Q^2} & \d{\eps'}/d{\nu}
\end{array}
\right| d \nu d Q^2 
= \pi (\eps')^2 d \nu d Q^2 \left|
\begin{array}{cc}
\d{Q^2}/d{\cos \theta} & \d{Q^2}/d{\eps'} \\
\d{\nu}/d{\cos \theta} & \d{\nu}/d{\eps'}
\end{array}
\right|^{-1} \non \\
&=& \pi {\eps' \over \eps} d\nu dQ^2 \,. 
\eeqs

\nid Therefore, the expression of the differential cross-section
eventually becomes
\beq
\label{eq:fw-dsigmaresult}
{ {d\sigma} \over {d\nu dQ^2} } = 4 \pi { {\al_{em}^2} \over Q^4}
{\eps' \over \eps} {M \over E_p}
(2 \W_1 \sin^2 {\theta \over 2} + \W_2 \cos^2 {\theta \over 2} ) \,.
\eeq

\nid In the laboratory frame where $\theta$ is the scattering angle,
it is possible to fix $x$ and $Q^2$ (or equivalently $Q^2$ and $\nu$) 
and vary $\theta$, thus experimentally $\W_1$ and $\W_2$ can be determined
separately. Since there are more small angle experimental data, $\W_2$
is better measured. 

Theoretically, what we usually do is to relate the cross-section, in
terms of the structure functions, to the imaginary part of a forward
scattering amplitude \cite{Christ&Mueller72} via an optical theorem. 

\nid To be more specific, define the forward scattering amplitude 
$\T_\mn$ as
\beq
\label{eq:fw-amp}
\T_\mn = i 4\pi^2 {E_p \over M} \half \sum_\sigma \int d^4x e^{i q \cdot x}
\bra{p\sigma} T j^\mu (x) j^\nu (0) \ket{p\sigma} \,,
\eeq

\nid we have (see \ref{eq:fw-w})
\beq
\label{eq:fw-optical}
\W_\mn = 2 Im \; \T_\mn \,.
\eeq

\begin{figure}[ptb]
\begin{center}
\epsfxsize=5.5in
\epsfysize=3in
\leavevmode
\hbox{ \epsffile{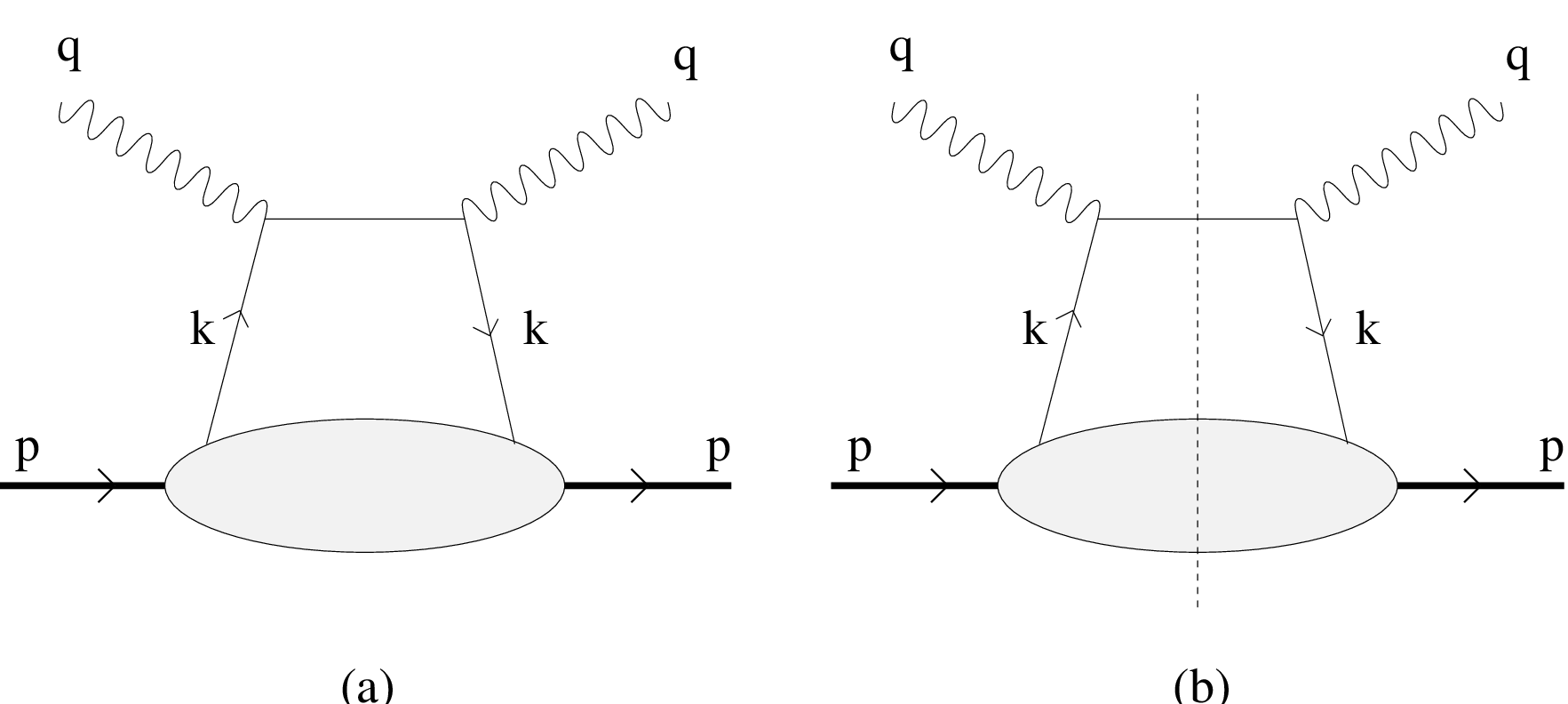} }
\end{center}
\caption{Forward Amplitude and Structure Function}
{This is the graphic representation of the DIS forward amplitude (a) and 
the DIS structure function (b). The difference lies in the cut (the 
dotted line) in the middle of (b) which puts all the intermediate 
particles it passes through on the mass shell.}
\label{fig:fw-amp}
\end{figure}

\nid The forward amplitude is shown in figure \ref{fig:fw-amp}(a) while
the structure function is shown in figure \ref{fig:fw-amp}(b), both
up to kinematic factors. The only difference in the graphs is the
cut in the middle in figure \ref{fig:fw-amp}(b), which means that all
the intermediate lines the cut goes through are put on the mass
shell. Mathematically this is equivalent to replacing the Feynman
propagators of all the intermediate particles with its corresponding
on-shell $\de$-function and the proper factors (see 
section \ref{sec:fw-gamma}).

\section{Operator Product Expansion Analysis}

We will use a straight forward operator product expansion (OPE) analysis 
to discuss the deeply inelastic scattering (DIS) of previous sections. 
While similar and more complete discussions can be found in
 \cite{Collinsbook}, we will present ours in a slightly different
notation, adopted from \cite{G8070}.

\subsection{General Statement}

Let $\hat{O}_1(x)$ and $\hat{O}_2(x)$ be local operators built 
out of fundamental
fields like $q, \tilde{q}$ (quark and anti-quark fields) or $A_\mu$
(gauge fields) and finite number of (covariant) derivatives at one
space-time point, for example, 
\beq
\label{eq:opeg}
j_\mu^f(x) = \tilde{q}^f(x) \ga_\mu q(x) \;\;\;\;\;\;\;\;; 
\hat{O}_\mu(x) = \tilde{q}D_\mu(x) q(x) \;; ... \;.
\eeq

\nid We have, in the short distance limit,
\beq
\label{eq:fw-ope}
\hat{O}_1(x) \hat{O}_2(0) 
\stackrel{x \rightarrow 0}{\longrightarrow} \sum_{r=1}^N
\hat{O}_{\mu_1 \mu_2 ... \mu_{n_r}}^{(r)}(0) 
\; E_{\mu_1 \mu_2 ... \mu_{n_r}}^{(r)}(x) + Remainder \,.
\eeq

\nid $\hat{O}^{(r)}$s are local operators which can include the
identity operator $\hat{I}$ and $E^{(r)}(x)$ are the so-called Wilson
coefficients, $c$-number functions that are usually singular in the
limit of $x \rightarrow 0$.

By a dimensional analysis we can obtain the asymptotic behavior of 
the Wilson coefficients $E^{(r)}(x)$ at short distance. The convention
we use is such that mass and energy has dimension $+1$, i.e., 
$[E] = dim (E) = [M] = dim (M) = +1$, 
while coordinate has dimension of $-1$, ie, $[x] = -1$. Thus,
we have under our convention, $[q]  = - {3 \over 2}$,
$[A_\mu] = -1$, and so on.

\nid Now let $d_r$ be the naive dimension of $\hat{O}^{(r)}$, $d_1$ and
$d_2$ be those of $\hat{O}_1$ and $\hat{O}_2$, respectively, $E^{(r)}(x)$ has
an $x$ behavior/dependence, for small $x$, given by
$(\sqrt{x^2})^{-\lm}$, where $\lm = d_r - d_1 - d_2$ since we must
have $d_1+d_2 = d_r - \lm$. That is, we have
\beq
\label{eq:fw-asym}
E_{\mu_1 \mu_2 ... \mu_{n_r}}^{(r)}(x) 
\stackrel{x \rightarrow 0}{\sim}(\sqrt{x^2})^{ d_1+d_2 - d_r} \,.
\eeq

\nid When we increase $N$, we can only generate new operators by
adding new fields or more derivatives into the operators, 
these new operators must have more negative dimensions, that is, 
$d_r$ decreases with increasing $N$, therefore, we may choose $N$
large enough such that the remainder becomes as small as desired,
namely, $remainder \sim (x^2)^R$ for any desired positive value $R$ in
the small $x$ limit.

\subsection{Renormalization Group Equations}

We work in the zero quark mass limit. The operators under
renormalization behaves like
\beq
\label{eq:fw-oprenorm}
\hat{O}_i(x, \mu^2, \al_\mu) = Z_i^{-1}( {\mu^2 \over \mu_0^2}, \al_{\mu_0} ) 
\; \hat{O}_i(x, \mu_0^2, \al_{\mu_0}) \,,
\eeq

\nid where we have explicitly shown the dependence of the operators on
the renormalization scale. This is the so-called multiplicative
renormalization and $Z_i$ more generally can be a matrix when there is
mixing among the operators.

\nid Taking the logarithmic derivative of the scale $\mu^2$ on both
sides of \ref{eq:fw-oprenorm} we have
\beqs
\mu^2 {d \over {d \mu^2}} \hat{O}_i(x, \mu^2, \al_\mu) &=&
[ \mu^2 {d \over {d \mu^2}} Z_i^{-1}( {\mu^2 \over \mu_0^2}, \al_{\mu_0})] \;
\hat{O}_i(x, \mu_0^2, \al_{\mu_0}) \non \\
&=& - {1 \over {Z_i}} \mu^2 {{d Z_i} \over {d \mu^2}} \;\, 
Z_i^{-1} \; 
\hat{O}_i(x, \mu_0^2, \al_{\mu_0} ) \non \\
&\equiv& \ga_i(\al_\mu) \; \hat{O}_i(x, \mu^2, \al_\mu) \;,
\eeqs

\nid where in the last step we have defined the so-called anomalous
dimension $\ga_i$ of the operator $\hat{O}_i$ as
\beq
\label{eq:fw-gadef}
\ga_i(\al_\mu) = - {1 \over {Z_i}} \mu^2 {{d Z_i} \over {d \mu^2}} \,.
\eeq

\nid Note that $\ga_i$ is a function of $\al_\mu$ only. This is
because from \ref{eq:fw-gadef} $\ga_i$ can only be a function of
$\mu^2$ and $\al_\mu$ and is dimensionless. Because we have taken all
the quark masses to be zero, there is no scale to set $\mu$, thus
$\ga_i$ is a function of the coupling only. On the other hand, if 
$m_q \neq 0$, then $\ga_i \to \ga_i(\al_\mu, {\mu^2 \over m_q^2})$,
complications will arise. 

The operators, at the same time, obey a renormalization group equation 
of the form
\beq
\label{eq:fw-rge}
\mu^2 {d \over {d \mu^2}} \hat{O}_i(x, \mu^2, \al_\mu) = 
\ga_i(\al_\mu) \; \hat{O}_i(x, \mu^2, \al_\mu) \;,
\eeq

\nid or more generally, when there is operator mixing,
\beq
\label{eq:fw-grge}
\mu^2 {d \over {d \mu^2}} \hat{O}_i(x, \mu^2, \al_\mu) = 
\sum_j \ga_{ij}(\al_\mu) \; \hat{O}_j(x, \mu^2, \al_\mu) \;.
\eeq

To establish the renormalization group equations of
the Wilson coefficients, let us look at a simple case where
only one term is kept in the operator product expansion, that is,
\beq
\hat{O}_1(x, \mu^2, \al_\mu) \hat{O}_2(0, \mu^2, \al_\mu) 
\stackrel{x \rightarrow 0}{\longrightarrow} 
\hat{O}^{(3)}(0, \mu^2, \al_\mu) \; E^{(3)}(x, \mu^2, \al_\mu) + Remainder \,.
\eeq

\nid Taking the logarithmic derivative of $\mu^2$ on both sides and 
omitting the contribution from the remainder, we have, suppressing the
$\mu^2$ and $\al_\mu$ dependence,
\beq
(\ga_1 + \ga_2) \hat{O}_1(x) \hat{O}_2(0) 
\stackrel{x \rightarrow 0}{\longrightarrow}
\ga_3 \hat{O}^{(3)}(0) E^{(3)} + \hat{O}^{(3)}(0) \mu^2 {d \over {d \mu^2}}
E^{(3)} \;.
\eeq

\nid Combined with the operator product expansion itself, we have
\beq
\mu^2 {d \over {d \mu^2}} E^{(3)}(x) \equiv 
(\mu^2 \d/d{\mu^2} + \be \d/d{\al_\mu})  E^{(3)}(x)
= (\ga_1 + \ga_2 - \ga_3) E^{(3)}(x) \;,
\eeq

\nid where $\be = \be(\al_\mu) = \mu^2 {{d \al_\mu} \over {d \mu^2}}$
is the QCD beta-function. 

\nid Therefore, the renormalization group equation for the Wilson
coefficient functions is
\beq
\label{eq:fw-wcrge}
\left(\mu^2 \d/d{\mu^2} + \be \d/d{\al_\mu} + \ga_3(\al_\mu)
- \ga_1(\al_\mu) - \ga_2(\al_\mu) \right) 
\; E^{(3)}(x, \mu^2, \al_\mu) = 0 \;.
\eeq

\subsection{OPE Description of DIS \label{sec:fw-ope}}

We now put the operator product expansion formalism to use on the deeply
inelastic scattering process in \ref{sec:fw-kinematics}. We work on
the forward amplitude $\T_\mn$ defined in \ref{eq:fw-amp} and the
structure functions (and thus the cross-sections) can be obtained
through \ref{eq:fw-optical}. As usual we work in the so-called
Bjorken limit where we have 
\beq
\label{eq:bjlimit}
\left\{
\begin{array}{lll}
-q^2 \!\!\!& = \, Q^2 \;\;\; &{\rm large} \non \\
m\nu \!\!\!& = \, p \cdot q \;\;\; &{\rm large} \\
x_{Bj} \!\!\!& = \, {1 \over \om} = { Q^2 \over {2 p \cdot q} } 
\;\;\; &{\rm fixed} \;. \non
\end{array}
\right.
\eeq

\subsubsection{Forward Operator Product Expansion}

We expand $T j_\mu(x) j_\nu(0)$ as $x_\mu \to 0$.
The tensor structure of $\T_\mn$ can be proven as \cite{G8070}
\beqs
\label{eq:fw-orig-ope}
T j_\mu(x) j_\nu(0) = \hat{A}_\mn E^{(0)}(x^2) \hat{I} 
+ \hat{A}_\mn \sum_{i,n} \, F_n^{(i)} (x^2)
\hat{O}_{\mu_1 ... \mu_n}^{(i,n)} (0) 
x^{\mu_1}x^{\mu_2} ...x^{\mu_n} 
\,\,\,\,\,\,\,\,\,\,\,\, \non \\ 
+ \,\, \hat{B}_{\mu \nu \al \be} \sum_{i,n} \, E_n^{(i)} (x^2)
\hat{O}_{\al \be;\mu_1 ... \mu_n}^{(i,n)} (0) 
x^{\mu_1}x^{\mu_2} ...x^{\mu_n} + Rem. \;,
\eeqs

\nid where $\hat{A}_{\mu \nu}$ and $\hat{B}_{\mu \nu \al \be}$ are 
conserved tensor structure operators and (see also section \ref{sec:ope})
\beqs
\label{eq:fw-op-ab}
\hat{A}_{\mu \nu} &=& g_{\mu \nu} \Box - \pdr_\mu \pdr_\nu \non \\
\hat{B}_{\mu \nu \al \be} &=& g_{\mu \al}g_{\nu \be} \Box 
+g_{\mu \nu} \partial_\al \partial_\be
-g_{\mu \al} \partial_\nu \partial_\be
-g_{\nu \be} \partial_\mu \partial_\al \;.
\eeqs

\nid We suppose symmetric combinations in $\mu$, $\nu$ (since
from section \ref{sec:fw-decomp}, $\T_\mn$, after all, is symmetric) 
and also we suppose that all indices in $\hat{O}$s are symmetrized 
(even $\al$, $\be$ with $\mu_i$s). We note that the $\hat{O}$s are the
same operator sets for both terms and that the label $n$ is actually
the angular momentum quantum number, or spin, of the corresponding
operator. The term {\sl twist} is defined as the difference between
the naive dimension of $\hat{O}$ and its spin, ie, {\sl twist} of 
$\hat{O}_n = [\hat{O}_n] - n$.

We need to evaluate the above OPE between
two symmetric external proton states (see \ref{eq:fw-amp}), however,
in practice in the kinematic region where high energy particle
experiments are conducted, especially in the small $x_{Bj}$ region where
most of our interest lies, we can put in some transverse momentum to make
the external states asymmetric and physics will not see the difference
at this level (note it is the main goal of this thesis to discuss what will
happen with asymmetric external states when we go to higher level of
accuracy). That is,
\beq
\bra{p}T j_\mu(x) j_\nu(0) \ket{p} = 
\displaystyle\lim_{r \to 0} \bra{p-r} T j_\mu(x) j_\nu(0) \ket{p} \,.
\eeq

\nid Therefore the identity operator $\hat{I}$ does not contribute to
$\T_\mn$ in the OPE. We need to work to the next terms in 
 \ref{eq:fw-orig-ope} and we have
\beqs
\label{eq:fw-matrixexp}
\bra{p}T j_\mu(x) j_\nu(0) \ket{p} 
&\stackrel{x \rightarrow 0}{\longrightarrow}&
(g_{\mu \nu} \Box - \pdr_\mu \pdr_\nu) 
\sum_{i,n} \, F_n^{(i)} (x^2) x^{\mu_1}x^{\mu_2} ...x^{\mu_n} 
\bra{p}\hat{O}_{\mu_1 ... \mu_n}^{(i,n)} (0) \ket{p} \non \\
&& + \; (g_{\mu \al}g_{\nu \be} \Box 
+g_{\mu \nu} \partial_\al \partial_\be
-g_{\mu \al} \partial_\nu \partial_\be
-g_{\nu \be} \partial_\mu \partial_\al) \non \\
&& \;\;\;\; \sum_{i,n} \, E_n^{(i)} (x^2)
x^{\mu_1}x^{\mu_2} ...x^{\mu_n}
\bra{p} \hat{O}_{\al \be;\mu_1 ... \mu_n}^{(i,n)} (0)\ket{p} \;.
\eeqs

For the matrix elements 
$\bra{p}\hat{O}_{\mu_1 ... \mu_n}^{(i,n)} (0) \ket{p}$, 
the indices can be made only from either $g_\mn$ or $p_\mu$ because $x_\mu$
is gone, however, the contribution of the two types of indices are
different. After the Fourier transformation, we know from dimensional
analysis that
\beq
x_{\mu_i} \Rightarrow { q_{\mu_i} \over {q}^2 } \,.
\eeq

\nid Thus we have, for the $g_\mn$ terms,
\beq
g_{\mu_i \mu_j} { q_{\mu_i} \over {q}^2 }{ q_{\mu_j} \over {q}^2 }
\sim {1 \over q^2} \,,
\eeq

\nid while at the same time,
\beq
p_{\mu_i} p_{\mu_j} { q_{\mu_i} \over {q}^2 }{ q_{\mu_j} \over {q}^2 }
= ( \frac {p \cdot q} {q^2} )^2 \sim {\rm O}(1) \,.
\eeq

\nid Therefore the $g_\mn$ contributions are small compared with those of
$p_\mu p_\nu$. So the only term we keep is the contribution from a
totally symmetric (in $\mu_i$) combination $p_{\mu_1} p_{\mu_2} ...
p_{\mu_n}$. It is clear that this approximation is good to 
$O({1 \over q^2})$. That is, we write
\beq
\label{eq:fw-reducedef}
\bra{p}\hat{O}_{\mu_1 ... \mu_n}^{(i,n)} (0) \ket{p}
= p_{\mu_1} p_{\mu_2} ... p_{\mu_n} \, \bra{p}|\hat{O}^{(i)}_n|\ket{p}
+ O({1 \over q^2}) \,,
\eeq

\nid where $\bra{p}|\hat{O}^{(i)}_n|\ket{p}$ is the so-called reduced
matrix element of the operator $\hat{O}_{\mu_1...\mu_n}^{(i,n)}$. 
Because the external states are on-shell protons,  
$\bra{p}|\hat{O}^{(i)}_n|\ket{p}$ does not depend on kinematic
variables and thus is a number depending only on $i$ and $n$, the
flavor and spin indices, respectively. 

Note indeed this approximation is the so-called leading twist
approximation because non-leading twist operators will have in their
contributions extra factors of ${1 \over q^2}$ following similar
discussion as the above. 

\nid We therefore obtain, after taking leading twist approximation, 
\beqs
\label{eq:fw-redmatrixexp}
&& \bra{p}T j_\mu(x) j_\nu(0) \ket{p} 
\stackrel{x \rightarrow 0}{\longrightarrow}
\sum_{i,n} \, \left\{ (g_{\mu \nu} \Box - \pdr_\mu \pdr_\nu)
(p \!\cdot\! x)^n  F_n^{(i)} (x^2) \right. \non \\
&& \;\;\;\;\;\;\;\;\;\;
+ \left. [g_\mn(p \!\cdot\! \pdr)^2 + p_\mn p_\nu \Box
     - (p_\mu \pdr_\nu + p_\nu \pdr_\mu) p \!\cdot\! \pdr]
    (p \!\cdot\! x)^{n\!-\!2}  E_{n\!-\!2}^{(i)} (x^2) \right\} \;.
\eeqs

The forward amplitude $\T_\mn$ is essentially the Fourier transform 
with momentum $q$ of the matrix elements (\ref{eq:fw-amp}). 
And under the Fourier transform, it is clear that 
$ x_\mu \Rightarrow - i \d{}/d{q_\mu}$. Thus we can rewrite the
part concerning the Fourier transform of the Wilson coefficients 
$E_n^{(i)} (x^2)$ in terms of logarithmic derivatives of $q^2$ as 
\beqs
\label{eq:fw-logderi}
\int d^4 x \, e^{i q \cdot x} (p \cdot x)^n E_n^{(i)}(x^2) 
&=& (- i p_\mu \d{}/d{q_\mu})^n \int d^4 x \, e^{i q \cdot x}  
E_n^{(i)}(x^2) \non \\
&=& ((-i 2 p \cdot q) \d{}/d{q^2})^n 
\; \tilde{e}_n^{(i)}(q^2) \non \\
&=& \left( { {-i 2 p \cdot q} \over {q^2} } \right)^n
\left(q^2 \d{}/d{q^2}\right)^n \tilde{e}_n^{(i)}(q^2) \,,
\eeqs

\nid where $\tilde{e}_n^{(i)}$ is the Fourier transform of 
$E_n^{(i)}$; and we have a similar set of equations for $F_n^{(i)}$. 
(Note in our notation, 
$\left(q^2 \d{}/d{q^2}\right)^n \tilde{e}_n^{(i)}(q^2)$ is exactly the
coefficient $\tilde{C}^{Ja}(Q)$ in \cite{Collinsbook} with the
replacement of $n \to J$ (spin index) and $i \to a$(flavor index). )

\nid We then define the Wilson coefficients in the momentum
space $\tilde{E}_n^{(i)}$ and $\tilde{F}_n^{(i)}$ as 
\beqs
\label{eq:fw-momwc}
\int d^4 x \, e^{i q \cdot x} (p \!\cdot\! x)^n E_n^{(i)}(x^2)
\!&\!=\!&\! { {\!-2i} \over {(Q^2)^2} } 
\left( { {2 p \!\cdot\! q} \over {Q^2} } \right)^n
\tilde{E}_n^{(i)} (Q^2) 
\equiv { {\!-2i} \over {Q^4} } \om^n \tilde{E}_n^{(i)} (Q^2)
\non \\
\int d^4 x \, e^{i q \cdot x} (p \!\cdot\! x)^n F_n^{(i)}(x^2)
\!&\!=\!&\! { i \over {Q^2} } 
\left( { {2 p \cdot q} \over {Q^2} } \right)^n
\tilde{F}_n^{(i)} (Q^2)  
\equiv { i \over {Q^2} }  \om^n \tilde{F}_n^{(i)} (Q^2)
\;,
\eeqs

\nid where we have used equation \ref{eq:x-bj}. Substitute these
definitions into the Fourier transform of the matrix elements 
and note that using integration by parts we can show that 
the derivatives in the conserved tensor operators
(\ref{eq:fw-op-ab}) become factors of momentum $q$ in the fashion
$\pdr_\mu \to i q_\mu$, we arrive at
\beqs
\label{eq:fw-melemft}
\int d^4 x \, e^{i q \cdot x} \bra{p}T j_\mu(x) j_\nu(0) \ket{p}
= - \sum_{i,n} i \bra{p}|\hat{O}^{(i)}_n|\ket{p} \left\{
-\left(g_\mn - { {q_\mu q_\nu} \over {q^2} } \right)
\om^n \tilde{F}_n^{(i)} (Q^2) \right. \non \\
+ \left. {2 \over Q^2} \left( p_\mu p_\nu
- { {p_\mu q_\nu + p_\nu q_\mu} \over q^2} p \cdot q
+ g_\mn { {(p \!\cdot\!q)^2} \over q^2} \right) \om^{n\!-\!2}
\tilde{E}_{n\!-\!2}^{(i)} (Q^2) \right\} \;. \;\;\; 
\eeqs

\nid Rewriting the last term in the parenthesis in front of
$\tilde{E}_{n\!-\!2}^{(i)}$ as
\beq
g_\mn { {(p \!\cdot\!q)^2} \over q^2} 
= (g_\mn - {{q_\mu q_\nu} \over {q^2}} + {{q_\mu q_\nu} \over {q^2}})
{{(p \!\cdot\!q)^2} \over q^2} \,,
\eeq

\nid we obtain
\beqs
&&\int d^4 x \, e^{i q \cdot x} \bra{p}T j_\mu(x) j_\nu(0) \ket{p}
\non \\
&&= - \sum_{i,n} i \bra{p}|\hat{O}^{(i)}_n|\ket{p} \left\{
-\left(g_\mn - { {q_\mu q_\nu} \over {q^2} } \right) \om^n 
\left(\tilde{F}_n^{(i)} (Q^2) - {2 \over {Q^2 \om^2}} 
{{(p \!\cdot\!q)^2} \over q^2} \tilde{E}_{n\!-\!2}^{(i)} (Q^2)\right)
\right. \non \\
&& \;\;\;\;\;\;\;\;\;\;\;\;
+ \left. \left( p_\mu p_\nu 
- { {p_\mu q_\nu + p_\nu q_\mu} \over q^2} p \cdot q
+ {{q_\mu q_\nu} \over {q^2}} { {(p \!\cdot\!q)^2} \over q^2}
\right) {2 \over Q^2} \om^{n\!-\!2}  
\tilde{E}_{n\!-\!2}^{(i)} (Q^2) \right\} \;.
\eeqs

\nid From equation \ref{eq:x-bj} we have
\beq
- {2 \over {Q^2 \om^2}}{{(p \!\cdot\!q)^2} \over q^2} = \half \;,
\eeq

\nid thus
\beqs
\label{eq:fw-opefinal}
&&\int d^4 x \, e^{i q \cdot x} \bra{p}T j_\mu(x) j_\nu(0) \ket{p}
\non \\
&&= - \sum_{i,n} i \bra{p}|\hat{O}^{(i)}_n|\ket{p} \left\{
-\left(g_\mn - { {q_\mu q_\nu} \over {q^2} } \right) \om^n 
(\tilde{F}_n^{(i)} (Q^2) +\half \tilde{E}_{n\!-\!2}^{(i)} (Q^2) )
\right. \non \\
&& \;\;\;\;\;\;\;\;\;\;\;\;
+ \left. {1 \over {M \nu}} \left( p_\mu p_\nu 
- { {p \cdot q} \over q^2} (p_\mu q_\nu + p_\nu q_\mu)
+ { {(p \!\cdot\!q)^2} \over (q^2)^2} q_\mu q_\nu
\right) \om^{n\!-\!1}  
\tilde{E}_{n\!-\!2}^{(i)} (Q^2) \right\} \;. \;\;\;\;
\eeqs

Recall the definition of $\T_\mn$, equation \ref{eq:fw-amp}, and the
tensorial decomposition of the structure function, equation 
 \ref{eq:fw-w12def}, we obtain the tensorial decomposition of the
forward amplitude as
\beqs
\label{eq:fw-t12def}
\T_\mn &=& i 4\pi^2 {E_p \over M} \int d^4x e^{i q \cdot x}
\bra{p} T j^\mu (x) j^\nu (0) \ket{p} \non \\
&=& \left( -g_\mn + {{q_\mu q_\nu} \over q^2} \right) \T_1
+ {1 \over M^2} \left( p_\mu p_\nu 
         - { {p \cdot q} \over q^2 } (p_\mu q_\nu + p_\nu q_\mu)
         + { {(p \cdot q)^2} \over {(q^2)^2} } q_\mu q_\nu
    \right) \T_2 \,, \non \\
\eeqs

\nid where the forward invariant amplitude $\T_{1,2}$ are related to
the structure functions $\W_{1,2}$ by
\beq
\label{eq:fw-tandw}
\W_1 = 2 Im \, T_1 \;; \;\;\;\;\;\;\; \W_2 = 2 Im \, T_2 \;.
\eeq

\nid Comparing with the expression of the operator product expansion
of $\T_\mn$, equation \ref{eq:fw-opefinal}, we obtain the formula for
the invariant amplitudes as
\beqs
\label{eq:fw-invamp}
\T_1 &=\!&\! { {4 \pi^2 E_p} \over M } \sum_n \om^n
\sum_i \bra{p}|\hat{O}^{(i)}_n|\ket{p} 
(\tilde{F}_n^{(i)} (Q^2) +\half \tilde{E}_{n\!-\!2}^{(i)} (Q^2) )
\non \\
\nu \T_2 &=& 4 \pi^2 E_p \, \sum_n \om^{n\!-\!1}
\sum_i \; \bra{p}|\hat{O}^{(i)}_n|\ket{p} \; 
\tilde{E}_{n\!-\!2}^{(i)} (Q^2) \;.
\eeqs

\subsubsection{Dispersion Relation and Optical Theorem}

Equation \ref{eq:fw-invamp} is essentially a power series expansion
of the invariant amplitudes. To further the computation we need the
analyticity properties of $\T_1$ and $\T_2$

Explicitly write out the time ordered product in $\T_\mn$ 
(suppressing the spin average) by inserting a summation over arbitrary
intermediate states $\ket{r}$ we get
\beqs
\T_\mn &=& i 4\pi^2 {E_p \over M} \int d^4x e^{i q \cdot x}
\bra{p} T j^\mu (x) j^\nu (0) \ket{p} \non \\
&=& 4\pi^2 {E_p \over M} i \sum_r \int d^4 x \, e^{i q \cdot x} 
\{ \; \theta (x_0) \bra{p} j_{\mu}(x) \ket{r} \bra{r} j_{\nu}(0) \ket{p} 
\non \\
&& \;\;\;\;\;\;\;\;\;\;\;\;\; 
+ \; \theta (-x_0) \bra{p} j_{\nu}(0) \ket{r} \bra{r} j_{\mu}(x) \ket{p}
\; \} \;.
\eeqs

\nid By applying the translation operators (see equation
 \ref{eq:translation}) we have
\beqs
\label{eq:fw-texpansion}
\T_\mn \!&\!=\!&\! i 4\pi^2 {E_p \over M} 
i \sum_r \int d^4 x \, e^{i q \cdot x}
\{ \, \theta (x_0) e^{i (p-p_r) \cdot x}
\bra{p} j_{\mu}(0) \ket{r} \bra{r} j_{\nu} (0)  \ket{p}
\non \\
&& \;\;\;\;\;\;\;\;\;\;\;\;\;\;\;\;\;\;\;\;
+ \; \theta (-x_0) e^{-i (p - p_r) \cdot x}
\bra{p} j_{\nu}(0) \ket{r} \bra{r} j_{\mu} (0)  \ket{p} \, \}
\non \\
\!&\!=\!&\! i 4 \pi^2 {E_p \over M} (2 \pi)^3 \sum_r \int d x_0
\{ \de^3(\vec{q} \!+\! \vec{p} \!-\! \vec{p}_r) 
e^{i (q_0 \!+\! {p}_0 \!-\! p_{r\!,\!0}) \cdot x_0} \theta (x_0)
\bra{p} j_{\mu}(0) \ket{r} \bra{r} j_{\nu} (0) \ket{p}  \non \\
&& \;\;\;\;\;\;\;\;\;\;\;\;\;
+ \; \de^3(\vec{q} - \vec{p} + \vec{p}_r)
e^{i (q_0 - p_0 + p_{r,0}) \cdot x_0} \theta (-x_0)
\bra{p} j_{\nu}(0) \ket{r} \bra{r} j_{\mu} (0)  \ket{p} \, \} \non \\
\!&\!=\!&\! - \,4\pi^2 {E_p \over M} (2 \pi)^3 \sum_r
\left( \; \frac {\de^3(\vec{q} + \vec{p} - \vec{p}_r)
           \bra{p} j_{\mu}(0) \ket{r} \bra{r} j_{\nu} (0)  \ket{p} }
   {q_0 + {p}_0 - p_{r,0} +i \eps} \right. \non \\
&& \;\;\;\;\;\;\;\;\;\;\;\;\;\;\;\;\;\;\;\;\;\;\;\;\;\;\;\;\;\;\; 
\left. - \;\frac {\de^3(\vec{q} - \vec{p} + \vec{p}_r)
              \bra{p} j_{\nu}(0) \ket{r} \bra{r} j_{\mu}(0) \ket{p}}
   {q_0 - p_0 + p_{r,0} - i \eps} \; \right) \;,
\eeqs     

\nid where in the second step we have performed the spacial
integration that yields the $3$-d momentum $\de$-function. In the
last step when we perform the time integration we have put in the
proper damping factors $\pm \eps$ in the exponent to assure the
convergence at the boundaries in an adiabatic approximation. 

To visualize the analytic properties of $\T_\mn$, again we go to the
rest frame of the proton where $p \cdot q = M q_0 = M \nu$. The
independent variables $\T_\mn$ depends on are $p$ and $q$ while for
the invariant amplitudes the independent variables are actually $Q^2$
and $\om$, with now (see equation \ref{eq:x-bj})
\beq
\om = { {2 p \cdot q} \over {Q^2} } = { {2 M} \over Q^2 } q_0 \,.
\eeq

\nid Thus, for fixed $Q^2$, singularities in $\om$ are the same as
singularities in $q_0$. Note this result is covariant--it is just for
the purpose of visualization that we need to choose a frame.

From equation \ref{eq:fw-texpansion} we have singularities in $q_0$ at
$(i) \; q_0 + p_0 = p_{r,0} \equiv E_r$ and 
$(ii) \; q_0 - p_0 = - E_r$. We shall discuss them one by one. 

\begin{description}

\item{(i) } The first set of poles in $q_0$ occur when

\beq
(q_0 + {p}_0)^2 = E_r^2 = M_r^2 + \vec{p}_r^2 
= M_r^2 + (\vec{q} + \vec{p})^2 \,, \non
\eeq

\nid where in the last step we have used the spacial $\de$-function. 
$M_r$ is the invariant mass of the intermediate state $\ket{r}$. 
We therefore have
\beq
M_r^2 = (q + p)^2 = -Q^2 + 2 p \cdot q + M^2 \,, \non
\eeq

and the corresponding poles in $\om$ are
\beq
\om_r = 1 + \frac {M_r^2 - M^2} {Q^2} \,. \non
\eeq

\item {(ii) } The second set of poles in $q_0$ occur when, after
similar discussion as above,
\beq
M_r^2 = (q-p)^2 = q^2 -2 p \cdot q + M^2 \,, \non
\eeq

\nid and the $\om$ poles are at
\beq
\om_r = - 1 - \frac {M_r^2 - M^2 } {Q^2} \,.
\eeq

\end{description}

\begin{figure}[ptb]
\begin{center}
\epsfxsize=4in
\epsfysize=3.2in
\leavevmode
\hbox{ \epsffile{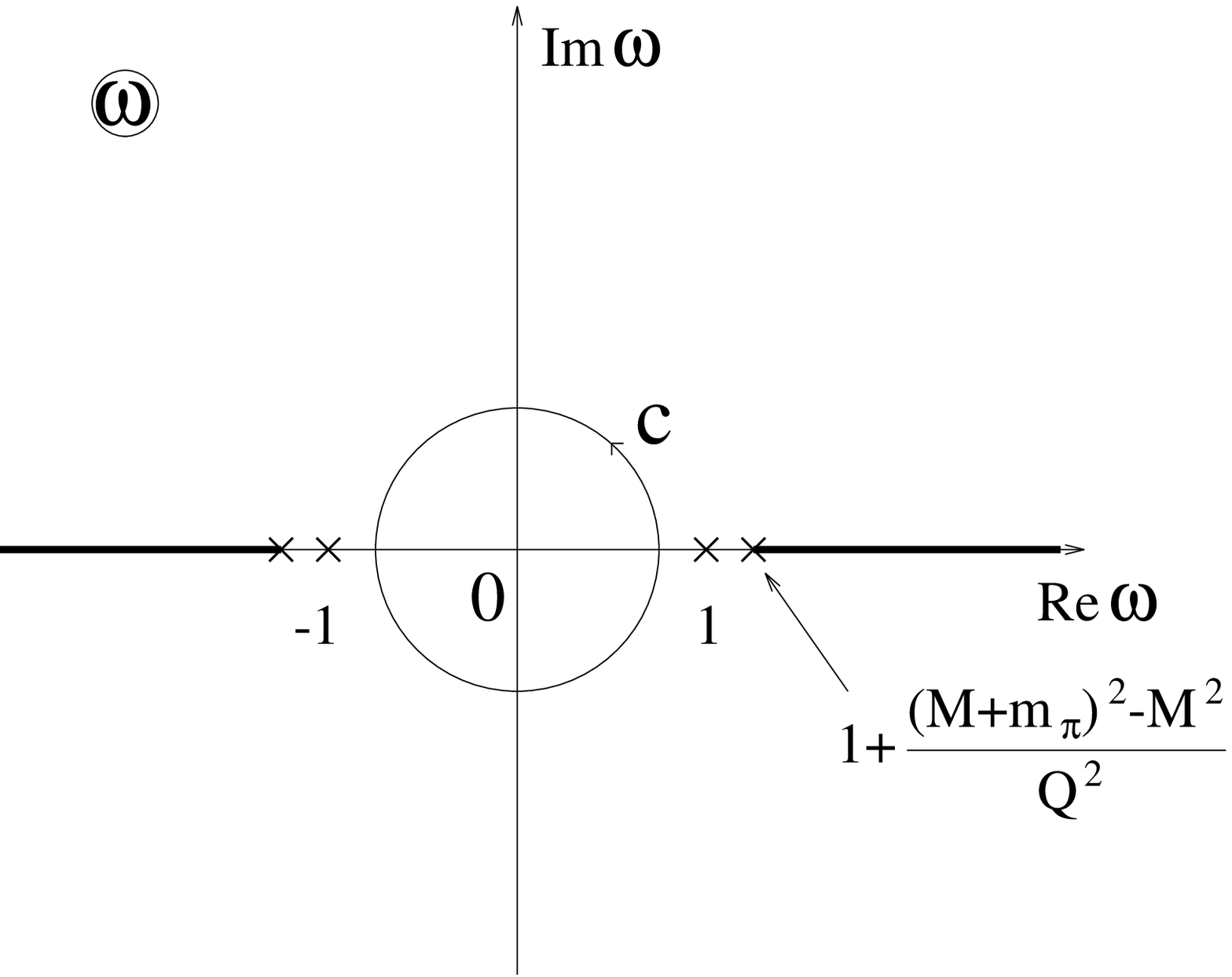}}
\end{center}
\caption{Analyticity of the Forward Amplitude}
{This figure shows the analytic properties of the forward amplitude on 
the complex $\om$ plane.
The crosses are poles and the bold regions on the real axis are the
branching cuts. The apmlitude has a convergence circle of unit radius.
$c$ is the original contour used in the dispersion integral. It lies
completely within the unit circle and thus encloses no singularity.}
\label{fig:T-analyticity}
\end{figure}

\nid The intermediate states $\ket{r}$ can be $\ket{p}$,
$\ket{p,\pi}$, $\ket{p,\pi,\pi}$, etc. Electro-magnetic current and
Baryon number conservation require that $M$, mass of the proton
($\ket{p}$ state), be the smallest invariant mass of all possible
states involving any baryon. This means we always have $M_r \ge M$. 
Therefore we can clearly visualize the analytic property of $\T_\mn$
on the complex $\om$-plane as shown in figure \ref{fig:T-analyticity}:
\begin{itemize}

\item $\T_\mn$ has a pole at $\om =1$ and another one at $\om = -1$. 
They correspond to an elastic scattering where $\ket{r} = \ket{p}$ and
thus $M_r = M$.

\item $\T_\mn$ also has two branch cuts. One on the positive real
$\om$-axis and starts from the point 
$\om = 1 + \frac { (M+m_\pi)^2 - M^2} {Q^2}$, which corresponds to the
lowest excited state $\ket{p,\pi}$ with a pion generated almost at
rest and as such $M_r = M + m_\pi$. It extends out to
infinity, or rather, as long as the collision center of mass energy is
big enough to generate the states. The other one is simply a mirror
reflection of this cut about the imaginary $\om$-axis.

\end{itemize}

\nid It is therefore obvious that $\T_\mn$ has an analytic circle of 
unit radius on the $\om$-plane and thus its Taylor expansion in $\om$
exists as long as we are inside the unit circle.

However, the region of $0 \le \om \le 1$ is not physical. The physics,
namely, deeply inelastic scattering in the Bjorken limit, is happening
in the kinematic region where $0 \le x_{Bj} \le 1$ (see 
section \ref{sec:fw-kinematics}) and corresponds to $\om \ge 1$. Thus the
operator product expansion of equation \ref{eq:fw-invamp}, although
called the OPE of the forward amplitude, can not be applied directly
as it is to the physical scattering, nor the computation of structure
functions and cross sections. 

It turns out \cite{Christ&Mueller72} 
that by deriving a dispersion relation for $\nu
\T_2$, we can indeed correctly apply the operator product expansion to
the physical situation and relate directly to the structure function
$\nu \W_2$. 

We start by writing a dispersion integral of $\nu \T_2$,
\beq
\nu \T_2 (\om, Q^2) = {1 \over {2 \pi i}} \int_c 
\frac {d \om'}{\om' - \om} \nu \T_2 (\om', Q^2) \,
\eeq

\nid where $c$ is any coutour enclosing the origin on $\om$-plane and
lying complete inside the unit circle (see figure
 \ref{fig:T-analyticity}). The analyticity of $\T_\mn$ within the unit
circle will guarantee its convergence.

\begin{figure}[ptb]
\begin{center}
\epsfxsize=4in
\epsfysize=3.5in
\leavevmode
\hbox{ \epsffile{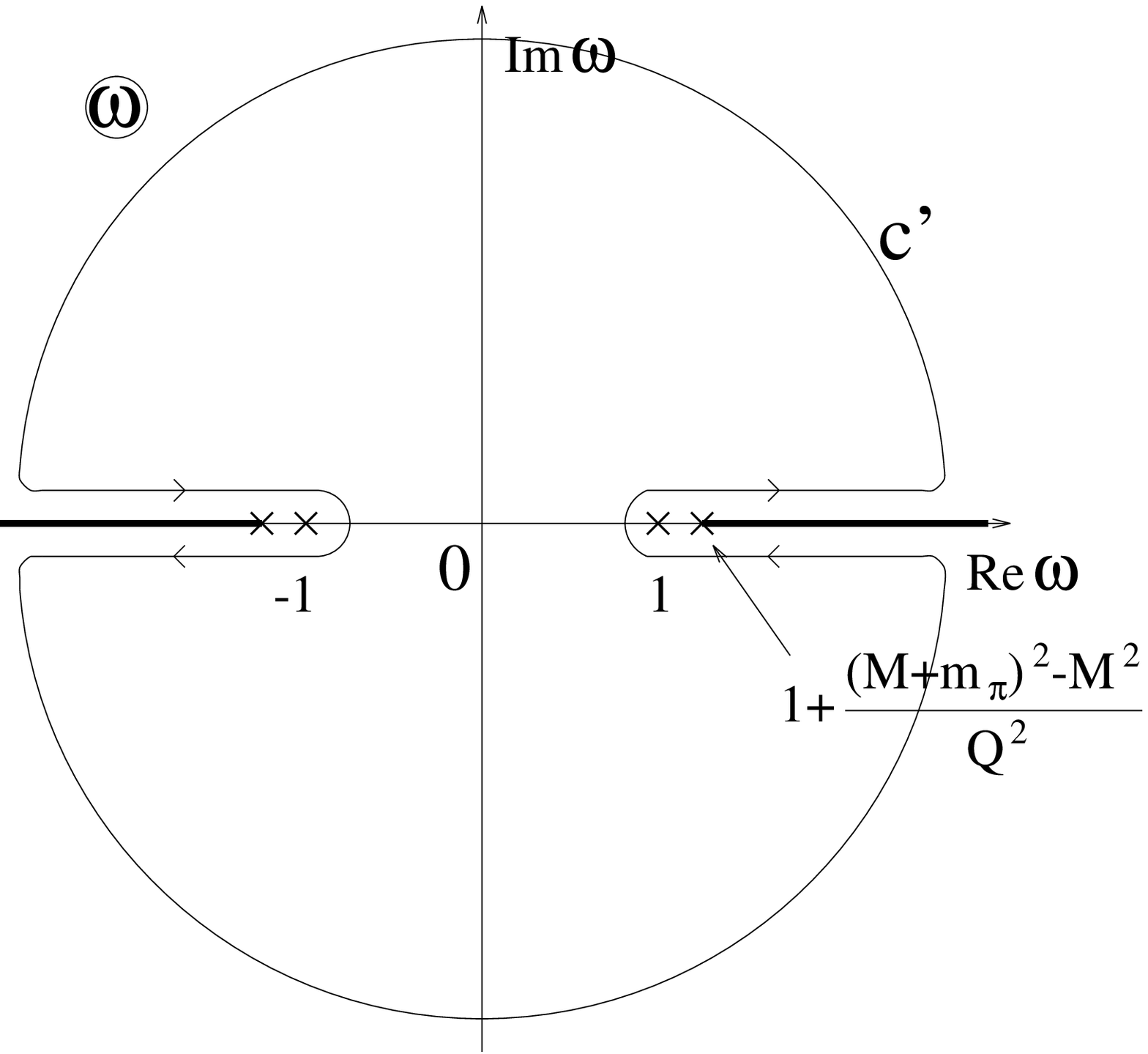} }
\end{center}
\caption{Dispersion Integral Contour}
{This figure shows the contour $c'$ of the dispersion integral of the 
forward amplitude. It is a continuous distortion from the $c$ contour of
figure \ref{fig:T-analyticity}.  
$c'$ does not cross any cut nor does it enclose any singularity on the
complex $\om$ plane.}
\label{fig:contour}
\end{figure}

\nid We can continuously and analytically distort the contour $c$ 
to $c'$ as shown in figure \ref{fig:contour}, where $c'$ crosses no
cut, encloses no pole and has its boundary segments pushed to
infinity. At infinity  although actually $\nu \T_2$ approaches a constant
 \cite{G8070}, there are cancellations between $\pm \infty$
resulting in extra convergent factors and as such we can drop the
integral region of the two semi-circles at infinity and obtain
\beqs
\label{eq:fw-distortion}
\nu \T_2 (\om, Q^2) &=& {1 \over {2 \pi i}} \int_{c \to c'}
\frac {d \om'}{\om' - \om} \nu \T_2 (\om', Q^2) \non \\
&=& {1 \over {2 \pi i}} \int_{1^-}^\infty \frac {d \om'}{\om' - \om}
(\nu \T_2 (\om' + i\eps, Q^2) -  \nu \T_2 (\om' - i\eps, Q^2)) \non \\
&& + {1 \over {2 \pi i}} \int_{-1^+}^{-\infty} \frac {d \om'}{\om' - \om}
(\nu \T_2 (\om' - i\eps, Q^2) -  \nu \T_2 (\om' + i\eps, Q^2)) \,,
\eeqs

\nid where $\eps$ is a infinitesimal positive number.

To simplify the above expression, we need the symmetry property of
$\T_\mn$ in $\om$, or equivalently, in $q_0$. $\T_\mn$ as a
time-ordered product can be represented systematically by a set of 
Feynman diagrams. Equation \ref{eq:fw-texpansion} is for real $q_0$
values. If we take $q_0$ to be complex and forget $i \eps$, we have an
extended $\T_\mn^c (q_0^c)$ on a complex plane. The usual time-ordered
product is obviously obtained when we take $q_0 \to |q_0| +i \eps$ or
$q_0 \to -|q_0| -i \eps$. This extended $\T_\mn^c (q_0^c)$ clearly
obeys
\beq
\T_\mn(q_0) = \T_\mn (-q_0) \,,
\eeq

\nid which can be obtained directly from equation
 \ref{eq:fw-texpansion}. When we go to real values, we actually have
\beq
\T_2 (\om+i\eps,Q^2) = \T_2(-\om - i\eps, Q^2) \,,
\eeq

\nid which means that for $\nu \T_2$ we have
\beq
\nu \T_2 ( \om \pm i \eps, Q^2) = - \nu \T_2 (-\om \mp i \eps, Q^2)
\,.
\eeq

\nid Therefore, by setting $\om' \to -\om'$ in the second integral of
equation \ref{eq:fw-distortion} and use the above we obtain
\beqs
\nu \T_2 (\om, Q^2) &=& {1 \over {2 \pi i}} \int_{1^-}^\infty 
\frac {d \om'}{\om' - \om}
(\nu \T_2 (\om' + i\eps, Q^2) -  \nu \T_2 (\om' - i\eps, Q^2)) \non \\
&& \;\;\;\;\;
+ \; {1 \over {2 \pi i}} \int_{1^-}^{\infty} \frac {d \om'}{\om' + \om}
(\nu \T_2 (-\om' - i\eps, Q^2) -  \nu \T_2 (-\om' + i\eps, Q^2)) \non
\\
&=& {1 \over {2 \pi i}} \int_{1^-}^\infty 
\frac {d \om'}{\om' - \om}
(\nu \T_2 (\om' + i\eps, Q^2) -  \nu \T_2 (\om' - i\eps, Q^2)) \non \\
&& \;\;\;\;\;
+ \; {1 \over {2 \pi i}} \int_{1^-}^{\infty} \frac {d \om'}{\om' + \om}
(- \nu \T_2 (\om' + i\eps, Q^2) +  \nu \T_2 (\om' - i\eps, Q^2)) \non
\\
&=& {1 \over {2 \pi i}} \int_{1^-}^\infty d \om'
\left( {1 \over {\om' \!-\!\om}} \!-\! {1 \over {\om' \!+\! \om}} \right) 
[\nu \T_2 (\om' \!+\! i\eps, Q^2) -  \nu \T_2 (\om' \!-\! i\eps, Q^2)]
\,. \non \\
&=& {1 \over {2 \pi i}} \int_{1^-}^\infty d \om'
\frac {2 \om} {\om'^2 - \om^2} 
[\nu \T_2 (\om' \!+\! i\eps, Q^2) -  \nu \T_2^* (\om' \!+\! i\eps, Q^2)]
\non \\
&=& {\om \over {\pi i}} \int_{1^-}^\infty 
\frac {d \om'} {\om'^2 - \om^2} 2 i \, Im \, \nu \T_2 (\om'+ i\ eps, Q^2)
\,,
\eeqs

\nid where in the last step we have used the Hermiticity of the
electro-magnetic current. Recalling equation \ref{eq:fw-optical}, the 
optical theorem, we obtain the dispersion relation of $\nu \T_2$
as
\beq
\label{eq:fw-dispersion}
\nu \T_2 (\om, Q^2) = {\om \over \pi} \int_{1^-}^\infty 
\frac {d \om'} {\om'^2 - \om^2} \nu \W_2 (\om', Q^2) \,.
\eeq

\nid Note that in the above, $\om$ should still be seen as inside the
convergence circle (the unit circle) of $\nu \T_2$ while only $\om'$ 
has the physical meaning of the inverse of Bjorken-$x$. Therefore we 
can expand the denominator of the dispersion integrand and get
\beqs
\nu \T_2 (\om, Q^2) &=& {\om \over \pi} \int_{1^-}^\infty
\frac {d \om'} {\om'^2} \nu \W_2 (\om', Q^2)
\frac {1} {1 - {\om^2 \over {\om'^2}} } \non \\
&=& {\om \over \pi} \int_{1^-}^\infty
\frac {d \om'} {\om'^2} \nu \W_2 (\om', Q^2) 
\sum_{n=0}^\infty \left({\om^2 \over {\om'^2}}\right)^n \non \\
&=& \sum_{n=0}^\infty \frac {\om^{2n \!+\!1}} {\pi}
\int_{1^-}^\infty \frac {d \om'}{ (\om')^{2n \!+\!2} }
\nu \W_2 (\om', Q^2) \,.
\eeqs

\nid Change the integration variable from $\om'$ to $x_{Bj} =
\om'^{-1} \equiv x$ we have
\beqs
&& d x = - \frac {d \om'} { \om'^2} \non \\
&& \Rightarrow \frac {d \om'} { (\om')^{2n \!+\!2} }
 = - x^{2n} d x \,,
\eeqs

\nid and thus
\beqs
\label{eq:fw-Texpfinal}
\nu \T_2 (\om, Q^2) &\stackrel{\om \eps O(0)} {=}& 
\sum_{n=0}^\infty \om^{2n \!+\!1}
{1 \over \pi} \int_0^1 dx \; x^{2n} \nu \W_2 (x, Q^2) \non \\
&=& \sum_{n \, even}^\infty \om^{n+1}
{1 \over \pi} \int_0^1 dx \; x^n \nu \W_2 (x, Q^2) \,.
\eeqs

\nid Again we would like to stress that $\om$ and $x$ are completely
different quantities. $\om$ is essentially a mathematical qunatity
introduced to Taylor expand the (invariant) amplitude(s) around the
origin in an operator product expansion. It is the so-called {\sl
moment variable} and has its value limited between $0$ and $1$. On the
other hand, $x$ is the Bjorken-$x$, $x_{Bj}$, a kinematic variable of
the actual physical scattering (see \ref{eq:x-bj}) with its value
also limited between $0$ and $1$. The convolution of the structure
function $\nu \W_2$ with the $n\!+\!1$-th power of $x$ is called
taking the $n$-th moment of the structure function. We can see that 
because of the symmetry properties of $\nu \T_2$, only odd moments 
of the structure function enters the expansion of the invariant amplitude. 

Comparing equations \ref{eq:fw-Texpfinal} and \ref{eq:fw-invamp} we
can relate the operator product expansion of the forward amplitude
with the moments of the (physical) structure function and we get 
\beq
4 \pi^2 E_p \, \sum_i \; \bra{p}|\hat{O}^{(i)}_n|\ket{p} \; 
\tilde{E}_n^{(i)} (Q^2) = 
{1 \over \pi} \int_0^1 dx \; x^n \nu \W_2 (x, Q^2) \;, \;\;\;\;\;
n \; even \,.
\eeq

\nid That is, although the structure function $\nu \W_2$ itself
can not be related directly to an expansion of local operators and
their corresponding Wilson coefficients, its moments in $x_{Bj}$ can
indeed be expressed as a product of local operators and Wilson
coefficients by application of operator product expansion and
dispersion relations. The result, after a trivial rewriting, is
\beq
\label{eq:fw-operesult}
\int_0^1 dx \; x^n \nu \W_2 (x, Q^2) = \frac { (2\pi)^3 E_p } {2}
\sum_i \; \bra{p}|\hat{O}^{(i)}_n|\ket{p} \; \tilde{E}_n^{(i)} (Q^2)
\;, \;\;\;\;\; n \; even \;.
\eeq

\subsubsection{The Operators}

Before we discuss the momentum evolution and renormalization group
properties of the operators and their corresponding Wilson coefficients
in equation \ref{eq:fw-operesult}, let us identify the operators
themselves. 

By recalling equation \ref{eq:fw-reducedef} we know that the dominant
operators should have the smallest negative dimension since those with
larger negative ones will be accompanied by extra suppressing factors of
${1 \over Q^2}$ at high energy. In QCD, the leading twist operators
are the quark operators $\hat{O}_{\mu_1 ... \mu_n}^f$ and the gluon
operators $\hat{O}_{\mu_1 ... \mu_n}^G$ as the following
\beq
\label{eq:fw-ops}
\left\{
\begin{array}{l}
\hat{O}_{\mu_1 ... \mu_n}^f = 2 \; \tilde{q}_f \ga_{\mu_1}
D_{\mu_2} ... D_{\mu_n} q_f  \\
\hat{O}_{\mu_1 ... \mu_n}^G = -2 F^\al_{\mu_1}
\cd_{\mu_2} ... \cd_{\mu_{n\!-\!1}} F_{\al \mu_n} \,.
\end{array}
\right.
\eeq

\nid In the quark operator, $D_\mu$ is the covariant derivative 
and $D_\mu = \pdr_\mu - i g A_\mu$ with $g$ the QCD coupling 
$A_\mu = \sum_i A_\mu^i \lm^i/2$ where $\lm^i$ is the fundamental 
representation of the $SU(3)$ color group and $A_\mu^i$ the gauge 
field. $f$ is the flavor label and we can form singlets and octets 
out of the flavor indices. The Dirac indices are all symmetrized. 
The dimension of the quark operator is $[O^f_n]= -n -2 $.

\nid The dimension of the gluon operator is also $[O^G_n] = -n-2$. The
Dirac indices are also symmetrized. In detail, $O^G_n$ is
\beq
\hat{O}_{\mu_1 ... \mu_n}^G = -2 \sum_{i_1,...,i_{n\!-\!1}}
F^{i_1, \al_{\mu_1}} (\cd_{\mu_2})^{i_1 i_2} (\cd_{\mu_3})^{i_2 i_3} 
... (\cd_{\mu_{n\!-\!1}})^{i_{n\!-\!2} i_{n\!-\!1}} 
F_{\al \mu_n}^{i_{n\!-\!1}} \,,
\eeq

\nid where the $i$'s are color indices and $\cd_\mu^{ij} = \pdr_\mu
+ \sum_l g f_{ilj}A_\mu^l$ with $f_{ilj}$ the adjoint representation
of the $SU(3)$ color group.  

We will not give an explicit proof of equation \ref{eq:fw-ops}, but 
rather present some motivations and explanations.
\begin{description}

\item{$\hat{O}_{\mu_1 ... \mu_n}^f$} Because the Dirac indices are
symmetrized, only one gamma matrix can be used since two of them will
give $g_\mn$ terms. The other indices have to be covariant derivatives
not only to preserve gauge symmetry but also to have the most
efficient way to get indices except for gamma matrices. For example,
in the case of $n=2$, we can have 
$\tilde{q}_f \ga_{\mu_1}q_f \tilde{q}_f \ga_{\mu_2}q_f$, which has
dimension of $-6$, or $\tilde{q}_f \ga_{\mu_1} D_{\mu_2} q_f$, which
has a dimension of $-4$. The Wilson coefficient ($E$ function) of the
former will have an extra power of ${1 \over Q^2}$ and thus the latter
dominates.

\item{$\hat{O}_{\mu_1 ... \mu_n}^G$} The gluon field $F_\mn$ has two
indices and dimension of $[F] = -2$, however, they are
anti-symmetric. To symmetrize the $\mu_i$ indices we have to
contract one of the two indices in each $F_\mn$. Thus they are not as
efficient as the (gluonic) covariant derivative $\cd_\mu$'s.

\end{description}

\section{Renormalization Group Analysis}

Recall that the operators need renormalization and they obey 
renormalization group equations \ref{eq:fw-rge} and \ref{eq:fw-grge}.  
Their corresponding Wilson coefficients obey a related equation
 \ref{eq:fw-wcrge} so that their products, which are the $n$ moments of
the physical structure function, do not depend on the renormalization
scale.

\subsection{Equations and Solutions}

We define the Parton distribution functions $x P^f(x,Q^2)$s by 
\beq
\label{eq:fw-pdfdef}
\sum_f e^2_f \!\int_0^1 \!d x \, x^n x P^f(x,Q^2) 
\equiv \int_0^1 \!dx \; x^n \nu \W_2 (x, Q^2) = \frac { (2\pi)^3 E_p } {2}
\sum_i \; \bra{p}|\hat{O}^{(i)}_n|\ket{p} \; \tilde{E}_n^{(i)} (Q^2)
\;.
\eeq

\nid At this point $x P^f$s seem more like mathematical objects rather
than physical ones. However, in the Parton Model of DIS, $x P^f$ are
indeed the (momentum fraction) distribution functions of quarks of
flavor $f$ inside the proton \cite{DESYreview}. 

In leading order of the QCD coupling $\al(Q^2)$ (when $Q^2$ is large), 
only quark operators come in because photons only couple to quarks 
directly. Thus in equation \ref{eq:fw-pdfdef} the $\hat{O}$'s are
quark operators $\hat{O}^f$ from equation \ref{eq:fw-ops}. We have
chosen the normalization of $\hat{O}^f$ such that
\beq
(2 \pi)^3 E_p \bra{p}  \hat{O}_{\mu_1 ... \mu_n}^f \ket{p} 
= 2 p_{\mu_1 ... \mu_n} 
\eeq

\nid for quarks of flavor $f$ in free field theory. On the other hand,
in QCD $\hat{O}^f$ requires renormalization and so a scale $\mu$ has
to be introduced. There will in general be complicated $Q^2$ and $\mu^2$
dependence in the reduced matrix elements and Wilson coefficients. To
be more specific, if $\mu^2/Q^2 << 1$ then $\tilde{E}_n^f$ would have terms
involving $\mu^2$ as 
\beq
\tilde{E}_n^f (Q^2, \mu^2)= \tilde{E}((\al(\mu^2)\log Q^2/\mu^2)^n)
= \cc_f + O ((\al(\mu^2)\log Q^2/\mu^2)^n) \,.
\eeq

\nid If we simply choose $\mu^2 = Q^2$, then all the $Q^2$ dependence
is in the reduced matrix elements obtained by equation
 \ref{eq:fw-reducedef} and we have 
\beq
\label{eq:fw-wc}
\tilde{E}_n^f (Q^2)= e^2_f + O(\al(Q^2)) \,
\eeq

\nid under the normalization convention we use. Therefore we have, in 
leading order, at the renormalization scale $\mu^2 = Q^2$, 
\beq
\sum_f e^2_f \int_0^1 d x \, x^n x P^f(x,Q^2) =
\frac { (2\pi)^3 E_p } {2} \sum_f \; \bra{p}|\hat{O}^f_n|\ket{p}
e^2_f \,,
\eeq

\nid which means that up to kinematic factors the moments of the 
parton distribution functions are the reduced matrix elements in 
operator product expansion, or explicitly, 
\beq
\label{eq:fw-pdf-ops}
\int_0^1 d x \, x^n x P^f(x,Q^2) = \frac { (2\pi)^3 E_p } {2}
\bra{p}|\hat{O}^f_n|\ket{p}_{Q^2} \,.
\eeq

\nid The subscript $Q^2$ means that the reduced matrix elements are
renormalized at the scale $\mu^2 = Q^2$.

Recall that the operators obey the renormalization group equation 
(see equations \ref{eq:fw-rge} and \ref{eq:fw-grge}), at an
arbitrary scale $\mu$,
\beq
\label{eq:fw-erge}
(\mu^2 \d/d{\mu^2} + \be \d/d{\al_\mu}) \; \hat{O}^f_n(\mu^2, \al_\mu)
= \ga_n^f (\al_\mu) \, \hat{O}^f_n(\mu^2, \al_\mu) \,,
\eeq

\nid where $\ga_n^f$ is the anomalous dimension of the operator.

\nid The matrix elements of the operators will obey the same
equation. Furthermore, because the differences between the matrix
elements and the reduced matrix elements are only kinematic factors
(see equation \ref{eq:fw-reducedef}), the reduced matrix elements and
thus the $n$-moments of the parton distribution functions also obey
the same renormalization equation, namely, we have
\beq
\label{eq:fw-pdfrge}
(\mu^2 \d/d{\mu^2} + \be \d/d{\al_\mu}) \; 
\int_0^1 d x \, x^n x P^f(x,Q^2) = \ga_n^f (\al_\mu)
\int_0^1 d x \, x^n x P^f(x,Q^2) \,.
\eeq

\nid The solution to the renormalization group equation
 \ref{eq:fw-erge} is (formally when we have operator mixing and an
anomalous dimension matrix)
\beq
\label{eq:fw-rge-solu}
\hat{O}^f_n(\mu^2, \al_\mu) = e^{- \int_{\al_\mu}^{\al_{\mu_0}}
d \al' \frac {\ga_n^f(\al')} {\be(\al')}} 
\hat{O}^f_n(\mu^2_0, \al_{\mu_0}) \,,
\eeq

\nid and the moments of parton distribution functions will obey
exactly the same evolution in momentum scale.

\nid The anomalous dimensions are calculable in perturbation 
theory. A lowest order calculation gives the dominant large $\mu^2$
dependence.

\subsection{Relationship to DGLAP Evolution \label{sec:DGLAP}}

The DGLAP equations (Dokthitze, Gribov, Lipatov, Altarelli, Parisi)
(for example, see \cite{A&P77}) state that the momentum scale $Q^2$
evolution of the parton distribution functions $P^f(x, Q^2)$ obeys
\beq
\label{eq:fw-DGLAP}
Q^2 \frac {d} {d Q^2} \, x P^f(x,Q^2) = \frac {\al (Q^2)} {2 \pi}
\int_x^1 \frac {d x'}{x'} \, \ga^f ({x \over {x'}}) \; x' P^f(x',Q^2) \,,
\eeq

\nid where $\ga^f(x)$ is the so-called Altarelli-Parisi splitting
function of the corresponding parton distribution(s). $\ga^f(x)$ is 
defined to be zero outside the range $(0,1)$. Again in general
$\ga^f(x)$ can be a matrix due to mixing among the parton distribution
functions. 

The DGLAP evolution equations and the renormalization group equations
 \ref{eq:fw-pdfrge} are actually equivalent, with the proper
identification of the splitting function with the anomalous
dimension. We will not explicitly prove this (for that see, for
example, \cite{Peskin, Cheng&Li}), but rather show that given the
DGLAP equation \ref{eq:fw-DGLAP} we can obtain equation \ref{eq:fw-pdfrge} 
and illustrate the explicit relationship between the splitting function 
and the anomalous dimension (in the leading order).

Taking the $n+1$th moment (here we adopt the convention that
$n$th moment is take by convoluting with $x^{n\!-\!1}$) of both 
sides of equation \ref{eq:fw-DGLAP} we obtain
\beq
Q^2 \frac {d} {d Q^2} \int_0^1 d x \, x^n x P^f(x,Q^2) = 
\frac {\al (Q^2)} {2 \pi}  \int_0^1 d x \, x^n
\int_x^1 \frac {d x'}{x'} \ga^f ({x \over {x'}}) x' P^f(x',Q^2) \,.
\eeq
\nid Change the integration variables in the right-hand-side (rhs) 
from $x$ and $x'$ to $x'$ and $x/x'$ by $dx = x' \, d (x/x')$ and 
$dx \, dx'/x' = d(x/x') \,dx'$, so that 
\beq
\int_0^1 dx \int_x^1 \frac {d x'}{x'} 
= \int_0^1 dx' \int_0^1 d \left(\frac {x}{x'}\right) \,,
\eeq

\begin{figure}[ptb]
\begin{center}
\epsfxsize=3in
\epsfysize=3in
\leavevmode
\hbox{ \epsffile{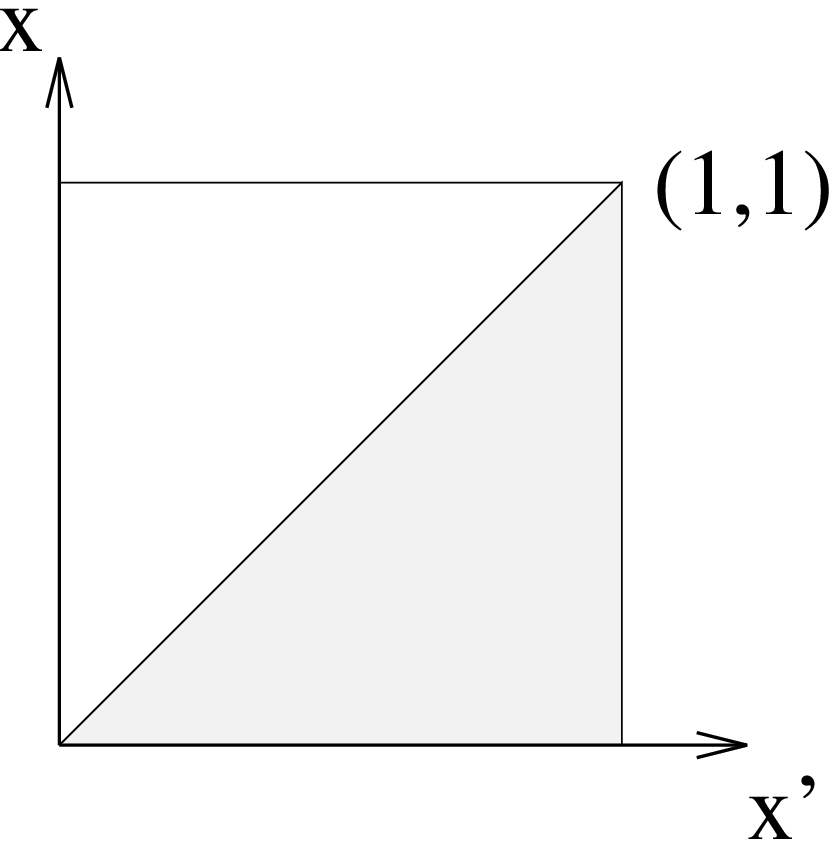} }
\end{center}
\caption{Integration Region and Change of Variables}
{This figure illustrates the change of variables and integration limits 
in section \ref{sec:DGLAP}, where we show the equivalence between the 
DGLAP equation and the RGE from OPE analysis. The shaded area is the 
integration region.
}
\label{fig:x-x}
\end{figure}

\nid where we have also changed the integration limits appropriately
(see figure \ref{fig:x-x} for illustration). Rewrite $x^n$ as 
$(x/x')^n (x')^n$ we have
\beqs
rhs &=& \frac {\al (Q^2)} {2 \pi} \int_0^1 dx' 
\int_0^1 d( \frac {x}{x'}) x' P^f(x',Q^2)  (x')^n (\frac {x}{x'})^n
\ga^f ({x \over {x'}}) \non \\
&=& \frac {\al (Q^2)} {2 \pi} \int_0^1 dx' x'^n x' P^f(x',Q^2)
\int_0^1 d( \frac {x}{x'}) (\frac {x}{x'})^n \ga^f ({x \over {x'}})
\non \\
&\equiv& \frac {\al (Q^2)} {2 \pi} \int_0^1 dx \, x^n \,x P^f(x,Q^2)
\ga^f_n \,,
\eeqs

\nid where we have defined the $n$-moment of the splitting function as
\beq
\ga^f_n \equiv \int_0^1 dx \, x^n \ga^f(x) \,.
\eeq

\nid By the proper identification between the $n(+1)$th moment of the
splitting function and the anomalous dimension in leading order as
\beq
\ga^f_n(\al(Q^2)) = \frac {\al (Q^2)} {2 \pi} \ga_n^f \,,
\eeq

\nid and noting that at the normalization scale $\mu^2 = Q^2$ we have
$\al_\mu = \al(Q^2)$ and thus 
\beq
Q^2 \frac {d} {d Q^2} \equiv Q^2 \d/d{Q^2} + \be \d/d{\al}
\eeq

\nid we have indeed recovered equation \ref{eq:fw-pdfrge}.

\subsection{The Anomalous Dimensions and Splitting Functions 
\label{sec:fw-gamma}}

\begin{figure}[ptb]
\begin{center}
\epsfxsize=5.5in
\epsfysize=3in
\leavevmode
\hbox{ \epsffile{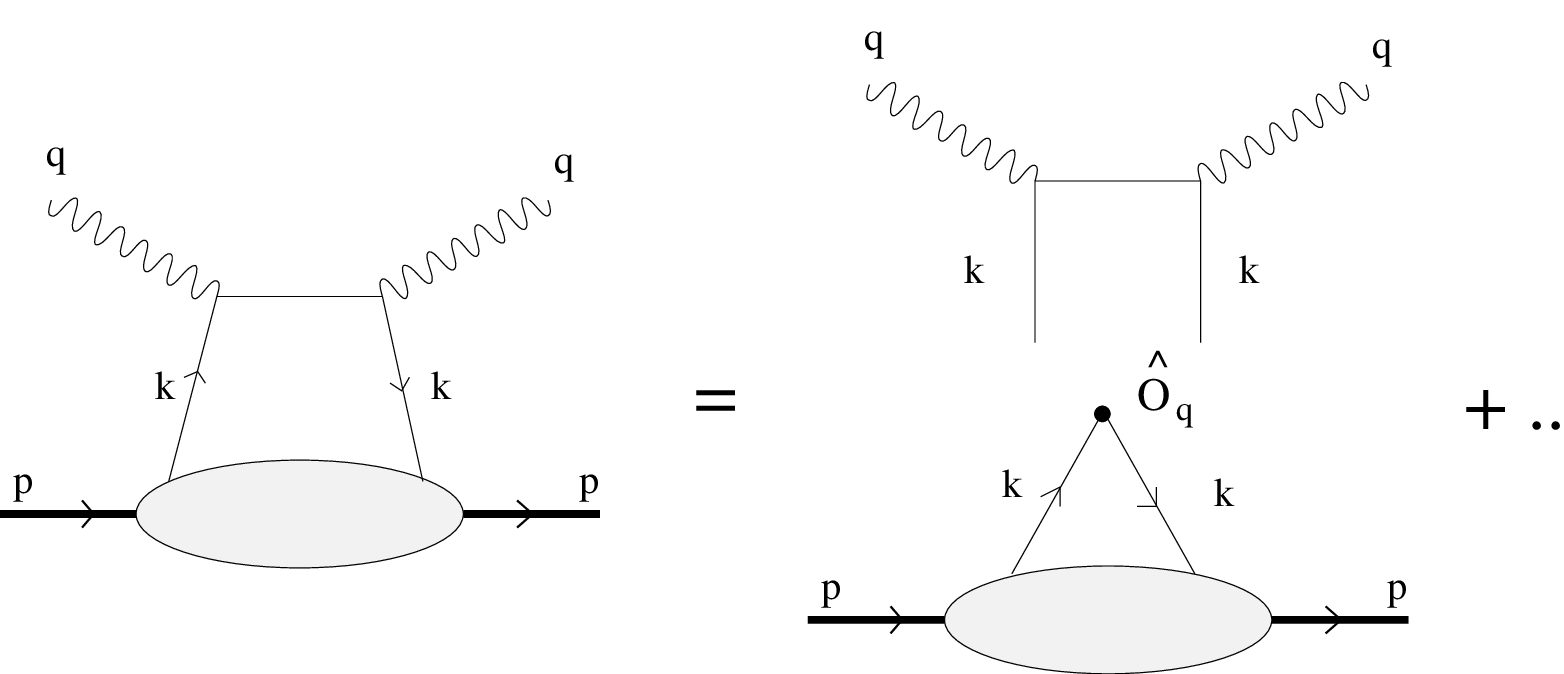} }
\end{center}
\caption{Forward Operator Product Expansion}
{The operator product of the forward amplitude to the lowest order. 
The upper part of the righ hand side is the Wilson coefficient. The 
lower part is the reduced matrix element. We have only shown the
leading quark-quark transition. 
}
\label{fig:fw-ope}
\end{figure}

In terms of diagrams, the operator product expansion of the
DIS forward amplitude into the product of reduced matrix elements and
Wilson coefficients is shown in figure \ref{fig:fw-ope}. If we put a
cut through both sides, it becomes the OPE of the structure function
into Wilson coefficient and parton distribution function. Only the
lowest order contribution is shown where the connecting part on the
right hand side is a quark operator.

\nid The parton distribution on the bottom evolves in the momentum
scale via an anomalous dimension until it reaches the scale of the
hard scattering $Q^2$. We will continue the computation and discussion
in the light cone (LC) gauge.

\subsubsection{Review of the Light Cone Gauge}

We take the convention that the $0$ component is the time component of
a four vector, and the $1,2,3$ components are the (spacial) $x, y, z$
components, respectively.

For any four vector $v$, we define the light cone components
\beq
v_{\pm} \equiv {1 \over \sqrt{2}} (v_0 \pm v_3) , \,\,\,\, 
\perb{v} \equiv \left(
\begin{array}{cc}
v_1 \\
v_2 
\end{array} \right) .
\eeq

\nid This means 
\beq
v^2 \equiv v \cdot v = 2 v_+ v_- - \perb{v}^2,
\eeq

\nid while for any two $4$-vectors $v_1$ and $v_2$,
\beq
v_1 \cdot v_2 = v_{1+}v_{2-} + v_{1-}v_{2+} - \perb{v}_1 \cdot \perb{v}_2.
\eeq

\nid We can then define the LC metric $g_{\al \be}$ by
\beqs
g_{+-} = g_{-+} = 1, \,\,\,\,\, 
g_{++} = g_{--} = 0, \,\,\,\,\, g_{\bot \bot} = -1 , \non \\
g_{+ \bot} = g_{\bot +} = g_{- \bot} = g_{\bot -} = 0 .\non
\eeqs

\nid such that for $\al, \be = +, -, \bot$
\beq 
v_1 \cdot v_2 = \sum_{\al, \be} g_{\al \be}
v_{1\al}v_{2\be}.
\eeq

The Dirac matrices $\ga_{\al}$ with $\al = 0, 1, 2, 3$ can be
viewed as a four vector and we can similarly define
\beq
\ga_{\pm} \equiv {1 \over \sqrt{2}} (\ga_0 \pm \ga_3), \,\,\,\, 
\perb{\ga} \equiv \left(
\begin{array}{cc}
\ga_1 \\
\ga_2 
\end{array} \right) .
\eeq

\nid It can be easily verified that the light cone components obey
the anti-commutation relationship
\beq
\label{eq:anticommudef}
\{\ga_{\al},\ga_{\be}\} = 2 g_{\al \be} \,.
\eeq

\nid In particular, we have
\beqs
\label{eq:gammasquare}
\ga_+^2 = \ga_-^2 = 0 \non \\
\ga_+ \ga_- + \ga_- \ga_+ = 2  \\
\ga_\pm \ga_{1,2} + \ga_{1,2} \ga_\pm = 0 \non
\eeqs

\nid and therefore,
\beqs
\label{eq:anticommu}
\ga_+ \ga_- \ga_+ = 2 \ga_+ \non \\
\ga_+ \! \not{\!v} \ga_+ = 2 v_+ \ga_+  
\eeqs

Introducing a vector $n_\mu$ such that for any $4$-vector $v$
\beq
n \cdot v = v_+ \,.
\eeq

\nid This means
\beq
\label{eq:n-def}
n_- = 1 \,, \;\;\;\;\; n_+ = n_\bot = 0 \,, \;\;\;\;\;
\& \;\, n^2 = n_\mu n^\mu = 0 \,.
\eeq

\nid $n_\mu$ is the so called LC null vector.

The light cone (LC) gauge is defined by requiring for the gauge
field $A$,
\beq
\label{eq:LC-def}
n \cdot A = A_+ = 0 \,.
\eeq
 
\nid The gluon propagator in the light cone gauge is 
\beq
\label{eq:LC-gluon}
\cd_\mn(k) = \frac {-i} {k^2 + i \eps} \left( g_\mn - { {n_\mu k_\nu + k_\mu
n_\nu} \over {n \cdot k}} \right)  
\equiv \frac {-i} {k^2 + i \eps}  D_\mn(k) \,,
\eeq

\nid where $D_\mn(k)$ is the so-called light cone gluon projector. We have
\beq
n_\mu \cd^\mn = n_\nu \cd^\mn = 0 \,.
\eeq

\nid The light cone gauge is also a {\sl ghostless} gauge.

\subsubsection{The Quark-Quark Anomalous Dimension}

\begin{figure}[ptb]
\begin{center}
\epsfxsize=5.5in
\epsfysize=2.4in
\leavevmode
\hbox{ \epsffile{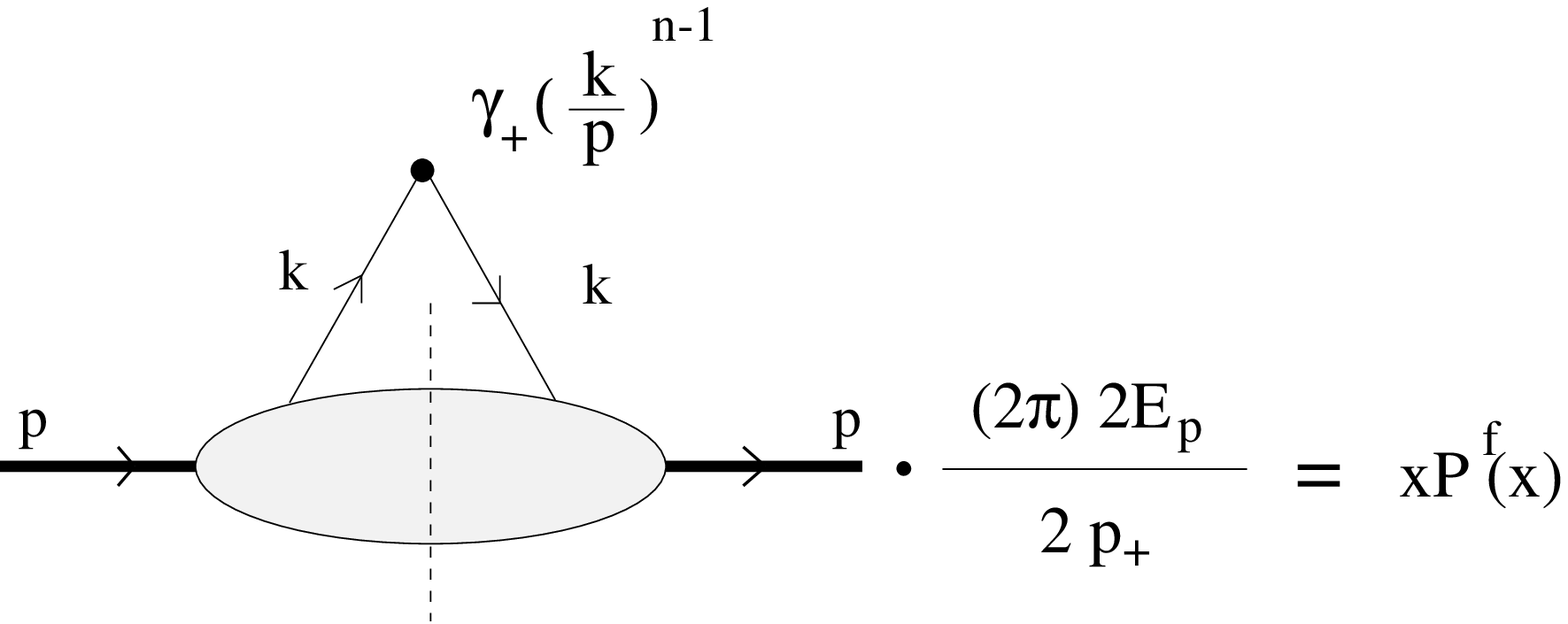} }
\end{center}
\caption{The Lowest Order Quark Distribution Function}
{This is the quark distribution function $P^f(x)$ where we have put in 
explicitly the lowest order quark vertex in the light cone gauge. The 
dotted line is the cut that puts all the intermediate lines it passes 
through on the mass shell.
}
\label{fig:fw-pdf}
\end{figure}

The diagram of the quark distribution function
$x P^f(x)$ is shown in figure \ref{fig:fw-pdf}, where we take
the lowest order quark-quark vertex function to be (\cite{G8070})
\beq
\label{eq:fw-ga0}
\Ga_+^{(0)} = \ga_+ \left( \frac {k_+}{p_+} \right)^{n\!-\!1} \,.
\eeq

\begin{figure}[ptb]
\begin{center}
\epsfxsize=5.5in
\epsfysize=3in
\leavevmode
\hbox{ \epsffile{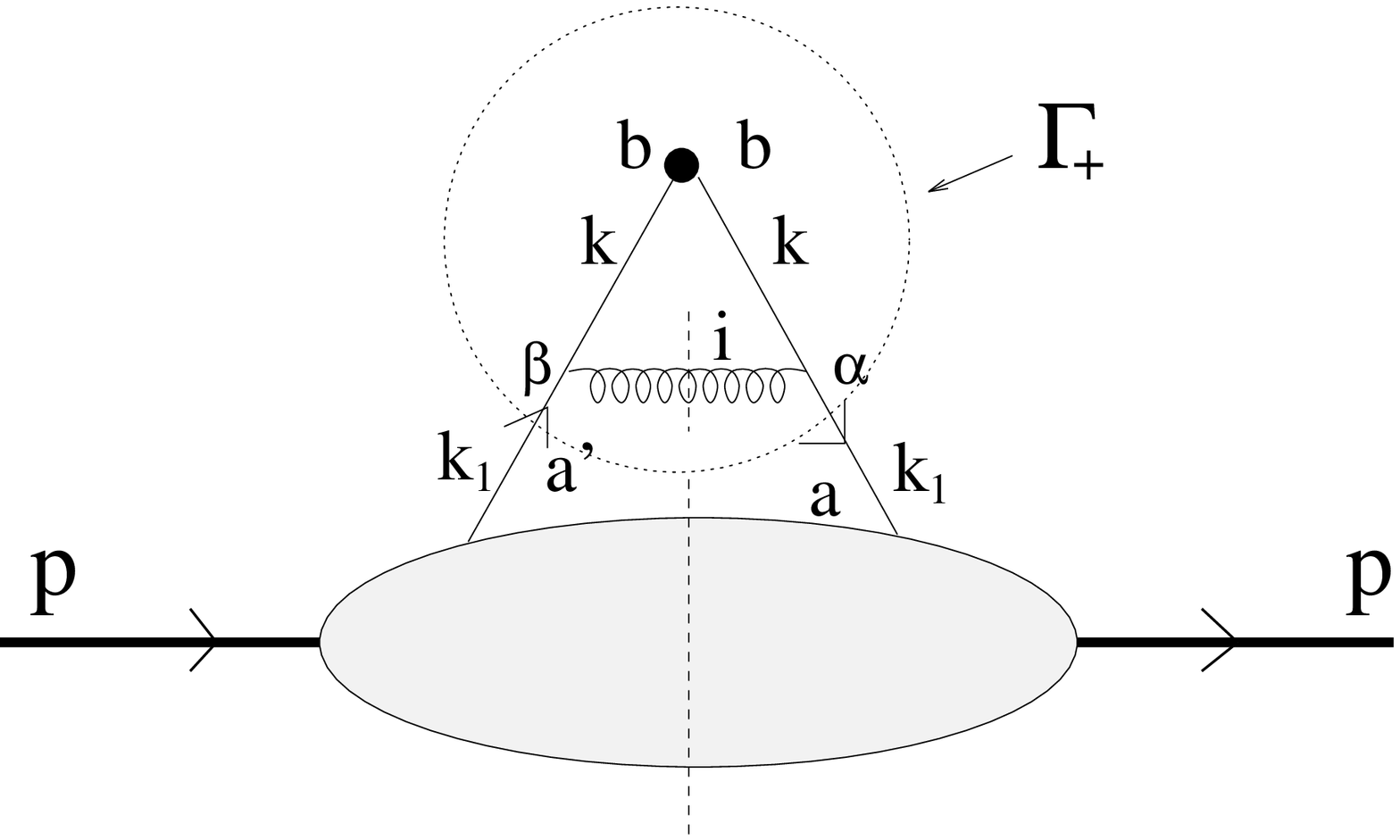} }
\end{center}
\caption{The Lowest Order Quark-Quark Diagram}
{This is the quark-quark diagram of the first order radiative correction to
the quark distribution function. $a,b$ are color indices and $i$ 
labels the on-shell gluon. 
}
\label{fig:fw-qq}
\end{figure}

Let us look at the first radiative corrections. The quark-quark
diagram is shown in figure \ref{fig:fw-qq}, where we have an extra
on-shell gluon line. 

\nid Using standard Feynman rules we can write the value of part of
the diagram within the dotted circle as
\beq
\label{eq:fw-ga1-def}
\Ga_+^{(1)} = C \int { {d^4 k} \over { (2 \pi)^4}}
\ga_\al {i \over \not{\!k}} \ga_+ 
\left( \frac {k_+}{p_+} \right)^{n\!-\!1} \!\!\!
(-2 \pi) \de^+[(k_1-k)^2] {i \over \not{\!k}} 
\ga_\be D^{\al \be} (k_1 -k) \,.
\eeq

\nid In the above, $C$ is the color factor and
\beq
C = (ig)^2 \sum_{i,b} (T^i_{ab}T^i_{ba'}) \,,
\eeq

\nid where $T^i_{ab} = \left(\frac {\lm^i} {2} \right)_{ab}$ is the
generator of the $SU(3)$ color group. Since we have \cite{Group}
\beq
\sum_i (T^i)^2 = C_F I \,,
\eeq

\nid where $C_F$ is the Casmir operator of the fundamental
representation of the gauge group and for $SU(N)$ 
\beq
C_F = \frac {N^2 -1} {2N} \,.
\eeq

\nid Defining the strong coupling constant $\al_s \equiv \al$ by 
$g^2 = 4 \pi \al$, we have
\beq
C = (ig)^2 C_F \de_{a'a} = - 4 \pi \al C_F \,,
\eeq

\nid where in the last step we have set $a = a'$. 

\nid In arriving at equation \ref{eq:fw-ga1-def} we have also used the
transition from the usual gluon propagator \ref{eq:LC-gluon} to the
on-shell gluon by
\beq
\frac {-i} {k^2 + i \eps} \to (-2 \pi) \de^+ [k^2] \,,
\eeq

\nid which is consistent with the optical theorem \ref{eq:fw-optical}.

\nid Therefore we obtain
\beq
\label{eq:fw-ga1}
\Ga_+^{(1)} = \frac {\al C_F} {2 \pi^2} 
\int \frac {d^4k \de^+[(k_1-k)^2]} { (k^2 + i \eps)^2}
\left( \frac {k_+}{p_+} \right)^{n\!-\!1} \!\!\!
\ga_\al \not{\!k} \ga_+ \not{\!k} \ga_\be D^{\al \be} (k_1 -k) \,.
\eeq

\nid Note in the light cone gauge, there is only one solution to 
$\de (k^2)$ because of the linearization of the light cone variable. 
We write the integral over $d^4k$ in light cone variables
\beq
d^4k = d k_0 d k_1 dk_2 dk_3 \equiv dk_+ dk_- d^2 \perb{k} \,,
\eeq

\nid and integrate over $k_-$ with the $\de$-function to get
\beq
d k_- \de [(k_1\!-\!k)^2] 
= d k_- \de [2 (k_1\!-\!k)_+(k_1 \!-\!k)_- \!-\! (\perb{k}_1 \!-\! \perb{k})^2]
= {1 \over {2 (k_1 \!-\! k)_+} } \,.
\eeq

\nid Define three momentum ratio variables
\beq
x = \frac {k_+}{p_+} \,, \;\;\;\;\;
x_1 = \frac {k_{1+}} {p_+} \,, \;\;\;\;\;
\om = \frac {k_+} {k_{1+}} = {x \over x_1} \,,
\eeq

\nid we evaluate $k^2$ with $(k_1 -k)^2 = 0$ and get
\beqs
k^2 &=& 2 k_+ k_- - \perb{k}^2 \non \\
&=& 2 k_+ ( k_{1-} - \frac {(\perb{k}_1 \!-\! \perb{k})^2}
{2 (k_1 \!-\! k)_+} ) - \perb{k}^2 \non \\
&=& \frac {k_+}{k_{1+}} (k_1^2 + \perb{k}_1^2) - \perb{k}^2 
- \frac {k_+}{(k_1 \!-\! k)_+} (\perb{k}_1 \!-\! \perb{k})^2 \,,
\eeqs

\nid where in the last step we have used
\beq
2 k_+ k_{1-} = \frac {k_+}{k_{1+}} 2 k_{1+}k_{1-} 
= \frac {k_+}{k_{1+}} (k_1^2 + \perb{k}_1^2) \,.
\eeq

\nid Expressing things in terms of $\om$ we have
\beqs
k^2 &=& \om k_1^2 + \om \perb{k}_1^2 - \perb{k}^2 
- \frac {\om} {1 \!-\! \om} (\perb{k}_1 \!-\! \perb{k})^2 \non \\
&=& - {1 \over {1 \!-\! \om}} [ -\om (1 \!-\! \om)k_1^2 
+ \om (\perb{k}_1 \!-\! \perb{k})^2 - \om (1 \!-\! \om) \perb{k}_1^2
+ (1 \!-\! \om) \perb{k}^2 \non \\
&=& - {1 \over {1 \!-\! \om}} [ ( \perb{k} - \om \perb{k}_1)^2
- \om (1 \!-\! \om)k_1^2] \,.
\eeqs

Our goal is not to calculate $\Ga_+$ completely, but only to calculate
the ultraviolet divergent part of it. In evaluating the numerator of 
 \ref{eq:fw-ga1} we keep only the terms of the highest $\perb{k}$
power, which turns out to be quadratic. Note because of the
$\de$-function, $(k_1 -k)_-$ also has a $\perb{k}^2$ contribution. 
The denominator is simple after we take the leading contribution, that
is,
\beq
k^2 = - \frac {\perb{k}^2} {1 \!-\! \om} \,.
\eeq

\nid As for the numerator, we now need
\beqs
\ga^\al \!\not{\!k} \ga_+ \!\not{\!k} \ga^\be D_{\al \be} &=&
\ga^\al (\{\not{\!k}, \ga_+ \} - \ga_+ \!\not{\!k} ) 
\not{\!k} \ga^\be D_{\al \be} \non \\
&=& 2 k_+ \ga^\al \! \not{\!k} \ga^\be D_{\al \be} 
- k^2 \ga^\al \ga_+ \ga^\be D_{\al \be} \,.
\eeqs

\nid The second term of the above is
\beqs
- k^2 \ga^\al \ga_+ \ga^\be \left( g_{\al \be}
- { {n_\al (k_1 \!-\!k)_\be + (k_1\!-\!k)_\al n_\be} 
\over {n \cdot (k_1 - k)}} \right) &=&
- k^2 \ga^\al \ga_+ \ga_\al + 0 \non \\
&=& 2 k^2 \ga_+ = - 2 \ga_+ \frac {\perb{k}^2} {1 \!-\! \om} \,, 
\;\;\;\;\;\;\;\;\;
\eeqs

\nid where in the first step we used the facts that $n \cdot \ga =
\ga_+$ and $\ga_+^2 = 0$ while in the last step we used the
anti-commuting relation \ref{eq:anticommudef}. The first term of
the numerator, on the other hand, can be expanded to be
\beq
2k_+ \{ \ga_\al \!\not{\!k} \ga^\al 
- { 1 \over { (k_1 \!-\!k)_+}} [ \ga_+ \!\not{\!k} 
(\not{\!k}_1-\!\not{\!k}) + (\not{\!k}_1-\!\not{\!k}) \!\not{\!k} 
\ga_+]\} \,.
\eeq

\nid We have from the $\de$-function
\beq
k_- \sim \frac {\perb{k}^2} {2 (k_1-k)_+} \,,
\eeq

\nid and thus
\beq
\ga_\al \!\not{\!k} \ga^\al = - 2 \!\not{\!k} = -2 \ga_+ k_-
= \ga_+ \frac {\perb{k}^2} { (k_1 \!-\!k)_+} \,.
\eeq

\nid At the same time,
\beqs
\ga_+ \!\not{\!k}(\not{\!k}_1-\!\not{\!k}) &=& 
- \ga_+ \!\not{\!k} \!\not{\!k} = - \ga_+ k^2 \non \\
&=& \ga_+ \frac {\perb{k}^2} {1 \!-\! \om} \,,
\eeqs

\nid while similarly
\beq
(\not{\!k}_1-\!\not{\!k}) \!\not{\!k} \ga_+ = 
\ga_+ \frac {\perb{k}^2} {1 \!-\! \om} \,.
\eeq

\nid Therefore the first term of the numerator is actually equal to
\beq
2 \ga \perb{k}^2 \frac {\om}{1 \!-\! \om}
\{ 1- \frac {2} {1 \!-\! \om} \} 
= - 2 \ga \perb{k}^2 \frac {\om}{1 \!-\! \om}
\frac {1 \!+\! \om} {1 \!-\! \om} \,.
\eeq

\nid Combined with the second term, we find that the divergent
contribution to the numerator is
\beq
-2 \ga_+ \frac {\perb{k}^2} {(1 \!-\! \om)^2}
[ \om (1 \!+\! \om) + (1 \!-\! \om)] 
= -2 \ga_+ \perb{k}^2 \frac {1 \!+\! \om^2} {(1 \!-\! \om)^2} \,.
\eeq

Rewrite the $\perb{k}$ integral as
\beq
\int d^2 \perb{k} = 2 \pi \int \perb{k} d\perb{k} = \pi \int d
\perb{k}^2 \,,
\eeq

\nid we arrive at the expression of the divergent part of
$\Ga_+^{(1)}$ as
\beqs
\Ga_+^{(1)} &=& \frac {\al C_F} {2 \pi} 
\int \frac {d k_+} {2 (k_1-k)_+} 
\left( \frac {k_+}{p_+} \right)^{n\!-\!1} 
\int \frac {d \perb{k}^2} {\perb{k}^2} \ga_+  
\left( \frac {k_+}{k_{1+}} \right)^{n\!-\!1}
\left( \frac {k_{1+}}{p_+} \right)^{n\!-\!1} 
2 (1 + \om^2) \non \\
&=& \ga_+ \left( \frac {k_{1+}}{p_+} \right)^{n\!-\!1} 
 \frac {\al C_F} {2 \pi} \int \frac {d \perb{k}^2} {\perb{k}^2}
\int d \om \, \om^{n\!-\!1} \frac {1 \!+\! \om^2} {1 \!-\! \om} \,.
\eeqs

Our original graph, one that is figure \ref{fig:fw-amp}(b) plus an
on-shell gluon running from the incoming quark to the outgoing quark,
is not divergent in $\perb{k}^2$. The divergence we encountered arises
because we let $Q^2$ become very large for fixed $\perb{k}^2$. This
means that the cutoff for the logarithmic $\perb{k}^2$ integral is at
$Q^2$.  That is, explicitly,
\beq
\Ga_+^{(1)} = \ga_+ \left( \frac {k_{1+}}{p_+} \right)^{n\!-\!1}
\frac {\al C_F} {2 \pi} \int^{Q^2} \frac {d \perb{k}^2} {\perb{k}^2}
\int_0^1 d \om \, \om^{n\!-\!1} \frac {1 \!+\! \om^2} {1 \!-\! \om}
\,,
\eeq

\nid where we have also explicitly written out the limit on the $\om$
integration. 

Taking the logarithmic derivative with respect to $Q^2$, and
recall equation \ref{eq:fw-ga0}, we obviously have
\beq
Q^2 {d \over {d Q^2}} \Ga_+^{(1)}(x,Q^2) 
= \frac {\al C_F} {2 \pi} 
\int_0^1 d \om \, \om^{n\!-\!1} \frac {1 \!+\! \om^2} {1 \!-\! \om}
\Ga_+^{(0)}(x, Q^2) \,.
\eeq

\nid Remember the graphic definition of the parton distribution
function and $\Ga_+^{(0),(1)}$ as the lowest and first order factors,
we realize that we indeed have
\beq
Q^2 {d \over {d Q^2}} \int_0^1 x^{n\!-\!1} x P^f(x, Q^2) 
= \frac {\al C_F} {2 \pi} \int^1_0 x^{n\!-\!1} x P^f(x, Q^2) 
\int_0^1 d \om \, \om^{n\!-\!1} \frac {1 \!+\! \om^2} {1 \!-\! \om} \,.
\eeq

\nid By exactly reversing the argument in section \ref{sec:DGLAP}
(with $x'$ now being $x_1$), it is straight forward to show that we 
indeed have the DGLAP evolution equation for the quark distribution 
function
\beq
\label{eq:fw-DGLAPqq}
Q^2 {d \over {d Q^2}} x P^f(x,Q^2) = \frac {\al C_F} {2 \pi}
\int_x^1 \frac {d x_1}{x_1} \ga ( {x \over x_1} )
x P^f (x_1, Q^2) \,,
\eeq

\nid where the Altarelli-Parisi splitting function is
\beq
\label{eq:fw-spltqq1}
\ga(y) = \frac {1 + y^2} {1 - y} \,.
\eeq

\begin{figure}[ptb]
\begin{center}
\epsfxsize=5.5in
\epsfysize=1.5in
\leavevmode
\hbox{ \epsffile{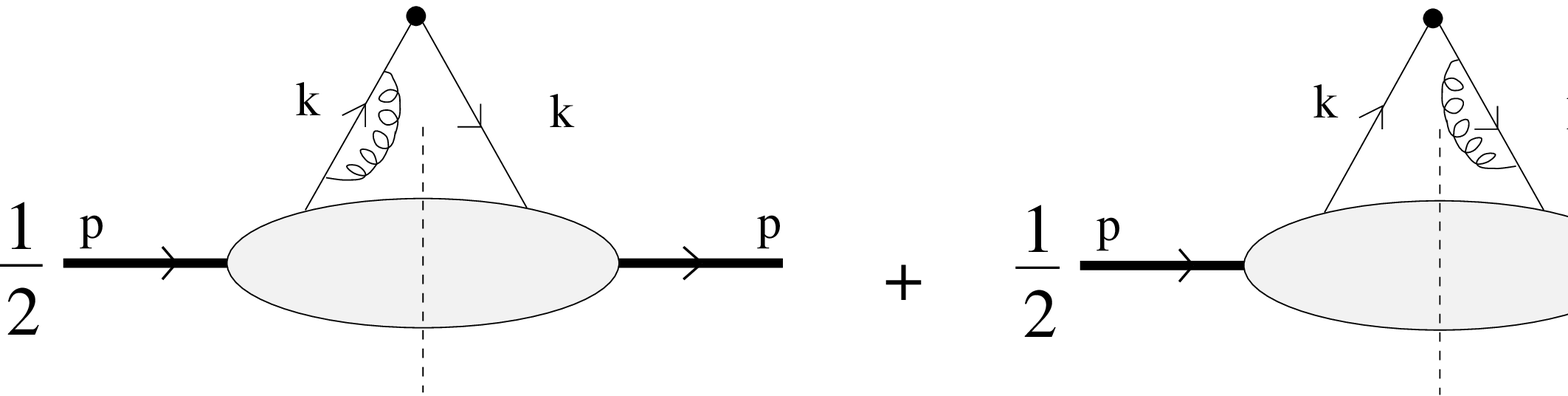} }
\end{center}
\caption{Quark Self Energy Diagram}
{These are the two quark self energy diagrams that are needed to be 
added to figure~\ref{fig:fw-qq} to cancel infrared divergence at the
end points of integration.
}
\label{fig:fw-selfE}
\end{figure}

The integral in equation \ref{eq:fw-DGLAPqq} is actually divergent at
the end point when $x_1 \to x$. This is an infrared divergence due to
emission of soft gluons with $(k_1 - k)_+ \to 0$. This divergence
should not be present in physical processes and indeed it is removed
once we add the quark self-energy graphs shown in figure 
 \ref{fig:fw-selfE}.

\nid After virtual corrections are included, the full quark-quark
splitting function that is free of singularities is 
\beq
\label{eq:fw-qqfinal}
\ga(x) = C_F [ \frac {1 + x^2} {(1 -x)_+} + {3 \over 2} \de (1-x)] \,,
\eeq

\nid where the special function $(1-x)_+^{-1}$ is defined by
\beq
\label{eq:fw-plusfcn}
\int_0^1 \frac {f(x) dx} {(1 -x)_+} 
= \int_0^1 dx  \frac {f(x) - f(1)} {(1 -x)} 
\eeq

\nid for any function $f(x)$ that has reasonable behavior.

\subsubsection{The Mixing of Evolution}

\begin{figure}[ptb]
\begin{center}
\epsfxsize=5.5in
\epsfysize=3in
\leavevmode
\hbox{ \epsffile{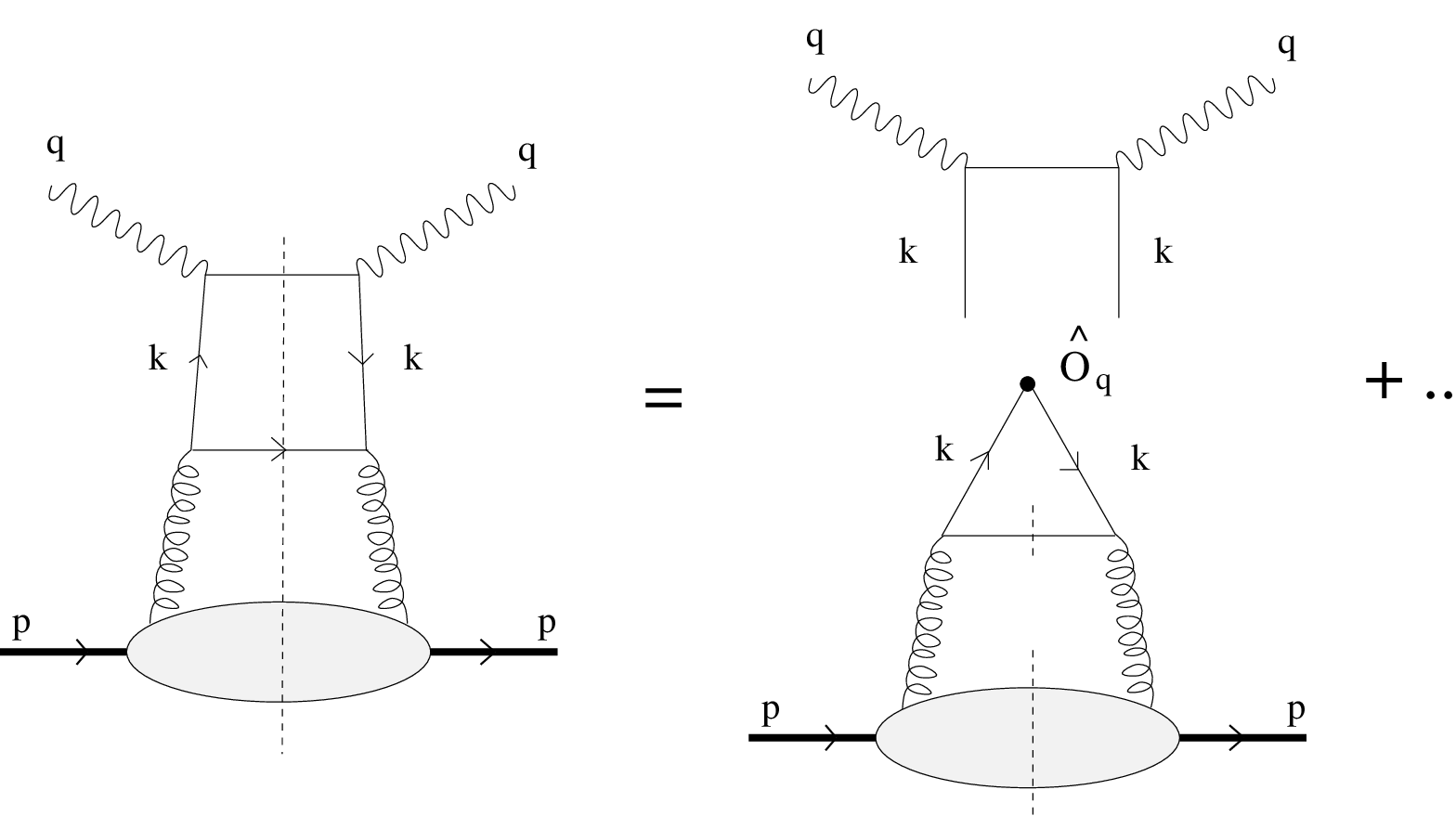} }
\end{center}
\caption{The Quark-Gluon Mixing Graph}
{This is the graph that generates the anomalous dimension for quark to
gluon mixing in forward DIS. 
}
\label{fig:fw-qg}
\end{figure}

There are additional order $\al$ terms in the $Q^2$-evolution equation
due to operator mixing, for example, see figure \ref{fig:fw-qg}. 
We thus have the necessity of considering both quark and gluon distributions.

To organize, define the flavor singlet quark distribution as
\beq
\label{eq:fw-singlet}
\Sigma (x,Q^2) = x \sum_f [ P^f (x, Q^2) + P^{\bar{f}} (x, Q^2)] \,,
\eeq

\nid while the flavor octet distribution is
\beq
\label{eq:fw-octet}
\Delta^{ff'}(x, Q^2) = x [ P^f (x, Q^2) -  P^{f'}(x, Q^2)] \,,
\eeq

\nid and a similarly defined function $\Delta^{\bar{ff'}}$ for
anti-quarks. 

\nid The octet distributions obey a generic DGLAP evolution equation
\beq
\label{eq:fw-DGLAPoct}
Q^2 {d \over {d Q^2}} \Delta^{ff'}(x, Q^2)
= \frac {\al(Q^2)} {2 \pi} \int_x^1 \frac {d x_1}{x_1} 
\ga_{qq'} \left( {x \over x_1} \right) \Delta^{ff'}(x_1, Q^2) 
\;+\; O(\al^2) \,.
\eeq

\nid However, the singlet distribution will mix under renormalization
with the gluon distribution $G (x, Q^2)$. The DGLAP equation has a
matrix form and explicitly
\beq
\label{eq:fw-DGLAPsin}
Q^2 {d \over {d Q^2}} \left(
\begin{array}{c}
\Sigma (x,Q^2) \\
G (x, Q^2)
\end{array} \right)
= \frac {\al(Q^2)}{2 \pi}
\int_x^1 \frac {d x_1}{x_1}
\left(
\begin{array}{cc}
\ga_{qq'} (x/x_1) & \ga_{qg} (x/x_1) \\
\ga_{gq} (x/x_1) & \ga_{gg} (x/x_1)
\end{array} \right)
\left(
\begin{array}{c}
\Sigma (x_1,Q^2) \\
G (x_1, Q^2)
\end{array} \right)
\;.
\eeq

\nid Let $C_F$ again be the Casmir operator in the fundamental
representation with $C_F = 4/3$ for $SU(3)$ color group, and let $C_A$
be the Casmir operator in the adjoint representation with $C_A = N_c =
3$ for the color group, we have the explicit expression of the four
splitting functions as (eg, see \cite{Peskin})
\beqs
\label{eq:fw-spltfcnfinal}
\ga_{qq}(x) \!&=&\! C_F \, \left[ \frac {1+x^2}{(1-x)_+} + 
{3 \over2} \de (1-x)\right] \\
\ga_{gq}(x) \!&=&\! C_F \; \frac { 1 + (1-x)^2 } {x} \\
\ga_{qg}(x) \!&=&\! \half \, \left[ x^2 + (1-x)^2 \right] \\
\ga_{gg}(x) \!&=&\! 2C_A \, \left[ \frac {x}{(1\!-\!x)_+} 
+ \frac {1\!-\!x}{x} + x(1\!-\!x) \right] 
+ \frac {11 C_A \!-\! 2 n_f}{6} \, \de (1\!-\!x)
\eeqs

\section{High Energy Behavior}

It is more convenient to discuss the high energy behavior of the
parton distribution functions when we go to the $n$-moment space by
taking the $n$-th moment of the DGLAP evolution equation, or
equivalently directly using the renormalization group equation from
operator product expansion. 

\subsection{Anomalous Dimension Matrix}

\nid To the leading order in $\al$ we obtain, after taking the moments
of the splitting functions, the anomalous dimension matrix for the
singlet-gluon mixing to be
\beq
\label{eq:fw-gamatrix}
\ga_n (\al (Q^2)) = \frac {\al(Q^2)} {2 \pi}
\left(
\begin{array}{cc}
\ga_n^{qq} & \ga_n^{qg} \\
\ga_n^{gq} & \ga_n^{gg}
\end{array}
\right)
\eeq

\nid where $\ga_n$ is the $n$-th moment of the corresponding $\ga(x)$ 
in equation \ref{eq:fw-spltfcnfinal} and explicitly
\beqs
\label{eq:fw-nmoments}
\ga_n^{qq} &=& -{2 \over 3} \left[1 + 4 \sum_2^n {1 \over j} 
- {2 \over {n(n+1)}} \right] \\
\ga_n^{qg} &=& 2 \frac {n^2+n+2}{n (n+1)(n+2)} \\
\ga_n^{gq} &=& {8 \over 3} \frac {n^2+n+2} {n (n^2-1)} \\
\ga_n^{gg} &=& - 3 \left[ {1 \over 3} + {2 \over 9} \,n_f 
+ 4 \sum_2^n {1 \over j} - \frac {4}{n(n-1)}
- \frac {4} {(n+1)(n+2)} \right] \,.
\eeqs

\subsection{Dominant Moment Contribution}

The anomalous dimension matrix elements are usually singular at some
values of the moment index $n$ (see equation
 \ref{eq:fw-nmoments}). In the high energy limit, the dominant
contribution at small $x$ is from the right most pole of the anomalous
dimension. We will illustrate this by looking again at the DIS
structure function $\nu \W_2$.

\nid Adopting the common terminology let us now call $\nu \W_2$
structure function $F_2$. The $n$-th moment of $F_2$ is defined as
\beq
\label{eq:fw-F2momdef}
F_2^{(n)}(Q^2) = \int_0^1 dx \, x^{n\!-\!1} F_2 (x, Q^2) \,.
\eeq

\nid As we already established $F_2^{(n)}$ can be factorized as the
product of parton distribution functions and perturbatively calculable
Wilson coefficients with well defined $Q^2$ evolution. To predict
cross-section, however, we do need directly $F_2(x,Q^2)$ itself. We
can reconstruct $F_2$ from its moments by a Mellin transformation
\beq
\label{eq:fw-mellin}
F_2(x,Q^2) = \int_c \frac {dn}{2 \pi i} e^{n \log {1 \over x}}
F_2^{(n)}(Q^2) \,,
\eeq

\nid where $c$ is a contour lies in the complex $n$ plane parallel to
the imaginary axis and to the right of all singularities.

Let us first check to see whether this is a self-consistent
definition. Substitute \ref{eq:fw-F2momdef} into equation
 \ref{eq:fw-mellin} we obtain for the right hand side (rhs)
\beqs
rhs &=& \int_c \frac {dn}{2 \pi i} \int_0^1 dx' x'^{n\!-\!1} 
F_2 (x',Q^2) e^{n \log {1 \over x}} \non \\
&=& \int_c \frac {dn}{2 \pi i} \int_0^1 \frac {dx'}{x'}
e^{n \log {1 \over x} - n \log {1 \over {x'}}} F_2 (x',Q^2) \non \\
&=& \int_0^1 \frac {dx'}{x'} F_2 (x',Q^2) \int_c \frac {dn}{2 \pi i}
e^{n [ \log {1 \over x} - \log {1 \over {x'}}]} \non \\
&=& \int_0^1 d \log x' \, F_2 (x',Q^2) \,
\de \left( \log {1 \over x} - \log {1 \over {x'}} \right) \non \\
&=& F_2 (x, Q^2) = lhs \,,
\eeqs

\nid where in the fourth step we have used the mathematical identity
\beq
\int_{L- i \infty}^{L+i \infty} \frac {dn}{2 \pi i} e^{(n-c)z} 
= \de (z) \,.
\eeq

\nid We can always distort the contour $c$ to the left (but not crossing
any poles) so that it'll pick out the residues of the $n$ poles. Each
pole $n_p$ will eventually translate to a power factor of the
Bjorken-$x$ as $(x)^{-n_p}$. Therefore, at high energy and small-$x$
the dominant contribution comes from the right most $n$ pole of the
anomalous dimension on the complex $n$-plane.

\subsection{Double Leading Logarithmic Approximation \label{sec:fw-DLLA}}

To illustrate the point that leading (right most) pole in moment space of the
anomalous dimension dominates in the high energy limit, we will
explicitly compute the high energy behavior of the structure function
$F_2$ of deeply inelastic scattering.

According to previous analysis, the moments of the structure
function obey the momentum evolution equation (see, eg,
 \ref{eq:fw-rge-solu}) 
\beq
F_2^{(n)}(Q^2) = F_2^{(n)}(Q_0^2) e^{\int_{Q_0^2}^{Q^2} 
\frac {d \lm^2}{\lm^2} \ga_n (\al(\lm^2))} \,.
\eeq

\nid Thus the structure function itself is given by
\beq
F_2 (x, Q^2) = \int_c \frac {dn}{2 \pi i} F_2^{(n)}(Q_0^2)
e^{n \log {1 \over x} + \int_{Q_0^2}^{Q^2}  \frac {d \lm^2}{\lm^2} 
\ga_n (\al(\lm^2))} \,.
\eeq

\nid From equation \ref{eq:fw-nmoments} it is obvious that the
gluon-gluon anomalous dimension has the right most pole among others
at $n=1$. Using the lowest order running coupling we can write the
leading contributing term of the anomalous dimension
\beq
\ga_n(\al(\lm^2)) = \frac {C_A}{\pi (n-1)} \frac{1}{b_0 \log
\lm^2/\Lm^2} \,,
\eeq

\nid where $b_0$ is the first coefficient of the QCD $\be$-function. 

\nid By substitution we have
\beqs
F_2 (x, Q^2) &=& \int_c \frac {dn}{2 \pi i} F_2^{(n)}(Q_0^2)
e^{n \log {1 \over x} + \frac {C_A}{\pi b_0}
\log \frac {\log Q^2/\Lm^2} {\log Q_0^2 /\Lm^2} \frac {1} {n\!-\!1}} 
\non \\
&\equiv& \int_c \frac {dn}{2 \pi i} F_2^{(n)}(Q_0^2)
e^{an+ \frac {b}{n-1}} \,,
\eeqs

\nid where we have defined at high energy,
\beqs
a &\equiv& \log {1 \over x} >> 1 \non \\
b &\equiv& \frac {C_A}{\pi b_0} \log 
\frac {\log Q^2/\Lm^2} {\log Q_0^2 /\Lm^2}
\eeqs

The high energy limit corresponds to $n\to 1$ and $a >>1$ so we can
make a saddle point approximation for the exponent function $f(n) = a n +
b {1 \over {n-1}}$. We have
\beq
f'(n) = a - \frac {b} {(n-1)^2} = 0 \Rightarrow
n_0 = 1 + \sqrt{{b \over a}} \,.
\eeq

\nid Thus at the saddle point $n_0$,
\beqs
f(n_0) = a (\sqrt{{b \over a}} +1 ) + {b \over \sqrt{{b \over a}}}
= a + 2 \sqrt{ab} \non \\
f''(n_0) = \frac {2b} {(\sqrt{b/a})^3} = \frac {2 a^{3/2}} 
{b^{1/2}} > 0 \,,
\eeqs

\nid which means $c$ is a fine contour. Let $i\nu = n-1$ we obtain
finally,
\beqs
\label{eq:fw-sad-hie}
F_2 (x, Q^2) &=& F_2^{(n)}(Q_0^2) \int_{-\infty}^{\infty} 
\frac {d \nu} {2 \pi} e^{(a + 2 \sqrt{ab})}
e^{- \frac {2 a^{3/2}} {b^{1/2}} \nu^2} \non \\
&=&  F_2^{(n)}(Q_0^2) e^{n \log {1 \over x}}
e^{2 \sqrt{\log {1 \over x} \frac {C_A}{\pi b_0}
\log \frac {\log Q^2/\Lm^2} {\log Q_0^2 /\Lm^2}}}
\sqrt{ \frac { \pi b^{1/2}} {a^{3/2}}} \non \\
&=&  F_2^{(n)}(Q_0^2) \left( {1 \over x} \right)^1
e^{2 \sqrt{\log {1 \over x} \frac {C_A}{\pi b_0}
\log \frac {\log Q^2/\Lm^2} {\log Q_0^2 /\Lm^2}}}
\sqrt{ \frac { \pi \sqrt{\frac {C_A}{\pi b_0}
\log \frac {\log Q^2/\Lm^2} {\log Q_0^2 /\Lm^2}} }
{(\log {1 \over x})^{3/2}}} \,. \non \\
\eeqs

This is the high energy limit of the forward DIS amplitude/structure
function. The dominate small-$x$ behavior is a power dependence on 
$\left({1 \over x}\right)^{n_0}$ with $n_0$ given by a saddle point 
approximation. It is obvious that the value of $n_0$, which determines
the leading high energy and small-$x$ behavior, is dictated by the
right most pole on the complex $n$ plane. In this forward case, it is
at $n=1$ and thus we arrive at the $x^{-1}$ leading behavior of the
gluon distribution under DLLA.

\chapter{Decomposition of the Non-forward Amplitude}
\thispagestyle{myheadings}
\markright{}

In this chapter, we discuss in detail the tensorial decomposition
of the non-forward Amplitude. First we define the amplitude under
discussion by laying out the kinematics of the process; then we
proceed to use current conservation and the symmetry properties of the
amplitude itself to decompose it into various invariant components and
define the corresponding invariant amplitudes in the non-forward case.

\section{Kinematics \label{sec:kinematics}}

The process we consider is $\ga^* + P \rightarrow \ga^* + P$ (virtual
photon + proton goes to virtual photon + proton) shown in 
fig.~\ref{fig:ampdef} with ${\rm T}_{\mu \nu}$ its amplitude. It is a 
double virtual Compton scattering in the sense that both the incoming 
photon $q'$ and the outgoing photon $q$ have, in general, non-vanishing 
invariant masses. There is a non-zero four-momentum transfer from the 
virtual photon to the target proton which we label as $r$. Thus the 
struck proton $p'$ has a slightly different four-momentum than 
the original proton $p$. And we have
\beq
\label{eq:p'q'def}
p' = p - r \,\,; \;\;\;\; q' = q - r \,\,.
\eeq

\nid Furthermore, the incoming virtual photon $q'$ has a different 
invariant mass than the outgoing virtual photon $q$ with 
\beq
\label{eq:invmasses}
q'^2 = -Q_1^2 , \,\,\,\, q^2 = -Q_2^2 \,.
\eeq 

We restrict our discussion to amplitudes. Cross sections are
obtained from the square of the amplitudes. In this non-forward case there 
is not a simple optical theorem relating the imaginary part of an amplitude
to a cross section \cite{Christ&Mueller72}. Our main interest is in the 
high energy limit, so similar to \cite{BrodskyVM} we take all but one
of the light cone (LC) components of the proton momentum to be zero. We do 
the same for the momentum transfer $r$. We will choose the non-zero 
component of both as the plus component, which means
\beq
p = (p_+, 0, \, \perb{0}) \,, \,\,\,\,\,\,\,\, r = (r_+, 0, \, \perb{0}) \,,
\eeq

\nid and

\beq
\label{eq:rsquare}
p^2 = 0 = r^2 \equiv -t \,.
\eeq

\nid That is, we are in the zero nucleon mass and zero $t$ limit.

The fact that the target remains a proton means that 
\beq
{p'}^2 = 0 = (p-r)^2 = p^2 - 2 p \cdot r + r^2 \,, 
\eeq

\nid and thus
\beq
\label{eq:pdotr}
p \cdot r = 0 \,,
\eeq

\nid which gives
\beq
\label{eq:pdotq'}
p \cdot q' = p \cdot q \,.
\eeq

\nid And since
\beq
{q'}^2 = (q - r)^2 = q^2 - 2 q \cdot r + r^2 = q^2 - 2 q \cdot r \,,
\eeq

\nid which results in
\beq
\label{eq:qdotr}
2 q \cdot r = q^2 - {q'}^2 \,.
\eeq

Together with the fact that $r$ is proportional to the external 
momentum $p$, we have,
\beq
\label{eq:zeta}
r = \zeta p \,, \,\,\,\,\,\,\,\,\,\,\,\, 
\zeta = \frac {\,\,\,\,q^2-q'^2} {2 p \cdot q} \, .  
\eeq

\nid We always suppose $-q^2 = Q_2^2 \leq Q_1^2 = q'^2$ so that $\zeta \ge 0$.

It is worth noting that the process of a double virtual Compton
scattering is not a physical one, however, it provides a general
framework in which one can move the discussion continuously from one
physical limit to another. In particular, DIS corresponds to $q = q'$ 
while DVCS corresponds to $q=0$.

\nid Also worth noting is that the kinematic limit we are discussing, 
namely, $r \neq 0$ but $t = 0$, is not a physical one, since for a
physical non-forward process, $t$ is bounded by 
\beq
\label{eq:tbound}
|t| \geq {{x^2 m^2} \over {1 - x}} \,,
\eeq

\nid where $x$ is the Bjorken $x$ variable and $m$ the target proton
mass. However, our main interest lies in the high energy and small-$x$
behavior of the amplitude and therefore $t=0$ is a good approximation.
In principle we can always discuss finite $t$ behavior by giving $r$ a
transverse component $\perb{r} \neq 0$. There the proportionality
between $r$ and $p$ (see  \ref{eq:zeta}) becomes a proportionality
between the plus components of the two momenta (i.e. $r_+ = \zeta
p_+$)(see, for example, \cite{Radyprd97}).

\section{The Amplitude \label{sec:ampdef}}


The amplitude of the process in fig.~\ref{fig:ampdef} in coordinate space
is a matrix element between two proton states of a time ordered
product of two electromagnetic currents. If we label the space time
point of the incoming interaction as $x+z$ and the outgoing one as
$x$, we have
\beq
{\rm T}_{\mu \nu}^{coord} = \bra{p'} T \, j_{\mu}(x)
j_{\nu} (x+z) \ket{p} \,.
\eeq

\nid As usual we want to Fourier transform the amplitude into the 
momentum space where the kinematics and the dynamics of the process can 
be much more readily discussed. The momentum space amplitude, which we
label as ${\rm T}_{\mu \nu}$, depends on all the relevant independent 
kinematic variables (in the momentum space). In this non-forward case 
there are three kinematic degrees of freedom and sets of variables
such as $(p,p',q)$ or $(p,q,r)$ or $(p, q', r)$ are valid choices and
are all equivalent to each other. We choose the set $(p, q', q)$ for 
convenience of later discussion. Therefore, using the convention that
an incoming momentum $p$ at space time point $x$ enters the Fourier
transformation as an exponential of negative exponent $e^{- i p \cdot x}$ 
and an outgoing momentum as one with a positive exponent, we can write 
explicitly the amplitude of fig.~\ref{fig:ampdef} in momentum space as 
\beqs
{\rm T}_{\mu \nu} &=& {\rm T}_{\mu \nu} (p,q',q) \non \\
&=& i \int d^4 x d^4 z \, e^{-i q' \cdot (x+z) + i q \cdot x} \,
\bra{p'} T \, j_{\mu}(x) j_{\nu} (x+z) \ket{p} \,.
\eeqs                                                                

Integration over $d^4 x$ gives only an overall four dimensional 
momentum space $\de$-function. Ignoring it we can rewrite the
amplitude, up to normalization factors like $(2 \pi)^4$, as
\beq
{\rm T}_{\mu \nu} (p,q',q) = i \int d^4 z \, e^{-i q' \cdot z} \,
\bra{p'} T \, j_{\mu}(0) j_{\nu} (z) \ket{p} \,.
\eeq                                                    

\nid It is obvious that this amplitude should be invariant under an
overall translation in coordinate space, for example, $(z \rightarrow
0 \; \& \; 0 \rightarrow -z)$. Explicitly, by using the translation 
operator $e^{i \hat{p} \cdot x}$ where $\hat{p}$ is the momentum
operator and for any operator in coordinate space $\hat{O}(z)$
(see equations \ref{eq:translation} and \ref{eq:p-hat})
\beq
e^{i \hat{p} \cdot x} \hat{O}(z-x) e^{-i \hat{p} \cdot x}
 = \hat{O}(z) \,,
\eeq

\nid we have
\beqs
\label{eq:z&-z}
{\rm T}_{\mu \nu} (p,q',q) &=& i \int d^4 z \, e^{-i q' \cdot z} \,
\bra{p'} T \, j_{\mu}(0) j_{\nu} (z) \ket{p} \non \\
&=& i \int d^4 z \, e^{-i q' \cdot z} \,
\bra{p'} T \, e^{i \hat{p} \cdot z} j_{\mu}(-z) e^{-i \hat{p} \cdot z}
e^{i \hat{p} \cdot z} j_{\nu} (0) e^{-i \hat{p} \cdot z}  \ket{p} \non
\\
&=& i \int d^4 z \, e^{-i q' \cdot z} e^{i p' \cdot z}
\bra{p'} T \,j_{\mu}(-z) j_{\nu} (0)  \ket{p} e^{-i p \cdot z} \non \\
&=& i \int d^4 z \, e^{-i (q' - p' + p) \cdot z}
\bra{p'} T \,j_{\mu}(-z) j_{\nu} (0)  \ket{p} \non \\
&=& i \int d^4 z \, e^{-i q \cdot z}
\bra{p'} T \,j_{\mu}(-z) j_{\nu} (0)  \ket{p} \, .
\eeqs

\nid Anticipating the operator product expansion analysis we want to
recast the amplitude into the so-called light cone expansion where
$(z \rightarrow \half z \,\,\, \& \,\,\, 0 \rightarrow - \half
z)$. Similar to the above process we have
\beqs
{\rm T}_{\mu \nu} (p,q',q) &=& i \int d^4 z \, e^{-i q' \cdot z} \,
\bra{p'} T \, j_{\mu}(0) j_{\nu} (z) \ket{p} \non \\
&=& i \int d^4 z \, e^{-i q' \cdot z} \,
\bra{p'} T \, e^{i \hat{p} \cdot \half z} j_{\mu}(- \half z) 
e^{-i \hat{p} \cdot \half z} e^{i \hat{p} \cdot \half z} 
j_{\nu} (\half z) e^{-i \hat{p} \cdot \half z}  \ket{p} \non
\\
&=& i \int d^4 z \, e^{-i q' \cdot z} e^{i p' \cdot \half z}
\bra{p'} T \,j_{\mu}(- \half z) j_{\nu} (\half z)  \ket{p} 
e^{-i p \cdot \half z} \non \\
&=& i \int d^4 z \, e^{-i (q' + \half (p - p')) \cdot z}
\bra{p'} T \,j_{\mu}(- \half z) j_{\nu} (\half z)  \ket{p} \non \\
&=& i \int d^4 z \, e^{-i (q' + \half r) \cdot z}
\bra{p'} T \,j_{\mu}(-z) j_{\nu} (0)  \ket{p} \, .
\eeqs   

\nid Therefore, we arrive at
\beq
\label{eq:ampdef}
{\rm T}_{\mu \nu} (p,q',q) = i \int d^4 z \, e^{-i \ovl{q} \cdot z} \, 
\bra{p'} T \, j_{\mu}(-{z \over 2}) j_{\nu} ({z \over 2}) \ket{p}\,
\eeq

\noindent where we have defined 
\beq
\label{eq:qbardef} 
\ovl{q} = q' + \half r = q - \half r = {1 \over 2} (q'+q) \,.
\eeq

\nid Because from (\ref{eq:qdotr}) we have
\beq
\label{eq:qdotq'}
q \cdot q' = q^2 - q \cdot r = \half (q'^2 + q^2) 
= - \half (Q_1^2 + Q_2^2) \,,
\eeq

\nid the square of this newly defined $\ovl{q}$ becomes
\beq
\label{eq:qbarsquare}
\ovl{q}^2 = {1 \over 4} (q'+q)^2 = \half (q'^2 + q^2) = q \cdot q' \,.
\eeq

\nid If we define
\beqs
\label{eq:Qbarsquare}
\ovl{Q}^2 &\equiv& - \ovl{q}^2 = - \half (q'^2 + q^2) \non \\ 
&=& \half (Q_1^2 + Q_2^2) \,,
\eeqs 

\nid we will see in later discussions that $\ovl{Q}^2$ is now the
{\it natural} scale of the scattering process and it characterizes the
hardness of the scattering. 

\section{Current Conservation \label{sec:currentconsv}}

Similar to the case shown in the previous chapter for the forward scattering 
we will use the conservation of the electro$\!-\!$magnetic(e\&m) current
(\ref{eq:e&mcurrentconsv}) to simplify the tensorial decomposition of 
the non-forward amplitude ${\rm T}_{\mu \nu}$. However, because now
the two  e\&m currents carries different momenta, we need to be more
careful in writing down the consequences of the current conservation
in momentum space.

\nid Because
\beq
\label{eq:e&mcurrentconsv}
\d/d{z^\mu} j_\mu(z) = 0 
\eeq

\nid from (\ref{eq:z&-z}) we have
\beqs
0 &=& i \int d^4 z \, e^{-i q' \cdot z} \,
\bra{p'} T \, j_{\mu}(0) \d/d{z^\nu} j_{\nu} (z) \ket{p} \non \\ 
&=& i \int d^4 z \, e^{-i q' \cdot z} \,  \d/d{z^\nu}
\bra{p'} T \, j_{\mu}(0) j_{\nu} (z) \ket{p} \non \\ 
&=& i \int d^4 z \, \d/d{z^\nu} (e^{-i q' \cdot z} \,  
\bra{p'} T \, j_{\mu}(0) j_{\nu} (z) \ket{p} ) \non \\       
&& - i \int d^4 z \, (-iq')_\nu e^{-i q' \cdot z}
\bra{p'} T \, j_{\mu}(0) j_{\nu} (z) \ket{p} ) \non \\ 
&=& 0 + iq'_\nu {\rm T}_{\mu \nu} \,
\eeqs

\nid and
\beqs
0 &=& i \int d^4 z \, e^{-i q \cdot z} \,
\bra{p'} T \, (\d/d{z^\mu} j_{\mu}(-z)) j_{\nu} (0) \ket{p} \non \\
&=& i \int d^4 z \, e^{-i q \cdot z} \,  \d/d{z^\mu}
\bra{p'} T \, j_{\mu}(-z) j_{\nu} (0) \ket{p} \non \\
&=& i \int d^4 z \, \d/d{z^\mu} (e^{-i q \cdot z} \,
\bra{p'} T \, j_{\mu}(0) j_{\nu} (z) \ket{p} ) \non \\
&& - i \int d^4 z \, (-iq)_\nu e^{-i q \cdot z}
\bra{p'} T \, j_{\mu}(-z) j_{\nu} (0) \ket{p} ) \non \\
&=& 0 + iq_\mu {\rm T}_{\mu \nu} \, .
\eeqs 

\nid Thus in momentum space the conservation of the 
electro$\!-\!$magnetic current now requires the amplitude to satisfy
(cf \ref{eq:fw-crtcsv})
\beq
\label{eq:currentconsv}
q^\mu {\rm T}_{\mu \nu} = 0; \,\,\,\,\, q'^\nu {\rm T}_{\mu \nu} = 0. 
\eeq

\section{Decomposition and Invariant Amplitudes}

\begin{figure}[ptb]
\begin{center}
\epsfxsize=5.5in
\epsfysize=3in
\leavevmode
\hbox{ \epsffile{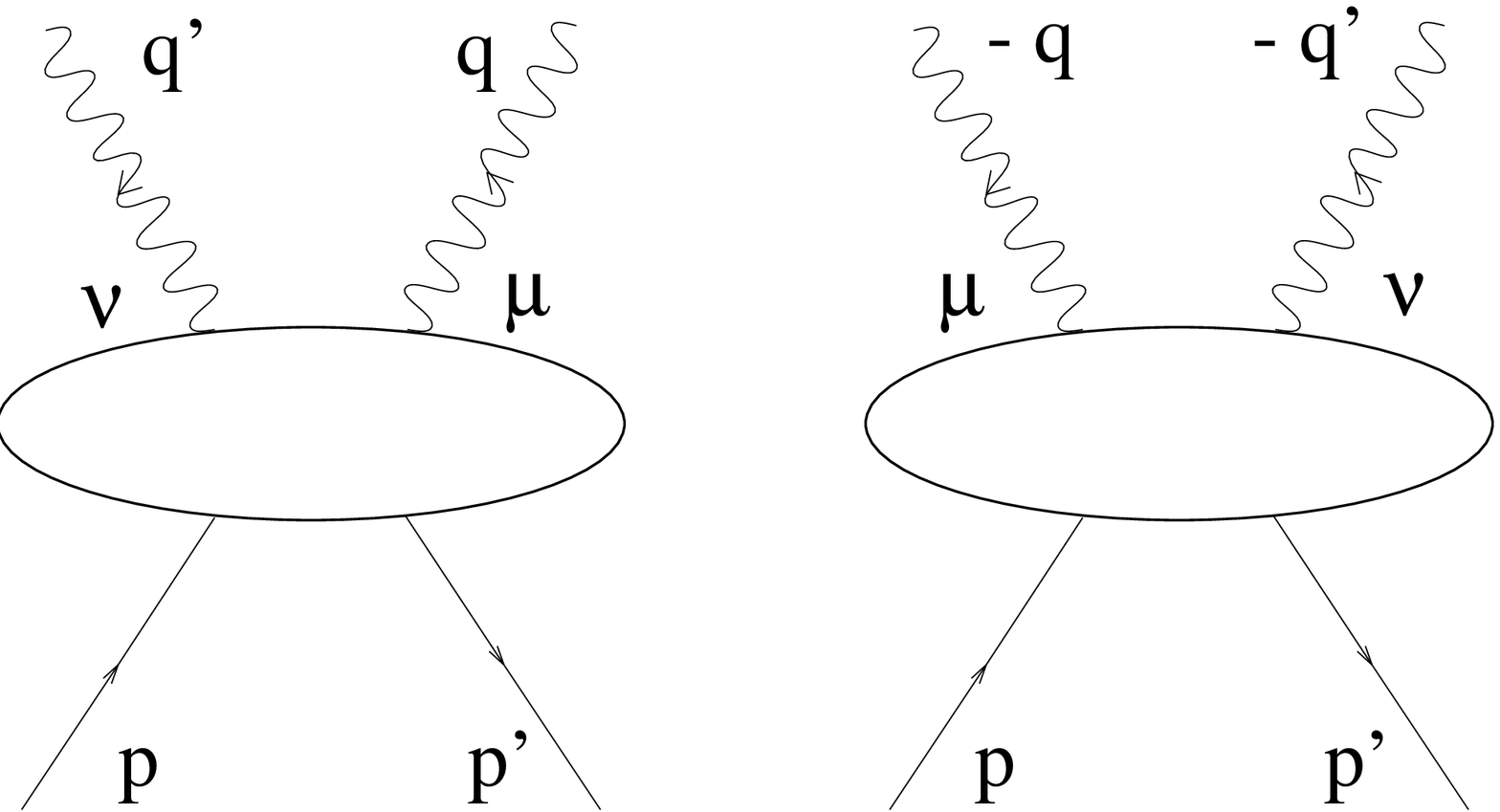}}
\end{center}
\caption{Symmetry of the Non-forward Scattering Amplitude}
{Because the two diagrams shown are indeed identical after the momentum
labeling change $\mu \leftrightarrow \nu, \,\,\,q \leftrightarrow
-q'$, the non-forward amplitude is invariant under the same 
transformation.} 
\label{fig:symmetry}
\end{figure}

To obtain the complete tensorial decomposition of ${\rm T}_{\mu \nu}$
we also need the symmetry properties of the amplitude. From 
fig.~\ref{fig:symmetry} we can see clearly that the second diagram is
simply the first diagram with a different momentum labeling and thus
the two have identical values, both equal to ${\rm T}_{\mu \nu}$. 
Note that to be completely explicit, each diagram in
fig.~\ref{fig:symmetry} should have its own corresponding cross
diagram, and the symmetry exists for the sum of the cross diagram and
the original one.
 
Therefore, while ${\rm T}_{\mu \nu}$ is no longer symmetric, it is 
invariant under the transformation $\mu \leftrightarrow \nu, 
\,\,\,q \leftrightarrow -q'$, that is,
\beq
\label{eq:symmetry}
{\rm T}_{\mu \nu} (p, q',q) = {\rm T}_{\nu \mu} (p, -q, -q') \,.
\eeq

From Section \ref{sec:ampdef} we know that ${\rm T}_{\mu \nu}$ depends
on the momenta $p$, $q$ and $q'$, similar to the last chapter we can
write the most general tensorial decomposition of ${\rm T}_{\mu \nu}$
as 
\beqs
\label{eq:decompdef}
{\rm T}_{\mu \nu} (p, q',q) &=& a_0 g_{\mu \nu} + a_1 p_\mu p_\nu
+ a_2 q'_\mu q'_\nu + a_3 q_\mu q_\nu + a_4 p_\mu q'_\nu 
+ a_5 q'_\mu p_\nu \non \\
&& + a_6 p_\mu q_\nu + a_7 q_\mu p_\nu
+ a_8 q'_\mu q_\nu + a_9 q_\mu q'_\nu \,.
\eeqs

\nid As in the previous chapter, this is a spin averaged
electron-scattering, thus we do not need to consider any helicity
issues and hence there are no $\eps$-tensor terms in the
decomposition. 

\nid The $a_i$s are invariant amplitudes depending again only on scalar
products of the momenta. Because we have equations
(\ref{eq:pdotr}), (\ref{eq:pdotq'}), (\ref{eq:qdotr}), and (\ref{eq:qdotq'}), 
we take the 
$a_i$s to only depend on $q'^2$, $q^2$ and $p \cdot q$, that is
\beq
a_i = a_i (q'^2, q^2, p \cdot q), \,\,\,\,\,\,\,\,\,\, i = 0, 1, 2,
..., 9 \,.
\eeq

From (\ref{eq:symmetry}) and (\ref{eq:decompdef}) we have
\beqs
\label{eq:tnumu}
{\rm T}_{\nu \mu} (p, -q, -q') &=& a_0(-) g_{\mu \nu} 
+ a_1(-) p_\mu p_\nu + a_2(-) q'_\mu q'_\nu + a_3(-) q_\mu q_\nu  
- a_4(-) p_\mu q'_\nu \non \\
&& \!\! - a_5(-) q'_\mu p_\nu 
- a_6(-) p_\mu q_\nu - a_7(-) q_\mu p_\nu
+ a_8(-) q'_\mu q_\nu + a_9(-) q_\mu q'_\nu  \non \\
&=& {\rm T}_{\mu \nu} (p, q',q)
\eeqs

\nid where we have used the short handed notation
\beq
\label{eq:a(-)def}
a_i(-) = a_i(q \leftrightarrow -q') = a_i(q'^2, q^2, -p \cdot q) \,.
\eeq

\nid Equation \ref{eq:tnumu} is true for any values of the momenta
$q'$, $q$ and $p$, in particular, we can fix the scalar products but
still vary arbitrarily the individual momentum. Thus the coefficients
of the same momentum combinations must be identical, which leads to the
following relationships among the invariant amplitudes:
\beqs
\label{eq:a&a(-)}
a_0 = a_0(-) \,; a_1 = a_1(-) \,; a_8 = a_8(-) \,; a_9 = a_9(-) \,;
\non \\
a_2 = a_3(-) \,; a_3 = a_2(-) \,; a_4 = -a_7(-) \,; a_5 = -a_6(-) \,.
\eeqs

By using the conservation of the e\&m current (\ref{eq:currentconsv})  
we can establish another two sets of equations of the invariants. We
have
\beqs
\label{eq:ccexpansion}
q^\mu {\rm T}_{\mu \nu} = 0 &=& a_0 q_\nu + a_1 p \cdot q p_\nu
+ a_2 q \cdot q' q'_\nu + a_3 q^2 q_\nu + a_4 p \cdot q q'_\nu
+ a_5 p \cdot q' p_\nu \non \\
&& + a_6 p \cdot q q_\nu + a_7 q^2 p_\nu
+ a_8 q \cdot q' q_\nu + a_9 q^2 q'_\nu  \,; \non \\
q'^\nu {\rm T}_{\mu \nu} = 0 &=& a_0 q'_\mu + a_1 p \cdot q' p_\mu
+ a_2 q'^2 q'_\mu + a_3 q \cdot q' q_\mu + a_4 q'^2 p_\mu
+ a_5 p \cdot q' q'_\mu \non \\
&& + a_6 q \cdot q' p_\mu + a_7 p \cdot q' q_\mu
+ a_8 q \cdot q' q'_\mu + a_9 q'^2 q_\mu \,.
\eeqs

\nid Because again this is always true when we vary arbitrarily the
momenta while keeping the scalar products fixed, the coefficients of
the same combinations of momenta must identically vanish, leading to 
\beqs
\label{eq:a-equations}
a_0 + a_3 q^2 + a_6 \, p \cdot q + a_8 \, q \cdot q' &=& 0 \non \\
      a_7 q^2 + a_1 \, p \cdot q + a_5 \, q \cdot q' &=& 0 \non \\
      a_9 q^2 + a_4 \, p \cdot q + a_2 \, q \cdot q' &=& 0 \non \\
a_0 + a_2 q'^2 + a_5 \, p \cdot q' + a_8 \, q \cdot q' &=& 0 \non \\
      a_4 q'^2 + a_1 \, p \cdot q' + a_6 \, q \cdot q' &=& 0 \non \\
      a_9 q'^2 + a_7 \, p \cdot q' + a_3 \, q \cdot q' &=& 0 \,.
\eeqs

\nid As in the previous chapter, we will use these equations to reduce
the number of independent invariant amplitudes. We first rewrite
(\ref{eq:a-equations}) by taking $a_{0,1,2,3,5}$ as given and expressing
the rest of $a_i's$ in terms of them. We obtain
\beqs
\label{eq:a_i-solution}
a_4 &=& - { {q \cdot q'} \over {p \cdot q} } a_2 
        - { q^2 \over {p \cdot q} } a_9 \non \\
&=& - { {p \cdot q'} \over {q'^2} } a_1 
    - { {q \cdot q'} \over {p \cdot q} } a_2
    + { {q \cdot q' q^2} \over {q'^2 p \cdot q} } a_3
    - { {q \cdot q'} \over {q'^2} } a_5 \,, \non \\
a_6 &=& - { {p \cdot q} \over {q \cdot q'} } a_1
        - { {q'^2} \over {q \cdot q'} } a_4 \non \\
&=& + { q'^2 \over {p \cdot q} } a_2 
    - { q^2 \over {p \cdot q} } a_3 + a_5 \, ,\non \\
a_7 &=& - { {p \cdot q} \over q^2 } a_1
        - { {q \cdot q'} \over q^2 } a_5 \,, \non \\
a_8 &=& - { 1 \over {q \cdot q'} } a_0     
        - { {q'^2} \over {q \cdot q'} } a_2
        - { {p \cdot q'} \over {q \cdot q'} } a_5 \,, \non \\
a_9 &=& - { {q \cdot q'} \over q'^2 } a_3
        - { {p \cdot q'} \over q'^2 } a_7 \non \\
&=& { {p \cdot q p \cdot q'} \over {q^2 q'^2} } a_1
    - { {q \cdot q'} \over q'^2 } a_3 
    + { {p \cdot q' q \cdot q'} \over {q^2 q'^2} } a_5 \,.
\eeqs

\nid We now use equation  \ref{eq:a&a(-)} to further simplify the
above result. We have
\beqs
\label{eq:a&a(-)result}
a_8 & = & a_8(-) \;\; \Rightarrow \non \\
 & & - { 1 \over {q \cdot q'} } a_0
 - { {q'^2} \over {q \cdot q'} } a_2 
 - { {p \cdot q'} \over {q \cdot q'} } a_5 =
 - { 1 \over {q \cdot q'} } a_0(-)
        - { {q'^2} \over {q \cdot q'} } a_2(-)
        + { {p \cdot q'} \over {q \cdot q'} } a_5(-) \non \\
& \Rightarrow & a_5 + a_5(-) = { {p \cdot q} \over {q'^2} } (a_3 - a_2)
\,, \non \\
a_9 & =& a_9(-) \;\; \Rightarrow \non \\
&& { {p \cdot q p \cdot q'} \over {q^2 q'^2} } a_1
    - { {q \cdot q'} \over q'^2 } a_3
    + { {p \cdot q' q \cdot q'} \over {q^2 q'^2} } a_5  \non \\
&& \;\;\;\;\;\;\;\;
= { {p \cdot q p \cdot q'} \over {q^2 q'^2} } a_1(-)
    - { {q \cdot q'} \over q'^2 } a_3(-)
    - { {p \cdot q' q \cdot q'} \over {q^2 q'^2} } a_5(-) \non \\
& \Rightarrow & a_5 + a_5(-) = { {p \cdot q} \over {q^2} } (a_3 - a_2) \,,
\eeqs

\nid where we have used $a_1 = a_1(-)$, $a_3(-) = a_2$ and 
$p \cdot q = p \cdot q'$. From the above, it is obvious that we must
have
\beq
\label{eq:a235first}
a_2 = a_3 \,,\,\,\,\,\,\, a_5 = - a_5(-) \,.
\eeq

\nid Furthermore,
\beqs
a_5 &=& -a_6(-) \;\; \Rightarrow \non \\
&& a_5 = - ( - { q'^2 \over {p \cdot q} } a_2(-)
   + { q^2 \over {p \cdot q} } a_3(-) + a_5(-) ) \non \\
&\Rightarrow& a_5 + a_5(-) = { {q'^2} \over {p \cdot q} } a_3
   - { {q^2} \over {p \cdot q} } a_2 \,,
\eeqs

\nid which, combined with (\ref{eq:a235first}), gives
\beq
\label{eq:a23}
a_2 = a_3 = 0 \,. 
\eeq

At the same time,
\beqs
a_4 &=& - a_7(-) \;\; \Rightarrow \non \\
&&  - { {p \cdot q'} \over {q'^2} } a_1
    - { {q \cdot q'} \over {p \cdot q} } a_2
    + { {q \cdot q' q^2} \over {q'^2 p \cdot q} } a_3
    - { {q \cdot q'} \over {q'^2} } a_5 
= - ( { {p \cdot q} \over q^2 } a_1 (-)
        - { {q \cdot q'} \over q^2 } a_5(-) ) \non \\
&\Rightarrow& a_5 = - { {p \cdot q} \over {q \cdot q'} } a_1 \,.
\eeqs

\nid Therefore, going back to (\ref{eq:a_i-solution}) we have
\beqs
a_4 &=& - { {p \cdot q'} \over {q'^2} } a_1
    - { {q \cdot q'} \over {q'^2} } 
(- { {p \cdot q} \over {q \cdot q'} } ) a_1 \non \\
&=& 0 \,, \non \\
a_6 &=& a_5 = - { {p \cdot q} \over {q \cdot q'} } a_1 \,, \non \\
a_7 &=& - { {p \cdot q} \over q^2 } a_1
        - { {q \cdot q'} \over q^2 } 
(- { {p \cdot q} \over {q \cdot q'} } ) a_1 \non \\
&=& 0 \,, \non \\
a_8 &=& - { 1 \over {q \cdot q'} } a_0
        - { {p \cdot q'} \over {q \cdot q'} } 
(- { {p \cdot q} \over {q \cdot q'} } ) a_1 \non \\
&=& - { 1 \over {q \cdot q'} } a_0 
    + { {(p \cdot q)^2} \over { (q \cdot q')^2 } } a_1 \,, \non \\
a_9 &=& { {p \cdot q p \cdot q'} \over {q^2 q'^2} } a_1
    + { {p \cdot q' q \cdot q'} \over {q^2 q'^2} } 
    (- { {p \cdot q} \over {q \cdot q'} } ) a_1 \non \\
&=& 0 \,.
\eeqs                                                        

It is now clear that only two independent invariant amplitudes, 
$a_0$ and $a_1$, remain in the end while all the other ones either
vanish identically or can be expressed in terms of $a_0$ and $a_1$.
In summary, we have
\beqs
\label{eq:a-results}
a_2 = a_3 = a_4 = a_7 = a_9 = 0 \non \\
a_5 = a_6 = - { {p \cdot q} \over {q \cdot q'} } a_1 \non \\
a_8 = - { 1 \over {q \cdot q'} } a_0
    + { {(p \cdot q)^2} \over { (q \cdot q')^2 } } a_1 \,.
\eeqs

\nid Thus, substitute these back into (\ref{eq:decompdef}) we have
\beqs
\label{eq:decompfinal}
{\rm T}_{\mu \nu} (p, q',q) &=& a_0 g_{\mu \nu} + a_1 p_\mu p_\nu
- { {p \cdot q} \over {q \cdot q'} } a_1 
(q'_\mu p_\nu + p_\mu q_\nu) \non \\
&&
+ ( - { 1 \over {q \cdot q'} } a_0
    + { {(p \cdot q)^2} \over { (q \cdot q')^2 } } a_1) \, q'_\mu q_\nu 
\non \\
&=& \left( g_{\mu \nu} - { {q'_\mu q_\nu} \over {q \cdot q'} } \right)
a_0 \non \\
&& + \left( p_\mu p_\nu 
- { {p \cdot q} \over {q \cdot q'} } (q'_\mu p_\nu + p_\mu q_\nu)
+ { {(p \cdot q)^2} \over { (q \cdot q')^2 } } q'_\mu q_\nu \right)
a_1 \,.
\eeqs     

By defining two invariant amplitudes ${\rm T_1}$ and ${\rm T_2}$ we 
finally obtain the general tensorial decomposition of ${\rm T}_{\mu \nu}$ as
\beqs
\label{eq:decomposition}
{1 \over {\sqrt{1-\zeta}}} {\rm T}_{\mu \nu} &=& \left(-g_{\mu \nu} + 
\frac {q'_\mu q_\nu}{\ovl{q}^2} \right){\rm T_1} \non \\
&& + \, {1 \over M^2} \left(p_\mu p_\nu - \frac {p \cdot q} {\ovl{q}^2}
(p_\mu q_\nu + q'_\mu p_\nu) + ( \frac {p \cdot q} {\ovl{q}^2})^2 
q'_\mu q_\nu \right){\rm T_2} \,.
\eeqs

\noindent ${\rm T_i}={\rm T_i}(q'^2, q^2, p \cdot q)$ are invariant 
amplitudes analogous to those in DIS (see section \ref{sec:fw-ope}).
They are even functions of $p \cdot q$ (because they are essentially
$a_0$ and $a_1$). 

\nid We have pulled out from ${\rm T_i}$ an explicit factor of 
$\sqrt{1-\zeta}$, as was done in, e.g.\ ,\cite{Radyphlt961} (see 
equation \ref{eq:zeta} for the definition of $\zeta$). This factor 
comes from the external spinors of the proton as the following:
The normalization convention we use for computing Feynman diagrams is 
such that an external incoming Fermion with momentum $p$ and spin
state $r$ enters a Feynman diagram expression as 
${ {U_r(p)} \over {\sqrt{p}} }$ where ${U_r(p)}$ is 
the standard $4$-component basis spinor for a Fermion. Thus there is 
an explicit factor of 
$\sqrt{p'}^{-1} \equiv (\sqrt{1 - \zeta} \sqrt{p})^{-1}$ in the
expression of ${\rm T}_{\mu \nu}$. We pull it outside explicitly
so that later on the expression of the lowest order Wilson coefficient 
will be simpler (see equation \ref{eq:1storderwcoeff} in 
section \ref{sec:wilsoncoeff} ).
 
It is worth noting that we still have only two independent invariant 
amplitudes even in the presence of three independent invariants, in
contrast to the case of forward scattering where we only have two
(see sections \ref{sec:fw-decomp} and \ref{sec:fw-ope}).

\chapter{Operator Product Expansion}
\thispagestyle{myheadings}
\markright{}

In this chapter we will discuss the operator product expansion (OPE)
of the non-forward amplitude ${\rm T}_{\mu \nu}$ 
(see section \ref{sec:ampdef}). We will identify the 
operators in the OPE, obtain their reduced matrix elements between 
(asymmetric) external states and define new moment variables. We will 
express ${\rm T}_{\mu \nu}$ and eventually the invariant amplitudes 
${\rm T}_1$ and ${\rm T}_2$ (see equation \ref{eq:decomposition})
in terms of the reduced matrix elements and the corresponding Wilson 
coefficients. 

\section{Non-forward Operator Product Expansion \label{sec:ope}}

In the short distance limit, we can perform an OPE for ${\rm T}_{\mu \nu}$ 
as a sum of products of local operators and their corresponding Wilson 
coefficients \cite{Collinsbook}: 
\beqs
\label{eq:orig-ope}
T \, j_\mu (-{z \over 2}) j_\nu ({z \over 2}) 
\stackrel{z_\mu \rightarrow 0}{\longrightarrow} 
\hat{A}_{\mu \nu} \, \sum_{J=1}^\infty
\sum_{n=0}^J \sum_{i=1}^{u_J} \, F_{J,n}^{(i)} (z^2)
\hat{O}_{\mu_1 ... \mu_J}^{(i)(J,n)} (0) 
{{z^{\mu_1}} \over 2}... {{z^{\mu_J}} \over 2}
\,\,\,\,\,\,\,\,\,\,\,\, \non \\ 
+ \,\, \hat{B}_{\mu \nu \al \be}
\sum_{J=1}^\infty \sum_{n=0}^{J\!+\!2} \sum_{i=1}^{u_J} \, E_{J,n}^{(i)}
(z^2) \hat{O}_{\al \be ; \mu_1 ... \mu_J}^{(i)(J\!+\!2,n)} (0)
{{z^{\mu_1}} \over 2}... {{z^{\mu_J}} \over 2} \,,
\eeqs

\noindent where $\hat{A}_{\mu \nu}$ and $\hat{B}_{\mu \nu \al \be}$ are 
conserved tensor structure operators. The summations should be
regarded as such that in addition to the literal summation over $J$, $n$ and
$i$, all the indices except $\mu, \nu$ are also symmetrized as usual.  

To obtain the explicit expression of $\hat{B}_{\mu \nu \al \be}$ we
start with the most general form that an tensor operator in the
position of $\hat{B}$ can have:
\beqs
\label{eq:opbgeneral}
\hat{O}_{\mu \nu \al \be} &=& a_1 g_{\mu \nu} g_{\al \be} \Box
+ a_2 g_{\mu \al} g_{\nu \be} \Box + a_3 g_{\mu \be} g_{\nu \al} \Box
\non \\
&& + a_4 g_{\mu \nu} \pdr_\al \pdr_\be 
+ a_5 g_{\mu \al} \pdr_\nu \pdr_\be + a_6 g_{\mu \be} \pdr_\al\pdr_\nu 
\non \\
&& + a_7 g_{\nu \be} \pdr_\mu \pdr_\al
+ a_8 g_{\nu \al} \pdr_\mu \pdr_\be +  a_9 g_{\al \be} \pdr_\mu \pdr_\nu
\,.
\eeqs

\nid Current conservation requires
\beq
\pdr_\mu \hat{O}_{\mu \nu \al \be} = \pdr_\nu \hat{O}_{\mu \nu \al \be}
= 0 \,.
\eeq

\nid We have
\beqs
0 &=& a_1 \pdr_\nu g_{\al \be} \Box + a_2 \pdr_\al g_{\nu \be} \Box 
+ a_3 \pdr_\be g_{\nu \al} \Box + a_4 \pdr_\nu \pdr_\al \pdr_\be
\non \\
&& + a_5 \pdr_\al \pdr_\nu \pdr_\be + a_6 \pdr_\be \pdr_\al\pdr_\nu
+ a_7 g_{\nu \be} \pdr_\al \Box + a_8 g_{\nu \al} \Box \pdr_\be 
+  a_9 g_{\al \be} \Box \pdr_\nu \non \\
&=& ( (a_1 + a_9) g_{\al \be} \pdr_\nu 
+ (a_2 + a_7) g_{\nu \be} \pdr_\al + (a_3 + a_8) g_{\nu \al} \pdr_\be)
\Box \non \\
&& +(a_4 + a_5 + a_6) \pdr_\nu \pdr_\al \pdr_\be \,,
\eeqs

\nid and 
\beqs
0 &=& a_1 \pdr_\mu g_{\al \be} \Box + a_2 g_{\mu \al} \pdr_\be \Box
+ a_3 g_{\mu \be} \pdr_\al \Box + a_4 \pdr_\mu \pdr_\al \pdr_\be
\non \\
&& + a_5 g_{\mu \al} \pdr_\be \Box + a_6 g_{\mu \be} \pdr_\al \Box
+ a_7 \pdr_\be \pdr_\mu \pdr_\al + a_8 \pdr_\al \pdr_\mu \pdr_\be
+  a_9 g_{\al \be} \Box \pdr_\mu \non \\
&=& ( (a_1 + a_9) g_{\al \be} \pdr_\mu
+ (a_2 + a_5) g_{\mu \al} \pdr_\be + (a_3 + a_6) g_{\mu \be} \pdr_\al)
\Box \non \\
&& +(a_4 + a_7 + a_8) \pdr_\mu \pdr_\al \pdr_\be \,.
\eeqs                                              

\nid Since the above two equations are tensor equations, the
coefficient of each (different) tensor structure operator must
vanish, which gives us
\beqs
a_1+a_9 = a_2 + a_7 = a_2 + a_5 = a_3 +a_8 = a_3 +a_6 = 0 \non \\
a_4 + a_5 + a_6 = a_4 + a_7 + a_8 = 0 \,.
\eeqs

\nid Thus we have
\beqs
a_9 &\!=\!& - a_1 \non \\
a_4 &\!=\!& a_2 + a_3 \non \\
a_5 &\!=\!& a_7 = - a_2 \non \\
a_6 &\!=\!& a_8 = - a_3 \,,
\eeqs

\nid which, after substitution back into (\ref{eq:opbgeneral}), leads us
to
\beqs
\hat{O}_{\mu \nu \al \be} &=& a_1 g_{\mu \nu} g_{\al \be} \Box
+ a_2 g_{\mu \al} g_{\nu \be} \Box + a_3 g_{\mu \be} g_{\nu \al} \Box
\non \\
&& + (a_2 + a_3) g_{\mu \nu} \pdr_\al \pdr_\be
- a_2 g_{\mu \al} \pdr_\nu \pdr_\be - a_3 g_{\mu \be} \pdr_\al \pdr_\nu
\non \\
&& - a_2 g_{\nu \be} \pdr_\mu \pdr_\al
- a_3 g_{\nu \al} \pdr_\mu \pdr_\be - a_1 g_{\al \be} \pdr_\mu \pdr_\nu
\non \\
&=& (g_{\mu \nu} g_{\al \be} \Box - g_{\al \be} \pdr_\mu \pdr_\nu) a_1
+ (g_{\mu \al} g_{\nu \be} \Box + g_{\mu \nu} \pdr_\al \pdr_\be \non \\
&& - g_{\mu \al} \pdr_\nu \pdr_\be - g_{\nu \be} \pdr_\mu \pdr_\al ) a_2
+ (g_{\mu \be} g_{\nu \al} \Box  \non \\
&&+ g_{\mu \nu} \pdr_\al \pdr_\be 
 - g_{\mu \be} \pdr_\al \pdr_\nu - g_{\nu \al} \pdr_\mu \pdr_\be) a_3
\,.
\eeqs                   

\nid If we symmetrize the indices $\al$ and $\be$ we have
\beqs
\hat{O}_{\mu \nu \al \be} &=& g_{\al \be} 
(g_{\mu \nu} \Box - \pdr_\mu \pdr_\nu) a_1 \non \\
&& + (g_{\mu \al} g_{\nu \be} \Box + g_{\mu \nu} \pdr_\al \pdr_\be 
- g_{\mu \al} \pdr_\nu \pdr_\be - g_{\nu \be} \pdr_\mu \pdr_\al ) a_2
\,.
\eeqs 

\nid It is now obvious that $\hat{B}_{\mu \nu \al \be}$ is the same as in
the forward case (see equation \ref{eq:fw-op-ab}), i.e.\ ,
\beq
\label{eq:op-b}
\hat{B}_{\mu \nu \al \be} = g_{\mu \al}g_{\nu \be} \Box 
+g_{\mu \nu} \partial_\al \partial_\be
-g_{\mu \al} \partial_\nu \partial_\be
-g_{\nu \be} \partial_\mu \partial_\al \, ,
\eeq

\noindent and corresponds to the tensor structure multiplying ${\rm T}_2$ 
in (\ref{eq:decomposition}).

We have not found the explicit form of $\hat{A}_{\mu \nu}$ which would 
generate the tensor structure corresponding to ${\rm T}_1$. But, as in 
the forward case, once we know ${\rm T}_2$, we can obtain ${\rm T}_1$ 
by a Callen-Gross relationship, at least in leading logarithmic level 
(see equation \ref{eq:Callen-Gross}). In the following discussion,
we will leave $\hat{A}_{\mu \nu}$ as an unspecified general conserved
tensor operator. Eventually we will obtain the necessary terms from
the Callen-Gross relationship.


\section{The Operators \label{sec:operators}}

The operators that enter into the OPE of an amplitude are composed of
the fields involved in the interaction and derivatives. In the most
general case, the derivatives can either be internal, bi-directional
derivatives like the ones in equation \ref{eq:fw-ops}, or overall
derivatives that act outside the fields. These overall derivatives, 
when evaluated between external states, give the difference in
momentum (in momentum space) between the incoming and outgoing states, 
or, equivalently, the net momentum inflow/outflow of the local vertex 
of the interaction. Thus operators with overall derivatives are 
identically zero in a forward scattering process, and we do not need to
consider them in forward OPE (see section \ref{sec:fw-ope}). However, 
in a non-forward case as the one we are discussing, because the 
momentum flowing into the local vertex is $r$ instead of zero, we have 
to include in the expansion new sets of operators that have overall 
derivatives.

In leading twist (the physical consequences of leading twist in this 
case will be clear later) the operators for QCD are (cf. 
equation \ref{eq:fw-ops})
\beqs
\label{eq:operators}
\hat{O}_{\mu_1 ... \mu_J}^{(q)(J,n)} & = & \partial_{\mu_1} ...
\partial_{\mu_n} \, \tilde{q}(0) \gamma_{\mu_{n+1}}i \stackrel
{\leftrightarrow}D_{\mu_{n+2}}...i \stackrel{\leftrightarrow}
D_{\mu_J}q(0)  \\
\hat{O}_{\mu_1 ... \mu_J}^{(g)(J,n)} & = & \partial_{\mu_1} ...\partial_
{\mu_n} \, F_{\mu_{n+1}}^{\,\,\,\,\,\,\,\,\nu}(0) i \stackrel{\leftrightarrow}
{\cal D}_{\mu_{n+1}}...i \stackrel{\leftrightarrow}{\cal D}_{\mu_J} 
F_{\mu_J \nu}(0) \,,
\eeqs

\nid where it is again understood that the indices are all
symmetrized.

\noindent After taking the matrix elements between the asymmetric external 
states, and after taking a Fourier transform, the external derivatives would 
eventually be turned into factors of $r_\mu$ while the internal derivatives 
into either $r_\mu$ or $(p+p')_\mu \equiv (2p-r)_\mu$. We thus define the 
two moment variables in the non-forward case as
\beq
\label{eq:momvardef}
\om = \frac {(2p-r) \cdot q} {{\ovl{Q}}^2}, \,\,\,\,\,\,\,\,\,\,\,\, 
\nu = \frac {r \cdot q} {{\ovl{Q}}^2} \,\, .
\eeq

\noindent The forward case would be the limit $\nu=0$ while DVCS corresponds 
to $\nu=1$.

In a QCD-Parton picture, the diagrams contributing to ${\rm T}_{\mu \nu}$
are the so-called {\it hand-bag} diagrams as shown in figure \ref{fig:handbag}.
If we parametrize the momentum of the scattered parton as $k=xp+yr$ (see,
e.g. \cite{Radyphlt961}), where $x$ and $y$ are two Bjorken type scaling
variables defined by
\beqs
\label{eq:x&ydef}
x &=& { {Q_1^2} \over {2p \cdot q} } \equiv {1 \over \tilde{\om}} \non \\ 
y &=& { {Q_1^2} \over {2r \cdot q} } \equiv {1 \over \tilde{\nu}} \,, 
\eeqs

\nid our moment variables are related to these quantities via
\beqs
\label{eq:momvarrelation}
\om &=& {Q_1^2 \over \ovl{Q}^2} \,\, (x \tilde{\om} + y \tilde{\nu} 
- {1 \over 2} \tilde{\nu}) \,\,\,\non \\
\nu &=& {Q_1^2 \over \ovl{Q}^2} \,\, ({1 \over 2} \tilde{\nu}) \,.
\eeqs

There will be mixing among operators with the same $J$ but different 
$n$ labels. Under evolution in the momentum scale, which as we shall
see later, is characterized by $\ovl{Q}^2$, the internal derivatives,
with which the dynamics of the process lies, can be turned into either
overall derivatives or again internal ones, and, as stated before,
eventually give rise to factors of either $2p-r$ or $r$. On the other
hand, the overall derivatives, which are in essence only involved with 
kinematics, can be turned into themselves only and eventually give 
factors of $r$. Therefore, evolution in $\ovl{Q}^2$ will lead to
mixing of these operators in only one direction in an upper-triangular 
fashion, namely,
\beq
\label{eq:mixing-dir}
\hat{O}^{(i)(J,n)} \rightarrow \hat{O}^{(i)(J,n')} \,,\;\;\;\;\;\;
0 \leq n \leq n' \leq J \,.
\eeq
  
\nid However, we do still have the freedom to choose, for simplification,
at a factorization scale $\mu_0$, that all internal derivatives give 
factors of $(2p-r)$. Thus we can write
\beq
\label{eq:reducedef}
{\bra{p'} \hat{O}_{\mu_1 ... \mu_J}^{(i)(J,n)} \ket{p}}_{(\mu_0)}  = 
r_{\mu_1}...r_{\mu_n}(2p-r)_{\mu_{n+1}}...(2p-r)_{\mu_J} \, 
{\bra{p'} | \hat{O}^{(i)(J,n)} | \ket{p}}_{(\mu_0)} \,.
\eeq

\noindent This is also the definition of the reduced matrix
elements of $\hat{O}^{(i)(J,n)}$. It is clear that from the above
discussion, these reduced matrix elements will depend only on $J-n$,
since the overall derivatives give (in certain sense) trivial kinematic 
factors of $r$, and we have
\beq
\label{eq:reducej-n}
{\bra{p'} | \hat{O}^{(i)(J,n)} | \ket{p}} = {\bra{p'} | \hat{O}^{(i)(J-n,0)} 
| \ket{p}} \,\,\, .
\eeq

The choice of (\ref{eq:reducedef}) is essentially the same as taking
the leading twist approximation. Analogous to the discussion in the 
forward case (\ref{sec:fw-ope}), the indices of the asymmetric matrix
elements of the operators in the OPE must be made from $g_{\mu_i
\mu_j}$ or $p_{\mu_i} (p'_{\mu_i}) $. After Fourier transformation the
factors of $z_{\mu_i}$ are turned into momentum factors, namely,
\beq
z_{\mu_i} \Rightarrow { {\ovl{q}_{\mu_i}} \over {\ovl{q}^2} } \,.
\eeq

\nid And we have 
\beq
g_{\mu_i \mu_j} { {\ovl{q}_{\mu_i}} \over {\ovl{q}^2} }
{ {\ovl{q}_{\mu_j}} \over {\ovl{q}^2} } \sim {1 \over \ovl{q}^2} \,,
\eeq

\nid while (see \ref{eq:pdotqbar})
\beq
\label{eq:ltwist}
p_{\mu_i} p_{\mu_j} { {\ovl{q}_{\mu_i}} \over {\ovl{q}^2} }
{ {\ovl{q}_{\mu_j}} \over {\ovl{q}^2} }
= ( \frac {p \cdot q} {\ovl{q}^2} )^2 \sim {\rm O}(1) \,.
\eeq

\nid Therefore the leading contribution from the operators in the OPE
should all give factors of the external momentum, while $g_{\mu_i
\mu_j}$ type of contributions are small (by factors of ${1 \over
{\ovl{q}^2} }$). It is clear that this also makes the leading twist
operators dominate (cf. \cite{Collinsbook}). Non-leading twist terms in 
this case are suppressed by at least a power of ${1 \over {\ovl{Q}^2}}$. 
And $\ovl{q}$ is now the measure of the momentum scale of the
scattering.

\section{Wilson Coefficients and Explicit OPE \label{sec:wc&ope}}

To obtain the explicit expression of the OPE of the amplitude, we
rewrite equation \ref{eq:orig-ope} by explicit substitution of the
tensor structure operators as
\beqs
\label{eq:jjexpansion}
T \, j_\mu (-{z \over 2}) j_\nu ({z \over 2})
& \!\stackrel{z_\mu \rightarrow 0}{\longrightarrow} \!&
\hat{A}_\mn
\, \sum_{J=1}^\infty
\sum_{n=0}^J \sum_{i=1}^{u_J} \, F_{J,n}^{(i)} (z^2)
\hat{O}_{\mu_1 ... \mu_J}^{(i)(J,n)} (0) {{z^{\mu_1}} \over 2}
... {{z^{\mu_J}} \over 2} \,\,\,\,\,\,\,\,\,\,\,\,\,\,\,\, \non \\
&+& (g_{\mu \al}g_{\nu \be} \Box
  +g_{\mu \nu} \partial_\al \partial_\be
  -g_{\mu \al} \partial_\nu \partial_\be
  -g_{\nu \be} \partial_\mu \partial_\al ) \non \\
&& \,\,\,\,\,\,\,\,\,\,\,\,\,\,\,\,\,\,\,\,\,\,\,\,
\sum_{J=1}^\infty \sum_{n=0}^{J\!+\!2} \sum_{i=1}^{u_J} 
\, E_{J,n}^{(i)} (z^2) 
\hat{O}_{\al \be ; \mu_1 ... \mu_J}^{(i)(J\!+\!2,n)} (0)
{{z^{\mu_1}} \over 2}... {{z^{\mu_J}} \over 2} \,. \non \\
\eeqs             

\nid We evaluate the above time-ordered product between the asymmetric
external proton states and expand it at an arbitrarily chosen
factorization scale $\mu_0$ into products of reduced matrix elements,
corresponding Wilson coefficients, and the kinematic factors coming
from the tensor structure operators. We obtain
\beqs
\label{eq:jjmatrixexp}
&&\bra{p'} T \, j_{\mu}(-{z \over 2}) j_{\nu} ({z \over 2}) \ket{p} 
\longrightarrow 
\non \\ 
&& \;\;\;\;\; 
 \sum_{J=1}^\infty \sum_{n=0}^{J} \sum_{i=1}^{u_J}
{\bra{p'} | \hat{O}^{(i)(J,n)} | \ket{p}}_{(\mu_0)} 
\hat{A}_{\mu \nu} F_{J,n}^{(i)} (z^2) 
(2)^{-J}((2p \!-\! r) \cdot z)^{J \!-\! n} (r \cdot z)^n \non \\
&& \;\;\;\;\;\;\;\;\;
 + \sum_{J=1}^\infty \sum_{n=0}^{J\!+\!2} \sum_{i=1}^{u_J}
{\bra{p'} | \hat{O}^{(i)(J\!+\!2,n)} | \ket{p}}_{(\mu_0)} 
\non \\ 
&& \;\;\;\;\;\;\; 
 \cdot \, \hat{B}_{\mu \nu}  E_{J,n}^{(i)} (z^2) 
(2)^{-J}((2p \!-\! r) \cdot z)^{J\!-\!n} (r \cdot z)^{n\!-\!2}
\cc_{\al \be}(2p\!-\!r, r) \,. 
\eeqs

\nid Because the $\al$ and $\be$ indices can either come from
$2p-r$ or $r$ when going to the reduced matrix elements we have 
included a factor of $\cc_{\al \be}$ that incorporates all the
possible combinations in a symmetric fashion. Explicitly, it is the
completely symmetrized sum of the following three terms: 
\beqs
(2p-r)_\al (2p-r)_\be r_{\mu_i} r_{\mu_j} \,\,, \non \\
(2p-r)_\al r_\be (2p-r)_{\mu_i} r_{\mu_j} \,\,, \\
 r_\al r_\be (2p-r)_{\mu_i} (2p-r)_{\mu_j} \,\,, \non
\eeqs
  
\nid together with a normalization factor that we need to put in to
compensate for the double counting of terms because, as stated before,
the summation signs have already implied a complete symmetrization of
all indices except $\mu$ and $\nu$.

\nid The first term of equation (\ref{eq:jjmatrixexp}) becomes
\beq
\sum_{J=1}^\infty \sum_{n=0}^{J} \sum_{i=1}^{u_J}
{\bra{p'} | \hat{O}^{(i)(J,n)} | \ket{p}}_{(\mu_0)}
(2)^{-J} (2 \!-\! \zeta)^{J \!-\! n} \zeta^n
\hat{A}_\mn \, F_{J,n}^{(i)} (z^2)
(p \cdot z)^J \,,
\eeq

\nid where we have used equation (\ref{eq:zeta}). The second term,
after a shift of labeling $J+2 \rightarrow J$, becomes

\beqs
&&\sum_{J=1}^\infty \sum_{n=0}^{J} \sum_{i=1}^{u_J} 
{\bra{p'} | \hat{O}^{(i)(J,n)} | \ket{p}}_{(\mu_0)} \,
 (g_{\mu \al}g_{\nu \be} \Box
  +g_{\mu \nu} \partial_\al \partial_\be
  -g_{\mu \al} \partial_\nu \partial_\be
  -g_{\nu \be} \partial_\mu \partial_\al ) 
\non \\
&& \;\;\;\;\;\;\;\;\;\;\;\;\;\;\;
\cdot \; E_{J\!-\!2,n}^{(i)} (z^2) 
2^{-(J\!-\!2)} (2-\zeta)^{J\!-\!n\!-\!2} \zeta^{n\!-\!2}
(p \cdot z)^{J\!-\!4} \cc_{\al \be}(2p\!-\!r, r) 
\non \\
&=& \sum_{J=1}^\infty \sum_{n=0}^{J} \sum_{i=1}^{u_J}
{\bra{p'} | \hat{O}^{(i)(J,n)} | \ket{p}}_{(\mu_0)} \;
2^{-(J\!-\!2)} \; (2-\zeta)^{J\!-\!n\!-\!2} \, \zeta^{n\!-\!2} 
\;\; \cc
\non \\
&& \;\;\;\;\;
\cdot \; ( ((2p\!-\!r)_\mu (2p\!-\!r)_\nu \Box 
+ g_{\mu \nu} ((2p\!-\!r) \cdot \pdr)^2 
\non \\
&& \;\;\;\;\;\;\;\;\;\;\;\;\;\;\;\;\;\;\;
- \; ((2p\!-\!r)_\mu \pdr_\nu \!+\! \pdr_\mu (2p\!-\!r)_\nu) \,
 (2p\!-\!r) \!\cdot\! \pdr ) \, (r \!\cdot\! z)^2 
\non \\
&& \;\;\;\;\;\;\;\;\;\;\;
+ \; (((2p\!-\!r)_\mu r_\nu \Box + g_{\mu \nu} (2p\!-\!r) \!\cdot\! \pdr \,
  r \!\cdot\! \pdr
  - (2p\!-\!r)_\mu \pdr_\nu \, r \!\cdot\! \pdr 
  - \pdr_\mu r_\nu (2p\!-\!r) \!\cdot\! \pdr)  
\non \\
&& \;\;\;\;\;\;\;\;\;\;\;\;\;\;\;\;\;\;\;
+ \; (r_\mu (2p\!-\!r)_\nu \Box + g_{\mu \nu} 
     (2p\!-\!r) \!\cdot\! \pdr \, r \!\cdot\! \pdr 
\non \\
&& \;\;\;\;\;\;\;\;\;\;\;\;\;\;\;\;\;\;\;\;\;\;\; 
- \pdr_\mu (2p\!-\!r)_\nu \, r \!\cdot\! \pdr 
  - r_\mu \pdr_\nu (2p\!-\!r) \!\cdot\! \pdr) ) 
\;\; (2p\!-\!r) \cdot z \,\, r \cdot z 
\non \\
&& \;\;\;\;\;\;\;\;\;\;\;
+ \; (r_\mu r_\nu \Box + g_{\mu \nu} (r \cdot \pdr)^2 
\non \\
&& \;\;\;\;\;\;\;\;\;\;\;\;\;\;\;\;\;\;\;  
- \; (r_\mu \pdr_\nu + \pdr_\mu r_\nu) \, r \cdot \pdr) 
\;\;  ((2p\!-\!r) \!\cdot\! z)^2 )
\;\; (p \cdot z)^{J\!-\!4} \; E_{J\!-\!2,n}^{(i)} (z^2) \,, 
\eeqs

\nid where $\cc$ is the normalization factor that will compensate for
the double counting. After using equation (\ref{eq:zeta}), it is
straight forward to see that the four terms coming from 
$\cc_{\al \be}(2p\!-\!r, r)$ all give the same value. This means that
we should set $\cc = {1 \over 4}$ in the above expression, which now
becomes 
\beqs
&&\sum_{J=1}^\infty \sum_{n=0}^{J} \sum_{i=1}^{u_J}
{\bra{p'} | \hat{O}^{(i)(J,n)} | \ket{p}}_{(\mu_0)}\;
2^{-(J\!-\!2)}\; (2-\zeta)^{J\!-\!n} \; \zeta^{n} \non \\ 
&& \;\;\;\;\;
( p_\mu p_\nu \Box + g_{\mu \nu} (p \cdot \pdr)^2
 - (p_\mu \pdr_\nu + \pdr_\mu p_\nu) \, p \cdot \pdr )
\; (p \cdot z)^{J\!-\!2} \; E_{J\!-\!2,n}^{(i)} (z^2) \,.
\eeqs

\nid Therefore we have, in summary, for the asymmetric matrix elements,
\beqs
\label{eq:jjmatrixfinal}
&&\bra{p'} T \, j_{\mu}(-{z \over 2}) j_{\nu} ({z \over 2}) \ket{p}
\longrightarrow  \non \\
&& \;\;\;\;
\sum_{J=1}^\infty \sum_{n=0}^{J} \sum_{i=1}^{u_J}
{\bra{p'} | \hat{O}^{(i)(J,n)} | \ket{p}}_{(\mu_0)}
\left( { {(2 \!-\! \zeta)^{J \!-\! n} \zeta^n} \over {2^J} }
\hat{A}_\mn 
\; F_{J,n}^{(i)} (z^2)
\; (p \cdot z)^J  \right. \non \\
&& \;\;\;\;\;\;
\left. + \; { {(2\!-\!\zeta)^{J\!-\!n} \zeta^{n}} \over {2^{J\!-\!2}} } 
( p_\mu p_\nu \Box + g_{\mu \nu} (p \!\cdot\! \pdr)^2
 - (p_\mu \pdr_\nu + \pdr_\mu p_\nu) \, p \!\cdot\! \pdr ) \; 
E_{J\!-\!2,n}^{(i)} (z^2) \; (p \!\cdot\! z)^{J\!-\!2} \right) \,.
\non 
\eeqs

The amplitude ${\rm T}_{\mu \nu}$ is the Fourier transform of the
matrix elements with momentum $\ovl{q}$ (see equation \ref{eq:ampdef}). 
Under the Fourier transform, we have, as in the forward case (see
section \ref{sec:fw-ope}, note the difference in sign convention) 
$ z_\mu \Rightarrow i \d{}/d{\ovl{q}_\mu}$. Therefore, by taking the
logarithmic derivative of $\ovl{q}^2$ of the Fourier transformed
Wilson coefficients, we have 
\beqs
\label{eq:logderivative}
&& {1 \over 2^J} \int d^4 z \, e^{-i \bar{q} \cdot z} (p \cdot z)^J
E_{J,n}^{(i)}(z^2) \non \\
&& \;\;\;\; = \; {1 \over 2^J} (i p_\mu \d{}/d{\ovl{q}_\mu})^J 
\int d^4 z \, e^{-i \bar{q} \cdot z}  E_{J,n}^{(i)}(z^2) \non \\
&& \;\;\;\; = \; ((p \cdot \ovl{q}) \d{}/d{\ovl{q}^2})^J 
\; \tilde{e}_{J,n}^{(i)}(\ovl{q}^2) \non
\\
&& \;\;\;\; = \; \left( { {p \cdot \ovl{q}} \over {\ovl{Q}^2} } \right)^J
\left(-i \ovl{q}^2 \d{}/d{\ovl{q}^2}\right)^J 
\tilde{e}_{J,n}^{(i)}(\ovl{q}^2) \,,
\eeqs

\nid where $\tilde{e}_{J,n}^{(i)}$ is the Fourier transform of 
$E_{J,n}^{(i)}$. We also have a similar equation for $F$.  

At the same time, by a integration by parts it is straight
forward to show that, as in the forward case, the derivatives in
the tensor structure operator will simply turn into factors of
the $\ovl{q}$ momentum, namely, $\pdr_\mu \Rightarrow -i \ovl{q}_\mu$.
We will drop the $\hat{A}_\mn$ terms and concentrate on the explicit
calculation of the structure function ${\rm T}_2$. Again the necessary
terms generated from $\hat{A}_\mn$ and thus the structure function 
${\rm T}_1$ will be obtained from a Callen-Gross relationship once we
know ${\rm T}_2$. We obtain the following results for the expression of the
amplitude in the short distance limit:
\beqs
\label{eq:amp-ope}
&& {\rm T}_{\mu \nu} (p,q',q) = i \int d^4 z \, e^{-i \ovl{q} \cdot z} 
\, \bra{p'} T \, j_{\mu}(-{z \over 2}) j_{\nu} ({z \over 2}) \ket{p}
\non \\
&& \;\;\;\;\;\;\;\; 
= -\; i \sum_{J,n,i} {\bra{p'} | \hat{O}^{(i)(J,n)} | \ket{p}}_{(\mu_0)} 
\; (2 \!-\! \zeta)^{J \!-\! n} \; \zeta^n \non \\
&& \;\;\;\;\;\;\;\;\;\;\;
\cdot \; ( \hat{A}_\mn terms 
+ \;\;( p_\mu p_\nu \ovl{q}^2 + g_{\mu \nu} (p \cdot \ovl{q})^2
 - (p_\mu \ovl{q}_\nu + \ovl{q}_\mu p_\nu) \; p \cdot \ovl{q} ) 
\;\;\;\;\;\;\;\;\;\;\;\;\;\;\;\;\;\;\;\;\;\;\; \non \\
&& \;\;\;\;\;\;\;\;\;\;\;\;\;\;\;\;\;\;\;\;\;\;\;\;\;\;\;\;\;\;\;\;\;
\;\;\;\;\;\;\; \cdot \; 
\left( { {p \cdot \ovl{q}} \over {\ovl{Q}^2} } \right)^{J\!-\!2}
\left(-i \ovl{q}^2 \d{}/d{\ovl{q}^2}\right)^{J\!-\!2}
\tilde{e}_{J\!-\!2,n}^{(i)}(\ovl{q}^2) \, ) \,.
\eeqs

\nid From equations \ref{eq:pdotr} and \ref{eq:qbardef} it is
obvious 
\beq
\label{eq:pdotqbar}
p \cdot q' = p \cdot \ovl{q} = p \cdot q \,.
\eeq

\nid Thus for the tensor structure generated by $\hat{B}_{\mu \nu}$
(\ref{eq:op-b}) in equation \ref{eq:amp-ope} we have
\beqs
\label{eq:2ndtensor}
&& p_\mu p_\nu \ovl{q}^2 + g_{\mu \nu} (p \cdot \ovl{q})^2
 - (p_\mu \ovl{q}_\nu + \ovl{q}_\mu p_\nu) p \cdot \ovl{q} \non \\
&& \;\;\;\;\;
=\ovl{q}^2 \left( p_\mu p_\nu 
+ g_{\mu \nu} { {(p \cdot q)^2} \over \ovl{q}^2 }
- { {p \cdot q} \over \ovl{q}^2 } \; 
  (p_\mu \ovl{q}_\nu + \ovl{q}_\mu p_\nu) \right) \non \\
&& \;\;\;\;\;
=\ovl{q}^2 \left( p_\mu p_\nu
- { {p \cdot q} \over \ovl{q}^2 }
  (p_\mu q_\nu - {\zeta \over 2} \; p_\mu p_\nu 
  + q'_\mu p_\nu + {\zeta \over 2} p_\mu p_\nu) \right. \non \\
&& \;\;\;\;\;\;\;\;\;\;\;\;\;\;\;\;\;\;\;\;\;\;\;
\left. + \;\; g_{\mu \nu} { {(p \cdot q)^2} \over \ovl{q}^2 } 
 + q'_\mu q_\nu { {(p \cdot q)} \over \ovl{q}^2 }^2
 - q'_\mu q_\nu { {(p \cdot q)} \over \ovl{q}^2 }^2 \right) \non \\
&& \;\;\;\;\;
= \ovl{q}^2 \left( p_\mu p_\nu
- { {p \cdot q} \over \ovl{q}^2 } \; (p_\mu q_\nu + q'_\mu p_\nu)
+ { {(p \cdot q)} \over \ovl{q}^2 }^2 q'_\mu q_\nu \right) 
+ { {(p \cdot q)^2} \over {(\ovl{q}^2)^2} }
  \left( g_{\mu \nu} - { {q'_\mu q_\nu} \over {\ovl{q}^2} } \right)
\,. \non \\
\eeqs

\nid Also from equation \ref{eq:momvardef} we have
\beq
\label{eq:omnu}
\om = (2 - \zeta) \frac {p \cdot q} {{\ovl{Q}}^2}, \,\,\,\,\,\,\,\,\,
\nu = \zeta \frac {p \cdot q} {{\ovl{Q}}^2} \,\, .
\eeq

\nid Therefore, we can rewrite the expression of ${\rm T}_{\mu \nu}$
(\ref{eq:amp-ope}) as
\beqs
\label{eq:ampopefinal}
&& {\rm T}_{\mu \nu} (p,q',q) = i \int d^4 z \, e^{-i \ovl{q} \cdot z}
\, \bra{p'} T \, j_{\mu}(-{z \over 2}) j_{\nu} ({z \over 2}) \ket{p}
\non \\   
&& \;\;\;\;\;
= - i \;\sum_{J,n,i} {\bra{p'} | \hat{O}^{(i)(J,n)} | \ket{p}}_{(\mu_0)}
\; \om^{J \!-\! n} \; \nu^n \;\; ( \;
\;\hat{A}_\mn terms \; \non \\
&& \;\;\;\;\;\;\;
+ \;\; ( { {\ovl{Q}^2} \over {p \cdot q} } )^2
\ovl{q}^2 \left( { {(p \cdot q)^2} \over {(\ovl{q}^2)^2} }
\left( g_{\mu \nu} - { {q'_\mu q_\nu} \over {\ovl{q}^2} } \right) 
\right. \non \\
&& \;\;\;\;\;\;\;\;\;\;\;
+ \; \left( p_\mu p_\nu
- { {p \cdot q} \over \ovl{q}^2 } (p_\mu q_\nu + q'_\mu p_\nu)
+ { {(p \cdot q)} \over \ovl{q}^2 }^2 q'_\mu q_\nu \right) 
\left(-i \ovl{q}^2 \d{}/d{\ovl{q}^2}\right)^{J\!-\!2}
\tilde{e}_{J\!-\!2,n}^{(i)}(\ovl{q}^2) \; ) \; . \non \\
\eeqs

\nid Note we could not obtain the explicit expression of the $\hat{A}_\mn$
terms because of its general form. However, there is a term of the
correct tensor structure (see equation \ref{eq:decomposition}) 
generated from the second term. This is another indication of the 
Callen-Gross relationship that relates the two invariant amplitudes.  

Now let us define the Wilson coefficients in momentum space $\tilde{E}$ as
\beq
\label{eq:wcoeffdef}
{ {p \cdot q} \over \ovl{Q}^2 } i ({1 \over \ovl{q}^2})^2
\tilde{E}_{J,n}^{(i)} = ( -i \ovl{q}^2 { \partial \over 
{\partial \ovl{q}^2}} )^J \int d^4 z \, e^{-i \bar{q} \cdot z}
E_{J,n}^{(i)},
\eeq 

\noindent where we have pulled out an explicit factor of ${{p \cdot q} 
\over \ovl{Q}^2}$ to make the form of $\tilde{E}$ simple 
(see equation \ref{eq:1storderwcoeff}). We will have
\beqs
\label{eq:opefinal}
{\rm T}_{\mu \nu} \!& \!= \!& \!i^2 \sum_{J,n,i}{\bra{p'}| \hat{O}^{(i)(J,n)} 
|\ket{p}}_{(\mu_0)} \om^{J\!-\!n} \nu^n
\left( ( -g_{\mu \nu} + \frac {q'_\mu q_\nu} {\ovl{q}^2} ) \,
{ {p \! \cdot \! q} \over \ovl{Q}^2} \, \tilde{F'}_{J,n}^{(i)}
(\al_s, {\ovl{Q} \over {\mu_0}}) \right. \;\;\;\;\;\;\; \non \\
& & \left. + \; {1 \over {p \! \cdot \! q}} 
\left( p_\mu p_\nu - \frac {p \! \cdot \! q} {\ovl{q}^2}
(p_\mu q_\nu + q'_\mu p_\nu) + (\frac {p \cdot q} {\ovl{q}^2})^2 
q'_\mu q_\nu \right) \tilde{E}_{J\!-\!2,n}^{(i)}
(\al_s,{\ovl{Q} \over \mu_0}) \right) \, .
\eeqs

\noindent $\tilde{F'}$ should be a linear combination of $\tilde{E}$
and a similarly defined $\tilde{F}$. We have explicitly indicated 
their dependence on the factorization scale $\mu_0$. 

By comparing with equation \ref{eq:decomposition} we finally 
obtain the explicit expression of the invariant amplitudes, in terms of
reduced matrix elements and their corresponding Wilson coefficients, 
as the following:
\beqs
\label{eq:invamp}
\sqrt{1-\zeta} \, {\rm T}_1 = - \sum_{J,n,i}
{\bra{p'}| \hat{O}^{(i)(J,n)} |\ket{p}}_{(\mu_0)} \om^{J\!-\!n} \nu^n 
\tilde{F'}_{J,n}^{(i)}(\al_s, {\ovl{Q} \over \mu_0}) 
\,\,\,\,\,\,\,\,\,\,\,\,\,\,\,\,\,\,\,\,\,\,\,\,\, \\
{ {p \! \cdot \! q} \over M^2} \sqrt{1-\zeta} \, {\rm T}_2 = - \sum_{J,n,i}
{\bra{p'}| \hat{O}^{(i)(J,n)}|\ket{p}}_{(\mu_0)} \om^{J\!-\!n} \nu^n 
\tilde{E}_{J-2,n}^{(i)}(\al_s, {\ovl{Q} \over \mu_0}) \,\,\,. \,\,
\,\,\,\,\,\,\,
\eeqs

We will concentrate on the discussion of ${\rm T}_2$ from now on. We
always regard ${\rm T}_1$ as being obtained from ${\rm T}_2$ from a 
Callen-Gross relationship that can readily be extracted from the
above discussion:
\beq
\label{eq:Callen-Gross}
{\rm T}_1 = {{p \cdot q} \over {M^2}} {\rm T}_2 \,.
\eeq    

\chapter{Renormalization Group Analysis}
\thispagestyle{myheadings}
\markright{}

In this chapter, We perform a renormalization group (RG) analysis of
the non-forward amplitude. We write down the renormalization
group equation (RGE) of the operators and their corresponding 
Wilson coefficients of the operator product expansion (OPE) 
of the amplitude, and find the formal solution to the RGE. Then 
we compute the evolution kernels that will generate the anomalous 
dimensions and discuss their properties. Finally we will calculate 
explicitly the lowest order Wilson coefficients.

\section{Renormalization Group Equation \label{sec:RGE}}

As stated in the last chapter, we will concentrate on the invariant
amplitude ${\rm T}_2$, and ${\rm T}_1$ can be obtained from the
Callen-Gross relationship (\ref{eq:Callen-Gross}). $\T_2$ is analogous
to the structure function $F_2$ ($\nu \W_2$) (see section 
\ref{sec:fw-strfcn}), however, the analogy is not complete because of 
the subtleties in the discussion of dispersion relationship later on. 

Recall that $\T_2$ is given, after OPE, as
\beq
{ {p \! \cdot \! q} \over M^2} \sqrt{1-\zeta} \, {\rm T}_2 
= - \sum_{J,n,i}
{\bra{p'}| \hat{O}^{(i)(J,n)}|\ket{p}}_{(\mu_0)} \om^{J\!-\!n} \nu^n
\tilde{E}_{J-2,n}^{(i)}(\al_s, {\ovl{Q} \over \mu_0}) \,\,\,.
\eeq

\nid It can be regarded as a double distribution function. In the
usual language it is the series sum of ``double'' moments in terms of the
two new moment variable $\om$ and $\nu$ that we defined earlier
(see equation \ref{eq:momvardef}): 
\beq
\label{eq:doubmomsum}
{\rm T}_2 =  \sum_{J,n} \om^{J-n} \nu^n {\rm T}_{2}^{(J,n)},
\eeq

\noindent with the double moments ${\rm T}_{2}^{(J,n)}$ defined as
\beq
\label{eq:doubmomdef}
\sqrt{1-\zeta} { {p \! \cdot \! q} \over M^2} {\rm T}^{(J,n)}_
{2,(\ovl{Q}, \mu_0)}=- \sum_i {\bra{p'}| \hat{O}^{(i)(J,n)}|\ket{p}}
_{(\mu_0)}\tilde{E}_{J-2,n}^{(i)}(\al_s, {\ovl{Q} \over \mu_0})
\eeq

We will eventually analytically continue in $J$, but leave $n$ 
as discrete. The discussion of analyticity properties of $\T_2$ is 
almost identical to that presented for the forward case (see 
section \ref{sec:fw-ope}) 
with minor adjustments. Explicitly writing out the spin sum of 
$\T_{\mu \nu}$ we have (see equations \ref{eq:z&-z} and \ref{eq:ampdef})
\beqs
{\rm T}_{\mu \nu} (p,q',q) &=& i \half \sum_s \int d^4 z 
\, e^{-i \ovl{q} \cdot z} \, 
\bra{p',s} T \, j_{\mu}(-{z \over 2}) j_{\nu} ({z \over 2}) \ket{p,s}
\non \\
&=& i \int d^4 z \, e^{-i q \cdot z}
\bra{p'} T \,j_{\mu}(-z) j_{\nu} (0)  \ket{p} \non \\
&\stackrel{z \rightarrow -z}{\equiv}&
i \int d^4 z \, e^{i q \cdot z}
\bra{p'} T \,j_{\mu}(z) j_{\nu} (0)  \ket{p} \non \\ 
&=& i \sum_r \int d^4 z \, e^{i q \cdot z} 
\{ \, \theta (z_0) \bra{p'} j_{\mu}(z) \ket{r} \bra{r} j_{\nu}(0) \ket{p} 
\non \\
&& + \theta (-z_0) \bra{p'} j_{\nu}(0) \ket{r} \bra{r} j_{\mu}(z) \ket{p}
\, \}
\non \\
&=& i \sum_r \int d^4 z \, e^{i q \cdot z}
\{ \, \theta (z_0) e^{i (p'-p_r) \cdot z}
\bra{p'} j_{\mu}(0) \ket{r} \bra{r} j_{\nu} (0)  \ket{p}
\non \\
&& + \theta (-z_0) e^{-i (p - p_r) \cdot z}
\bra{p'} j_{\nu}(0) \ket{r} \bra{r} j_{\mu} (0)  \ket{p} \, \}
\,,      
\eeqs

\nid where we have used the translation operators (see 
equation \ref{eq:translation}) in the last step. Performing the spacial 
integration yields a $3$-d $\de$-function and we have, similar 
to the forward case,
\beqs
{\rm T}_{\mu \nu} &=& i (2 \pi)^3 \sum_r \int d z_0
( \de^3(\vec{q} + \vec{p}\,' - \vec{p}_r) 
e^{i (q_0 + {p'}_0 - p_{r,0}) \cdot z_0} \theta (z_0)
\bra{p'} j_{\mu}(0) \ket{r} \bra{r} j_{\nu} (0)  \ket{p}  \non \\
&& +  \de^3(\vec{q} - \vec{p} + \vec{p}_r)
e^{i (q_0 - p_0 + p_{r,0}) \cdot z_0} \theta (-z_0)
\bra{p'} j_{\nu}(0) \ket{r} \bra{r} j_{\mu} (0)  \ket{p} ) \non \\
&=& -(2 \pi)^3 \sum_r
\left(  \frac {\de^3(\vec{q} + \vec{p}\,' - \vec{p}_r)
           \bra{p'} j_{\mu}(0) \ket{r} \bra{r} j_{\nu} (0)  \ket{p} }
   {q_0 + {p'}_0 - p_{r,0} +i \eps} \right. \non \\
&& \left. - \frac {\de^3(\vec{q} - \vec{p} + \vec{p}_r)
           \bra{p'} j_{\nu}(0) \ket{r} \bra{r} j_{\mu} (0)  \ket{p}}
   {q_0 - p_0 + p_{r,0} - i \eps} \right) \,.
\eeqs

Again by looking at the poles in $q_0$ we can discuss the
analytic properties of the amplitude. The first term gives rise to poles
at 
\beq
q_0 + {p'}_0 = p_{r,0} \equiv E_r \,,
\eeq

\nid which means that (cf \ref{sec:fw-ope})
\beq
(q_0 + {p'}_0)^2 = E_r^2 = M_r^2 + \vec{p}_r^2 
= M_r^2 + (\vec{q} + \vec{p}\,')^2 \,, \non
\eeq

\nid where in the last step we have used the spacial $\de$-function. $M_r$
is again the invariant mass of the intermediate state $r$. Thus we
have, recall equations \ref{eq:p'q'def} and \ref{eq:pdotq'},
\beq
M_r^2 = (q + p')^2 = (q'+ p)^2 = q'^2 + 2 p \cdot q \,. \non
\eeq

\nid Rewrite $q'^2$ as
\beq
q'^2 = \half (q'^2 + q^2) - \half (q^2 - q'^2) \non
\eeq

\nid and use equations \ref{eq:qdotr} and \ref{eq:momvardef}, we
arrive at
\beqs
\label{eq:q'+psquare}
M_r^2 &=& (q'+p)^2 = \ovl{q}^2 + (2 p - r) \cdot q \non \\ 
&=& \ovl{q}^2 (1 - \om) \,, 
\eeqs

\nid and thus the poles in $\om$ are at 
\beq
\om_r = 1 + \frac {M_r^2} {\ovl{Q}^2} \,.
\eeq

\nid The second term of the spectral expansion of the amplitude gives
poles at
\beq
q_0-p_0 = - p_{r,0} = - E_r \, \non
\eeq

\nid and after similar discussion results in
\beqs
\label{eq:q-psquare}
M_r^2 &=& (q-p)^2 = q^2 -2 p \cdot q = \ovl{q}^2 - (2p-r) \cdot q
\non \\
&=& \ovl{q}^2 (1+\om) \,,
\eeqs

\nid which means that there are also poles in $\om$ at
\beq
\om_r = - 1 - \frac {M_r^2} {\ovl{Q}^2} \,.
\eeq

\nid We see again that the momentum scale is set naturally by 
$\ovl{Q}^2 = - \ovl{q}^2$. 
 
It is clear that in this non-forward case, $\T_\mn$ also has an
analytic circle of unit radius in the $\om$-plane and has a Taylor
expansion in $\om$. The discussion on the dispersion relationship 
for $\T_2$ then follows almost identically to that of the forward 
case (see section \ref{sec:fw-ope})

\nid We can indeed use the dispersion relationship to analytically 
extend the discussion from the region of $\om \leq 1$ where $\T_\mn$ 
is analytic but kinematically unphysical to the physical region of the
Bjorken limit where $0 \leq (\om)^{-1} \leq 1$ in the same manner as
in the forward case. And we find that the inverse of the 
expansion \ref{eq:doubmomsum} is
\beqs
\label{eq:inverse}
{\rm T}_{2}^{(J,n)} = \int_c {{d \om} \over \om} \int_c {{d \nu} \over
\nu} \, ({1 \over \om})^{J\!-\!n} ({1 \over \nu})^n {\rm T}_2
(\ovl{Q}^2, \om, \nu) \non \\
= 2i \int_1^\infty {{d \om} \over \om} \int_c {{d \nu} \over
\nu} \, ({1 \over \om})^{J\!-\!n} ({1 \over \nu})^n {\rm W}_2
(\ovl{Q}^2, \om, \nu) \,,
\eeqs

\noindent where $c$ is any contour in the $\om (\nu)$ plane that encloses 
the origin and ${\rm W}_2= {\rm Im \,T}_2$. Note that in obtaining the 
second equality we have used the fact that ${\rm T}_2$ is even in $\om$. 
Also note that the moment index of $\om$ in this case is, in our
notation, $J-n$.

$\nid {\rm T}_2$ now as a double distribution function is given by a
Mellin transformation:
\beq
\label{eq:mellin}
{\rm T}_2(\ovl{Q}^2, \om, \nu) = \int_c {{dJ} \over {2 \pi i}} 
\sum_{n=0}^{\infty} e^{(J-n) \log \om} \, \nu^n  \, {\rm T}_{2}^{(J,n)}
\eeq

\noindent with the contour $c$ lying parallel to the imaginary axis in the 
$J-n$ plane and to the right of all singularities in that plane.

Under renormalization, the operators scale according to a renormalization
group equation 
\beq
\label{eq:rge-op}
\mu^2 { d \over {d \mu^2}} \hat{O}^{(i)(J,n)} = \sum_{n'=n}^{J}\sum_{i'}
\tilde{\ga}_{nn'}^{ii'} \hat{O}^{(i')(J,n')}
\eeq

\noindent where $\tilde{\ga}$ is the anomalous dimension matrix which 
acts in the product space $J \otimes u_J$. Because the mixing of these
operators is such that $n \rightarrow n' \geq n$ (see 
equation \ref{eq:mixing-dir} the matrix $\tilde{\ga}$ is upper 
triangular in the $J$ dimensional $n$-space.

\nid Since $\T_2$ is a physical quantity it should not depend on the
factorization scale. The Wilson coefficients obey a similar 
renormalization group equation thus cancelling out the scale
dependence of the operators:
\beq
\label{eq:rge-wcoeff}
\mu^2 { d \over { d \mu^2}} \tilde{E}^{(i)}_{J,n}(\al_s(\mu), 
{\ovl{Q} \over \mu}) = - \sum_{n'=0}^n \sum_{i'} \tilde{E}^
{(i')}_{J,n'} \,\,\, \tilde{\ga}_{n'n}^{i'i}.
\eeq

\noindent We can write the solution to this RG equation as
\beq
\label{eq:formalsolu}
\tilde{E}^{(i)}_{J,n}(\al_s(\mu), {\ovl{Q} \over \mu}) = 
\sum_{n'=0}^n \sum_{i'} \tilde{E}^{(i')}_{J,n'}
(\al_s(\ovl{Q}), 1) \,\,\, {\rm M}_{n'n}^{i'i}(\al 
(\ovl{Q}), {\ovl{Q} \over \mu}),
\eeq

\noindent where ${\rm M}$ is a path ordered exponential of the anomalous 
dimension matrix, formally given as
\beq
\label{eq:formalexpo}
{\rm M}_{n'n}^{i'i}(\al_s(\ovl{Q}), {\ovl{Q} \over \mu}) = 
\left( {\cal P} \exp (- \int_{\mu^2}^{\bar{Q}^2} \tilde{\ga}(\al_s 
(\lambda^2)) { {d \lambda^2} \over \lambda^2}) \right)_{n'n}^{i'i} \,.
\eeq

\noindent Therefore by substitution the double moments are given by
\beq
\label{eq:doubmomsolu}
\sqrt{1\!-\!\zeta} { {p \! \cdot \! q} \over M^2}{\rm T}^{(J,n)}_
{2,(\ovl{Q}, \mu_0)}=- \sum_{n'=0}^n \sum_{i'i} \tilde{E}_
{J-2,n'}^{(i')}(\al_s(\ovl{Q}), 1) \,\, 
{\rm M}_{n'n}^{i'i}(\al_s (\ovl{Q}), {\ovl{Q} \over \mu_0}) 
\,\, {\bra{p'}| \hat{O}^{(i)(J,n)}|\ket{p}}_{(\mu_0)} \,.
\eeq 

In principle, for the high energy scattering process, if we know 
the reduced matrix elements at a given momentum scale $\mu$, we can 
calculate the Wilson coefficients order by order in perturbation 
theory at some high momentum scale $Q$, at which the scattering
actually takes place and perturbative QCD is valid, then use the 
anomalous dimensions, also computed to some fixed order in 
perturbation theory, to evolve along the momentum scale to $\mu$ so
that we can use these reduced matrix elements and obtain (theoretical 
predictions of) the invariant amplitude, the amplitude and eventually 
the cross section.

\nid The matrix elements can come from two sources in general. They can
either come from a first-principle calculation in lattice QCD, or from
a high energy experiment. The former is phenomenologically not quite 
feasible yet. For the latter, we actually measure the cross section of 
the high energy scattering at a certain high momentum scale, use the 
above mentioned procedure in reverse and phenomenological models to 
extract the matrix elements, and then by RG analysis we can predict 
the cross section at any other high momentum scale where factorization 
is valid and compare the results with actual experimental data at the
scale.

In the next two sections, we calculate the anomalous dimension
(or rather, evolution kernels from which they can be extracted) and
Wilson coefficients in the lowest non-trivial order.

\section{Anomalous Dimensions \label{sec:adims}}

\begin{figure}[ptb]
\begin{center}
\epsfxsize=5.5in
\epsfysize=2in
\leavevmode
\hbox{ \epsffile{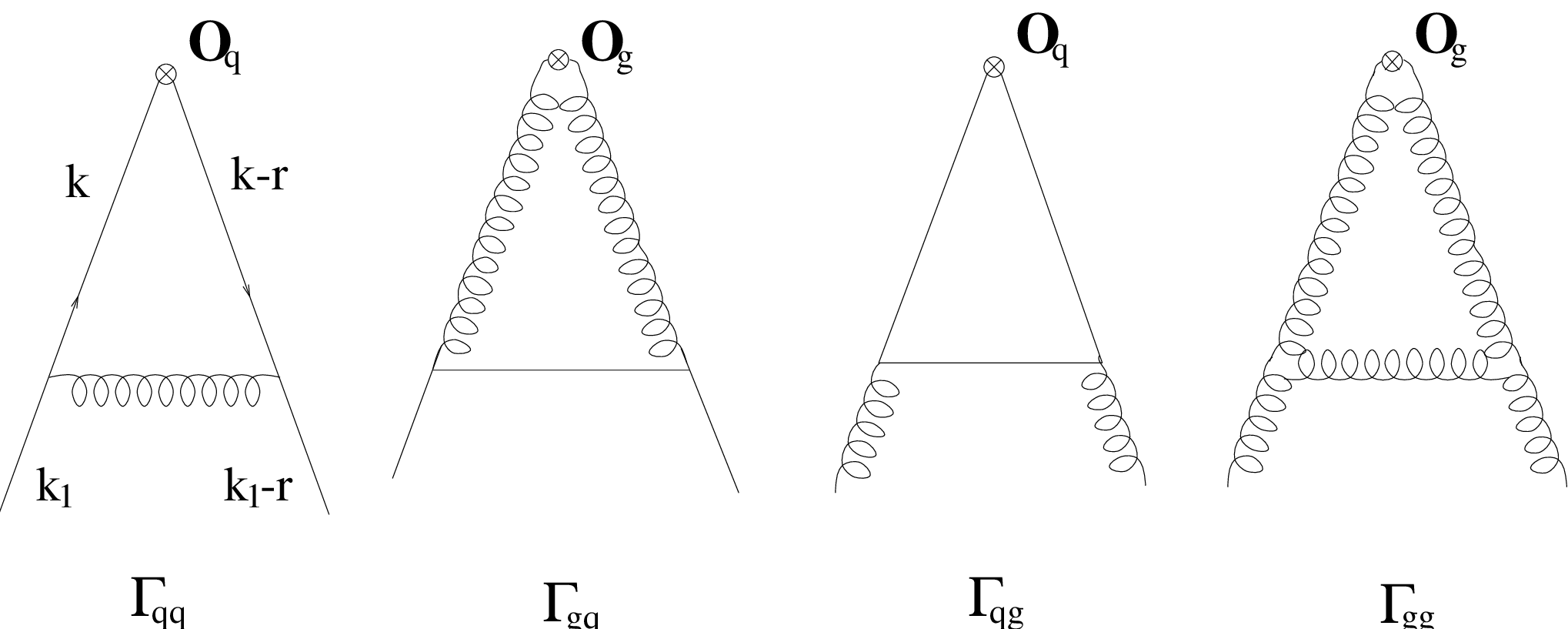}}
\end{center}
\caption{Triangle Graphs for the Anomalous Dimensions}
{These are the four lowest order ``triangle'' diagrams contributing to the
anomalous dimensions. The momentum labels of the four are the same
(only the quark-quark diagram shows them explicitly).}
\label{fig:triangle}
\end{figure}

We will use the light cone (LC) gauge and compute the lowest order
anomalous dimensions in terms of light cone variables. In the LC
gauge, the lowest order anomalous dimensions of the quark and 
gluon operators are generated by the usual triangle diagrams. These
diagrams are shown collectively in figure \ref{fig:triangle}. The
graphs where we have more than two quark/gluon lines meeting at the vertex 
are zero because of the gauge choice. The price we pay is the presence
of extra end point colinear divergences in the integration of light cone
variables, which, as we will see, will cancel out eventually after we
include the self-energy graphs.

\noindent We define the momentum fraction variables (which are the
moment variables) as
\beq
\label{eq:mtvardef}
\om = { k_+ \over k_{1+}}, \,\,\,\,\, \nu = {r_+ \over k_{1+}} \,.
\eeq

\nid  The conventions (cf. ref. \cite{Mueller78})
we use for the quark and gluon vertices are 
\beqs
\label{eq:vertex}
O_q^{(J,n)(+)}(k,k-r) & = & \ga_+(2k-r)_+^{J\!-\!n\!-\!1}r_+^n  \non \\
O_g^{(J,n)\al \be}(k,k-r) & = & 2 g_{\al \be} \, n \cdot k 
\, n \cdot {(k-r)} \, (2k-r)_+^{J\!-\!n\!-\!2}r_+^n \,, \,\,\,\,\,\,
\eeqs

\noindent where $n$ is the LC null vector (see equation \ref{eq:n-def}).

\subsection{Quark-Quark Anomalous Dimension \label{sec:qq}}

\begin{figure}[ptb]
\begin{center}
\epsfxsize=3in
\epsfysize=3.8in
\leavevmode
\hbox{ \epsffile{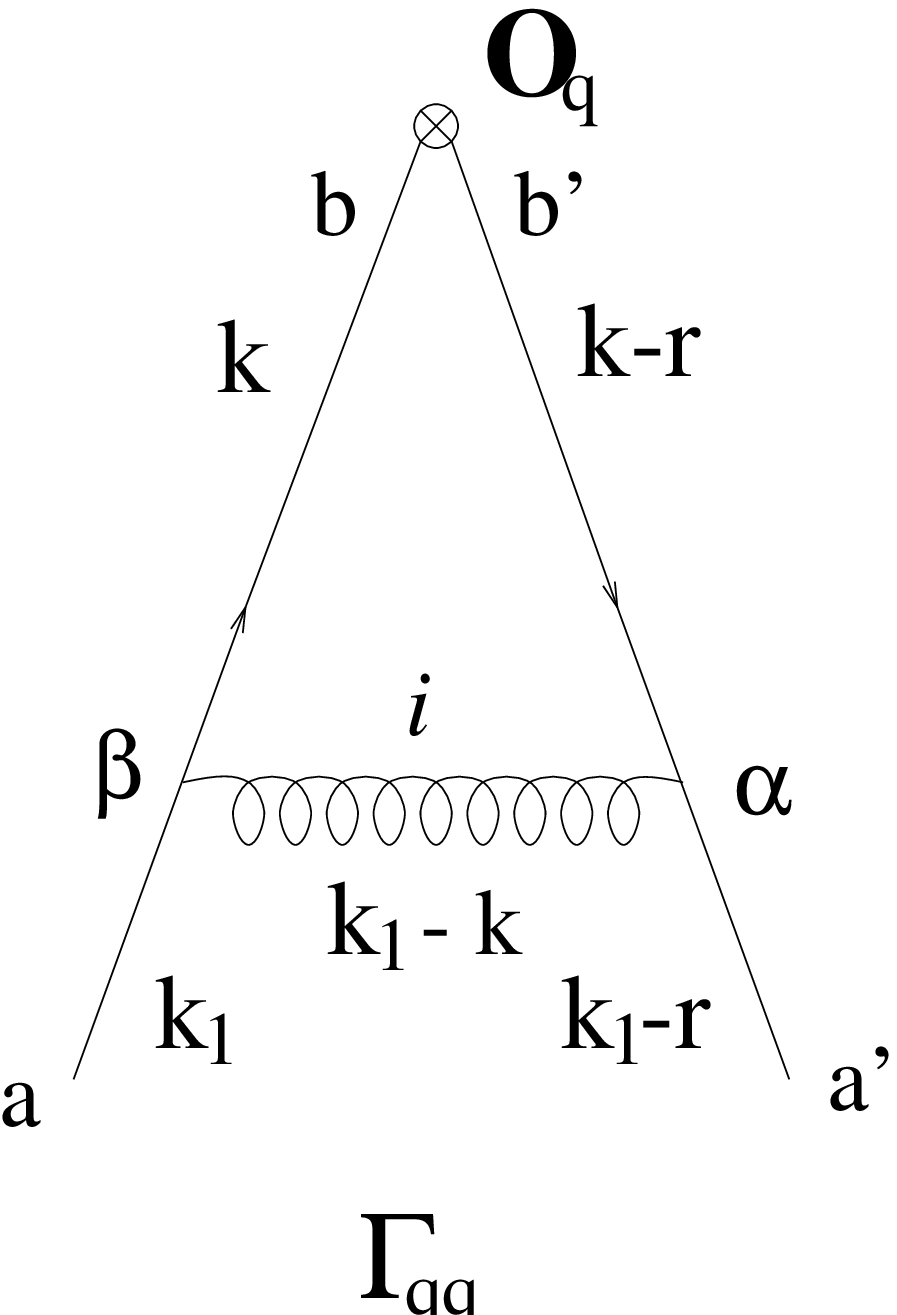}}
\end{center}
\caption{The Quark-Quark Diagram of the Anomalous Dimensions}
{This is the detailed version of the first triangle graph, the
quark-quark diagram. $\al$ and $\be$ are Lorentz indices; $a$, $a'$,
$b$ and $b'$ are color indices for the quarks while $i$ labels the
qluon line. The difference between this diagram and
Figure \ref{fig:fw-qq} in the forward case, besides the
non-forwardness, is that we are not putting the gluon on-shell.}
\label{fig:qq}
\end{figure}

The detailed version of the first diagram in figure \ref{fig:triangle} 
is shown in figure \ref{fig:qq}, where $\al$ and $\be$ are Lorentz 
indices; $a$, $a'$, $b$ and $b'$ are color indices for the quarks 
while $i$ labels the gluon line. In contrast to figure \ref{fig:fw-qq} 
of the forward case, the gluon is not on-shell. Using standard Feynman 
rules and the LC gauge we can write down its value as
\beq
\ga_{qq}^{+,1} = (ig)^2 \sum_{i,b} (T^i_{ab}T^i_{ba'}) 
\int { {d^4 k} \over { (2 \pi)^4}}
\ga_\al {i \over { \not{\!k}-\not{r}}} O_q^{(J,n)(+)}(k,k-r)
{i \over \not{\!k}} \ga_\be { {-i D^{\al \be} (k_1 \!-\! k)} 
\over { (k_1 -k)^2} } \,,
\eeq

\nid which, after the color algebra (see section \ref{sec:fw-gamma})
gives (where we have taken the incoming and outgoing quark to have the
same color index)
\beq
\ga_{qq}^{+,1} = (ig)^2 C_F \int { {d^4 k} \over { (2 \pi)^4}}
\ga_\al {i \over { \not{\!k}-\not{r}}} O_q^{(J,n)(+)}(k,k-r)
{i \over \not{\!k}} \ga_\be { {-i D^{\al \be} (k_1 -k)}
\over { (k_1 -k)^2}} \,,
\eeq

\noindent where the quark vertex $O_q$ is given in (\ref{eq:vertex}) and the
LC gluon projector (numerator of the LC gluon propagator) (see 
equation \ref{eq:LC-gluon})
\beq
D^{\al \be}(k) = g^{\al \be} - { {n_\al k_\be + k_\al
n_\be} \over {n \cdot k}} \,.
\eeq

\nid Again $n$ is the LC null vector.
Writing the integral of the loop momentum in terms of light-cone 
variables (see section \ref{sec:fw-gamma}) we get
\beq
\ga_{qq}^{+,1}= -i { {g^2 C_F} \over { (2 \pi)^4}} \int 
d^2 \perb{k} \, dk_+ dk_- { {(2k-r)_+^{J\!-\!n\!-\!1}r_+^n A(k,r)} 
\over { (k^2 + i \eps) ((k-r)^2 + i \eps) 
((k_1-k)^2 + i \eps)}} \,,
\eeq

\nid where 
\beq
A(k,r)= \ga_\al (\not{\!k}-\not{r}) \ga_+ \not{\!k} \ga_\be
D^{\al \be}(k_1-k) \,.
\eeq

We perform the $k_-$ integration first, in terms of the poles of the
integrand, which come from the denominator factors
\beqs
k^2 + i \eps & = & 2 k_+ k_- - \perb{k}^2 + i \eps \, , \non \\ 
(k-r)^2 + i \eps & = & 2 (k - r)_+ k_- - \perb{k}^2 + i \eps \, ,\\
(k_1-k)^2 + i \eps & = & 2 (k-k_1)_+ k_- - \perb{k}^2 + i \eps \,. \non
\eeqs

However, the positions of poles in $k_-$ are different when 
$k_+$ is in different regions. More specifically,
there are four regions we need to consider:

\nid 1) $k_+ < 0$, by setting the denominators to zero we find that the
poles are at, respectively,
\beq
k_- = { \perb{k}^2 \over {2 k_+} } + i\eps \, , \,\,\,\,\,
k_- = { \perb{k}^2 \over {2 (k-r)_+} } + i\eps \, , \,\,\,\,\,
k_- = { \perb{k}^2 \over {2 (k-k_1)_+} } + i\eps \, .
\eeq

\begin{figure}[ptb]
\begin{center}
\epsfxsize=5.5in
\epsfysize=5.7in
\leavevmode
\hbox{ \epsffile{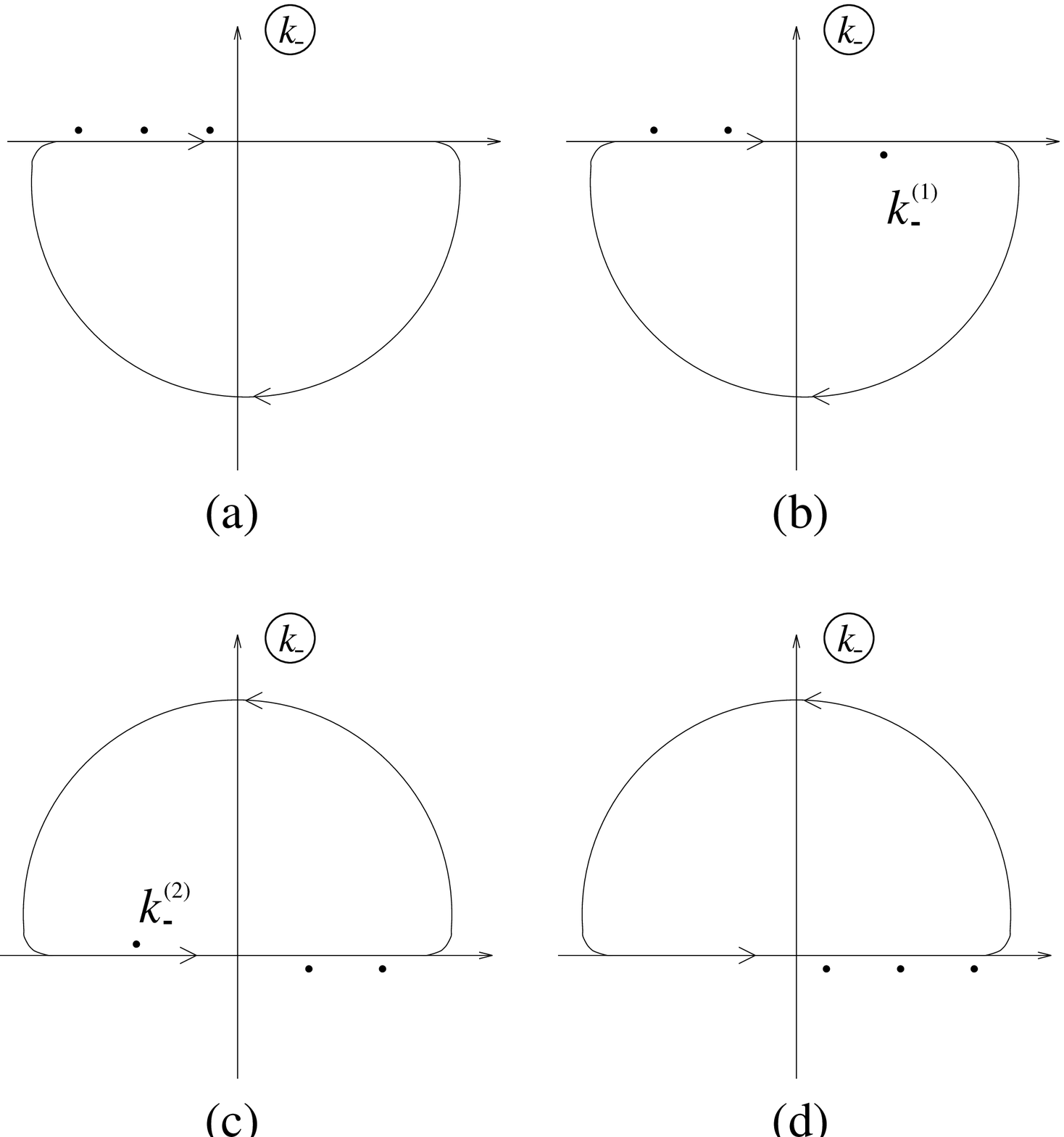}}
\end{center}
\caption{The Poles and Contours of $k_{-}$ Integration}
{The figures show, on the $k_-$ complex plane, the poles and different 
contours of $k_-$ integration in the four regions of its value 
(see section \ref{sec:qq}). The dots are the pole positions with their 
distances to the real $k_-$ axis exaggerated.}
\label{fig:poles}
\end{figure}

\nid All three poles lie above the real axis of the complex $k_-$
plane (see figure \ref{fig:poles} (a)). We can complete the contour 
of integration of $k_-$ in the 
lower-half-plane and clockwise, enclosing no poles and thus the 
integration gives zero value.  

\nid 2) $0 < k_+ < r_+ (< k_{1+}) $, the poles are at, respectively, 
\beq
k_- = { \perb{k}^2 \over {2 k_+} } - i\eps \, , \,\,\,\,\,
k_- = { \perb{k}^2 \over {2 (k-r)_+} } + i\eps \, , \,\,\,\,\,
k_- = { \perb{k}^2 \over {2 (k-k_1)_+} } + i\eps \, .
\eeq

\nid The first pole is below the real axis while the other two are
above it (see figure \ref{fig:poles} (b)). We can complete the 
integration contour again in the
lower-half-plane and clockwise, picking up the pole
at $k_-= {{\perb{k}^2} \over {2 k_+}} - i \eps 
\equiv k_-^{(1)}$. The value of the integration would then be the
residue of the $k_-^{(1)}$ pole with a factor of $-2 \pi i$.

\nid 3) $r_+ < k_+ < k_{1+} $, the poles are at, respectively, 
\beq
k_- = { \perb{k}^2 \over {2 k_+} } - i\eps \, , \,\,\,\,\,
k_- = { \perb{k}^2 \over {2 (k-r)_+} } - i\eps \, , \,\,\,\,\,
k_- = { \perb{k}^2 \over {2 (k-k_1)_+} } + i\eps \, .
\eeq

\nid The first two poles are below the real axis while the third one is 
above it (see figure \ref{fig:poles} (c)). We can complete the 
integration contour now in the
upper-half-plane and counterclockwise, picking up the pole
at $k_-= { {\perb{k}^2} \over {2 (k_+ - k_{1+})}} + i \eps 
\equiv k_-^{(2)}$.  The value of the integration would then be the
residue of the $k_-^{(2)}$ pole with a factor of $2 \pi i$.

\nid 4) $k_+ > k_{1+}$, the poles are at, respectively,
\beq
k_- = { \perb{k}^2 \over {2 k_+} } - i\eps \, , \,\,\,\,\,
k_- = { \perb{k}^2 \over {2 (k-r)_+} } - i\eps \, , \,\,\,\,\,
k_- = { \perb{k}^2 \over {2 (k-k_1)_+} } - i\eps \, .
\eeq

\nid All three poles lie below the real axis of the $k_-$ plane (see
figure \ref{fig:poles} (d)). We can 
complete the contour of integration of $k_-$ in the upper-half-plane and
counterclockwise, again enclosing no poles and the integration gives
zero value.  

Thus the $k_+$ integration should be performed in two regions: 
\beqs
\ga_{qq}^{+,1} \!&\! = \!&\! -i { {g^2 C_F} \over { (2 \pi)^4}} \int \!
d^2 \perb{k} \, \left( \int_0^{r_+} \!\! dk_+ \!+\! 
\int_{r_+}^{k_{1+}} \!\! dk_+
\right) \int \! { {  dk_- \,\, (2k-r)_+^{J\!-\!n\!-\!1}r_+^n \,\, A(k,r)} 
\over { (k^2 + i \eps) ((k-r)^2 + i \eps) 
((k_1-k)^2 + i \eps)}}  \non \\
\!&\! = \!&\! -i { {\al_s C_F} \over { (2 \pi)^2}} \int 
d \perb{k}^2 \, \left( \int_0^{r_+} dk_+ + \int_{r_+}^{k_{1+}} dk_+
\right) (2k-r)_+^{J\!-\!n\!-\!1}r_+^n \non \\
& & \!\! \cdot \, \int dk_- { {A(k,r)} 
\over { (2 k_+ k_- \!- \perb{k}^2 + i \eps) 
(2 (k \!-\! r)_+ k_- \!- \perb{k}^2 + i \eps) 
(2 (k \!-\! k_1)_+ k_- \!- \perb{k}^2 + i \eps)}} \,. \non \\
\eeqs

\nid Evaluating the contour integral of $k_-$ at the poles specified
above, we have 
\beqs
\ga_{qq}^{+,1} \!\!&\!\! = \!\! & \!\! 
\!-i { {\al_s C_F} \over { (2 \pi)^2}} 
\int \! d \perb{k}^2 \! \left[ \int_0^{r_+} \!\! dk_+ 
{ {(-2 \pi i)} \over {2 k_+} } 
\left( { {(2k-r)_+^{J\!-\!n\!-\!1}r_+^n \, A(k,r) } 
\over { (2 (k \!-\! r)_+ k_- \!- \perb{k}^2 ) 
(2 (k \!-\! k_1)_+ k_- \!- \perb{k}^2) }} 
\right) \! \left|_{k_- = k^{(1)}_-} \right. \right. \non \\
& & + \int_{r_+}^{k_{1+}} dk_+
{ {(2 \pi i)} \over {2(k-k_1)_+} } \,
\left. \left( { {(2k-r)_+^{J\!-\!n\!-\!1}r_+^n \, A(k,r) } 
\over { (2 k_+ k_- \!- \perb{k}^2 ) 
(2 (k \!-\! r)_+ k_- \!- \perb{k}^2 )}} 
\right) \! \left|_{k_- = k^{(2)}_-} \right. \right] \, .
\eeqs

\nid Substituting the pole values into the denominators and simplify
them first, we get, for the pole $k_-^{(1)} = {{\perb{k}^2} 
\over {2 k_+}} \, $,
\beqs
&& (2 (k \!-\! r)_+ k_- \!- \perb{k}^2 ) 
(2 (k \!-\! k_1)_+ k_- \!- \perb{k}^2) 
\! \left|_{k_- = k^{(1)}_-} \right. \non \\
&& = \left( \perb{k}^2 { {(k \!-\! r)_+} \over {k_+}} - \perb{k}^2 \right)
\!\!\left( \perb{k}^2 { {(k \!-\! k_1)_+} \over {k_+}} - \perb{k}^2 \right) 
= (\perb{k}^2)^2 \! \left( 1 -  { {(k \!-\! r)_+} \over {k_+} } \right) \!\!
\left(1 + { {(k_1 \!-\! k)_+} \over {k_+} } \right) \non \\
&& = { 1 \over {\perb{k}^4} } \, { k_+ \over { (k_+ - (k \!-\! r)_+)
(k_+ + (k \!-\! r)_+) } } =  { 1 \over {\perb{k}^4} } \, 
{ k_+ \over {r_+ k_{1+}} } \, ,
\eeqs

\nid while for the pole $k_-^{(2)} = { {\perb{k}^2} \over {2 (k_+ -
k_{1+})}} \, $,
\beqs
&&(2 k_+ k_- \!- \perb{k}^2 ) 
(2 (k \!-\! r)_+ k_- \!- \perb{k}^2 ) 
\! \left|_{k_- = k^{(2)}_-} \right. \non \\
&& = ( \perb{k}^2 { {k_+} \over {(k \!-\! k_1)_+}} - \perb{k}^2 )
( \perb{k}^2 { {(k \!-\! r)_+} \over {(k \!-\! k_1)_+} } - \perb{k}^2)
= (\perb{k}^2)^2 \left( 1 + { {k_+} \over {(k_1 \!-\! k)_+} } \right)
\left( 1 +  {{(k \!-\! r)_+} \over {(k_1 \!-\! k)_+} } \right) \non \\
&& = { 1 \over {\perb{k}^4} } { {(k_1 \!-\! k)_+} \over
{ ((k_1 \!-\! k)_+ + k_+) ((k_1 \!-\! k)_+ + (k \!-\! r)_+)} }
=  { 1 \over {\perb{k}^4} } { {(k_1 \!-\! k)_+} \over
{(k \!-\! r)_+ k_{1+}} } \,.
\eeqs

\nid Thus we arrive at
\beqs
\label{eq:gamma3}
\ga_{qq}^{+,1}= -i {{\al_s C_F} \over { (2 \pi)^2}} 
\int { {d^2 \perb{k}} \over {\perb{k}^4}} \left(
{ { -2 \pi i} \over {2 r_+ k_{1+}}} \int_0^{r_+}
dk_+ k_+ A^{(1)} (2k-r)_+^{J\!-\!n\!-\!1}r_+^n \, \right. 
\,\,\,\,\,\,\,\,\,\,\,\,\,\,\,\,\,\,\,\,\,\,\,\,\,\, \nonumber \\
\,\,\,\,\,\,\,\,\,\,\,\,\,\,\,\,\,\,\,\,\,\,\,\,\,\,\,\,\,\, \left. 
+ { {2 \pi i} \over { -2 k_{1+} (k_1-r)_+}} \int_{r_+}^{k_{1+}}
dk_+ (k_1-k)_+ A^{(2)} (2k-r)_+^{J\!-\!n\!-\!1}r_+^n \right) ,
\eeqs

\noindent where $A^{(1)}$ and $A^{(2)}$ are $A(k,r)$ evaluated at 
the poles $k_-^{(1)}$ and $k_-^{(2)}$, respectively.

We are interested in the logarithmic divergence in the graphs so in 
computing the Dirac spinors we keep only the leading powers of 
$\perb{k}$. This leading power turns out to be quadratic, which
means we will keep in the following calculation of $A^{(1)}$ and
$A^{(2)}$ only terms that are proportional to $\perb{k}^2$.

Defining $\om$ and $\nu$ as before in (\ref{eq:mtvardef}), we have, 
at the poles,
\beqs
\label{eq:poles}
k_-^{(1)} = {{\perb{k}^2} \over {2 k_+}} 
= { {\perb{k}^2} \over {2 k_{1+} \om} } \,\,\,\;\;\;\;\; & \& & \,\,
k^2 = 2 k_+ k_- - \perb{k}^2 =  0 , \non \\
k_-^{(2)} = { {\perb{k}^2} \over {2 (k_+ \!-\! k_{1+})}} 
= { {-\perb{k}^2} \over { 2 k_{1+} (1 \!-\! \om)} } \,\,\,
& \& & \,\, k^2 = - { {\perb{k}^2} \over { (1 - \om) } }. \non
\eeqs

\nid Recall that $A(k,r)$ can be written into two terms:
\beqs
\label{eq:A-11}
A(k,r) & = & \ga_\al (\not{\!k}-\not{r}) \ga_+ \not{\!k} \ga_\be
D^{\al \be}(k_1-k) \non \\
&=& \ga_\al \not{\!k} \ga_+ \not{\!k} \ga_\be
D^{\al \be}(k_1-k) 
- \ga_\al \not{r} \ga_+ \not{\!k} \ga_\be
D^{\al \be}(k_1-k) \non \\
&\equiv& A_I - A_{II} 
\eeqs

Using the anti-commutation relationships of the Dirac matrices in
light cone variables (see equations \ref{eq:anticommudef} and 
\ref{eq:anticommu}) we can reduce the first term ($A_I$) to
\beqs
\ga_\al \not{\!k} \ga_+ \not{\!k} \ga_\be
D^{\al \be}(k_1-k) 
&=& 2 k_+ \ga_\al \not{\!k} \ga_\be D^{\al \be}(k_1-k)
- k^2 \ga_\al \ga_+ \ga_\be D^{\al \be}(k_1-k) \,.
\eeqs

\nid At the pole $k_-^{(1)}$, because $k^2 = 0$, we have
\beq 
A_I = 2 k_+ \ga_\al \not{\!k} \ga_\be D^{\al \be}(k_1-k)
\equiv A_I^{(1)} \,.
\eeq

\nid Expanding $D^{\al \be}(k_1-k)$ and again using anti-commutation
rules and the fact that $k^2 = 0$  we have
\beqs
\label{eq:A-12}
A_I^{(1)}
\!&=\!& \!2 k_+ \ga_\al \not{\!k} \ga^\al
- { {2 k_+} \over { k_{1+}(1  -  \om)} }
(\ga_+ \not{\!k} (\not{\!k}_1-\not{\!k}) + (\not{\!k}_1-\not{\!k})\not{\!k}
\ga_+) \non \,\,\,\,\,\,\, \\
\!&=\!& \! -4 k_+ \not{\!k} - { {2 \om} \over {1 \! - \! \om} }
(\ga_+\not{\!k}\not{\!k}_1 + \not{\!k}_1\not{\!k}\ga_+ - 2 \ga_+ k^2)
\non \\
\!&=\!& \!-4k_+(k_+ \ga_- + k_- \ga_+ - \perb{k} \cdot \perb{\ga})
- { {2 \om} \over {1 \! - \! \om} }
(\ga_+ \! \not{\!k}\not{\!k}_1 + \not{\!k}_1 \!\not{\!k}\ga_+) \,. 
\eeqs

\nid As stated before, we want to keep only the leading power in
$\perb{k}$. It is clear that in (\ref{eq:A-12}) the leading power of
$\perb{k}$ is quadratic and can only come from a term that contains 
the factor $k_-$. In addition, since we are computing the quark-quark
anomalous dimension, only terms that have similar vertex structure as
the original quark vertex $O_q$ in (\ref{eq:vertex}) contribute. Thus,
we keep in (\ref{eq:A-12}) only terms that are proportional to 
$\ga_+ k_-$. By expanding the dot products and inspecting the
result, it is straight forward to see that only the second term in
the first parenthesis survives and we have
\beq
A_I^{(1)} = -4 k_+ k_- \ga_+ = -2 \perb{k}^2 \ga_+ \,,
\eeq

\nid where we have also used (\ref{eq:gammasquare}).

The second term of (\ref{eq:A-11}) can be similarly reduced by keeping
only the terms proportional to $\ga_+ k_-$:
\beqs
A_{II}^{(1)} &=& \ga_\al \not{r} \ga_+ \not{\!k} \ga_\be
D^{\al \be}(k_1-k) \non \\
&=& \ga_\al \not{r} \ga_+ \not{\!k} \ga^\al 
- {1 \over { (k_1 - k)_+ }} 
(\ga_+ \not{r} \ga_+ \not{\!k} (\not{\!k}_1 - \not{\!k})
+ (\not{\!k}_1 - \not{\!k}) \not{r} \ga_+ \not{\!k} \ga_+) \non \\
&=& 0 - {1 \over { (k_1 - k)_+ } }
(2 r_+ \ga_+ ( \not{\!k} \not{\!k}_1 - k^2 ) 
+ (\not{\!k}_1 \not{r} - \not{\!k} \not{r} ) 2 k_+ \ga_+) \non \\
&=& - {1 \over { (k_1 - k)_+ }}
(2 r_+ \ga_+ ( \ga_- k_+ \not{\!k}_1 - 0 )
+ (0 -  \not{\!k} \not{r} ) 2 k_+ \ga_+) \non \\
&=& - {1 \over { (k_1 - k)_+ }} ( 0 - 2 k_+ \not{\!k} \not{r} \ga_+)
= {1 \over { (k_1 - k)_+ }} 2 k_+ k_- r_+ \ga_+ \ga_- \ga_+
\non \\
&=& { { 2 r_+ } \over { (k_1 - k)_+ } } \perb{k}^2 \ga_+
= { {2 \nu} \over { 1 - \om} } \perb{k}^2 \ga_+ \,. 
\eeqs 

\nid Note some of the zeros above do not mean numerically zero but
rather the contributions of those terms are zero. We use zero ($0$) 
this way through out the discussion unless otherwise stated.

\nid Combining the results we have
\beq
A^{(1)} = -2 \perb{k}^2 \ga_+ (1 + {\nu \over {1 -\om}}) \,.
\eeq

The evaluation of $A^{(2)}$ goes parallel but is a little bit more
involved since we no longer have $k^2 = 0$ but rather
$ k^2 = - { {\perb{k}^2} \over { (1 - \om) } } $. We outline the
main steps in the following:
\beqs
A^{(2)}_{I} &=& \ga_\al \not{\!k} \ga_+ \not{\!k} \ga_\be
D^{\al \be}(k_1-k) \non \\
&=& 2 k_+ \ga_\al \not{\!k} \ga_\be D^{\al \be}(k_1-k)
- k^2 \ga_\al \ga_+ \ga_\be D^{\al \be}(k_1-k) 
\non \\
&=& 2 k_+ \ga_\al \not{\!k} \ga^\al
- { {2 k_+} \over { k_{1+}(1  -  \om)} }
(\ga_+ \not{\!k} (\not{\!k}_1-\not{\!k}) + (\not{\!k}_1-\not{\!k})\not{\!k}
\ga_+) \non \\
&& - k^2 \ga_\al \ga_+ \ga^\al
+ { k^2 \over  { k_{1+}(1  -  \om)} }
(\ga_+ \ga_+  (\not{\!k}_1-\not{\!k}) +  (\not{\!k}_1-\not{\!k})
\ga_+ \ga_+) \non \\
&=& -4 k_+ \not{\!k} - { {2 \om} \over {1 \! - \! \om} }
(\ga_+\not{\!k}\not{\!k}_1 + \not{\!k}_1\not{\!k}\ga_+ - 2 \ga_+ k^2)
+ 2 k^2 \ga_+ + 0 \,. \non 
\eeqs

\nid Using identical arguments as those in calculation of $A^{(1)}$, plus
the fact that now the leading contribution (of $\perb{k}^2$) can also 
come from a $k^2$ factor in addition to $k_-$, we arrive at
\beqs
A^{(2)}_{I} &=& -4 k_+ k_- \ga_+ - 
{ {2 \om} \over {1 \! - \! \om} } (-2 \ga_+ k^2)
+  2 k^2 \ga_+ \non \\
&=& - 4 k_+  { {- \perb{k}^2} \over {2 (k_+ - k_{1+})}} \ga_+
+ ({{4 \om} \over {1 \! - \! \om}} + 2) \ga_+ 
(- { {\perb{k}^2} \over { (1 - \om) } }) \non \\
&=& { {\perb{k}^2} \over { (1 - \om)^2 } } \ga_+
(2 \om ( 1 - \om) -4 \om -2 (1-\om)) \non \\
&=& -{ {2 \perb{k}^2 \ga_+} \over { (1 - \om)^2 } }
(1 + \om^2) \,.
\eeqs

\nid On the other hand,
\beqs
A^{(2)}_{II} &=& \ga_\al \not{r} \ga_+ \not{\!k} \ga_\be
D^{\al \be}(k_1-k) \non \\
&=& \ga_\al \not{r} \ga_+ \not{\!k} \ga^\al 
- {1 \over { (k_1 - k)_+ }} 
(\ga_+ \not{r} \ga_+ \not{\!k} (\not{\!k}_1 - \not{\!k})
+ (\not{\!k}_1 - \not{\!k}) \not{r} \ga_+ \not{\!k} \ga_+) \non \\
&=& 0 - {1 \over { (k_1 - k)_+ } }
(2 r_+ \ga_+ ( \not{\!k} \not{\!k}_1 - k^2 ) 
+ (\not{\!k}_1 \not{r} - \not{\!k} \not{r} ) 2 k_+ \ga_+) \non \\
&=& - {1 \over { (k_1 - k)_+ }}
(2 r_+ \ga_+ ( \ga_- k_+ \not{\!k}_1 - k^2 )
+ (0 -  \not{\!k} \not{r} ) 2 k_+ \ga_+) \non \\
&=& - {1 \over { (k_1 - k)_+ }} ( -2 r_+ \ga_+ k^2
 - 2 k_+ k_- r_+ \ga_+ \ga_- \ga_+) \non \\
&=& { {2 r_+} \over { (k_1 - k)_+ }} \ga_+ 
 { {- \perb{k}^2} \over { (1 - \om) } }
+ { { 4 k_+} \over  { (k_1 - k)_+ }} \ga_+ r_+
{ {\perb{k}^2} \over {2 (k_+ - k_{1+})}} \non \\
&=& -2 \perb{k}^2 \ga_+ \left( {{r_+} \over {k_{1+}(1-\om)^2}}
+ {{k_+ r_+} \over {k_{1+}^2(1-\om)^2}} \right) \non \\
&=& - { {2 \nu \perb{k}^2 \ga_+} \over {(1-\om)^2} } ( 1 + \om ) \,.
\eeqs 

\nid Again combining the results we have
\beq
A^{(2)}  = - { {2 \perb{k}^2 \ga_+} \over {(1 - \om)^2}}
(1 + \om^2 - \nu (1 + \om)) \,.
\eeq

\nid Thus we have
\beqs
\label{eq:qq}
\ga_{qq}^{+,1}  =  { \al_s \over {2 \pi}} C_F \int \,
{{d \perb{k}^2} \over \perb{k}^2} \ga_+ k_{1+}^{J-1}\nu^n \,
\left( {1 \over \nu} \int_0^\nu \, d\om \, (2\om-\nu)^{J\!-\!n\!-\!1}
\, \om \, (1+ {\nu \over {1-\om}}) \right. 
\,\,\,\,\,\,\,\,\,\,\,\,\,\,\, \nonumber \\
 \left. \,\,\,\,\,\,\,\,\,\,\,\,\,\,  
+ {1 \over {1\!-\!\nu}} \int_\nu^1 \, d\om \, 
(2\om-\nu)^{J\!-\!n\!-\!1} \, { {1+\om^2-\nu(1+\om)} \over 
{1-\om}} \right) \,.
\eeqs

The $\om$ integration is divergent at $\om=1$. This 
divergence is cancelled by the self-energy graphs, the value of which
in LC gauge can be readily taken from \cite{Curci80} as
\beq
\label{eq:selfE}
Z_F(x)=1+ {{\al_s C_F} \over {2 \pi}} { 2 \over \eps}
( -2 I_0 - 2 \log{|x|} + {3 \over 2}) \,,
\eeq

\noindent where $I_0$ is the colinear divergence
\beq
\label{eq:diver}
I_0 = \int_0^1 { {dz} \over z} = \int_0^1 { {dz} \over {1-z}} \,.
\eeq

\nid This divergence can be viewed as essentially an artifact coming 
from the choice of using light cone calculations. For a detailed
treatment please refer to \cite{Curci80}. Here we simply state and use
their conclusions.

\noindent $Z_F$ depends on the longitudinal momentum fraction $x$. 
So in our case the self-energy contribution from the $k_1-r$ line 
should have $x= \frac {n \cdot (k_1 -r)} {n \cdot k_1} = 1 -\nu$ while the
$k_1$ line has simply $x=1$. Adding $ {1 \over 2} ( Z_F(1) + Z_F(1-\nu))$ 
to (\ref{eq:qq}) and identifying the logarithmic divergence 
$\int { {d \perb{k}^2} \over { \perb{k}^2}} = 
{ 2 \over \eps}$ we will obtain the expression of $\ga_{qq}$ as in 
(\ref{eq:gammaresult}), namely,
\beqs
\label{eq:qqf}
\ga_{qq}^{(\!+\!)}\!& = & \!{ \al_s \over {2 \pi}} C_F \, \ga_+ 
k_{1+}^{J-1}\nu^n \, \int_0^1 d \om \, (2\om-\nu)^{J\!-\!n\!-\!1}
\left( \Theta(\nu \! - \! \om) \, {1 \over \nu} \, \om \, 
(1+ {\nu \over {1-\om}}) \right. \nonumber \\
& & \,\,\,\,\,\,\,\,\,\,\,\,
\left. + \, \Theta( \om \! - \! \nu ) {1 \over {1\!-\!\nu}} 
{ {1+\om^2-\nu(1+\om)} \over {1-\om}} 
+ \de(1\!-\!\om) (-\! 2 I_0 - \log(1\!-\!\nu) + {3 \over 2})
\right) \,. \non \\
\eeqs

\nid Note there are no other diagrams contributing in LC gauge. Recall
this is because the gauge choice makes diagrams that have more than
two lines meeting at the vertex vanish.

\subsection{Anomalous Dimension Kernels \label{sec:kernels}}

While the details of the computation of the other three anomalous
dimensions are shown in the Appendices, we list the result in the
following:
\beqs
\label{eq:gammaresult}
\ga_{qq}^{(\!+\!)}\!& = & \!{ \al_s \over {2 \pi}} C_F \, \ga_+ 
k_{1+}^{J-1}\nu^n \, \int_0^1 d \om \, (2\om-\nu)^{J\!-\!n\!-\!1}
\left( \Theta(\nu \! - \! \om) \, {1 \over \nu} \, \om \, 
(1+ {\nu \over {1-\om}}) \right. \nonumber \\
& & \,\,\,\,\,\,\,\,\,
\left. + \, \Theta( \om \! - \! \nu ) {1 \over {1\!-\!\nu}} 
{ {1\!+\!\om^2\!-\!\nu(1\!+\!\om)} \over {1-\om}} 
+ \de(1\!-\!\om) (-\! 2 I_0 - \log(1\!-\!\nu) + {3 \over 2})
\right) \,, \non \\
\ga_{gq}^{(\!+\!)}\!& = & \!{ \alpha_s \over {2 \pi}} C_F \,
\ga_+
k_{1+}^{J-1}\nu^n \! \int_0^1 \! d \om (2
\om\!-\!\nu)^{J\!-\!n\!-\!2}
\left( \Theta(\nu \! - \! \om) {1 \over \nu} (\om^2\!-\!2\om)
\right. \,\,\,\,\,\,\,\,  \non \\
& & \,\,\,\,\,\,\,\,\,\,\,\,\,\,\,\,\,  \left.
\!- \Theta( \om \! - \! \nu ){1 \over {1\!-\!\nu}}
(1\!+\!(1\!-\!\om)^2\!-\!\nu ) \right) \,, \non \\
\ga_{qg}^{ij} \! &=& \!{ \al_s \over {2 \pi}} {1 \over 2} 2g^{ij} 
k_{1\!+}^J\nu^n \! \int_0^1 \! d \om (2 \om\!-\!\nu)^{J\!-\!n\!-\!1}
\!\left(\!- \Theta(\nu \! - \! \om){1 \over \nu} \om 
(2\om\!-\!1\!-\!\nu)\! \right.  \\
&& \,\,\,\,\,\,\,\,\,\,\,\,\,\,\,\,\,\,\,\,\,\,\,\,\,\,  \left. 
+ \Theta( \om \! - \! \nu ){1 \over {1\!-\!\nu}} 
(\om^2 \!+\! (1\!-\!\om)^2\!-\!\om \nu)\! \right) \,, \non \\
\ga_{gg}^{ij} \!&\! =\!& \!{ {\al_s C_A} \over {2 \pi}} 2g^{ij} 
k_{1\!+}^J\nu^n \!\int_0^1\!\! d \om (2\om\!-\!\nu)^{J\!-\!n\!-\!2}
\!\left(\!\Theta(\nu \! - \! \om) {1 \over 2} 
({1 \over \nu}(4\om^3\!\!-\!\om^2\!\!+\!4\om) \non \right.\\
\!&\!+\!&\!\!\! ({{\om^3\!\!+\!\om^2\!\!-\!2\om^2\nu} \over {1-\om}}\!-\!
3(\om^2\!+\!\om))) + { {\Theta( \om \! - \! \nu )} \over {1\!-\!\nu}} 
({{2(1\!-\!\om\!+\!\om^2)^2} \over {1-\om}}\!+\! \nu \!
{{\nu(1\!+\!\om^2)\!-\!2(1\!+\!\om^3)} \over {1-\om}}) \non \\
& &  \,\,\,\,\,\,\,\,\,\,\,\,\,\,\,\,\,\,\,\,\,\,\,
\left. \!+ (1\!-\!\nu) \delta(1\!-\!\om) 
( \!-\!2I_0\!-\!\log(1\!-\!\nu)\!+\!{ b_0 \over {2C_A}}) \right) \,. \non
\eeqs

\noindent We have included virtual corrections coming from the 
self-energy graphs that will cancel the colinear singularities at
end point in the $\om$ integration ($I_0$ terms, see 
equation \ref{eq:diver}
for the definition of $I_0$) and give correct constant terms in
the anomalous dimensions. $\Theta$ is the usual step function and
\beq
\label{eq:b0def}
b_0= 11- {2 \over 3} n_f
\eeq

\nid  is the leading coefficient of the QCD  $\be$-function. 

\subsection{Comments on Anomalous Dimensions}

Some comments are in order. It is straight forward to show that in the 
forward limit where $\nu=0$, equation \ref{eq:gammaresult} reduces to the 
conventional 
Altarelli-Parisi splitting functions \cite{A&P77}. However, because of the 
non-forwardness, there is no simple probability interpretation for these 
evolution kernels as splitting functions.

The reason we have two terms for each triangle graph is that for different 
$k_+$ integration regions pinching of the $k_-$ pole is different (see 
section \ref{sec:qq}). Thus there is no straight forward optical-theorem type 
dispersion relationship between the cross section and the imaginary part 
of the amplitude. Nonetheless, we can still do analytic continuation in 
$J$ and relate the matrix elements to ``double parton distributions'' 
\cite{Christ&Mueller72} which now may not have a direct probability 
interpretation. 
The second terms of our $\ga$ differ from those of \cite{Frankf&Strikman98} 
only because the factor $n \cdot k \, n \cdot (k-r)$ in our convention
for the gluon vertex (see equation \ref{eq:vertex})
is not included in the definition of their gluon vertex. Once we
take this into account and use the same vertices, their results are
identical to the second terms of the corresponding kernels of ours. 
\footnote{
Simultaneous to our work, Blumlein {\it et al} \cite{Blumlein97} also 
calculated the evolution kernels of what they call ``twist 2 light-ray 
operators for unpolarized and polarized DIS''. It would be useful to make 
a detailed comparison to see whether the seeming deviation between our 
results is a nontrivial disagreement or mainly due to different notations 
and approaches. Because of these differences in notation and approaches,
together with the fact that their main results $eqs.(15)-(18)$ were 
stated but not derived, we are not able to make such a comparison at the 
present time.}

After we perform the $\om$ ( i.e.\ $k_+$) integral, we can in principle
obtain the corresponding anomalous dimension matrix in the moment
space. At first sight, the $(1-\nu)^{-1}$ factor in the second terms
may generate a series of infinite sums over powers of $\nu$, which will 
spoil the locality 
of the vertex thus invalidate the operator product expansion. Detailed 
calculations show that because the lower bound of integration is now $\nu$ 
instead of zero, there will also always be at least one power of $(1-\nu)$ 
coming out from the integral. Furthermore, the highest powers of $\nu$  
cancel completely between the first two terms of each $\ga$, leaving the 
highest surviving $\nu$ powers to be $\nu^{J-1}$ for the quark sector and 
$\nu^J$ for the gluonic sector, as they must be to make the OPE valid.
All the $I_0$ and $\log(1-\nu)$ terms that might potentially spoil the
operator product expansion also cancel completely between each 
triangle graph and its corresponding self-energy graphs. 

Because of the mixing among different $n$ moments, as well as between quark
and gluon sectors, it is very difficult to read off the anomalous dimensions
of the mixing between two operators (with same $J$) of definite $n$ moments.
\footnote{After the completion of this work, through private
communications with Professors X. Ji and A. V. Radyushkin, the author
learned that by adopting a basis of linear combinations of the operators in
\ref{eq:operators} using Gegenbauer polynomials as coefficients we
might be able to explicitly diagonalize the anomalous dimension matrix in the
moment space.} However, the form of the anomalous dimension matrix 
simplifies greatly when the high energy limit is taken, and we will be able to 
make connection between this general formalism and the conventional 
leading logarithmic approximation (LLA) analysis results.

\section{Lowest Order Wilson Coefficients \label{sec:wilsoncoeff}}

\begin{figure}[ptb]
\begin{center}
\epsfxsize=5.5in
\epsfysize=1.8in
\leavevmode
\hbox{ \epsffile{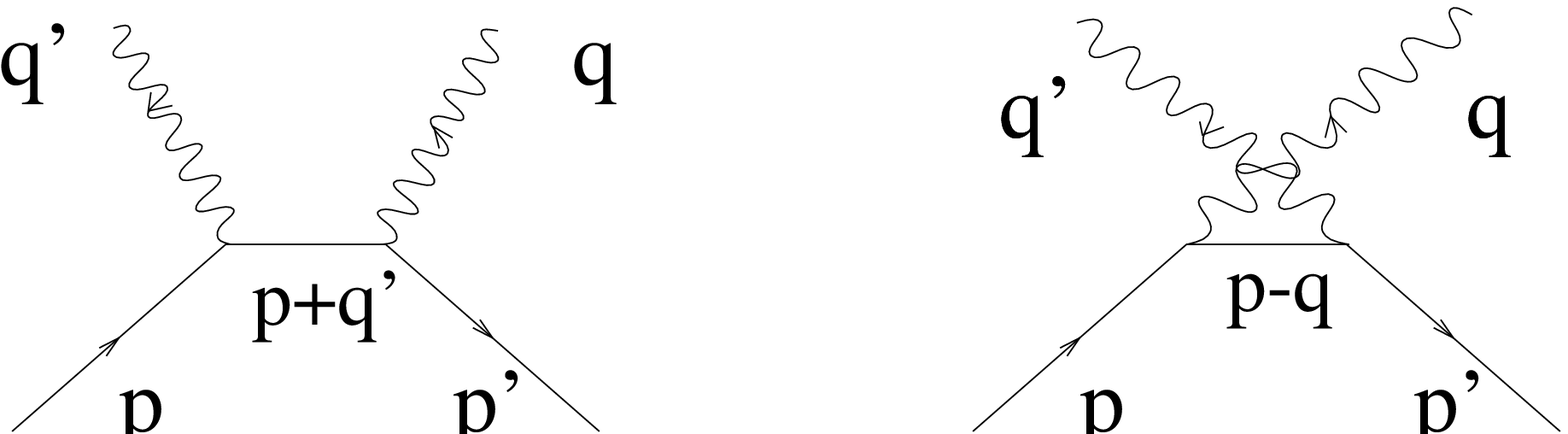}}
\end{center}
\caption{The Tree Level Born Diagrams of the Non-forward Amplitude}
{These are the tree level (Born) terms contributing to the
non-forward amplitude from which we can calculate the lowest order
Wilson coefficients. We have explicitly included the cross term.}
\label{fig:born}
\end{figure}

The lowest order Wilson coefficients can be calculated in perturbation
theory from tree level Born diagrams as shown in figure \ref{fig:born}, where
we have explicitly included the cross diagram. 

Using standard Feynman rules we can write the value of the Born
diagrams, after spin average, as
\beqs
\label{eq:borndef}
{\rm T}_{\mu \nu}^{(0)} & = & \half \sum_s (i e_q)^2 \ovl{U}_s(p') 
\left( 
\ga_\mu \frac {i (\not{p} + \not{q}')} {(p + q')^2 + i \eps} \ga_\nu
\right. \non \\
&& \left. 
+ \ga_\nu \frac {i (\not{p} - \not{q})} {(p - q)^2 + i \eps} \ga_\mu 
\right) U_s(p) \,.
\eeqs

\nid By using equations \ref{eq:zeta}, \ref{eq:q'+psquare} and
\ref{eq:q-psquare} we find
\beqs
{\rm T}_{\mu \nu}^{(0)} & = & -i e_q^2 \frac {\sqrt{1-\zeta}} {2 \ovl{q}^2}
\left( {1 \over {1 - \om}} 
\sum_s \ovl{U}_s(p) \ga_\mu (\not{p} + \not{q}') \ga_\nu U_s(p) \right. 
\non \\
&& \left. + {1 \over {1 + \om}}
\sum_s \ovl{U}_s(p) \ga_\nu (\not{p} - \not{q}) \ga_\mu U_s(p) \right)
\non \\
&=& -i e_q^2 \frac {\sqrt{1-\zeta}} {2 \ovl{q}^2} \sum_{l=0}^\infty \om^l
(Tr(\not{p}\ga_\mu (\not{p} + \not{q}') \ga_\nu)
+ (-1)^l Tr(\not{p}\ga_\nu (\not{p} - \not{q}) \ga_\mu) ) \non \\
&=& -2i e_q^2 \frac {\sqrt{1-\zeta}} {\ovl{q}^2} \sum_{l=0}^\infty \om^l
(p_\mu (p+q')_\nu + p_\nu (p+q')_\mu - g_\mn p \cdot (p+q')) \non \\
&& + (-1)^l (p_\mu (p-q)_\nu + p_\nu (p-q)_\mu - g_\mn p \cdot (p-q))
\non \\
&=& -2i e_q^2 \frac {\sqrt{1-\zeta}} {\ovl{q}^2} \sum_{l=0}^\infty
\om^l
( (p_\mu p_\nu + p_\nu p_\mu + p_\mu q'_\nu + p_\nu q'_\mu 
- g_\mu p \cdot q ) \non \\
&& + (-1)^l (p_\mu p_\nu + p_\nu p_\mu
- p_\mu q_\nu - p_\nu q_\mu + g_\mn p \cdot q)) \non \\
&=& -2i e_q^2 \frac {\sqrt{1-\zeta}} {\ovl{q}^2} \sum_{l=0}^\infty
\om^{2l}
(2(p_\mu p_\nu + p_\nu p_\mu) - p_\mu r_\nu - r_\mu p_\nu \non \\
&& + \om (p_\mu (q' + q)_\nu + (q' + q)_\mu p_\nu - 2 g_\mn p \cdot q) )
\non \\
&\equiv& - 2i e_q^2 \frac {\sqrt{1-\zeta}} {\ovl{Q}^2} 
\sum_{l=0}^\infty \om^{2l} C_\mn \,,
\eeqs

\nid where
\beq
C_\mn = (2p-r)_\mu r_\nu + r_\mu (2p-r)_\nu
+ 2 \om (p_\mu \ovl{q}_\nu + \ovl{q}_\mu p_\nu - g_\mn p \cdot q) \non \,.
\eeq

\nid By similar derivations that lead to equation (\ref{eq:2ndtensor}),
as well as the definition of the moment variables 
(see equation \ref{eq:momvardef}) we find
\newpage
\beqs
C_\mn &=& 2 (2-\zeta) p_\mu p_\nu 
- 2 (2-\zeta) { {p \cdot q} \over \ovl{q}^2 }
(p_\mu \ovl{q}_\nu + \ovl{q}_\mu p_\nu - g_\mn p \cdot q) \non \\
&=& 2 (2-\zeta)(p_\mu p_\nu  
- { {p \cdot q} \over \ovl{q}^2 } (p_\mu q_\nu + q'_\mu p_\nu) 
+ { {(p \cdot q)^2} \over \ovl{q}^2 } g_\mn) \non \\
&=& 2 (2-\zeta) \left( p_\mu p_\nu 
- { {p \cdot q} \over \ovl{q}^2 } (p_\mu q_\nu + q'_\mu p_\nu)
+ ({ {p \cdot q} \over \ovl{q}^2 })^2 q'_\mu q_\nu 
+ { {(p \cdot q)^2} \over \ovl{q}^2 } 
  (g_\mn - { {q'_\mu q_\nu} \over \ovl{q}^2}) \right) \non \\
&=& 2 \om \left( { \ovl{Q}^2 \over {p \cdot q} }
(p_\mu p_\nu
- { {p \cdot q} \over \ovl{q}^2 } (p_\mu q_\nu + q'_\mu p_\nu)
+ ({ {p \cdot q} \over \ovl{q}^2 })^2 q'_\mu q_\nu )
+ p \cdot q ( - g_\mn + { {q'_\mu q_\nu} \over \ovl{q}^2}) \right) \,.
\non \\
\eeqs

\nid Thus we obtain the value of the Born diagrams as
\beqs
\label{eq:bornresult}
{\rm T}_{\mu \nu}^{(0)} & = & - 4 i e_q^2 \sqrt{1-\zeta} \sum_{l=0}^\infty 
\om^{2l\!+\!1} \left( {{p \! \cdot \! q} \over {\ovl{Q}^2}}
( -g_{\mu \nu} + \frac {q'_\mu q_\nu} {\ovl{q}^2} ) \right. \non \\
& & \left. + {1 \over {p \cdot q}} \left(p_\mu p_\nu - \frac {p \cdot q} 
{\ovl{q}^2}(p_\mu q_\nu + q'_\mu p_\nu) + ( \frac {p \cdot q} 
{\ovl{q}^2})^2 q'_\mu q_\nu \right) \right) \, .
\eeqs

\noindent Comparing to (\ref{eq:opefinal}) we obtain
\beqs
\label{eq:1storderwcoeff}
\tilde{E}^{(q)(0)}_{J,n=0}(\al_s(\ovl{Q}),1) &=& 4ie_q^2 \,, 
\,\,\,\,\,\,\,\,\, J \,\,\, odd, \, J \geq 3 \,, \,\,\,\,\,\,\,\, \non \\
\tilde{E}^{(q)(0)}_{J,n}(\al_s(\ovl{Q}), 1) &=& 0 \,, \,\,\,\,\,
\,\,\,\,\,\,\,\,\,\,\, J,n \,\,\, otherwise \,. \,\,\,\,\,\,\,\,
\eeqs

\noindent Thus for each $J$ value, only the Wilson coefficient of
$n=0$ has a non-zero value at the tree level. This means that at 
leading order there is no dependence on the non-forwardness in the 
Wilson coefficients.

\chapter{High Energy Limit}
\thispagestyle{myheadings}
\markright{}

In this chapter we will solve the renormalization group (RG) equation for 
the Wilson coefficient (\ref{eq:rge-wcoeff}) in the high energy limit, 
and discuss the
relationship between our general results and that of the usual forward
double leading logarithmic approximation (DLLA) \cite{BrodskyVM}.

\section{Reduction of the Path Ordered Exponential}

Since we have computed the lowest order Wilson coefficients
(\ref{eq:1storderwcoeff}), and the reduced matrix elements are
regarded as input from experimental data, we need now to evaluate the
path ordered exponential 
${\rm M}_{n'n}^{i'i}(\al(\ovl{Q}), {\ovl{Q} \over \mu})$ in equation
(\ref{eq:formalsolu}) in order to obtain the double moments 
${\rm T}^{(J,n)}_{2,(\ovl{Q}, \mu_0)}$ and then the invariant
amplitude ${\rm T}_2$. 

\nid Recall that formally we have (\ref{eq:formalexpo})
\beq
{\rm M}_{n'n}^{i'i}(\al_s(\ovl{Q}), {\ovl{Q} \over \mu}) =
\left( {\cal P} \exp (- \int_{\mu^2}^{\bar{Q}^2} \tilde{\ga}(\al_s
(\lm^2)) { {d \lm^2} \over \lm^2}) \right)_{n'n}^{i'i} \,,
\eeq           

\nid thus to explicitly solve the RG equation we need to diagonalize
the anomalous dimension matrix in both flavor ($i$) and moment 
($n$) indices. However, in the high energy limit the situation
simplifies. 

Similar to the forward case, after analytical continuation in
$J$, at high energy, the dominant contributions to ${\rm T}_2$ come 
from the leading (right most) poles in the $J$-plane, which, as we will
see, means that the diagonal elements in the moment space ($n$ space)
of the anomalous dimension matrix dominate the evolution. And because
the gluon anomalous dimension has the right most pole (one unit higher
than that of the quark sector) (see equation \ref{eq:leadingpoles}), 
together with
the fact that gluons have bigger color charge, in flavor space ($i$
space) at high energy the gluon anomalous dimension dominates. 

\nid However, gluons only have color charge and thus can not interact 
directly with 
the color-neutral virtual photon. The scattering must happen via a 
quark loop, which means we have to force at least one gluon-quark 
transition at the end of the evolution. Thus in each term of the 
expansion of the path ordered exponential (\ref{eq:formalexpo}), 
all the factors of the product of anomalous dimensions are 
$\tilde{\ga}^{gg}$ except the first one, which should be 
$\tilde{\ga^{qg}}$ due to the quark-gluon transition.

\nid Therefore we need only to evaluate ${\rm M}$ between a 
quark and a gluon state and thus collapse the two flavor sums in
(\ref{eq:doubmomsolu}) to $i'=q$ and $i=g$. Formally, if we now label 
$\ga^{ii'}$ as the anomalous dimension matrix in the moment space, 
by expanding the path ordered exponential,
we have
\beqs
\label{eq:reduction}
{\rm M}_{n'n}^{i'i}(\al_s(\ovl{Q}), {\ovl{Q} \over \mu}) 
&\stackrel{i'=q \,,i=g} {\longrightarrow}&
\left( {\cal P} \exp (- \int_{\mu^2}^{\bar{Q}^2} \tilde{\ga}(\al_s
(\lm^2)) { {d \lm^2} \over \lm^2}) \right)_{n'n}^{qg} 
\non \\
&=& \left( \sum_{l=0}^\infty {1 \over {l!}} 
{\cal P} \int_{\mu^2}^{\bar{Q}^2} 
\prod_{k=l}^1 ( { {d \lm_k^2} \over \lm_k^2} 
(-\tilde{\ga}(\al_s(\lm_k^2))) ) \right)_{n'n}^{qg} \non \\
&=& \left(  \sum_{l=0}^\infty
\int_{\mu^2}^{\bar{Q}^2} { {d \lm_l^2} \over \lm_l^2}
(-\tilde{\ga}(\al_s(\lm_l^2)))
\int_{\mu^2}^{\lm_l^2} { {d \lm_{l\!-\!1}^2} \over \lm_{l\!-\!1}^2}
(-\tilde{\ga}(\al_s(\lm_{l\!-\!1}^2)))
\cdot ...  \right. \non \\
&& \left. \,\,\,\,\,\,\,\,\,\,\,\,\,\,\, \cdot
\int_{\mu^2}^{\lm_2^2} { {d \lm_1^2} \over \lm_1^2}
(-\tilde{\ga}(\al_s(\lm_1^2))) \right)_{n'n}^{qg} \non \\ 
&\stackrel{hi-E}{\rightarrow}& \!\!\!\! \left( \sum_{l=0}^\infty
\int_{\mu^2}^{\bar{Q}^2} { {d \lm_l^2} \over \lm_l^2}
(-\tilde{\ga}^{qg}(\al_s(\lm_l^2)))
\int_{\mu^2}^{\lm_l^2} { {d \lm_{l\!-\!1}^2} \over \lm_{l\!-\!1}^2}
(-\tilde{\ga}^{gg}(\al_s(\lm_{l\!-\!1}^2)))
\cdot ... \right. \non \\
&& \left. \,\,\,\,\,\,\,\,\,\,\,\,\,\,\, \cdot
\int_{\mu^2}^{\lm_2^2} { {d \lm_1^2} \over \lm_1^2}
(-\tilde{\ga}^{gg}(\al_s(\lm_1^2))) \right)_{n'n} \non \\  
&\equiv& \!\!\!\! \left( 
- \int_{\mu_0^2}^{\bar{Q}^2} {{d \lm^2} \over \lm^2}
{\ga}^{qg}(\alpha_s (\lm^2)) {\cal P}
\exp (- \int_{\mu_0^2}^{\lm^2} {{d \lm'^2} \over \lm'^2}
{\ga}^{gg}(\alpha_s (\lm'^2))) \right)_{n'n} \,, \non \\
\eeqs

\nid which can be rewritten as
\beq
\label{eq:redfinal}
{\rm M}_{n'n}^{qg}(\alpha_s (\overline{Q}), {\overline{Q} \over \mu_0}) 
= \left( - \int_{\mu_0^2}^{\bar{Q}^2} {{d \lm^2} \over \lm^2}
{\ga}^{qg}(\alpha_s (\lm^2)) \right)_{n'l} \left({\cal P}
\exp (- \int_{\mu_0^2}^{\lm^2} {{d \lm'^2} \over \lm'^2}
{\ga}^{gg}(\alpha_s (\lm'^2))) \right)_{ln} \,.
\eeq

\noindent The leading pole terms of the relevant anomalous dimensions 
can be parametrized as (suppressing the $J$ label), after integration 
of (\ref{eq:gammaresult}), 
\beqs
\label{eq:leadingpoles}
\ga^{qg}_{nn'}(\lm^2) &=& -{n_f \over {2 \pi}} \alpha_s(\lm^2)
\,\,\, \ovl{\ga}^{qg}_{nn'} \,, \,\,\,\,\,\,\,\,\,\,\,\,\,\,\,\, 
{\rm with} \,\,\,\,\,\,\, \ovl{\ga}^{qg}_{nn'} = 
{{\,\,\,\, 1 + c^{qg}_{n',n}} \over {J-n}} \,\,; \,\, \\
\ga^{gg}_{nn'}(\lm^2) &=& -{C_A \over \pi} \alpha_s(\lm^2)  
\,\,\, \ovl{\ga}^{gg}_{nn'} \,, \,\,\,\,\,\,\,\,\,\,\,\,\,\,\,
{\rm with} \,\,\,\,\,\,\, \ovl{\ga}^{gg}_{nn'} = 
{ { 1 + c^{gg}_{n',n}} \over {J-n-1}}\,\, . 
\eeqs

\noindent where $0 \leq n \leq n' \leq J$ for $\ovl{\ga}_{nn'}$ and
more importantly
\beq   
c^{i'i}_{n'=n,n} = 0 \,.
\eeq

\nid This means that only the diagonal elements have leading pole
contributions.

\nid Because of the upper triangular structure and the leading pole 
positions (recall that the right most pole dominates) of these 
$\ovl{\ga}$ matrices, we have, for $m \leq l \leq n$,
\beq
\label{eq:gammasum}
\sum_l \ovl{\ga}_{ml} \,\, \ovl{\ga}_{ln} = \ovl{\ga}_{mn}
\,\, \ovl{\ga}_{nn} \equiv \ovl{\ga}_{mn} \, a_n \,,
\eeq

\noindent where we have labeled the diagonal terms of the anomalous
dimensions (leading pole terms) $a_n$. This means in a product 
of $\ovl{\ga}$ matrices we can, in leading logarithmic 
approximation (LLA), take only diagonal entries in all the factors 
except for the first one (see equation \ref{eq:exporeduction}).

On the other hand, because we have the QCD running coupling 
(see, eg, \cite{Peskin})
\beq
\label{eq:alphadef} 
\al(\lm^2) = { {4 \pi} \over {b_0 \log (\lm^2 /\Lm^2)} } \,,
\eeq

\nid where $\Lm \equiv \Lm_{QCD}$ is the QCD scale and $b_0$ the
leading QCD $\be$-function coefficient (see (\ref{eq:b0def})), the 
integration of the anomalous dimension at high energy gives
\beqs
\label{eq:gammaint}
- \int_{\mu_0^2}^{\lm^2} {{d \lm'^2} \over \lm'^2}
{\ga}^{gg}(\alpha_s (\lm'^2))) &=&
{C_A \over \pi} \ovl{\ga}^{gg} 
\int_{\mu_0^2}^{\lm^2} {{d \lm'^2} \over \lm'^2} \al(\lm'^2)
\non \\
&=& { {4 C_A} \over {\pi} } \ovl{\ga}^{gg}
\int_{\mu_0^2}^{\lm^2} 
{ {d \log (\lm'^2 /\Lm^2)} \over {\log (\lm'^2 /\Lm^2)} } \non \\
&\equiv&  A(\lm^2) \ovl{\ga}^{gg} \,,
\eeqs

\nid where we define the usual double-log factor
\beq
\label{eq:Adef}
A(\lm^2)= A \equiv { {4 C_A} \over b_0} \log \left({{\log
(\lm^2 \!/ \Lm^2)} \over {\log (\mu_0^2 / \Lm^2)}} \right) \,.
\eeq
 
Thus the path ordered exponential of $\ga^{gg}$ (the second factor on
the right hand side of \ref{eq:redfinal}) becomes
\beqs
\label{eq:exporeduction}
\left({\cal P}
\exp (- \int_{\mu_0^2}^{\lm^2} {{d \lm'^2} \over \lm'^2}
{\ga}^{gg}(\alpha_s (\lm'^2))) \right)_{ln} 
= \left( {\cal P} e^{A{\lm^2} \ovl{\ga}^{gg}} \right)_{ln} \non \\
= (1 + A \ovl{\ga} + {1 \over {2!}} A^2 \ovl{\ga} \ovl{\ga} + ...)_{ln}
\non \\
= ( 1 + A \ovl{\ga}_{ln} + {A^2 \over {2!}} a_n \ovl{\ga}_{ln} + ...)
\non \\
= {\rm I} + (A + {A^2 \over {2!}} a_n + ...) \ovl{\ga}^{gg}_{ln} \non \\
= {\rm I} + {1 \over a_n^g}(e^{A(\lm^2)a_n^g} -1) \,\, \ovl{\ga}^
{gg}_{ln} \non \\
= {1 \over a_n^g} e^{A(\lm^2)a_n^g} \,\, \ovl{\ga}^{gg}_{ln} 
\,\,+\,\, {\rm non\!-\!leading \,\, terms} \,.
\eeqs

\nid Therefore, (\ref{eq:redfinal}) reduces to
\beqs
\label{eq:redresult}
{\rm M}_{n'n}^{qg}(\alpha_s (\overline{Q}), {\overline{Q} \over \mu_0})
&=& \left(- \int_{\mu_0^2}^{\bar{Q}^2} {{d \lm^2} \over \lm^2}
{\ga}^{qg}(\alpha_s (\lm^2)) \right)_{n'l}
{1 \over a_n^g} e^{A(\lm^2)a_n^g} \,\, \ovl{\ga}^{gg}_{ln} \non \\ 
&=& {n_f \over {2 \pi}} \int_{\mu_0^2}^{\bar{Q}^2} { {d \lm^2}
\over \lm^2} \alpha_s (\lm^2) \,\, e^{A(\lm^2)a_n} 
\ovl{\ga}^{qg}_{n'm} \,\ovl{\ga}^{gg}_{ln} {1 \over a_n^g} \non \\
&=& {n_f \over {2 \pi}} \int_{\mu_0^2}^{\bar{Q}^2} { {d \lm^2}
\over \lm^2} \alpha_s (\lm^2) \,\, e^{A(\lm^2)a_n} \,\,
\ovl{\ga}^{qg}_{n'n} \,,
\eeqs

\nid where in the last step we have again used (\ref{eq:gammasum}).

The $n'$ summation in the expression for the double moments 
(\ref{eq:doubmomsolu}) also collapses in LLA to the diagonal element of 
$\ovl{\ga}_{qg}$, and we have the final expression of the double moments 
as
\beqs
\label{eq:doubmomfinal}
{ {p \! \cdot \! q} \over M^2} \sqrt{1\!-\!\zeta} {\rm T}^{(J,n)}_
{2,(\overline{Q}, \mu_0)} \!&\!=\!&\!
- \sum_{n'=0}^n \sum_{i'i} \tilde{E}_{J-2,n'}^{(i')}(\al_s(\ovl{Q}), 1) 
\,\,{\rm M}_{n'n}^{i'i}(\al_s (\ovl{Q}), {\ovl{Q} \over \mu_0})
\,\,{\bra{p'}| \hat{O}^{(i)(J,n)}|\ket{p}}_{(\mu_0)} \non \\
\!&\!=\!&\! - \sum_{n'=0}^n \tilde{E}_{J-2,n'}^{(q)}(\al_s(\ovl{Q}), 1) \,\,
{\rm M}_{n'n}^{qg}(\al_s (\ovl{Q}), {\ovl{Q} \over \mu_0})
\,\, {\bra{p'}| \hat{O}^{(g)(J,n)}|\ket{p}}_{(\mu_0)} \non \\ 
\!&\!=\!&\! - \sum_{n'=0}^n \tilde{E}_{J-2,n'}^{(q)}(\al_s(\ovl{Q}), 1) \,\,
{n_f \over {2 \pi}} \int_{\mu_0^2}^{\bar{Q}^2} { {d \lm^2}
\over \lm^2} \alpha_s (\lm^2) \,\, e^{A(\lm^2)a_n} \non \\
&& \,\,\,\,\,\,\,\,\,\,\,\,\,\,\,\,\,\,\,\,\,\,\,\,\,\,
\cdot \,\, \ovl{\ga}^{qg}_{n'n}
\,\,{\bra{p'}| \hat{O}^{(g)(J,n)}|\ket{p}}_{(\mu_0)} \non \\ 
\!&\!=\!&\! - \tilde{E}_{J-2,n}^{(q)}(\al_s(\ovl{Q}), 1) \,\,
{n_f \over {2 \pi}} \int_{\mu_0^2}^{\bar{Q}^2} { {d \lm^2}
\over \lm^2} \alpha_s (\lm^2) \,\, e^{A(\lm^2)a_n} \,\, \non \\
&& \,\,\,\,\,\,\,\,\,\,\,\,\,\,\,\,\,\,\,\,\,\,\,\,\,\,\,\,\,
\cdot \,\, \ovl{\ga}^{qg}_{nn}
\,\,{\bra{p'}| \hat{O}^{(g)(J,n)}|\ket{p}}_{(\mu_0)} \non \\
\!&\!=\!&\! - \tilde{E}_{J-2,n}^{(q)}
(\alpha_s(\overline{Q}), 1) \,\, {\bra{p'}| \hat{O}^{(g)(J,n)}|
\ket{p}}_{(\!\mu_0\!)} \, \non \\
&& \,\,\,\,\,\,\,\,\,\,\,\,\,\,\,\,\,\,\,\,\,\,\,\,\,\,\,\,\,\,
\cdot \,\, {n_f \over {2 \pi}} 
\int_{\mu_0^2}^{\bar{Q}^2} 
{ {d \lm^2} \over \lm^2} \alpha_s (\lm^2) \,\, 
e^{{A(\lm^2)} \over {J-n-1}} {1 \over {J-n}} \,.
\eeqs

Now the invariant amplitude ${\rm T}_2$ (\ref{eq:mellin}), after 
analytically continuation in $J$, becomes
\beqs
\label{eq:t2integral}
-{ {p \! \cdot \! q} \over M^2} \sqrt{1-\zeta} {\rm T}_2( \overline{Q}^2,
\omega,\nu) =\int {{dJ} \over {2 \pi i}} \sum_{n=0}^\infty \nu^n  
\int_{\mu_0^2}^{\bar{Q}^2} { {d \lm^2} \over \lm^2} 
\,\alpha_s (\lm^2) {n_f \over {2 \pi}}
\,\, \tilde{E}_{J-2,n}^{(q)}(\alpha_s(\overline{Q}), 1)  
\, \non \\
\cdot 
\,\, {\bra{p'}| \hat{O}^{(g)(J,n)}|\ket{p}}_{(\!\mu_0\!)} 
{1 \over {J\!-\!n}} \exp \left( (J\!-\!n) \log \omega 
+ {{A(\lm^2)} \over {J\!-\!n\!-\!1}} \right) \,. \;\;\;
\eeqs

In the high energy limit we can make a saddle point approximation for
the $J$ integration because the structure of the exponent in the
expression. Define
\beq
\label{eq:saddlefcn}
f(J) = (J-n) \log \om + { A \over {J-n-1} } \,,
\eeq

\nid we obtain the first and second order derivative of $f$ as
\beqs
\label{eq:derivatives}
f'(J) &=& \log \om - {A \over {(J-n-1)^2}} \non \\
f''(J) &=& { {2A} \over {(J-n-1)^3} } \,.
\eeqs

\nid Setting $f'(J) = 0$ we obtain the saddle point $J_s$ as
\beq
\label{eq:saddlept}
J_s = 1 + n + \sqrt{{A \over {\log \omega}}} \,,
\eeq

\nid and at the saddle point we have
\beqs
\label{eq:sadptvalue}
f(J_s) &=& \log \om + 2 \sqrt{A \log \om} \non \\
f''(J_s) &=& { {2 \log^{3/2}\omega} \over {A^{1/2}} } \,,
\eeqs

\nid where indeed we can see that at $J_s$ we have $f'' > 0$. Now
the invariant amplitude becomes
\beqs
\label{eq:t2saddle}
-{ {p \! \cdot \! q} \over M^2} \sqrt{1-\zeta}{\rm T}_2 
= \sum_{n=0}^\infty \nu^n  
\int_{\mu_0^2}^{\bar{Q}^2} { {d \lm^2} \over \lm^2} 
\,\alpha_s (\lm^2) {n_f \over {2 \pi}}
\,\, \tilde{E}_{J-2,n}^{(q)}(\alpha_s(\overline{Q}), 1)  
\, \non \\
\cdot 
\,\, {\bra{p'}| \hat{O}^{(g)(J,n)}|\ket{p}}_{(\!\mu_0\!)} 
{ 1 \over {1 + \sqrt{ {A(\lm^2)} \over {\log \omega}}}} 
\sqrt{ { {\pi A(\lm^2)^{1/2}} \over {\log^{3/2}\om} } }
e^{\log \om} e^{2 \sqrt{A(\lm^2) \log \omega}} \non \\
= {n_f \over {2 \pi}} 
\sqrt{ \pi \over {\log^{3/2}\omega}} \, \omega \int_{\mu_0^2}^
{\bar{Q}^2} { {d \lm^2} \over \lm^2} \alpha_s (\lm^2) 
A(\lm^2)^{1/4} { 1 \over {1 + \sqrt{ {A(\lm^2)} 
\over {\log \omega}}}} \,\, \nonumber \\
\,\,\,\, \cdot e^{2 \sqrt{A(\lm^2) \log \omega}} \,\,\sum_
{n=0}^\infty \nu^n \, \tilde{E}_{J-2,n}^{(q)} (\alpha_s(\overline{Q}), 1) 
{\bra{p'}| \hat{O}^{(g)(J,n)}|\ket{p}}_{(\mu_0)} 
\,\,. 
\eeqs

\noindent Because the saddle point fixes the value $J-n=J_s-n=1+ \sqrt{{A 
\over {\log \omega}}}$, which fixes the number of internal derivatives 
inside the gluon operator, recall (\ref{eq:reducej-n}), it is clear that 
the reduced matrix element in (\ref{eq:t2saddle}) is the conventional 
(forward) gluon distribution function in moment space with a shifted 
moment label $J_s-n$. Despite the formal summation over $n$, which is 
essentially the only difference introduced in this limit by the 
non-forwardness, only one non-perturbative input is needed, which 
is the gluon density. This means that although in general there will 
be complicated dependence on $\nu$ in the double distributions, 
(in high energy and hard scattering limit) it is nontheless 
perturbative. 

\nid At leading order the $n$ summation collapses when we take the 
lowest order Wilson coefficients given in (\ref{eq:1storderwcoeff}) 
with the flavor averaged quark charge $n_f <e_q^2> =\sum_f e_q^2$,
which actually could be done even before taking the saddle point
approximation. We went to the saddle point first in order to make manifest 
the fact that there would be only one soft input, the forward gluon
density, even with the formal $n$ summation. Thus, subsitute
the result from equation \ref{eq:1storderwcoeff} we have
\beqs
\label{eq:t2result}
- {{p \! \cdot \! q} \over M^2} \sqrt{1\!-\!\zeta} \,
{\rm T}_2(\overline{Q}^2\!,\omega\!, \nu)\!=\!
{{2 i n_f  <e_q^2>} \over {\pi}} 
\sqrt{ \pi \over {\log^{3/2}\omega}} \, \omega
\, {\bra{p'} | \hat{O}^{(g)(J_s\!-\!n,0)}|\ket{p}}_{(\mu_0)} \non \\
\cdot \, \int_{\mu_0^2}^{\bar{Q}^2} 
{ {d \lm^2} \over \lm^2} \alpha_s (\lm^2) A(\lm^2)^{1/4} 
{ 1 \over {1 + \sqrt{ {A(\lm^2)} \over {\log \omega}}}}
\, e^{2 \sqrt{A(\lm^2) \log \omega}} \,.
\eeqs

Because of the pinching of $J_s-n$, the final evaluation of
(\ref{eq:t2result}) is the same in both forward and non-forward cases,
in the high energy limit under leading logarithmic approximation. Define
the integral
\beq
\label{eq:integral}
I = \int_{\mu_0^2}^{\bar{Q}^2} 
{ {d \lm^2} \over \lm^2} \alpha_s (\lm^2) A(\lm^2)^{1/4} 
{ 1 \over {1 + \sqrt{ {A(\lm^2)} \over {\log \omega}}}}
\, e^{2 \sqrt{A(\lm^2) \log \omega}} \, ,
\eeq

\nid with $A(\lm^2)$ given in equation \ref{eq:Adef}, which is not
very big even for very hard processes due to the double
logarithm. Therefore in the high energy limit where $\om$ is very
big, we have
\beq
\label{eq:almost1}
{ 1 \over {1 + \sqrt{ {A(\lm^2)} \over {\log \omega}}}} \rightarrow 1
\,.
\eeq

\nid We make the change of variable
\beq
\label{eq:udef}
\log(\lm^2/\Lm^2) = \log(\ovl{Q}^2/\Lm^2) - u \,,
\eeq

\nid which implies that
\beq
\label{eq:uvalue}
u = \log(\ovl{Q}^2/\Lm^2) - \log(\lm^2/\Lm^2) = \log(\ovl{Q}^2/\lm^2)
\,.
\eeq

\nid It is clear that $u > 0$ and is not very large either. To
simplify the notation, let
\beqs
\label{eq:abc}
a &=& { {4C_A} \over b_0 }   \non \\
b &=& \log(\ovl{Q}^2/\Lm^2)  \non \\
c &=& \log \left({{\log(\ovl{Q}^2 \!/ \Lm^2)} 
\over {\log (\mu_0^2 / \Lm^2)}} \right)  \non \,, 
\eeqs

\nid where we make the note that $a$ and $b$ can be seen as moderate
compared with the double logarithms $u$ and $c$, which are not big. 
Then
\beqs
\label{eq:doublog}
A (\lm^2) &=& \log \left({{\log(\lm^2 \!/ \Lm^2)} \over 
{\log (\mu_0^2 / \Lm^2)}} \right)
= \log \left( {{\log(\ovl{Q}^2 \!/ \Lm^2)} \over {\log (\mu_0^2 / \Lm^2)}}
\, (1 - {u \over {\log(\ovl{Q}^2 \!/ \Lm^2)}}) \right) \non \\
&=& c + \log (1 - {u \over b}) \non \\
&=& c - {u \over b} = c \left( 1 - {u \over {bc}} \right) \,.
\eeqs

\nid And the integral $I$ becomes
\beqs
\label{eq:Iabc}
I &=& \int_0^{\log \!{{\ovl{Q}^2} \over {\mu_0^2} }}
{{du} \over {b_0 (b-u)}} (ac)^{1/4} (1 - {u \over {4 b c}})
\, e^{2 \sqrt{\log \om \,a c\, (1 - {u \over {bc}})}} \non \\
 &=& \int_0^{\log \!{{\ovl{Q}^2} \over {\mu_0^2} }}
{{du} \over {b_0 (b-u)}} (ac)^{1/4} (1 - {u \over {4 b c}})
\, e^{2 \sqrt{\log \om \,a c} (1 - {u \over {2bc}})} \non \\
&\equiv& { {(ac)^{1/4}} \over {b_0}} e^{2 \sqrt{ac\log \om}}
\cdot \,II \,,
\eeqs

\nid where the second integral $II$ is defined as
\beq
\label{eq:IIdef}
II = \int_0^{\log \!{{\ovl{Q}^2} \over {\mu_0^2} }} du
{1 \over {b-u}} (1 - {u \over {4 b c}}) \,
e^{- {u \over b} \sqrt{ {{\log \om a} \over c}}} \,.
\eeq

\nid As we have discussed earlier, because the relative size of $a$ and $c$
and that $\log \om$ is considered big, 
$\sqrt{ {{\log \om a} \over c}}$ is actually a big number. 

\nid Rewrite (\ref{eq:IIdef}) as
\beqs
\label{eq:IIexpr}
II &=& {1 \over b} \int_0^{\log \!{{\ovl{Q}^2} \over {\mu_0^2} }} du
{1 \over { 1 - {u \over b}} } ({1 \over {4c}} (1 -  {u \over b})
+ 1 - {1 \over {4c}}) e^{- {u \over b} \sqrt{ {{\log \om a} \over c}}}
\non \\
&=& {1 \over b} \int_0^{\log \!{{\ovl{Q}^2} \over {\mu_0^2} }} du
({1 \over {4c}} + {1 \over {1 - {u \over b}}} (1 - {1 \over {4c}}))
e^{- {u \over b} \sqrt{ {{\log \om a} \over c}}} \,,
\eeqs

\nid and let
\beq
z = {u \over b} = {{\log(\ovl{Q}^2/\lm^2)} \over {\log(\ovl{Q}^2/\Lm^2)}}
\,,
\eeq

\nid we have
\beq
\label{eq:IIresult}
II = \int_0^{ {\log \!{{\ovl{Q}^2}\! /\! {\mu_0^2}}} \over b} dz
\left({1 \over {4c}} e^{- z \sqrt{ {\log \om a / c}}}
+ ( 1 - {1 \over {4c}}) { { e^{- z \sqrt{ {\log \om a / c}}}}
\over {1-z}} \right) \,.
\eeq 

\nid The $z$ integral can be evaluated using the formula of special
integral exponential function $IE_i(z)$
\beq
\label{eq:IEiform}
\int_0^\al { {dz} \over {1-z} } e^{-\lm z} = e^{-\lm}
(IE_i(\lm) - IE_i(\lm - \al \lm)) \,,
\eeq

\nid where the $IE_i(z)$ function has a branch cut discontinuity at
the negative real axis in the $z$-plane (i.e., 
cut at $z \in (-\infty,0]$), and has the following properties:
\beqs
\label{eq:IEippt}
IE_i(0) \;\; &=& - \infty \,, \non \\
IE_i(0.3725) \!\!&=& 0 \non \,,
\eeqs

\nid and while
\beq
\pdr_z IE_i(z) = {e^z \over z} \,, \non 
\eeq

\nid we have
\beq
{{IE_i(z)} \over {e^z}} << 1 
\eeq

\nid when $z \geq {\rm O}(1)$.

\nid Therefore the second term in equation \ref{eq:IIresult} does
not contribute in the leading logarithmic approximation, and we have
\beqs
\label{eq:IIfinal}
II &=& {1 \over {4c}} {1 \over {\sqrt{ \log \om a /c}}}
\left( 1 - e^{-\sqrt{ \log \om a /c} { {\log \!{{\ovl{Q}^2}\! /\! {\mu_0^2}}}
\over  {\log \!{{\ovl{Q}^2}\! /\! {\Lm^2}}} } } \right) \non \\
&=& {1 \over {4 \sqrt{\log \om a c} } } \,,
\eeqs

\nid where in the last step we have used the fact that $\sqrt{ \log \om a/c}$ 
is a big number.

Subsititute these results into the expression of the invariant
amplitude (\ref{eq:t2result}) we have
\beqs
\label{eq:t2}
- {{p \! \cdot \! q} \over M^2} \sqrt{1\!-\!\zeta} \,
{\rm T}_2(\overline{Q}^2\!,\omega\!, \nu)&\!=\!&
{{2 i n_f  <e_q^2>} \over {\pi}} 
\, {\bra{p'} | \hat{O}^{(g)(J_s\!-\!n,0)}|\ket{p}}_{(\mu_0)} \non \\
&& \sqrt{ \pi \over {\log^{3/2}\omega}} \, \omega
{ {(ac)^{1/4}} \over {b_0}} e^{2 \sqrt{ac\log \om}}
{1 \over {4 \sqrt{\log \om a c} } } \,.
\eeqs

\nid Since $ac = {{4C_A} \over b_0} A(\ovl{Q}^2)$, the final
expression of the double distribution ${\rm T}_2$ is
\beqs
\label{eq:t2final}
{{p \! \cdot \! q} \over M^2} \sqrt{1\!-\!\zeta} \,
{\rm T}_2(\overline{Q}^2\!,\omega\!, \nu)\!=\!
- { {i n_f \! <\!e_q^2\!>} \over {2 \sqrt{\pi} b_0 }} \! 
\left(\!{1 \over { { {4C_A} \over b_0 } \log^5 \omega A(\overline{Q}^2)}} 
\!\right)^{1 \over 4}\! \,\, \nonumber \\
\cdot \, {\bra{p'} | \hat{O}^{(g)(J_s\!-\!n,0)}|\ket{p}}_{(\mu_0)} \,\,\,
\omega \,\,\, e^{\! 2 \sqrt{{ {4C_A} \over b_0 } 
A(\bar{Q}^2) \log \! \omega}} \,.
\eeqs

\nid In the very high energy and hard scattering limit, the asymmetric
reduced matrix element becomes the forward gluon density
\beqs
\label{eq:nonftof}
{\bra{p'} | \hat{O}^{(g)(J_s\!-\!n,0)}|\ket{p}} &\rightarrow&
\sqrt{1\!-\!\zeta} \; {\bra{p} | \hat{O}^{(g)(J_s\!-\!n,0)}|\ket{p}} 
\non \\
&\propto& \sqrt{1\!-\!\zeta} \; (x g(x)) \,,
\eeqs

\nid with $x = {1 \over \om}$. We can see clearly that the leading 
high energy (small-$x$) behavior is exactly the same as that is given 
by a forward direction leading logarithmic analysis 
(see section \ref{sec:fw-DLLA}).

\chapter{Conclusions and Open Problems}
\thispagestyle{myheadings}
\markright{}

We have formulated a general operator product expansion of a non-forward 
and unequal mass virtual Compton scattering amplitude. We have found that, 
because of the non zero momentum transfer, the expansion now should be 
done in double moments with respect to the moment variables defined 
in equation \ref{eq:momvardef}. The double moments can be parametrized 
as products of the Wilson coefficients, which can be computed 
perturbatively at the
hard scattering scale, and the non-forward matrix elements of new sets 
of quark and gluon operators, namely, the double distribution
functions. These operators in general have total derivatives and they 
mix among themselves under renormalization. They obey a well-defined 
renormalization group equation that can be solved formally. We
have calculated the (equivalent to) evolution kernels of the double 
distribution from which the anomalous dimension matrix of these 
operators can be extracted.  

In the high energy limit, we have found the leading contributions of the
anomalous dimension and used them to solve the resulting renormalization 
group equation. We have recovered the conventional double leading
logarithmic (DLL) analysis results and have made the connection that 
in fact the double distributions are proportional to the conventional 
forward gluon density in this limit. This justifies previous analyses 
using DLL approximation and forward gluon density on non-forward 
processes like the exclusive diffractive vector meson production (e.g.\ 
\cite{BrodskyVM}). Furthermore, we have developed the formalism to
proceed in principle beyond DLL to next to leading order and/or
relaxing the leading logarithmic approximations.

Our analysis is quite general. We can go from deeply inelastic
scattering (DIS) (the forward case, corresponding to $\nu =0$) to 
deeply virtual Compton scattering (DVCS) (corresponding to $\nu=1$) in
principle by continuously varying the moment variable $\nu$. However,
the analyticity/analytical continuation in $\nu$, as well as the
relevant dispersion relationships, still need further investigation. 
At the same time, diffractive vector meson production ($\nu$ close to $1$)
can also be studied by replacing the second quark-photon vertex 
with effectively a (light cone) wave function of the vector
meson \cite{BrodskyVM, Brodsky&Lepage}. Also, here because the time-like 
four-momentum of the vector
meson, further study is needed in extending the discussion to the
time-like region.

There are also other details of the analysis still need to be filled in. The 
explicit form of the anomalous dimension matrix is not yet written
down, although as mentioned earlier, a change of base to a linear
combination of the quark and gluon operators \ref{eq:operators} using
Gegenbauer polynomial as coefficient might simplify the situation.
The dispersion relationship between the invariant amplitudes ${\rm T}_i$ 
and the double structure functions need also to be clarified. For example,
the exact physical meaning of the fact that different $k_-$ poles are being
picked out at different $k_+$ integration regions. The relationship
between this and the transition/interpolation between Altarelli-Parisi
evolution and Brodsky-Lepage evolution is also an important topic to
study in depth.

More experimental data are to come from DESY, Jefferson Lab, SLAC,
CERN and RHIC, the study of the physics of non-forward processes will
continue to be an interesting and important field.

\addcontentsline{toc}{chapter}{Bibliography}
\thispagestyle{myheadings}
\markright{}

\clearpage
\addcontentsline{toc}{chapter}{Appendices}
 
\par
\setcounter{chapter}{0}
\setcounter{section}{0}
\setcounter{subsection}{0}
\appendix

\chapter{Quark-Gluon Anomalous Dimension}
\thispagestyle{myheadings}
\markright{}

In this appendix we show the details of the calculation that leads to
our form of the quark-gluon anomalous dimension in equation
\ref{eq:gammaresult}.  

\begin{figure}[ptb]
\begin{center}
\epsfxsize=3in
\epsfysize=3.8in
\leavevmode
\hbox{ \epsffile{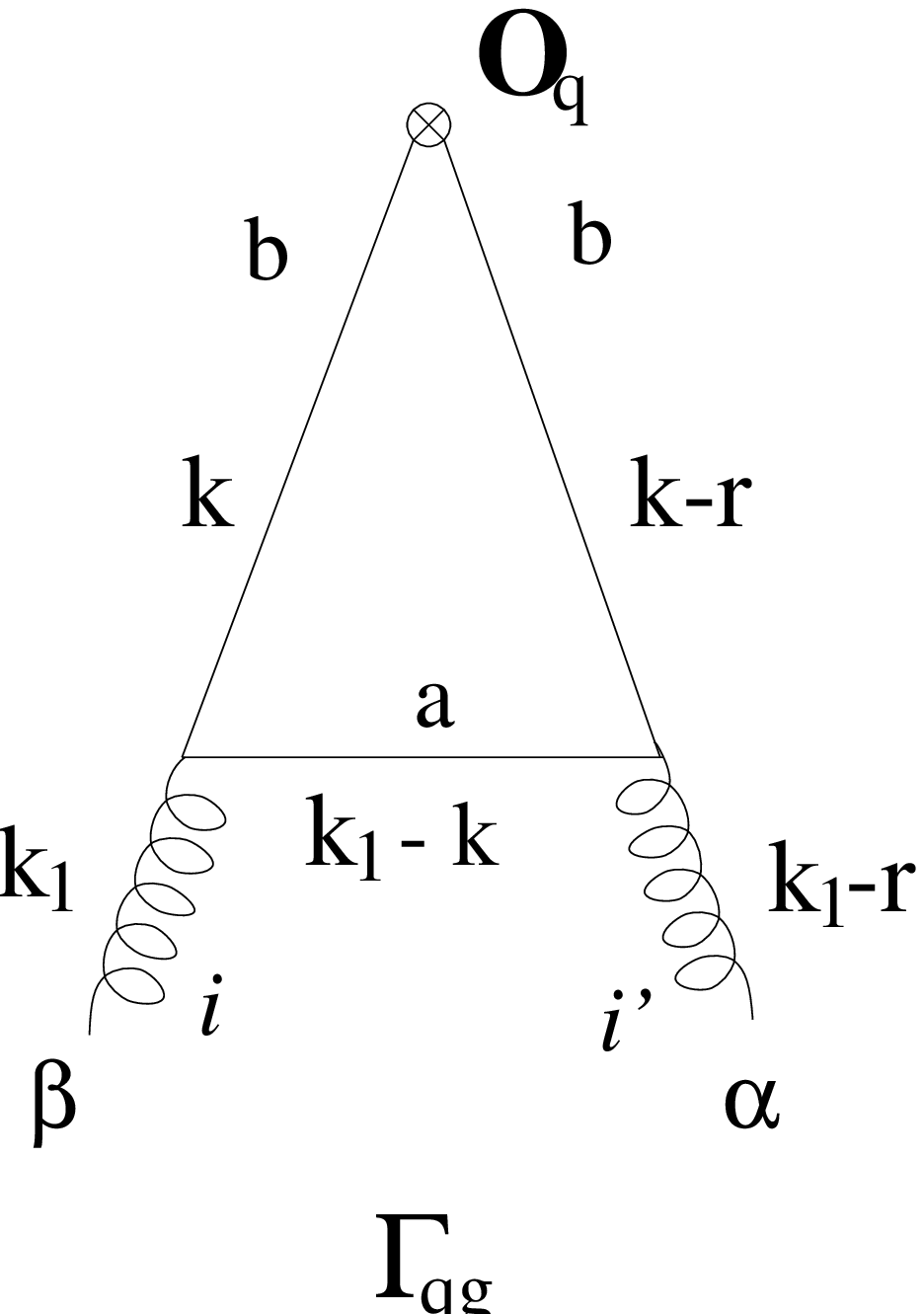}}
\end{center}
\caption{The Quark-Gluon Diagram of the Anomalous Dimensions}
{This is the detailed version of the third triangle graph, the
quark-gluon diagram. $\al$, $\be$ are Lorentz indices; 
$a$, $b$ are color indices for the quarks while $i$ and $i'$
label the gluon lines.}
\label{fig:qg}
\end{figure}

In the LC gauge, the value of the quark-gluon diagram, shown in
figure \ref{fig:qg}, is 
\beq
\label{eq:qg1}
\ga_{qg}^{\al \be} = (-ig)^2 \sum_{a,b} (T^i_{ab}T^{i'}_{ba})
\int { {d^4 k} \over { (2 \pi)^4}} (-1)
Tr \left( \ga_\al {i \over { \not{\!k}-\not{r}}} O_q^{(J,n)(+)}(k,k-r)
{i \over \not{\!k}} \ga_\be { i \over {\not{\!k}_1 - \not{\!k}} } \right)
\,,
\eeq

\nid where $O_q$ is the quark vertex from (\ref{eq:vertex}) and there is 
an extra factor of $(-1)$ coming from the fermion loop. After the
color algebra we obtain
\beq
\label{eq:qg2}
\ga_{qg}^{\al \be} = -ig^2 {1 \over 2} \de_{ii'} 
\int { {d^4 k} \over { (2 \pi)^4}}
{ {(2k-r)_+^{J\!-\!n\!-\!1}r_+^n B_{\al \be}(k,r)}
\over
{(k^2 + i\eps) ( (k_1 \!-\!k)^2 + i\eps) ( (k \!-\! r)^2 + i\eps) } }
\,,
\eeq

\nid where we have defined
\beq
\label{eq:Bqdef}
B_{\al \be}(k,r) = Tr(\ga_\al (\not{\!k} - \not{r}) \ga_+
\not{\!k} \ga_\be (\not{\!k}_1 - \not{\!k}) ) \,\,.
\eeq

\nid Almost identical to the calculation of the gluon-quark section,
after the $k_-$ integral we have
\beqs
\label{eq:qg3}
\ga_{qg}^{\al \be} &=& - {{\al_s} \over {2 \pi}}
{1 \over 4} \int { {d\perb{k}^2} \over {\perb{k}^4}} \left(
\int_0^{r_+} { {dk_+ k_+} \over { r_+ k_{1+}} }
(2k-r)_+^{J\!-\!n\!-\!1}r_+^n B^{(1)}_{\al \be} \,\right. \non \\
&& \,\,\,\,\,\,\,\,\,\,\,\,\,\,\,\,\,\,\,\,\,\,\,\,\,\,\,\,\,\,\,\,
\left. + \int_{r_+}^{k_{1+}}
{ {dk_+ (k_1-k)_+} \over { k_{1+} (k_1-r)_+}}
(2k-r)_+^{J\!-\!n\!-\!1}r_+^n B^{(2)}_{\al \be} \right) \,\,.
\eeqs 

\nid where $B^{(1)}$ and $B^{(2)}$ are $B_{\al \be}(k,r)$ evaluated at
the poles $k_-^{(1)}$ and $k_-^{(2)}$, respectively, and again with
\beqs
k_-^{(1)} = {{\perb{k}^2} \over {2 k_+}}
= { {\perb{k}^2} \over {2 k_{1+} \om} } \,\,\,\;\;\;\;\;\; & \& & \,\,
k^2 = 2 k_+ k_- - \perb{k}^2 =  0 , \non \\
k_-^{(2)} = { {\perb{k}^2} \over {2 (k_+ \!-\! k_{1+})}}
= { {-\perb{k}^2} \over { 2 k_{1+} (1 \!-\! \om)} } \,\,\,
& \& & \,\, k^2 = - { {\perb{k}^2} \over { (1 - \om) } }. \non
\eeqs       

\nid We have
\beqs
B_{\al \be}(k,r) &=& Tr(\ga_\al (\not{\!k} - \not{r}) \ga_+
\not{\!k} \ga_\be (\not{\!k}_1 - \not{\!k}) ) \non \\
&=& Tr(\ga_\al \not{\!k} \ga_+ \not{\!k} \ga_\be (\not{\!k}_1 - \not{\!k}) )
- Tr(\ga_\al \not{r} \ga_+ \not{\!k} \ga_\be (\not{\!k}_1 - \not{\!k}) )
\non \\
&\equiv& B_{I,\al \be} - B_{II, \al \be} \,\,.
\eeqs

Same as in the main text, we will pick up only the terms that are 
logarithmic divergent in the transverse momentum ($\perb{k}^2$ terms) 
and thus we need only look at the transverse components 
($\al \be = i,j = 1, 2$ terms). 

For the first term, we have
\beqs
B_{I,\al \be} &=& Tr(\ga_\al \not{\!k} \ga_+ \not{\!k} \ga_\be 
(\not{\!k}_1 - \not{\!k}) ) \non \\
&=& Tr(\ga_\al ({\ga_+, \not{\!k}} - \ga_+ \not{\!k}) \not{\!k} \ga_\be 
(\not{\!k}_1 - \not{\!k}) ) \non \\
&=& 2 k_+ Tr(\ga_\al \not{\!k} \ga_\be (\not{\!k}_1 - \not{\!k}) ) 
- k^2 Tr(\ga_\al \ga_+ \ga_\be (\not{\!k}_1 - \not{\!k}) ) \non \\
&=& 8 k_+ (k_\al (k_1 -k)_\be + k_\be (k_1 -k)_\al - g_{\al \be}
k \cdot (k_1 -k)) \non \\
&& - k^2 Tr(\ga_- \ga_+ \ga_- \ga_+ (k_1 - k)_- )
 - k^2 Tr( \perb{\ga} \ga_+ \perb{\ga} \ga_- (k_1 -k)_+) \,,
\eeqs

\nid where we have used the identity of gamma matrices \cite{Peskin}
\beq
Tr(\not{\!\!A} \not{\!\!B} \not{\!\!C} \not{\!\!D}) 
= 4 ( (A \cdot B) (C \cdot D) - (B \cdot D) (A \cdot C)
+ (A \cdot D) (B \cdot C) ) \, .
\eeq

\nid At the pole $k_-^{(1)}$ where $k^2 = 0$, we have, 
for the $ij$ component,
\beqs
B_{I,ij}^{(1)} &=& 8k_+(-k_i k_j - k_j k_i - g_{ij} k \cdot k_1) 
\non \\
&=& 8k_+ (g_{ij} \perb{k}^2 - g_{ij} { {\perb{k}^2} \over {2 \om} })
\non \\
&=& 8 k_+ g_{ij} \perb{k}^2 (1 - { 1 \over {2 \om} }) \non \\
&=& 4 k_{1+} g_{ij} \perb{k}^2 (2 \om -1) \, ,
\eeqs

\nid while at the pole $k_-^{(2)}$, where factors of transverse
momentum $\perb{k}$ can come from either $k_-$, $k^2$ or $k_i k_j$
terms, we have
\beqs
B_{I,ij}^{(2)} &=& 8k_+(-k_i k_j - k_j k_i - g_{ij} k \cdot (k_1 -k ))
 + 4 k^2 g_{ij} (k_1-k)_+ \non \\
&=& 8 k_+ (g_{ij} \perb{k}^2 - {1 \over 2} g_{ij} 
{ {\perb{k}^2} \over {1 - \om} })
- { {4 \perb{k}^2} \over {1 - \om} } g_{ij} k_{1+} (1 - \om) \non \\
&=& 4 k_{1+} g_{ij} \perb{k}^2 (2\om - { {\om} \over {1-\om} } - 1)
\non \\
&=& - 4 k_{1+} g_{ij} \perb{k}^2 { {\om^2 + (1-\om)^2} \over {1-\om} }
\,.
\eeqs

On the other hand, for the second term we have, by using the
anti-commutation relationships of the Dirac matrices (\ref{eq:anticommu}),
\beqs
B_{II, \al \be} &=& Tr(\ga_\al \not{r} \ga_+ \not{\!k} \ga_\be 
(\not{\!k}_1 - \not{\!k}) ) \non \\
&=& 2 \ga_\al Tr(\ga_+ \not{\!k} \ga_\be (\not{\!k}_1 - \not{\!k}) ) 
- Tr (\not{r} \ga_\al \ga_+ \not{\!k} \ga_\be (\not{\!k}_1 - \not{\!k}) )
\non \\
&=& 2 \ga_\al Tr(\ga_+ \not{\!k} \ga_\be (\not{\!k}_1 - \not{\!k}) ) 
- 2 g_{\al +} Tr(\not{r}\not{\!k}\ga_\be (\not{\!k}_1 - \not{\!k}) ) \non \\
&& + Tr(\not{r} \ga_+ \ga_\al \not{\!k} \ga_\be (\not{\!k}_1 - \not{\!k}) )
\non \\
&=& ... \non \\
&=& 2 \ga_\al Tr(\ga_+ \not{\!k} \ga_\be (\not{\!k}_1 - \not{\!k}) )
- 2 g_{\al +} Tr(\not{r} \not{\!k} \ga_\be (\not{\!k}_1 - \not{\!k}) ) \non \\
&& + 2 k_\al Tr(\not{r} \ga_+ \ga_\be (\not{\!k}_1 - \not{\!k}) ) 
- 2 g_{\al \be} Tr(\not{r} \ga_+ \not{\!k} (\not{\!k}_1 - \not{\!k}) ) 
\non \\
&& + 2(k_1 - k)_\al Tr(\not{r} \ga_+ \not{\!k} \ga_\be)
- Tr(\not{r} \ga_+ \not{\!k} \ga_\be (\not{\!k}_1 - \not{\!k}) \ga_\al) \, ,
\eeqs

\nid which gives
\beqs
\label{eq:BqII}
B_{II, \al \be} &=& 
\ga_\al Tr(\ga_+ \not{\!k} \ga_\be (\not{\!k}_1 - \not{\!k}) )
- g_{\al +} Tr(\not{r}\not{\!k}\ga_\be (\not{\!k}_1 - \not{\!k}) ) \non \\
&& + k_\al Tr(\not{r} \ga_+ \ga_\be (\not{\!k}_1 - \not{\!k}) )
- g_{\al \be} Tr(\not{r} \ga_+ \not{\!k} (\not{\!k}_1 - \not{\!k}) ) \non \\
&& + (k_1 - k)_\al Tr(\not{r} \ga_+ \not{\!k} \ga_\be) \,.
\eeqs

\nid At the pole $k_-^{(1)}$, the first term of (\ref{eq:BqII}), 
after taking only the logarithmic divergent contribution and only the
transverse component, becomes
\beq
\ga_\al Tr(\ga_+ (\ga_- k_+ - \perb{\ga} \cdot \perb{k}) \ga_\be
(\ga_+ (k_1 \!-\! k)_- + \ga_- (k_1 \!-\! k)_+ - \perb{\ga} \cdot 
(\perb{k_1} \!-\! \perb{k}) ) ) \stackrel{i,j}{\longrightarrow} 0 \,;
\eeq

\nid the second term is always zero because $g_{i +} = 0$; the third
term now becomes
\beqs
4 k_\al (r_+ (k_1 \!-\! k)_\be + g_{\be +} r \cdot (k_1 \!-\! k)
- r_\be (k_1 \!-\! k)_+ \stackrel{i,j}{\longrightarrow} -4 k_i k_j r_+ 
= 2 g_{ij} \perb{k}^2 r_+ \,;
\eeqs

\nid the fourth term is now
\beqs
&& - 4 g_{ij} (r_+ k \cdot (k_1 \!-\! k) + r \cdot (k_1 \!-\! k) k_+
- k \cdot r (k_1 \!-\! k)_+) \,\,\,\,\,\,\,\,\,\,\,\,\,\, \non \\
&& \,\,\, = \,\,\,- 4 g_{ij} (r_+ { \perb{k}^2 \over {2 \om} } 
- r_+ { \perb{k}^2 \over 2 } 
- { \perb{k}^2 \over {2 k_+} } r_+ (k_1 \!-\! k)_+ ) \non \\
&& \,\,\, = \,\,\, - 2 g_{ij} r_+ \perb{k}^2 ( {1 \over \om} - 1 - 
{ {1 \!-\! \om} \over \om } ) = 0 \, ;
\eeqs

\nid and the last term is
\beq
- k_i Tr(\not{r} \ga_+ \not{\!k} \ga_j) = - 4 k_i (r_+ k_j) 
= 2 g_{ij} \perb{k}^2 r_+ \, .
\eeq

\nid Thus we obtain
\beq
B_{II,ij}^{(1)} = 4 g_{ij} \perb{k}^2 r_+ = 4 k_{1+} g_{ij} \perb{k}^2 \nu
\,.
\eeq

\nid At the pole $k_-^{(2)}$, the first and second terms are both zero
for the same reason; the third and fifth terms will take the same
value as they do at the $k_-^{(1)}$ pole; for the fourth term we
now have
\beqs
&& - 4 g_{ij} (r_+ k \cdot (k_1 \!-\! k) + r \cdot (k_1 \!-\! k) k_+
- k \cdot r (k_1 \!-\! k)_+) \,\,\,\,\,\,\,\,\,\,\,\,\,\, \non \\ 
&& \,\,\, = - 4 g_{ij}
(r_+ (- {\perb{k}^2 \over {2 (1 \!-\! \om)} } 
+ { \perb{k}^2 \over {(1 \!-\! \om)} } ) 
+ { {r_+ k_+ \perb{k}^2} \over {2 k_{1+}(1 \!-\! \om)} } 
- r_+ (- { \perb{k}^2 \over {2 k_{1+}(1 \!-\! \om)} }) (k_1 \!-\! k)_+
\non \\
&& \,\,\, = -2 g_{ij} \perb{k}^2 r_+
( {1 \over {1 \!-\! \om}} + {\om \over {1 \!-\! \om}} +1 ) \non \\
&& \,\,\, = - 4  g_{ij} \perb{k}^2 r_+ {1 \over {1 \!-\! \om} }
\,.
\eeqs

\nid And we obtain
\beqs
B_{II,ij}^{(2)} &=& 4 k_{1+} g_{ij} \perb{k}^2 \nu 
- 4 k_{1+} g_{ij} \perb{k}^2 { \nu \over {1 \!-\! \om} } \non \\
&=& - 4 k_{1+} g_{ij} \perb{k}^2 { {\om\nu} \over {1 \!-\! \om} } \,.
\eeqs

Collecting terms we have
\beqs
B^{(1)}_{ij} &=& B_{I,ij}^{(1)} - B_{II,ij}^{(1)} 
= 4 k_{1+} g_{ij} \perb{k}^2 (2 \om \!-\! 1 \!-\! \nu) \non \\
B^{(2)}_{ij} &=& B_{I,ij}^{(2)} - B_{II,ij}^{(2)}
= - 4 k_{1+} g_{ij} \perb{k}^2 \,
{ {\om^2 \!+\! (1\!-\!\om)^2 \!-\! \om \nu} \over {1 - \om} } \,.
\eeqs

\nid Substitute this back into (\ref{eq:qg3}) we have
\beqs
\ga_{qg}^{ij} &=& - {{\al_s} \over {2 \pi}} g_{ij} 
\int { {d\perb{k}^2} \over {\perb{k}^2}} \left(
\int_0^{r_+} { {dk_+ k_+} \over { r_+} } 
(2k-r)_+^{J\!-\!n\!-\!1}r_+^n (2 \om \!-\! 1 \!-\! \nu) \right. 
\,\,\,\,\, \non \\
&& \,\,\,\,\,\,\,\,\,\,\,
\left. - \int_{r_+}^{k_{1+}}
{ {dk_+ (k_1-k)_+} \over {(k_1-r)_+}} (2k-r)_+^{J\!-\!n\!-\!1}r_+^n 
{ {\om^2 \!+\! (1\!-\!\om)^2 \!-\! \om \nu} \over {1 - \om} } \right)
\non \\
&=& {{\al_s} \over {2 \pi}} {1 \over 2} 2 g_{ij} k_{1+}^J \nu^n
\int { {d\perb{k}^2} \over {\perb{k}^2}} \left( \right.
\int_0^{\nu} { {d \om \om} \over {\nu} } 
(2 \om - \nu)^{J\!-\!n\!-\!1} (2 \om \!-\! 1 \!-\! \nu)
\,\,\,\,\, \non \\ 
&& \,\,\,\,\,\,\,\,\,\,\,
\left. - \int_{\nu}^{1} { {d \om} \over {1 \!-\! \nu} } 
(2 \om - \nu)^{J\!-\!n\!-\!1}  
(\om^2 \!+\! (1 \!-\! \om)^2 \!-\! \om \nu )  \right) \,.
\eeqs

\nid After factoring out the logarithmic divergence in $\perb{k}^2$ we
again recover the result from (\ref{eq:gammaresult})
\beqs
\ga_{qg}^{ij} \! &=& \!{ \al_s \over {2 \pi}} {1 \over 2} 2g^{ij} 
k_{1\!+}^J\nu^n \! \int_0^1 \! d \om (2 \om\!-\!\nu)^{J\!-\!n\!-\!1}
\!\left(\!- \Theta(\nu \! - \! \om){1 \over \nu} \om 
(2\om\!-\!1\!-\!\nu)\! \right. \non \\
&& \,\,\,\,\,\,\,\,\,\,\,\,\,\,\,\,\,\,\,\,\,\,\,\,\,\,  \left. 
+ \Theta( \om \! - \! \nu ){1 \over {1\!-\!\nu}} 
(\om^2 \!+\! (1\!-\!\om)^2\!-\!\om \nu)\! \right) \,.
\eeqs

\chapter{Gluon-Gluon Anomalous Dimension}
\thispagestyle{myheadings}
\markright{}

The computation of the gluonic sector follows similarly. In this
appendix we will show in some detail the calculation of the gluon-gluon
anomalous dimension.

\begin{figure}[ptb]
\begin{center}
\epsfxsize=3in
\epsfysize=3.8in
\leavevmode
\hbox{ \epsffile{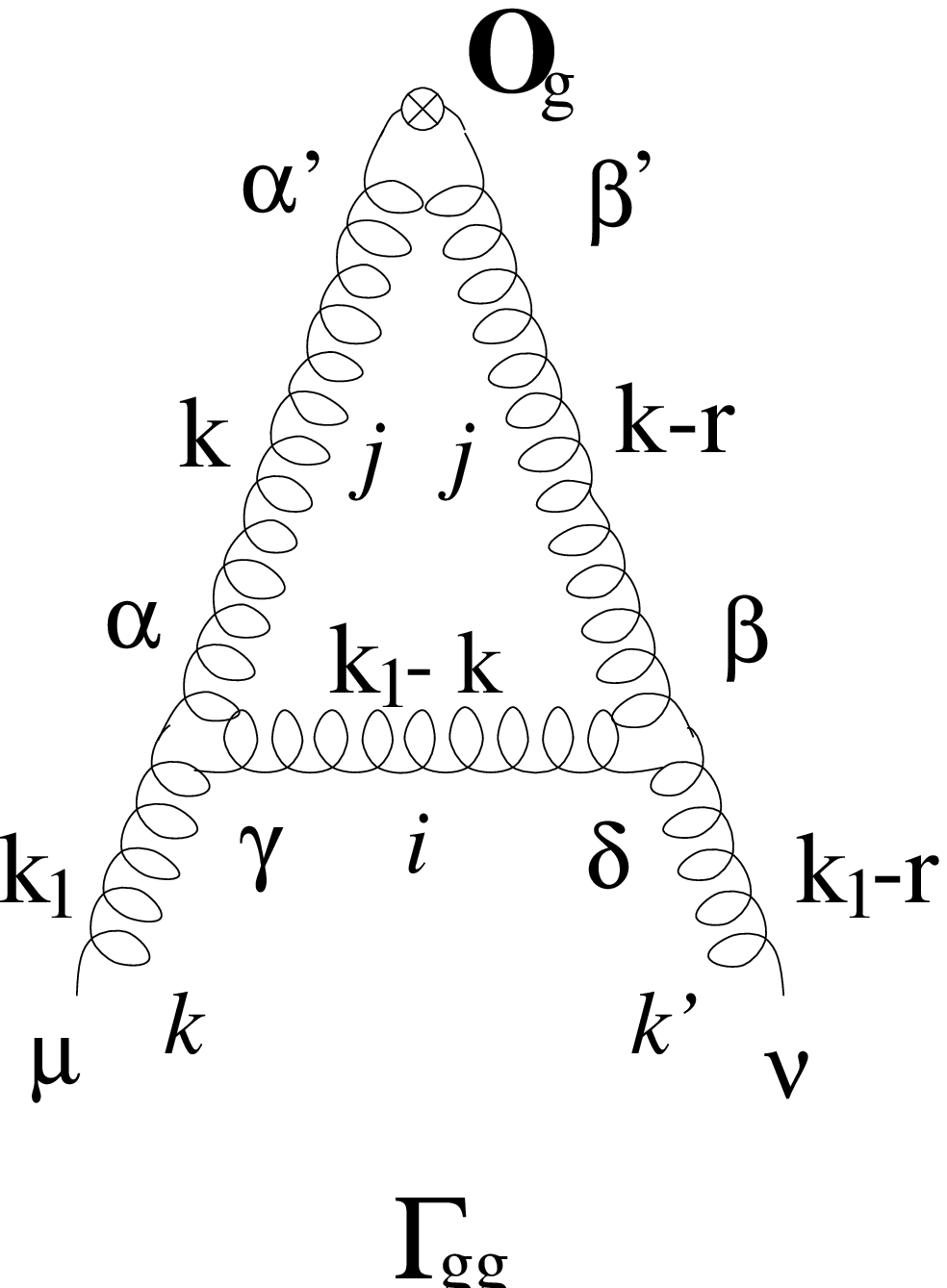}}
\end{center}
\caption{The Gluon-Gluon Diagram of the Anomalous Dimensions}
{This is the detailed version of the last triangle graph--the
gluon-gluon diagram. $\al$, $\al'$, $\be$, $\be'$, $\ga$, $\de$, 
$\mu$ and $\nu$ are Lorentz indices; $i$, $j$, $k$ and $k'$ 
label the gluon lines.}
\label{fig:gg}
\end{figure}

Using standard Feynman rules and light-cone (LC) gauge we can write down 
the value of 
the gluon-gluon triangle diagram (figure \ref{fig:gg}) as
\beqs
\ga_{gg}^{+,1} \!&=&\!  (-i)^3 g^2 \sum_{ij} (f_{ijk}f_{ik'j}) 
 \int { {d^4 k} \over { (2 \pi)^4}}
{ {\Ga_{\ga \al \de} \, D_{\ga \de}(k_1 - k) \, \Ga_{\de \nu \be} }
\over
{(k^2 + i\eps) ( (k_1 \!-\!k)^2 + i\eps) ( (k \!-\! r)^2 + i\eps) } } 
\non \\ 
&& \,\,\,\,\,\,\,\,\,\, \cdot D_{\be \be'}(k - r) \, 
O^{(J,n)}_{g, \be' \al'}(k,k - r) \, D_{\al' \al}(k) \,\, 
\eeqs

\nid where the gluon vertex $O^{(J,n)}_g$ is again taken from
(\ref{eq:vertex}) and $\Ga$ denotes the triple-gluon vertex.

\nid Performing the color algebra with
\beq
\sum_{ij} (f_{ijk}f_{ik'j}) = -C_A \de_{k,k'}
\eeq

\nid and setting the index $k=k'$ we arrive at
\beq
\label{eq:gg1}
\ga_{gg}^{+,1} = -ig^2 C_A \int { {d^4 k} \over { (2 \pi)^4}}
{ {2 (2k-r)_+^{J\!-\!n\!-\!2}r_+^n A^g_{\mu \nu}(k,r)}
\over 
{(k^2 + i\eps) ( (k_1 \!-\!k)^2 + i\eps) ( (k \!-\! r)^2 + i\eps) } } 
\eeq

\nid where
\beq
\label{eq:Agdef}
A^g_{\mu \nu} = \Ga_{\ga \al \de} \, D_{\ga \de}(k_1 - k) \, 
\Ga_{\de \nu \be} D_{\be \be'}(k - r) \, 
V^g_{\be' \al'}(k,k - r) \, D_{\al' \al}(k)
\eeq

\nid with 
\beq
D_{\ga \de}(k_1-k) = g_{\ga \de} - { {n_\ga (k_1-k)_\de + (k_1-k)_\ga
n_\de} \over {n \cdot (k_1-k)}}
\eeq

\nid and similar $ D_{\be \be'}(k - r) $ and 
$D_{\al' \al}(k)$. Again the LC null four vector $n$ is defined 
with properties $n^2=0$ and $n \cdot v = v_+$ for any four 
vector $v$. 

\nid The one point we need to pay extra attention to is the form of the 
gluon vertex in (\ref{eq:vertex}) and thus in (\ref{eq:Agdef}). The full 
tensorial structure of the LC gluon vertex in the non-forward case, 
generalized from that in \cite{Mueller78}, is  
\beq
V^g_{\be' \al'}(k,k-r)= g_{\be' \al'} n \! \cdot \! k 
n \! \cdot \! (k-r) - n \! \cdot \! (k-r) k_\be' n_\al'
- n \! \cdot \! k \, n_\be' (k-r)_\al'
+ k \! \cdot \! (k-r) n_\be' n_\al' \,, \;
\eeq

\nid and it is to be contracted with gluon lines 
$D_{\al' \al}(k)$ and $D_{\be \be'}(k-r)$. 

\nid We are now going to show explicitly that
\beqs
\label{eq:gvtxsimp}
D_{\be \be'}(k-r) \, V^g_{\be' \al'}(k, k-r) \, D_{\al' \al}(k)
\equiv V^g_{\be \al}(k, k-r)  \,\,\,\,\,\,  \non \\
\,\,\,\,\,\, D_{\be \be'}(k-r) \, g_{\be' \al'} n \! \cdot \! k 
n \! \cdot \! (k-r) \, D_{\al' \al}(k) \equiv V^g_{\be \al}(k, k-r),
\eeqs

\nid thus we can substitute the full gluon vertex by its first term
because of the LC projectors of the two gluon lines connected to 
the vertex and use (\ref{eq:vertex}) rather than the full expression
in (\ref{eq:Agdef}).

We split $V^g_{\be' \al'}(k, k-r)$ into two parts, with the $ g_{\be'
\al'}$ term being one part and the remaining three terms being the other.
That is, we write
\beq
D_{\be \be'}(k-r) \, V^g_{\be' \al'}(k, k-r) \, D_{\al' \al}(k)
\equiv V^{(g,1)}_{\be \al} - V^{(g,2)}_{\be \al} \,,
\eeq

\nid where
\beqs
V^{(g,1)}_{\be \al} \!&=&\!  D_{\be \be'}(k-r) \, g_{\be' \al'} 
n \! \cdot \! k n \! \cdot \! (k-r) \,  D_{\al' \al}(k) \,, \non \\
V^{(g,2)}_{\be \al} \!&=&\! D_{\be \be'}(k-r) \, 
(n \! \cdot \! (k-r) k_{\be'} n_{\al'} + n \! \cdot \! k n_{\be'} (k-r)_{\al'} 
- k \! \cdot \! (k-r) n_{\be'} n_{\al'} ) \, D_{\al' \al}(k) \, .\non
\eeqs

\nid We have
\beqs
V^{(g,1)}_{\be \al} \!&=&\! \left( g_{\be \be'} - 
{ {n_\be (k \!-\! r)_{\be'} + (k \!-\!r)_\be n_{\be'}} 
\over {n \cdot (k-r)}} \right)
(g_{\be' \al'} \, n \! \cdot \! k n \! \cdot \! (k \!-\! r)) \,
D_{\al' \al}(k) \non \\
\!&=&\! \left( n \! \cdot \! k n \! \cdot \! (k \!-\! r)  g_{\be \al'}
- n \! \cdot \! k (n_\be (k \!-\! r)_{\al'} + (k \!-\!r)_\be n_{\al'})
\right) \left( g_{\al' \al} - { {n_{\al'} k_\al + k_{\al'} n_\al} 
\over {n \cdot k}} \right) \non \\
\!&=&\!  n \! \cdot \! k n \! \cdot \! (k \!-\! r)  g_{\be \al}
- n \! \cdot \! k (n_\be (k \!-\! r)_\al + (k \!-\!r)_\be n_\al)
- n \! \cdot \! (k \!-\! r) (n_\be k_\al + k_\be n_\al) \non \\
& & \! + \, n \! \cdot \! (k \!-\! r) n_\be k_\al 
+ k \! \cdot \! (k-r) n_\be n_\al + n^2 (k \!-\! r)_\be k_\al
+ n \! \cdot \! k (k \!-\!r)_\be n_\al \non \\
\!&=&\! g_{\be \al} n \! \cdot \! k 
n \! \cdot \! (k-r) - n \! \cdot \! (k-r) k_\be n_\al
- n \! \cdot \! k n_\be (k-r)_\al
+ k \! \cdot \! (k-r) n_\be n_\al \non \\
& \equiv & V^g_{\be \al}(k, k-r) \,,
\eeqs

\nid and
\beqs
V^{(g,2)}_{\be \al} \!\! &=&\! \! \left( g_{\be \be'} -
{ {n_\be (k \!-\! r)_{\be'} + (k \!-\!r)_\be n_{\be'}} 
\over {n \cdot (k-r)}} \right)
(n \! \cdot \! (k-r) k_{\be'} n_{\al'} + n \! \cdot \! k n_{\be'} (k-r)_{\al'}
\non \\
& & \,\,\,\,\,\,\,\,\,\,
- k \! \cdot \! (k \!-\! r) n_{\be'} n_{\al'} ) \, D_{\al' \al}(k) \non \\
\!&=&\! ( n \! \cdot \! (k \!-\! r) k_\be n_{\al'}
+ n \! \cdot \! k n_\be (k \!-\! r)_{\al'}
- 2 k \! \cdot \! (k \!-\! r) n_{\be} n_{\al'} 
- n \! \cdot \! k (k \!-\! r)_\be n_{\al'} \non \\
& & \,\,\,\,\,  -{ {n \! \cdot \! k} \over {n \! \cdot \! (k \!-\! r)} }
(n \! \cdot \! (k \!-\! r) n_\be (k \!-\! r)_{\al'} 
  + n^2 (k \!-\! r)_\be (k \!-\! r)_{\al'}) \non \\
& & \,\,\,\,\, 
+ { {k \! \cdot \! (k \!-\! r)} \over {n \! \cdot \! (k \!-\! r)} }
(n \! \cdot \! (k \!-\! r) n_\be n_{\al'} + n^2  (k \!-\! r)_\be n_{\al'})
) \, D_{\al' \al}(k) \non \\
\!&=&\! ( n \! \cdot \! (k \!-\! r) k_\be n_{\al'}
+ n \! \cdot \! k n_\be (k \!-\! r)_{\al'}
- k \! \cdot \! (k \!-\! r) n_{\be} n_{\al'} 
- k \! \cdot \! (k \!-\! r) n_{\be} n_{\al'} \non \\
&& - n \! \cdot \! k (k \!-\! r)_\be n_{\al'}
- n \! \cdot \! k n_\be (k \!-\! r)_{\al'}
+ k \! \cdot \! (k \!-\! r)  n_\be n_{\al'} ) \, D_{\al' \al}(k) \non \\
\!&=&\! (n \! \cdot \! (k \!-\! r) k_\be n_{\al'}
- 2 k \! \cdot \! (k \!-\! r) n_{\be} n_{\al'} 
- n \! \cdot \! k (k \!-\! r)_\be n_{\al'}
+ k \! \cdot \! (k \!-\! r)  n_\be n_{\al'} ) \non \\ 
& & \,\,\,\,\,\,\,\,\,\,\,\,\,\,\,\,\, \cdot
\left( g_{\al' \al} - { {n_{\al'} k_\al + k_{\al'} n_\al} 
\over {n \cdot k}} \right) \non \\
\!&=&\! (n \! \cdot \! (k \!-\! r) k_\be n_\al
- 2 k \! \cdot \! (k \!-\! r) n_{\be} n_\al
- n \! \cdot \! k (k \!-\! r)_\be n_\al
+ k \! \cdot \! (k \!-\! r)  n_\be n_\al \non \\
&& - { {n \! \cdot \! (k \!-\! r)} \over {n \! \cdot \! k} }
(n^2 k_\be k_\al + n \! \cdot \! k  k_\be k_\al)
+ { {2 k \! \cdot \! (k \!-\! r)} \over {n \! \cdot \! k} }
(n^2 k_\be k_\al + n \! \cdot \! k  k_\be k_\al) \non \\
&& + ( n^2 (k \!-\! r)_\be k_\al 
       + n \! \cdot \! k (k \!-\! r)_\be n_\al)
- { {k \! \cdot \! (k \!-\! r)} \over {n \! \cdot \! k} }
(n^2  n_{\be} n_\al + n \! \cdot \! k n_{\be} n_\al) ) \non \\
\!&=&\! n \! \cdot \! (k \!-\! r) k_\be n_\al 
- k \! \cdot \! (k \!-\! r) n_{\be} n_\al
- n \! \cdot \! k (k \!-\! r)_\be n_\al
- n \! \cdot \! (k \!-\! r) k_\be n_\al \non \\
&& + 2 k \! \cdot \! (k \!-\! r) n_{\be} n_\al 
+  n \! \cdot \! k (k \!-\! r)_\be n_\al
- k \! \cdot \! (k \!-\! r) n_{\be} n_\al \non \\
& \equiv & 0 \,\,\, .
\eeqs

\nid It is now clear that we have established (\ref{eq:gvtxsimp}), and
explicitly we have
\beqs
& & D_{\be \be'}(k-r) V^g_{\be' \al'}(k, k-r) \, D_{\al' \al}(k)
= V^g_{\be \al}(k, k-r) \non \\
& & \,\,\,\,\,\,\, = g_{\be \al} n \! \cdot \! k 
n \! \cdot \! (k-r) - n \! \cdot \! (k-r) k_\be n_\al
- n \! \cdot \! k n_\be (k-r)_\al
+ k \! \cdot \! (k-r) n_\be n_\al \,. \non \\
\eeqs

The other factors in the definition of $A^g_{\mu \nu}$ 
(equation \ref{eq:Agdef}) are the triple-gluon vertices involved in the
diagram:
\beqs
\Ga_{\ga \al \mu} \!& = &\! g_{\ga \al} (2k-k_1)_\mu 
-g_{\al \mu}(k_1+k)_\ga + g_{\mu \ga} (2k_1 - k)_\al 
\equiv -\Ga_{\ga \mu \al} \,, \non \\
\Ga_{\de \nu \be} \!& = &\! g_{\de \nu} (2k_1\!-\!k\!-\!r)_\be 
-g_{\nu \be}(2r\!-\!k_1\!-\!k)_\de + g_{\be \de} (2k\!-\!k_1\!-\!r)_\nu 
\equiv -\Ga_{\be \nu \de} \, . \;\;
\eeqs

Thus we can rewrite (\ref{eq:Agdef}) as, since $D_{\ga \de} = D_{\de \ga}$,
\beqs
A^g_{\mu \nu} &=& \Ga_{\ga \al \mu} \, D_{\ga \de}(k_1 - k) \, 
\Ga_{\de \nu \be} D_{\be \be'}(k - r) \, 
V^g_{\be' \al'}(k,k - r) \, D_{\al' \al}(k) \non \\
&=&  \Ga_{\ga \al \de} \, D_{\ga \de}(k_1 - k) \, 
\Ga_{\de \nu \be} V^g_{\be \al}(k,k - r) \non \\
&=& (-\Ga_{\de \nu \be})  D_{\de \ga} (- \Ga_{\ga \al \mu})  V^g_{\be \al}
\equiv \Ga_{\be \nu \de} D_{\de \ga} \Ga_{\ga \mu \al} V^g_{\be \al}
\,.
\eeqs 

\nid Expanding the above we obtain the full expression of the ``gluon
triangle'' as
\beqs
\label{eq:ggfull}
&& \!\!\!\!\!\!\!\!\!\!\!\!\!
A^g_{\mu \nu} \equiv \Ga_{\be \nu \de} D_{\de \ga} \Ga_{\ga \mu \al} 
V^g_{\be \al} \non \\
&=& \!\!\!\! n \dt k n \dt (k \!-\! r) \,
[ \, ( \, (k \!-\! 2k_1) \dt (k \!-\! 2k_1 \!+\! r) 
+ (k_1 \!+\! k) \dt (k_1 \!+\! k \!-\! 2r) \, ) \, g_\mn \non \\
&& \!\!\!\! + \, 
  (k_1 \!+\! k \!-\! 2r)_\mu (k \!-\! 2k_1)_\nu 
+ (k \!-\! 2k_1 \!+\! r)_\mu (k_1 \!+\! k)_\nu \non \\
&& \!\!\!\! + \,
  (k \!-\! 2k_1)_\mu (k_1 \!-\! 2k \!+\! r)_\nu 
+ (k_1 \!+\! k)_\mu (k_1 \!-\! 2k \!+\! r)_\nu \non \\
&& \!\!\!\! + \,
  (k_1 \!-\! 2k)_\mu (k_1 \!+\! k \!-\! 2r)_\nu
+ (k_1 \!-\! 2k)_\mu (k \!-\! 2k_1 \!+\! r)_\nu
+ 4 (k_1 \!-\! 2k)_\mu (k_1 \!-\! 2k \!+\! r)_\nu \, ] \non \\
&-& \!\!\!\!
\frac {n \dt k n \dt (k \!-\! r)} {n \dt (k_1 \!-\! k)} \,
[ ( (k_1 \!+\! k) \dt (k_1 \!-\! k) n \dt (k_1 \!+\! k \!-\! 2r) 
\!+\! (k_1 \!-\! k) \dt (k_1 \!+\! k \!-\! 2r) \, n \dt (k_1 \!+\! k) )g_\mn
\non \\
&& \!\!\!\! + \,
n \dt (k_1 \!+\! k \!-\! 2r) \, ( \, (k_1 \!-\! k)_\mu (k \!-\! 2k_1)_\nu
+ (k_1 \!-\! 2k)_\mu (k_1 \!-\! k)_\nu \,) \non \\
&& \!\!\!\! + \,
n \dt (k_1 \!+\! k) \, ( \, (k_1 \!-\! k)_\mu (k_1 \!-\! 2k \!+\! r)_\nu
+ (k \!-\! 2k_1 \!+\! r)_\mu (k_1 \!-\! k)_\nu \,) \non \\
&& \!\!\!\! + \,
n \dt (k \!-\! 2k_1) (k_1 \!-\! k)_\mu (k_1 \!-\! 2k \!+\! r)_\nu
+ n \dt (k \!-\! 2k_1 \!+\! r) (k_1 \!-\! 2k)_\mu (k_1 \!-\! k)_\nu
\non \\
&& \!\!\!\! + \,
2 n \dt (k_1 \!-\! k)  (k_1 \!-\! 2k)_\mu  (k_1 \!-\! 2k \!+\! r)_\nu
\non \\
&& \!\!\!\! + \,
( \, (k_1 \!-\! k) \dt (k_1 \!+\! k) 
+ (k_1 \!-\! k) \dt (k \!-\! 2k_1) \,) \, n_\mu (k_1 \!-\! 2k \!+\! r)_\nu
\non \\
&& \!\!\!\! + \,
(k_1 \!-\! k) \dt (k_1 \!+\! k \!-\! 2r) n_\mu (k \!-\! 2k_1)_\nu
+ (k \!-\! 2k_1) \dt (k \!-\! 2k_1 \!+\! r) n_\mu (k_1 \!-\! k)_\nu
\non \\
&& \!\!\!\! + \,
( \, (k_1 \!-\! k) \dt (k_1 \!+\! k \!-\! 2r)
+ (k_1 \!-\! k) \dt (k \!-\! 2k_1 \!+\! r) \,) \, (k_1 \!-\! 2k)_\mu n_\nu 
\non \\
&& \!\!\!\! + \,
(k \!-\! 2k_1) \dt (k \!-\! 2k_1 \!+\! r) (k_1 \!-\! k)_\mu n_\nu 
+ (k_1 \!-\! k) \dt (k_1 \!+\! k) (k \!-\! 2k_1 \!+\! r)_\mu n_\nu \,]
\non \\
&-& \!\!\!\! 
n \dt (k \!-\! r) \; [ \; n \dt (k \!-\! 2k_1) \; k \dt (k \!-\! 2k_1 \!+\! r) 
g_\mn \non \\
&& \!\!\!\! + \,
n \dt (k \!-\! 2k_1) 
( k_\mu (k_1 \!-\! 2k \!+\! r)_\nu + (k_1 \!+\! k \!-\! 2r)_\mu k_\nu )
\non \\
&& \!\!\!\! + \,
n \dt (k_1 \!+\! k \!-\! 2r) (k_1 \!-\! 2k)_\mu k_\nu
+ n \dt k (k_1 \!-\! 2k)_\mu (k_1 \!-\! 2k \!+\! r)_\nu \non \\
&& \!\!\!\! + \,
(k_1 \!+\! k) \dt (k_1 \!+\! k \!-\! 2r) \, n_\mu k_\nu 
\!+\! k \dt (k \!-\! 2k_1 \!+\! r) n_\mu (k_1 \!+\! k)_\nu
\!+\! k \dt (k_1 \!+\! k) n_\mu (k_1 \!-\! 2k \!+\! r)_\nu \non \\
&& \!\!\!\! + \,
k \dt (k \!-\! 2k_1 \!+\! r) (k_1 \!-\! 2k)_\mu n_\nu \;] \non \\
&+& \!\!\!\!
\frac {n \dt (k \!-\! r)} {n \dt (k_1 \!-\! k)} 
\; [ \; n \dt (k_1 \!+\! k \!-\! 2r) \;
(\; n \dt (k \!-\! 2k_1) (k_1 \!-\! k)_\mu k_\nu
+ n \dt (k_1 \!-\! k) (k_1 \!-\! 2k)_\mu k_\nu \; ) \non \\
&& \!\!\!\! + \,
\;(\; n \dt (k \!-\! 2k_1) (k_1 \!-\! k) \dt (k_1 \!+\! k \!-\! 2r)
+ n \dt (k_1 \!+\! k \!-\! 2r) (k_1 \!-\! k) \dt (k_1 \!+\! k) \non \\
&& + \,
n \dt (k_1 \!+\! k) (k_1 \!-\! k) \dt (k_1 \!+\! k \!-\! 2r) \;)\;
n_\mu k_\nu \non \\
&& \!\!\!\! + \,
k \dt (k \!-\! 2k_1 \!+\! r) \;(\;
(k_1 \!-\! k) \dt (k_1 \!+\! k) \, n_\mu n_\nu 
+ (\, n \dt (k_1 \!+\! k) + n \dt (k \!-\! 2k_1) \,) n_\mu (k_1 \!-\! k)_\nu
\non \\
&& + \,
n \dt (k \!-\! 2k_1) (k_1 \!-\! k)_\mu n_\nu 
+ n \dt (k_1 \!-\! k) (k_1 \!-\! 2k)_\mu n_\nu \;) \non \\
&& \!\!\!\! + \,
(\; (\, n \dt (k \!-\! 2k_1) \, k \dt (k_1 \!-\! k)
 + n \dt k \, (k_1 \!-\! k) \dt (k_1 \!+\! k)
 + n \dt (k_1 \!+\! k) \, k \dt (k_1 \!-\! k) \,) \, n_\mu
\non \\
&& + \,
n \dt k \,(\, n \dt (k \!-\! 2k_1) (k_1 \!-\! k)_\mu
 + n \dt (k_1 \!-\! k) (k_1 \!-\! 2k)_\mu \,) 
\;) \; (k_1 \!-\! 2k \!+\! r)_\nu \;] \non \\
&-& \!\!\!\!
n \dt k \;[\; n \dt (k \!-\! 2k_1 \!+\! r) 
\, (k \!-\! 2k_1) \dt (k \!-\! r) \, g_\mn
+ (k \!-\! 2k_1) \dt (k \!-\! r) \, n_\mu (k_1 \!-\! 2k \!+\! r)_\nu
\non \\
&& \!\!\!\! + \,
n \dt (k \!-\! 2k_1 \!+\! r) \, ( \, 
(k_1 \!-\! 2k)_\mu (k \!-\! r)_\nu + (k \!-\! r)_\mu (k_1 \!+\! k)_\nu \,)  
\non \\ 
&& \!\!\!\! + \,
(\, n \dt (k_1 \!+\! k) \, (k \!-\! r)_\mu 
+ n \dt (k \!-\! r) \, (k_1 \!-\! 2k)_\mu \,) \, (k_1 \!-\! 2k \!+\! r)_\nu
\non \\
&& \!\!\!\! + \,
(\, (k \!-\! 2k_1) \dt (k \!-\! r) \, (k_1 \!+\! k \!-\! 2r)_\mu
+ (k_1 \!+\! k) \dt (k_1 \!+\! k \!-\! 2r) \, (k \!-\! r)_\mu \non \\
&& + \,
(k \!-\! r) \dt (k_1 \!+\! k \!-\! 2r) \, (k_1 \!-\! 2k)_\mu \,) 
\, n_\nu \;] \non \\
&+& \!\!\!\!
\frac {n \dt k} {n \dt (k_1 \!-\! k)} \;[\;
(\; (k \!-\! 2k_1) \dt (k \!-\! r) \,(\,
  n \dt (k_1 \!+\! k \!-\! 2r) \, (k_1 \!-\! k)_\mu
+ (k_1 \!-\! k) \dt (k_1 \!+\! k \!-\! 2r) \, n_\mu \,) \non \\
&& + \,
(\, n \dt (k_1 \!+\! k \!-\! 2r) \, (k_1 \!-\! k) \dt (k_1 \!+\! k)
 + n \dt (k_1 \!+\! k) \, (k_1 \!-\! k) \dt (k_1 \!+\! k \!-\! 2r) \,)
\, (k \!-\! r)_\mu \non \\
&& + \,
(\, n \dt (k_1 \!+\! k \!-\! 2r) \, (k_1 \!-\! k) \dt (k \!-\! r)
\!+\! n \dt (k \!-\! r) \, (k_1 \!-\! k) \dt (k_1 \!+\! k \!-\! 2r) \,)
\, (k_1 \!-\! 2k)_\mu \,)\; n_\nu \non \\
&& \!\!\!\! + \,
n \dt (k \!-\! 2k_1 \!+\! r) \;(\;
(\, n \dt (k_1 \!+\! k) \, (k \!-\! r)_\mu 
+ n \dt (k \!-\! r) \, (k_1 \!-\! 2k)_\mu \non \\
&& \hspace*{6.5cm} + \,
(k \!-\! 2k_1) \dt (k \!-\! r) \, n_\mu \,)\, (k_1 \!-\! k)_\nu \non \\
&& + \,
(\, (k \!-\! 2k_1) \dt (k \!-\! r) \, (k_1 \!-\! k)_\mu
+ (k_1 \!-\! k) \dt (k_1 \!+\! k) \, (k \!-\! r)_\mu \non \\
&& \hspace*{6.5cm} + \,
(k_1 \!-\! k) \dt (k \!-\! r) \, (k_1 \!-\! 2k)_\mu \,)\, n_\nu \;)
\non \\
&& \!\!\!\! + \,
n \dt (k_1 \!-\! k) 
\,(\, n \dt (k_1 \!+\! k) \, (k \!-\! r)_\mu 
  +   n \dt (k \!-\! r) \, (k_1 \!-\! 2k)_\mu \non \\
&& + \,
(k \!-\! 2k_1) \dt (k \!-\! r) \, n_\mu \,) 
\, (k_1 \!-\! 2k \!+\! r)_\nu \;] \non \\
&+& \!\!\!\!
k \dt (k \!-\! r) 
\;[\; n \dt (k \!-\! 2k_1) \, n \dt (k \!-\! 2k_1 \!+\! r) \, g_\mn
+ (k_1 \!+\! k) \dt (k_1 \!+\! k \!-\! 2r) \, n_\mu n_\nu \non \\
&& \!\!\!\! + \,
n \dt (k \!-\! 2k_1 \!+\! r) \, n_\mu (k_1 \!+\! k)_\nu
+ (\, n \dt (k \!-\! 2k_1) + n \dt (k_1 \!+\! k) \,) \, n_\mu 
(k_1 \!-\! 2k \!+\! r)_\nu \non \\
&& \!\!\!\! + \,
n \dt (k \!-\! 2k_1) \, (k_1 \!+\! k \!-\! 2r)_\mu n_\nu
+ (\, n \dt (k_1 \!+\! k \!-\! 2r) + n \dt (k \!-\! 2k_1 \!+\! r) \,)
\, (k_1 \!-\! 2k)_\mu n_\nu \;] \non \\
&-& \!\!\!\!
\frac {k \dt (k \!-\! r)} {n \dt (k_1 \!-\! k)} \;[\;
(\, n \dt (k \!-\! 2k_1) + n \dt (k_1 \!+\! k) \,) \,
n \dt (k \!-\! 2k_1 \!+\! r) \, n_\mu (k_1 \!-\! k)_\nu \non \\
&& \!\!\!\! + \,
(\, n \dt (k \!-\! 2k_1) + n \dt (k_1 \!+\! k) \,)\,
n \dt (k_1 \!-\! k) \; n_\mu (k_1 \!-\! 2k \!+\! r)_\nu \non \\
&& \!\!\!\! + \,
(n \dt (k_1 \!+\! k \!-\! 2r) \!+\! n \dt (k \!-\! 2k_1 \!+\! r) ) \,
(n \dt (k \!-\! 2k_1) \, (k_1 \!-\! k)_\mu 
\!+\! n \dt (k_1 \!-\! k) \, (k_1 \!-\! 2k)_\mu ) \, n_\nu \non \\
&& \!\!\!\! + \,
(\; (\, n \dt (k \!-\! 2k_1) + n \dt (k_1 \!+\! k) \,)\,
(k_1 \!-\! k) \dt (k_1 \!+\! k \!-\! 2r) \non \\
&& + \,
(\, n \dt (k_1 \!+\! k \!-\! 2r) + n \dt (k \!-\! 2k_1 \!+\! r) \,)\,
(k_1 \!-\! k) \dt (k_1 \!+\! k) \;) \; n_\mu n_\nu \;] \; ,
\eeqs

\nid where we have used $n \dt n = n^2 = 0$.

Similar to the calculation of $\ga_{qq}$ we perform the $k_-$ integral
in (\ref{eq:gg1}) first. The discussion of different regions of $k_+$ and
the corresponding pole values goes almost identical. We have
\beqs
\label{eq:gg2}
\ga_{gg}^{+,1} &=& -ig^2 C_A \int { {d^4 k} \over { (2 \pi)^4}}
{ {2 (2k-r)_+^{J\!-\!n\!-\!2}r_+^n A^g_{\mu \nu}(k,r)}
\over 
{(k^2 + i\eps) ( (k_1 \!-\!k)^2 + i\eps) ( (k \!-\! r)^2 + i\eps) } } 
\non \\
&=& -i {{\al_s C_A} \over { (2 \pi)^2}} 
\int { {d \perb{k}^2} \over {\perb{k}^4}} \left(
{ { -2 \pi i} \over { r_+ k_{1+}}} \int_0^{r_+}
dk_+ k_+ A^{g,(1)}_{\mu \nu} (2k-r)_+^{J\!-\!n\!-\!2}r_+^n \, \right. 
\,\,\,\,\,\,\,\,\,\,\,\,\,\,\,\,\,\,\,\,\,\,\,\,\,\, \nonumber \\
&& \,\,\,\,\,\,\,\,\,\,\,\,\,\,\,\,\,\,\,\,\,\,\,\,\,\,\,\,\,\, \left. 
+ { {2 \pi i} \over { - k_{1+} (k_1-r)_+}} \int_{r_+}^{k_{1+}}
dk_+ (k_1-k)_+ A^{g,(2)}_{\mu \nu} (2k-r)_+^{J\!-\!n\!-\!2}r_+^n \right)
\non \\
&=& - {{\al_s C_A} \over {2 \pi}}
\int { {d\perb{k}^2} \over {\perb{k}^4}} \left(
\int_0^{r_+} { {dk_+ k_+} \over { r_+ k_{1+}} }
(2k-r)_+^{J\!-\!n\!-\!2}r_+^n A^{g,(1)}_{\mu \nu} \,\right. \non \\
&& \,\,\,\,\,\,\,\,\,\,\,\,\,\,\,\,\,\,\,\,\,\,\,\,\,\,\,\,\,\,\,\,
\left. + \int_{r_+}^{k_{1+}} 
{ {dk_+ (k_1-k)_+} \over { - k_{1+} (k_1-r)_+}}
(2k-r)_+^{J\!-\!n\!-\!2}r_+^n A^{g,(2)}_{\mu \nu} \right) \non \\
&\equiv& \ga_{gg}^{1,(1)} + \ga_{gg}^{1,(2)} \,,
\eeqs

\nid where $A^{g,(1)}$ and $A^{g,(2)}$ are $A^g_{\mu \nu}(k,r)$ evaluated at 
the poles $k_-^{(1)}$ and $k_-^{(2)}$, respectively, and again we have
\newpage
\beqs
\label{eq:poles2}
k_-^{(1)} = {{\perb{k}^2} \over {2 k_+}} 
= { {\perb{k}^2} \over {2 k_{1+} \om} } \,\;\;\;\;\;\;\; & \& & \,\,
k^2 = 2 k_+ k_- - \perb{k}^2 =  0 , \non \\
k_-^{(2)} = { {\perb{k}^2} \over {2 (k_+ \!-\! k_{1+})}} 
= { {-\perb{k}^2} \over { 2 k_{1+} (1 \!-\! \om)} } \,\,\,
& \& & \,\, k^2 = - { {\perb{k}^2} \over { (1 - \om) } }. 
\eeqs

To extract the gluon-gluon anomalous dimension, we do not need to
calculate further the full expression \ref{eq:ggfull}.
Rather, the projection of $A^g_{\mu \nu}$ onto the transverse
directions will suffice. That is, by effectively treating the gluon 
vertex to be 
\beq
V^g_{ij} = g_{ij} n \! \cdot \! k \, n \! \cdot \! (k \!-\!r) \;.
\eeq

\nid we only need to keep in equation \ref{eq:ggfull}
those terms proportional to $g_{ij}$ and $k_i k_j \leftrightarrow 
-{1 \over 2} g_{ij} \perb{k}^2$. Thus we define 
\beq
A^g_{ij} \equiv g_{ij} A_I + k_i k_j A_{II}
\eeq

\nid and calculate $A_I$ and $A_{II}$ separately.  Note as stated in
the main text, we may choose $k_1$ and $r$ such that only their plus
components are non-vanishing, that is,
\beq
k_1 = (k_+, 0, \, \perb{0}) \, , \,\,\,\,\,\,\,\,
r = (r_+, 0, \, \perb{0}) \,,
\eeq

\nid which also leads to
\beq
k_1 \cdot r = k_1^2 = r^2 =0 \;.
\eeq

\nid Reading off (and simplifying) from the full expression of 
$A^g_{\mu \nu}$ we arrive at
\beqs
A_I &=& n \!\cdot\! k n \!\cdot\! (k\!-\!r) 
(2k^2 \!- \!2k_1 \!\cdot\! k \!-\! k \!\cdot\! r) \non \\
&& + \, { {n \!\cdot\! k n \!\cdot\! (k\!-\!r) } 
\over {n \!\cdot\! (k_1\!-\!k)} }
( 2 k^2 n \!\cdot\! (k_1 \!+\! k) - 2 k^2 n \!\cdot\! r 
- 2 n \!\cdot\! (k_1 \!+\! k) k \!\cdot\! r) \non \\
&& - \, n \!\cdot\! (k\!-\!r) n \!\cdot\! (k\!-\!2k_1) 
(k^2 \!- \!2k_1 \!\cdot\! k \!+\! k \!\cdot\! r)
-n \!\cdot\! k n \!\cdot\! (k\!-\!2k_1) (k^2 \!-\! 2k_1 \!\cdot\! k) \non \\
&& - \, n \!\cdot\! k ( n \!\cdot\! r (k^2 \!- \!2k_1 \!\cdot\! k
\!-\! k \!\cdot\! r)
-n \!\cdot\! (k\!-\!2k_1) k \!\cdot\! r) \non \\
&& + \, (k^2 \!-\! k \!\cdot\! r) ( (n \!\cdot\! (k\!-\!2k_1))^2 
+ n \!\cdot\! r (k\!-\!2k_1)) \,\, .
\eeqs

\nid Same as the discussion of the quark-quark anomalous dimension, we
are only interested in the leading powers of $\perb{k}$, which is also
quadratic, and, from the pole values, can only come from terms
proportional to either $k^2$ or $k_-$. Keeping only such terms we
obtain
\beq
A_I = k^2 A_I^a + (k^2 - 2 k_1 \!\cdot\! k) A_I^b + r_+k_- A_I^c \;,
\eeq

\nid where
\beqs
A_I^a &=& 2 \left( k_+^2 - 2 k_{1+}k_+ + 2 k_{1+}^2 - k_{1+}r_+ 
+ { {k_+^2 k_{1+} \!+\! k_+^3 \!-\! k_{1+}k_+ r_+ \!-\!2 k_+^2r_+ 
\!+\! k_+ r_+^2} \over {(k_1 -k)_+} } \right) \non \\
&=& 2 k_{1+}^2 \left (\om^2 -2 \om +2 - \nu +
{ {\om^2 + \om^3 -\om \nu - 2 \om^2 \nu + \om \nu^2} 
\over {1 - \om} } \right) \non \\
A_I^b &=& 4k_{1+}k_+ - k_+^2 - 2 k_{1+}r_+ - k_+r_+ 
\,=\, k_{1+}^2 (4 \om - \om^2 - 2 \nu - \om \nu) \non \\
A_I^c &=& -2 \left(k_+^2 - k_+ r_+ - 2 k_{1+}k_+ + 2 k_{1+}^2 
+ { { k_+^3 + k_+^2 k_{1+} - k_+^2r_+ - k_{1+}k_+ r_+}
\over {(k_1-k)_+} } \right) \non \\
&=& -2k_{1+}^2 (\om^2 - \om \nu - 2 \om +2 +
{ {\om^3 + \om^2 - \om^2 \nu - \om \nu} 
\over {1 - \om} }) \,\, .
\eeqs

\nid At the $k_-^{(1)}$ pole, from (\ref{eq:poles2}) we have
\beq
k^2 = 0 \,, \,\,\,\,\, 2k_1 \!\cdot\! k = { {\perb{k}^2} \over \om} \, ,
\eeq

\nid thus
\beqs
A_I^{(1)} &=& -{ {\perb{k}^2} \over \om } A_I^b 
+ { {\nu \perb{k}^2} \over {2 \om} } A_I^c \non \\
&=& -{ {\perb{k}^2} \over \om } 
k_{1+}^2 (4 \om - \om^2 - 2 \nu - \om \nu) \non \\
&& - { {\nu \perb{k}^2} \over \om } 
k_{1+}^2 (\om^2 - \om \nu - 2 \om +2
+ { {\om^3 + \om^2 - \om^2 \nu - \om \nu} \over {1 - \om} }) \non \\
&=& -{ {\perb{k}^2} \over \om } k_{1+}^2
( 4 \om - \om^2 + \nu (\om^2 -3 \om - \om \nu 
+ { {\om^3 + \om^2 - \nu (\om^2 + \om)} \over {1 -\om} } )) \,\, .
\eeqs

\nid At the $k_-^{(2)}$ pole, on the other hand, from (\ref{eq:poles2})
we have
\beq
k^2 = 2k_1 \!\cdot\! k = - { {\perb{k}^2} \over {1-\om}} \, ,
\eeq

\nid and thus
\beqs
A_I^{(2)} &=& - { {\perb{k}^2} \over {1-\om}} A_I^a 
- { { \nu \perb{k}^2} \over {2 (1-\om)}} A_I^c \non \\
&=& - { {\perb{k}^2} \over {1-\om}} k_{1+}^2
( 2 (\om^2 -2 \om +2 - \nu +
{ {\om^2 + \om^3 -\om \nu - 2 \om^2 \nu + \om \nu^2} 
\over {1 - \om} } ) \non \\
&& - \nu (\om^2 - \om \nu - 2 \om +2 +
{ {\om^3 + \om^2 - \om^2 \nu - \om \nu} 
\over {1 - \om} }) ) \non \\
&=& -{ {\perb{k}^2} \over {1 \!-\!\om} } 2 k_{1+}^2
( \om^2 - 2 \om +2 - 2 \nu 
+ { { \om^3 + \om^2 - 2 \om \nu (2 \om -\nu)} 
\over {1 - \om} }) \,\, .
\eeqs

On the other hand, from the full expression of $A^g_{\mu \nu}$ 
(\ref{eq:ggfull}) we have 
\beqs
A_{II} &=& 10 n \!\cdot\! k n \!\cdot\! (k\!-\!r)
- 2  n \!\cdot\! k n \!\cdot\! (k\!-\!r) 
- n \!\cdot\! (k\!-\!r) n \!\cdot\!(k \!+\!4r)
- n \!\cdot\! k n \!\cdot\!(k \!-\!5r) \non \\
&& + { {n \!\cdot\! k n \!\cdot\! (k\!-\!r)} \over {n \!\cdot\!(k_1\!-\!k)} }
( n \!\cdot\!(k_1 \!-\! k \!-\!2r) + n \!\cdot\!(k_1 \!-\! k \!+\!2r)) 
\non \\
&=& 8 k_+^2 - 8k_+r_+ + 4 r_+^2 \non \\
&=& 4 k_{1+}^2 ( 2 \om^2 - 2\om \nu + \nu^2) \,\, .
\eeqs

\nid Note that $A_{II}$ has the same value at both poles.

So now we have
\beqs
A^g_{ij} &\equiv& g_{ij} A_I + k_i k_j A_{II} \non \\
&=& g_{ij} ( A_I - {1 \over 2}\perb{k}^2 A_{II} ) \non \\
&=& g_{ij} (A_I - \perb{k}^2 k_{1+}^2 2 ( 2 \om^2 - 2\om \nu + \nu^2))
\,\, .
\eeqs

\nid Substitute the above result with the values of $A_I$ at different
poles into (\ref{eq:gg2}), we have
\beqs
\ga_{gg,ij}^{1,(1)} &=& - {{\al_s C_A} \over {2 \pi}}
\int { {d \perb{k}^2} \over {\perb{k}^4}}
\int_0^{r_+} { {dk_+ k_+} \over {r_+ k_{1+}} }
(2k-r)_+^{J\!-\!n\!-\!2}r_+^n A^{g,(1)}_{ij} \non \\
&=& - {{\al_s C_A} \over {2 \pi}}
\int { {d \perb{k}^2} \over {\perb{k}^4}} \int_0^{\nu}
{ {d \om \om} \over \nu } k_{1+}^{J\!-\!2}
(2 \om - \nu)^{J\!-\!n\!-\!2} \nu^n g_{ij} ( A_I^{(1)} -
{1 \over 2} \perb{k}^2 A_{II} ) \non \\
&=&  - {{\al_s C_A} \over {2 \pi}}
\int { {d \perb{k}^2} \over {\perb{k}^2}} 2 g_{ij} k_{1+}^J                   
\int_0^{\nu} d\om (2 \om - \nu)^{J\!-\!n\!-\!2} \nu^n 
{ \om \over {2 \nu}} \non \\
&& \,\,\,\,
\cdot (- {1 \over \om} ( 4 \om - \om^2 + \nu (\om^2 -3 \om - \om \nu 
+ { {\om^3 + \om^2 - \nu (\om^2 + \om)} \over {1 -\om} } )) \non \\
&& \,\,\,\,\,\,\,\,\,\,\,\,\,\,\,\,\,\,\,\,\,\,
-2  (2 \om^2 - 2\om \nu + \nu^2)) \non \\
&=& {{\al_s C_A} \over {2 \pi}} 
\int { {d \perb{k}^2} \over {\perb{k}^2}} 2 g_{ij} k_{1+}^J
\int_0^{\nu} d\om (2 \om - \nu)^{J\!-\!n\!-\!2} \nu^n  \non \\
&& \,\,\,\,\,\,\,\,
\cdot {1 \over 2} \left( {1 \over \nu} (4 \om^3 - \om^2 + 4 \om)
+ { {\om^3 + \om^2 - 2 \om^2 \nu} \over { 1 - \om} } - 3(\om^2 + \om) 
\right)
\eeqs

\nid and 
\beqs
\ga_{gg,ij}^{1,(2)} &=& - {{\al_s C_A} \over {2 \pi}}
\int { {d \perb{k}^2} \over {\perb{k}^4}}
\int_{r_+}^{k_1{1+}} { {dk_+ (k_1-k)_+} \over {k_{1+}(k_1\!-\!r)_+} }
(2k-r)_+^{J\!-\!n\!-\!2}r_+^n A^{g,(2)}_{ij} \non \\
&=& - {{\al_s C_A} \over {2 \pi}}
\int { {d \perb{k}^2} \over {\perb{k}^4}} \int_{\nu}^1
{ {d \om (1\!-\!\om)} \over {1 \!-\!\nu} } k_{1+}^{J\!-\!2}
(2 \om - \nu)^{J\!-\!n\!-\!2} \nu^n g_{ij} ( A_I^{(2)} -
{1 \over 2} \perb{k}^2 A_{II} ) \non \\
&=& - {{\al_s C_A} \over {2 \pi}}
\int { {d \perb{k}^2} \over {\perb{k}^2}} 2 g_{ij} k_{1+}^J
\int_{\nu}^1 d \om (2 \om - \nu)^{J\!-\!n\!-\!2} \nu^n
{ {1\!-\!\om} \over { 2 (1 \!-\!\nu)} } \non \\
&& \cdot (- { 2 \over {1 - \om}} ( \om^2 - 2 \om +2 - 2 \nu 
+ { {\om^3 + \om^2 - 2 \om \nu (2 \om -\nu)} \over {1 - \om} } ) \non\\
&& \,\,\,\,\,\,\,\,
-  2  (2 \om^2 - 2\om \nu + \nu^2)) \non \\
&=& {{\al_s C_A} \over {2 \pi}}
\int { {d \perb{k}^2} \over {\perb{k}^2}} 2 g_{ij} k_{1+}^J
\int_{\nu}^1 d\om (2 \om - \nu)^{J\!-\!n\!-\!2} \nu^n  \non \\
&& \,\,\,\,\,\,\,\,\,\,\,\,
\cdot {1 \over {1 \!-\! \nu} } \left(
{ { 2(1-\om+\om^2)^2 + \nu ( \nu(1+\om^2)-2(1+\om^3) ) }
\over {1 - \om} } \right) \,\, .
\eeqs

\begin{figure}[ptb]
\begin{center}
\epsfxsize=4.5in
\epsfysize=4in
\leavevmode
\hbox{ \epsffile{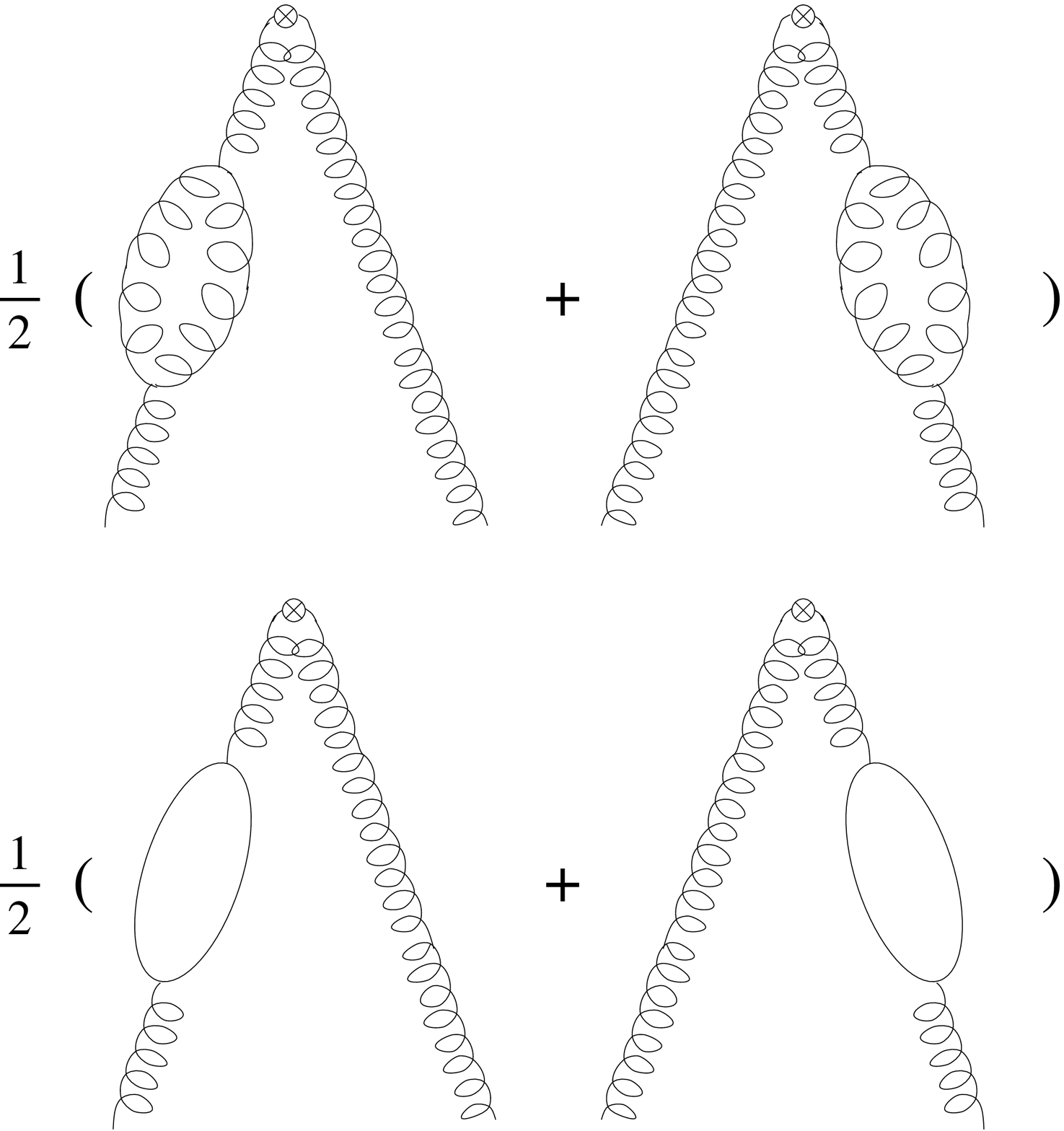}}
\end{center}
\caption{The Gluon Self-Energy Diagram}
{These are the gluon self-energy diagrams that are needed to cancel
the end-point divergences in the LC calculation. We need to include
both the gluon loop and the quark loop contributions.}
\label{fig:gselfE}
\end{figure}

Same as in the case of the quark-quark anomalous dimension, the $\om$
integration is divergent at the end point $\om =1$. This divergence is
again cancelled by self-energy graphs. However, in this case for each
gluon propagator connecting to the vertex, there are two of them, a
gluon loop and a quark loop (figure \ref{fig:gselfE}). Their values in the
LC gauge can again be taken readily from \cite{Curci80} as
\beqs 
Z_g(x) & = & 1 + { \al_s \over {2 \pi} } 2 C_A {2 \over \eps}
(- I_0 - \log{|x|} + { 11 \over 12 }) \non \\
Z_q(x) & = & 1 + { \al_s \over {2 \pi} } 2 C_A {2 \over \eps}
( - { {n_f} \over 18 } ) \,\, .
\eeqs

\nid Again the self-energies depend on the longitudinal momentum
fraction $x$, with $I_0$ the LC colinear divergence (see 
equation \ref{eq:diver}). Using the same 
argument as that was used for the quark-quark graph, we need add to the 
gluon-gluon anomalous dimension the following contribution from 
self-energy diagrams:
\beq
{1 \over 2} ( Z_g(1) + Z_q(1) + Z_g(1-\nu) + Z_q(1-\nu) )= 
{ \al_s \over {2 \pi} } 2 C_A {2 \over \eps} 
(-I_0 - {1 \over 2} \log(1-\nu) + { 11 \over 12 } -  
{ {n_f} \over 18}) \,\, .
\eeq

\nid Identifying $2 \over \eps$ as the logarithmic divergence in
transverse momentum and that
\beq
{ 11 \over 12 } - { {n_f} \over 18} = { 1 \over {4 N_c} } 
(11 - {2 \over 3} n_f) \equiv { 1 \over {4 N_c} } b_0
\eeq

\nid with $b_0$ the leading coefficient of the QCD $\be-$function, we
can see that we have shown that the gluon-gluon anomalous dimension is
indeed given by the same expression of $\ga_{gg}$ as 
in (\ref{eq:gammaresult}), that is,
\beqs
\ga_{gg}^{ij} \!&\! =\!& \!{ {\al_s C_A} \over {2 \pi}} 2g^{ij} 
k_{1\!+}^J\nu^n \!\int_0^1\!\! d \om (2\om\!-\!\nu)^{J\!-\!n\!-\!2}
\!\left(\!\Theta(\nu \! - \! \om) {1 \over 2} 
({1 \over \nu}(4\om^3\!\!-\!\om^2\!\!+\!4\om) \non \right.\\
\!&\!+\!&\!\! ({{\om^3\!\!+\!\om^2\!\!-\!2\om^2\nu} \over {1-\om}}\!-\!
3(\om^2\!+\!\om))) + { {\Theta( \om \! - \! \nu )} \over {1\!-\!\nu}} 
({{2(1\!-\!\om\!+\!\om^2)^2} \over {1-\om}}\!+\! \nu \!
{{\nu(1\!+\!\om^2)\!-\!2(1\!+\!\om^3)} \over {1-\om}}) \non \\
& &  \,\,\,\,\,\,\,\,\,\,\,\,\,\,\,\,\,\,\,\,\,\,\,
\left. \!+ (1\!-\!\nu) \delta(1\!-\!\om) 
( \!-\!2I_0\!-\!\log(1\!-\!\nu)\!+\!{ b_0 \over {2C_A}}) \right).
\eeqs

\chapter{Gluon-Quark Anomalous Dimension}
\thispagestyle{myheadings}
\markright{}

\begin{figure}[ptb]
\begin{center}
\epsfxsize=3in
\epsfysize=3.8in
\leavevmode
\hbox{ \epsffile{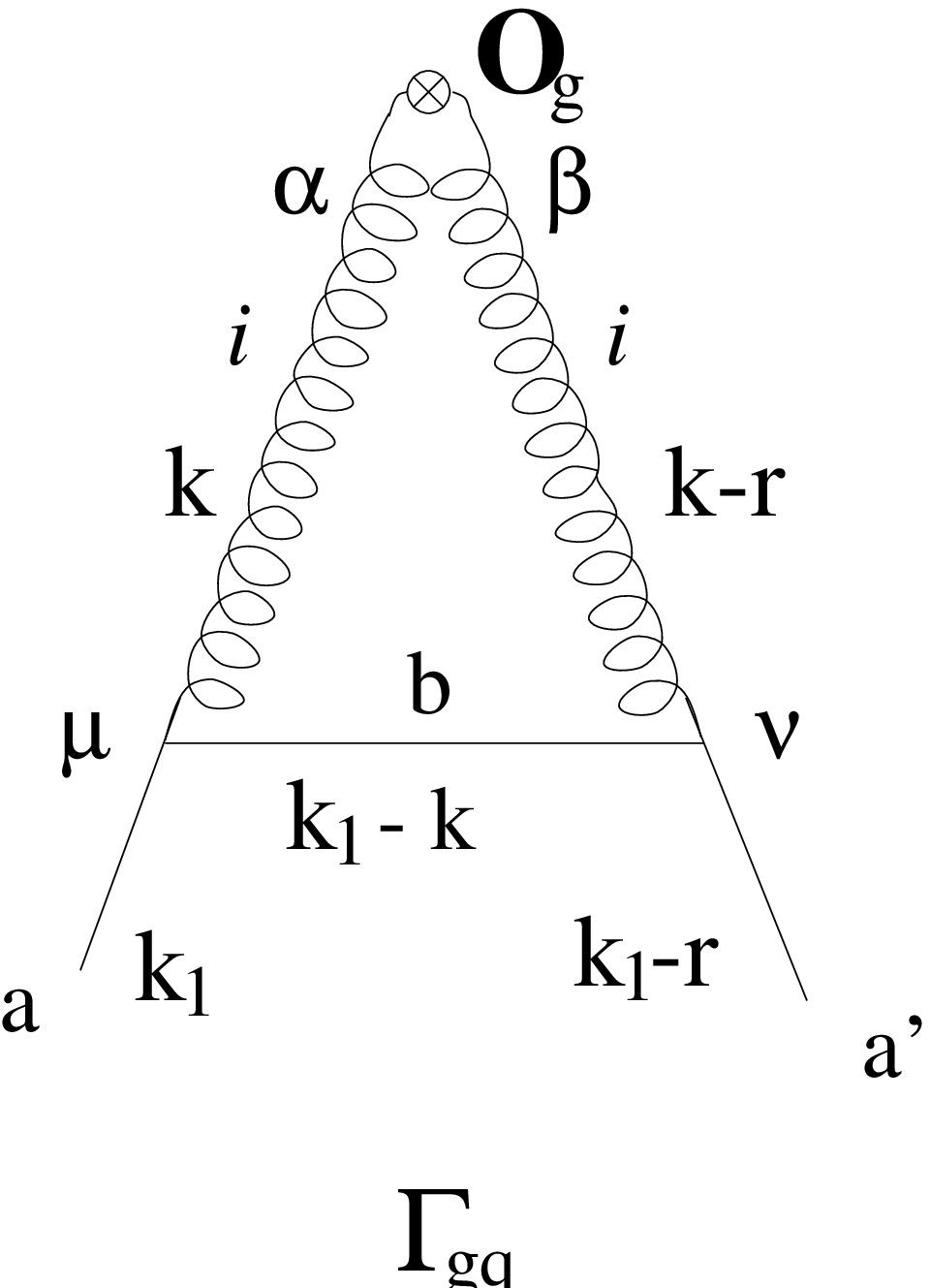}}
\end{center}
\caption{The Gluon-Quark Diagram of the Anomalous Dimensions}
{This is the detailed version of the second triangle graph, the
gluon-quark diagram. $\al$, $\be$, $\mu$ and $\nu$ are Lorentz
indices; $a$, $a'$, and $b$ are color indices for the quarks 
while $i$ labels the gluon lines.}
\label{fig:gq}
\end{figure}

In this appendix we show the details of computing the gluon-quark
anomalous dimension. The gluon-quark transition comes from the 
contribution of the second triangle diagram, as shown in detail 
in figure \ref{fig:gq}. By the same Feynman rules and in the LC gauge, 
we can write the value of the gluon-quark diagram as
\beqs
\label{eq:gq1}
\ga_{gq}^{+,1} &=& (-ig)^2 \sum_{i,b} (T^i_{ab}T^i_{ba'})
\int { {d^4 k} \over { (2 \pi)^4}}
\ga_\mu {i \over { \not{k_1}-\not{k}}} \ga_\nu  
{ {-i D_{\nu \be} (k-r)} \over { (k-r)^2 + i \eps} } \non \\
&& \,\,\,\,\,\,\,\,\,\,\,\,\,\,\,\,\,\,\,\,\,\,\,\,\,\,\,\, \cdot
O_{g,\be \al}^{(J,n)(+)}(k,k-r)
{ {-i D_{\al \mu} (k)} \over { k^2 + i \eps} } \non \\
&=& ig^2 C_F \de_{aa'} \int { {d^4 k} \over { (2 \pi)^4}}
{ {2 (2k-r)_+^{J\!-\!n\!-\!2}r_+^n B^g(k,r)}
\over
{(k^2 + i\eps) ( (k_1 \!-\!k)^2 + i\eps) ( (k \!-\! r)^2 + i\eps) } }
\,.      
\eeqs

\nid Similar to the calculation of the quark-quark diagram (see
section \ref{sec:qq}) we have set the color label of the two quark
lines to be equal ($a =a'$), and defined
\beq
\label{eq:Bgdef}
B^g(k,r) = \ga_\mu (\not{\!k}_1-\not{\!k}) \ga_\nu D_{\nu \be}(k-r)
V^g_{\be \al} D_{\al \mu}(k) \,.
\eeq

\nid The LC gluon projectors are 
\beqs
D_{\nu \be}(k-r) &=& g_{\nu \be} - { {n_\nu (k-r)_\be + (k-r)_\nu
n_\be} \over {n \cdot (k-r)}} \,,\non \\ 
D_{\al \mu}(k) &=& g_{\al \mu} - { {n_\al k_\mu + k_\al
n_\mu} \over {n \cdot k}} \,.
\eeqs

\nid From our convention of the gluon vertex (\ref{eq:vertex}),
by exactly the same argument as that for the gluon-gluon diagram 
(see the derivation of equation \ref{eq:gvtxsimp} in appendix B) 
we have 
\beqs
&& \! \!D_{\nu \be}(k-r) \, V^g_{\be \al}(k, k-r) \, D_{\al \mu}(k)
\equiv V^g_{\nu \mu}(k, k-r) \non \\
&& \,\,\,\,\,\, = g_{\nu \mu} n \! \cdot \! k n \! \cdot \! (k-r) 
- n \! \cdot \! (k-r) k_\nu n_\mu - n \! \cdot \! n_\nu (k-r)_\mu
+ k \! \cdot \! (k-r) n_\nu n_\mu \,. \non \\ 
\eeqs

By the same steps we followed before in computing the $k_-$ integral
and picking up different poles at different $k_+$ regions, we have
\beqs
\label{eq:gq2}
\ga_{gq}^{+,1} &=& {{\al_s C_F} \over {2 \pi}}
\int { {d\perb{k}^2} \over {\perb{k}^4}} \left(
\int_0^{r_+} { {dk_+ k_+} \over { r_+ k_{1+}} }
(2k-r)_+^{J\!-\!n\!-\!2}r_+^n B^{g,(1)} \,\right. \non \\
&& \,\,\,\,\,\,\,\,\,\,\,\,\,\,\,\,\,\,\,\,\,\,\,\,\,\,\,\,\,\,\,\,
\left. + \int_{r_+}^{k_{1+}}
{ {dk_+ (k_1-k)_+} \over { k_{1+} (k_1-r)_+}}
(2k-r)_+^{J\!-\!n\!-\!2}r_+^n B^{g,(2)} \right) \non \\
&\equiv& \ga_{gq}^{1,(1)} + \ga_{gq}^{1,(2)} 
\eeqs

\nid where $B^{g,(1)}$ and $B^{g,(2)}$ are $B^g(k,r)$ evaluated at
the poles $k_-^{(1)}$ and $k_-^{(2)}$, respectively, with
\beqs
k_-^{(1)} = {{\perb{k}^2} \over {2 k_+}}
= { {\perb{k}^2} \over {2 k_{1+} \om} } \,\,\,\;\;\;\;\; & \& & \,\,
k^2 = 2 k_+ k_- - \perb{k}^2 =  0 , \non \\
k_-^{(2)} = { {\perb{k}^2} \over {2 (k_+ \!-\! k_{1+})}}
= { {-\perb{k}^2} \over { 2 k_{1+} (1 \!-\! \om)} } \,\,\,
& \& & \,\, k^2 = - { {\perb{k}^2} \over { (1 - \om) } }. \non
\eeqs

For the evaluation of $B^g(k,r)$ we have
\newpage
\beqs
B^g(k,r) &=& \ga_\mu (\not{k_1}-\not{k}) \ga_\nu D_{\nu \be}(k-r)
V^g_{\be \al} D_{\al \mu}(k) \non \\
&=& \ga_\mu (\not{k_1}-\not{k}) \ga_\nu V^g_{\nu \mu}(k, k-r) 
= \ga_\mu (\not{k_1}-\not{k}) \ga_\nu (g_{\nu \mu} 
n \! \cdot \! k n \! \cdot \! (k-r) \non \\
&& - n \! \cdot \! (k-r) k_\nu n_\mu - n \! \cdot \! n_\nu (k-r)_\mu 
+ k \! \cdot \! (k-r) n_\nu n_\mu ) \non \\
&=& n \! \cdot \! k n \! \cdot \! (k-r) \ga_\mu (\not{k_1}-\not{k}) \ga_\mu 
- n \! \cdot \! k (\not{k}-\not{r}) (\not{k_1}-\not{k}) \not{n} \non \\
&& - n \! \cdot \! (k-r) \not{n} (\not{k_1}-\not{k}) \not{k}
+ k \! \cdot \! (k-r) \not{n} (\not{k_1}-\not{k}) \not{n} \non \\
&=& -2 n \! \cdot \! k n \! \cdot \! (k-r) (\not{k_1}-\not{k})
- n \! \cdot \! k (\not{k}\not{k_1} - \not{r}\not{k_1} +
\not{r}\not{k} -k^2)\not{n} \non \\
&& -n \! \cdot \! (k-r) \not{n} (\not{k_1} \not{k} -k^2)
+ k \! \cdot \! (k-r) \not{n} (\not{k_1}-\not{k}) \not{n} \non \\ 
&=& -2 k_+ (k-r)_+ (\not{k_1}-\not{k})
- k_+ (\not{k}\not{k_1} - \not{r}\not{k_1} +\not{r}\not{k} -k^2) \ga_+
\non \\
&& - (k-r)_+ \ga_+ (\not{k_1} \not{k} -k^2) + (k^2 - k \! \cdot \! r)
\ga_+ (\not{k_1}-\not{k}) \ga_+ \,.
\eeqs

\nid  As before we are interested in picking up the leading
logarithmic divergence terms which are again quadratic and come from
terms proportional to $k_-$ (and $k^2$ for the pole $k_-^{(2)}$). Since 
$\ga_{gq}$ is related to the gluon-quark transition in evolution, the
contributing terms will also have to have the structure of a quark
vertex, that is, be proportional to $\ga_+$ as well. We obtain
\beqs
B^g(k,r) &=& -2 k_+ (k-r)_+ (-k_- \ga_+) 
- k_+ (\ga_+\ga_-\ga_+ (k_- k_{1+}) - 0 - 0 - k^2 \ga_+) \non \\
&& - (k-r)_+ (\ga_+\ga_-\ga_+ k_{1+} k_- - k^2 \ga_+) 
+ (k^2 - k_- r_+)\ga_+\ga_-\ga_+ (k_1 -k)_+ \non \\
&=& 2 k_+ (k-r)_+ k_- \ga_+ - 2 \ga_+ k_+ k_- k_{1+} + \ga_+ k_+ k^2
- 2 (k-r)_+ \ga_+ k_{1+} k_- \non \\
&& + (k-r)_+ k^2 \ga_+ + 2 k^2 \ga_+ (k_1 -k)_+
- 2 (k_1 -k)_+ k_- r_+ \ga_+ \,\, .
\eeqs

Thus, for the $k_-^{(1)}$ pole we have
\newpage
\beqs
B^{g,(1)} &=& \ga_+ \left( 
{ {2 k_+ (k-r)_+} \over {2 k_+} } - { {2 k_+k_{1+}} \over {2 k_+} }
+ 0 - { {2 (k-r)_+ k_{1+}} \over {2 k_+} } \right. \non \\
&& \left. - { {2 (k_1 -k)_+ r_+} \over {2 k_+} } + 0 + 0 \right)
\perb{k}^2 \non \\
&=& \ga_+ \perb{k}^2 ( (k-r)_+ - k_{1+} - (k-r)_+ { k_{1+} \over k_+ }
- r_+ { {(k_1 -k)_+} \over k_+ } \non \\
&=& \ga_+ \perb{k}^2 k_{1+} (\om - \nu - 1 - { {\om - \nu} \over \om }
- \nu { {1-\om} \over \om} ) \non \\
&=& { \perb{k}^2 \over \om } \ga_+ k_{1+} ( \om^2 - 2 \om ) \,\, ,
\eeqs

\nid while for the $k_-^{(2)}$ pole we obtain
\beqs
B^{g,(2)} &=& \ga_+ \left(
- { {2 k_+ (k-r)_+} \over {2 (1- \om) k_{1+}} }
+ { {2 k_+k_{1+}} \over {2 (1- \om) k_{1+}} } - { {k_+} \over {1-\om}}
+ { {2 (k-r)_+ k_{1+}} \over {2 (1- \om) k_{1+}} } \right. \non \\
&& \left. - { {(k-r)_+} \over {1-\om} }
- { {2 (k_1 -k)_+} \over {1-\om} } 
+ { {2 (k_1 -k)_+ r_+} \over {2 (1- \om) k_{1+}} } 
\right) \perb{k}^2 \non \\
&=& { {\ga_+ \perb{k}^2} \over {1\!-\!\om} }
( - \om (k\!-\!r)_+ \!+ k_+ \!- k_+ \!+ (k\!-\!r)_+ \!- (k\!-\!r)_+ 
\!- 2 k_{1+} (1\!-\!\om) \!+ r_+ (1 \!-\!\om) ) \non \\
&=& - { {\ga_+ \perb{k}^2 k_{1+} } \over {1-\om} }
(\om^2 - 2\om + 2 - \nu) \non \\
&=& - { {\perb{k}^2} \over {1-\om} } k_{1+} \ga_+ 
(1 + (1 - \om)^2 - \nu) \,\,.
\eeqs

\nid Subsitute these back into (\ref{eq:gq2}) we have
\beqs
\ga_{gq}^{+,1} &=& {{\al_s C_F} \over {2 \pi}}
\int { {d\perb{k}^2} \over {\perb{k}^4}} \left(
\int_0^{r_+} { {dk_+ k_+} \over { 2 r_+ k_{1+}} }
(2k-r)_+^{J\!-\!n\!-\!2}r_+^n 
{ \perb{k}^2 \over \om } \ga_+ k_{1+} ( \om^2 - 2 \om ) \,\right. \non \\
&& \,\,\,\,\,\,\,\,\,\,
\left. + \int_{r_+}^{k_{1+}}
{ {dk_+ (k_1\!-\!k)_+} \over { 2 k_{1+} (k_1\!-\!r)_+}}
(2k-r)_+^{J\!-\!n\!-\!2}r_+^n 
( - { {\perb{k}^2} \over {1-\om} } k_{1+} \ga_+ 
(1 + (1 \!-\! \om)^2 - \nu)) \right) \non \\   
&=& {{\al_s C_F} \over {2 \pi}}
\int { {d\perb{k}^2} \over {\perb{k}^2}} \left(
\int_0^\nu { {d \om \om} \over \nu } { {\om^2 - 2 \om} \over \om }
k_{1+} k_{1+}^{J\!-\!2} (2 \om -\nu)^{J\!-\!n\!-\!2} \nu_+^n 
\ga_+ \right. \non \\
&& \left. - \int_\nu^1 { {d \om (1 -\om)} \over {1 - \nu} }
{ {1 + (1 - \om)^2 - \nu} \over {1 - \om} }
k_{1+} k_{1+}^{J\!-\!2} (2 \om -\nu)^{J\!-\!n\!-\!2} \nu_+^n
\ga_+ \right) \,\, ,
\eeqs

\nid upon which we have recovered the result listed in 
(\ref{eq:gammaresult}), that is, after factoring out the logarithmic 
divergence in the transverse momentum, we obtain 
\beqs
\ga_{gq}^{(\!+\!)}\!& = & \!{ \alpha_s \over {2 \pi}} C_F \,
\ga_+
k_{1+}^{J-1}\nu^n \! \int_0^1 \! d \om (2
\om\!-\!\nu)^{J\!-\!n\!-\!2}
\left( \Theta(\nu \! - \! \om) {1 \over \nu} (\om^2\!-\!2\om)
\right. \,\,\,\,\,\,\,\,  \non \\
& & \,\,\,\,\,\,\,\,\,\,\,\,\,\,\,\,\,  \left.
\!- \Theta( \om \! - \! \nu ){1 \over {1\!-\!\nu}}
(1\!+\!(1\!-\!\om)^2\!-\!\nu ) \right) \,\, .
\eeqs

\chapter*{Figures}
\addcontentsline{toc}{chapter}{Figures}
\thispagestyle{myheadings}
\markright{}

\centering

\input{intro.fig}
\clearpage

\clearpage

\clearpage

\clearpage

\clearpage

\clearpage

\clearpage

\clearpage

\clearpage

\clearpage

\clearpage

\clearpage

\input{c3.fig}
\clearpage

\clearpage

\clearpage

\clearpage

\clearpage

\clearpage
 
\clearpage
 
\clearpage

\end{document}